%% file: thesis.tex
\newcommand{\bq}{\begin{equation}}
\newcommand{\eq}{\end{equation}}
\newcommand{\bqa}{\begin{eqnarray}}
\newcommand{\eqa}{\end{eqnarray}}

\newcommand{\threev}[4]{\left(\begin{tabular}{c} $#1$ \\ $#2$ \\ $#3$ \end{tabular}\right)}

\newcommand{\f}{\varphi}
\newcommand{\h}{{1\over2}}

\newcommand{\fc}{\f^{\mathrm{cl}}}
\newcommand{\ind}[1]{\index{#1}#1}
\newcommand{\m}[2]{#1\;\mathrm{mod}\;#2}

\documentclass[12pt,a4paper]{book}

\usepackage{epsfig}
\usepackage{axodraw}
\usepackage{graphicx}
\usepackage{amsmath}
\usepackage{makeidx}
\usepackage[top=1in, bottom=1in, left=1in, right=1in]{geometry}

\makeindex

\begin{document}

\pagestyle{empty}
\include{titlesimple}

\include{title}

\pagestyle{headings}
\pagenumbering{roman}
\tableofcontents

\chapter{Introduction}

\pagenumbering{arabic}

For the last couple of decades a theory that goes by the unassuming name of `The Standard Model', has been the generally accepted theory of fundamental physics. This Standard Model has been very successful in describing experiments in particle physics. All particles that were theoretically predicted by it have been detected, except for one: the \ind{Higgs particle}. By now this Higgs particle, often called `the holy grail of high-energy physics', has become so important that billions of euros are spent to build large particle colliders, hoping to produce these Higgs particles. In Europe the LHC is being built, mainly for this purpose, and this 27 km long accelerator is expected to become operational in 2008.

Knowing this it is clear that the Higgs sector of the Standard Model is very important and interesting. The Higgs mechanism was proposed in the 60's by Brout and Englert \cite{Englert}, Higgs \cite{Higgs, Higgs2} and Guralnik, Hagen and Kibble \cite{Guralnik} to give masses to the gauge bosons and the fermions, while keeping the theory renormalizable. The main feature of this Higgs mechanism is the mechanism of spontaneous symmetry breaking (SSB), which was introduced into quantum field theory by Nambu \cite{Nambu, NambuJona-Lasinio}, in analogy to the BCS theory of superconductivity. This mechanism of SSB will be the \emph{main topic} of this thesis.

A nice introduction to SSB and the Higgs mechanism can be found in a review article by Bernstein \cite{Bernstein}.

\section{Spontaneous Symmetry Breaking}

How does SSB\index{spontaneous symmetry breaking} work in quantum field theory, and what is it? The \emph{canonical} approach\index{canonical approach} to SSB, which one finds in most textbooks (e.g.\ \cite{Peskin, Weinberg, Itzykson}), is as follows. One starts with a (bare) Lagrangian, obeying some symmetry in the fields (e.g. reflection or rotational symmetry), of which the bare (classical) potential has \emph{more} than one minimum. The most common, and most important, example is the `Mexican hat' potential. This means that the set of minima must also obey the symmetry, which means again that in any given minimum the fields cannot all be zero. Writing all fields into the single \emph{vector} $\f$ we have at the minima: $\f\neq0$. Therefore the classical lowest energy states, or vacua, are \emph{degenerate} and have a \emph{non-zero} field value, $\f_{\textrm{VAC}}\neq0$. In a quantum field theory the lowest energy state, or \ind{vacuum} $|\textrm{VAC}\rangle$, should be calculated from the \ind{Schr\"odinger equation}:
\bq
H|\textrm{VAC}\rangle = E_{\textrm{VAC}}|\textrm{VAC}\rangle \;.
\eq
Clearly, because of the very complicated form of the Hamiltonian $H$ in a quantum field theory, this equation can not be solved. Inspired by the classical minimum-energy states, one therefore \emph{postulates} that also the quantum vacuum is degenerate, and that:
\bq \label{vacpostulate}
\langle\textrm{VAC}|\f|\textrm{VAC}\rangle \neq 0 \;.
\eq
So there are \emph{multiple} vacuum states. But we can only live in one of these, and nature has chosen one of these vacuum states. Which one has been chosen, cannot be determined, and is therefore unimportant, because all theories built on one of these states have exactly the same physics.

This is called \ind{spontaneous symmetry breaking}, i.e.\ the vacuum state of the theory does \emph{not} have the same symmetry as the Lagrangian. So the dynamics of the theory obey a certain symmetry, which is not respected by the vacuum state.

Having postulated (\ref{vacpostulate}) one can then derive, via the equations of motion, the Schwinger-Dyson equations and the Feynman rules, that this gives a mass-like term for all particles coupling to the (Higgs) field $\f$. The fluctuation in this (Higgs) field around the constant value it has in the chosen vacuum is the Higgs particle.

After this one can calculate all Green's functions of the theory. Also one can construct the 1PI Green's functions and sum them, in the appropriate way, to obtain the effective potential. As we shall see this effective potential comes out to be \emph{complex} and can be \emph{non-convex} in certain domains. This is the well known \emph{convexity problem}\index{convexity problem}, i.e.\ the canonical perturbative calculation gives a non-convex effective potential, whereas general arguments show that this effective potential is \emph{convex}. The precise meaning of convex will be discussed in chapter \ref{chapeffact}. Also its convexity will be proven there.

\section{The Path Integral}

Now this mechanism of SSB can also be studied from the viewpoint of the \ind{path integral}. We know that the \ind{path-integral approach}, by Feynman, is just another way of formulating quantum mechanics, or quantum field theory. Like the Feynman rules in the canonical approach, the path integral is also a solution to the Schwinger-Dyson equations of the theory.

In this path-integral approach the path integral gives the Green's functions of the theory:
\bq
\frac{\int \mathcal{D}\f \; f(\f) \; \exp\left({i\over\hbar}S(\f)\right)}{\int \mathcal{D}\f \; \exp\left({i\over\hbar}S(\f)\right)} = \langle\textrm{VAC}|f(\f)|\textrm{VAC}\rangle \;.
\eq
Here $f(\f)$ is some expression built from the $\f$-fields, like $\f_i(x)$, or $\f_i(x)\f_j(y)$.

So from the path-integral viewpoint one should also be able to see whether we have SSB:
\bq
\langle\textrm{VAC}|\f|\textrm{VAC}\rangle \neq 0 \;,
\eq
as is postulated in the canonical approach.

It is exactly the \emph{path-integral} approach to SSB that we shall study in this thesis. The two simplest models to study SSB are of course the $N=1$ and $N=2$ \emph{Euclidean} linear sigma models, and these models we will consider.

It will appear that, although the canonical results and the path integral are solutions to the \emph{same} Schwinger-Dyson equations, the two approaches do \emph{not} give the same results in the case of a classical potential with more than one minimum, i.e.\ a non-convex classical potential.

Also it will appear that this path-integral approach cures the complexity and non-convexity that were obtained in the canonical approach.

\section{Literature}

What has been discussed in the literature about the \ind{convexity problem}? In \cite{Symanzik, Iliopoulos} Symanzik and Iliopoulos et al.\ were the first to realize that the effective potential is convex. A nice proof of this convexity property is given by Haymaker et al\ \cite{Haymaker}. Note that this proof is based on the path-integral formalism. The fact that there is a convexity problem (i.e.\ the perturbatively calculated effective potential, in case of a non-convex classical potential, is \emph{not} convex, despite general proofs that it should be) was first realized by O'Raifeartaigh et al.\ \cite{ORaifeartaigh}. After this there were several attempts to modify the computation of the effective potential to find a proper, convex effective potential. These attempts can be found in \cite{Fujimoto, Callaway, Bender, Cooper, Hindmarsh, Rivers}. Indeed these attempt were successful, in all of these articles a convex, well-defined effective potential is found for several models. All these attempts come down to the same idea, to get a convex and well-defined effective potential one should take the path integral seriously and calculate from there. This means one should include \emph{all} minima of the classical potential in the calculation, i.e.\ do perturbation theory around \emph{each} of the minima and add the generating functionals around each of the minima to obtain the \emph{complete} generating functional. If one then computes the effective potential from this complete generating functional one finds the result to be convex and well-defined for all field values.

However, in this new (path-integral) approach, SSB is lost in the strict sense, i.e.\ all of the convex effective potentials that are calculated in the articles above have their minimum at zero for finite space-time volume. For infinite volume the bottom of the effective potentials becomes flat (\ind{Maxwell construction}) and one is left with an infinite set of minima, living between the classical minima. What, then, is the true vacuum? Can one still determine what the vacuum is from these effective potentials? In \cite{Fujimoto} one can find a short remark about this. There the authors state that in the case of a non-convex classical potential, maybe the effective potential is \emph{not} the proper thing to look at to find the true vacuum. Or alternatively one might define SSB not as a non-zero vacuum expectation value, but as the sensitivity of the effective potential to small external sources. In this sense the new, convex effective potential is just as sensitive to a non-zero source as the old, non-convex effective potential.

However, besides these few vague remarks, no clear explanation is given as to what the new path-integral approach means for the physics of the theory.

O'Raifeartaigh et al.\ \cite{Wipf}, inspired by \cite{Fukuda}, introduce a constraint effective potential. This constraint effective potential is calculated from a path integral, in which a constraint that keeps the space-time averaged field to a non-zero value is included. Simply because of the constraint, there is SSB in the strict sense now. However in the infinite volume limit the constraint effective potential converges to the convex effective potential again, leaving one again with a flat bottom of minima. Again it is unclear what this means for the physics. Also Ringwald et al.\ \cite{Ringwald} define a constraint effective potential, however now the constraint keeps the average of the field over a certain limited domain of space-time to a non-zero value. Again the constraint effective potential converges to the convex effective potential in the infinite volume limit. Nothing is said about the physics behind this theory.

Branchina et al.\ \cite{Branchina} do go into more details about the physics. Here they also include all minima of the path integral (no constraint) and find a flat bottom. Their approach is essentially based on the canonical formalism and they find explicitly the ground states in a Gaussian approximation. They find two pure Gaussian states, which means that all linear superpositions of these states are also ground states. These correspond to the flat bottom of the effective potential. They calculate the probability to be in one of these states. This probability is only non-zero for the pure Gaussian states. This is their interpretation of SSB. However, for the rest nothing is said about the physics that follows from this approach.

Weinberg et al.\ \cite{Wu} further analyze the complex, non-convex effective potential one finds when only including \emph{one} minimum (i.e.\ canonical approach). They define the vacuum states of the theory to be states that, of course, minimize the Hamiltonian, but are also localized around some field value. It appears that the imaginary part of the complex effective potential is related to the decay rate of the (unstable) vacuum states which are localized around a point between the classical minima.

Dannenberg \cite{Dannenberg} further analyzes and resolves the \ind{convexity problem}. The point is that the convex effective potential, as calculated from the path integral, and the complex effective potential, as calculated in the canonical way (the sum of all 1PI diagrams), are simply not the same thing. In the path-integral approach one includes \emph{all} minima, in the canonical approach one includes only \emph{one} minimum. Although both ways are solutions to the same Schwinger-Dyson equations, they are \emph{not} equal. In this way it is completely understandable that the canonical approach gives a non-convex effective potential, even though one can prove from the path integral that the effective potential is convex. Both approaches are simply different and therefore give different results and physics.

Wiedemann \cite{Wiedemann} further analyzes what the non-convex complex effective potential and the convex effective potential tell one about the physics of the theory. It is shown that the flat section of the convex effective potential corresponds to the ground states of the theory. The complex effective potential gives one the boundaries of the flat section.

Having considered all of this literature one can conclude the following. The convexity problem is \emph{not} really a problem, it originates only because one compares two different things, at first thought to be the same. The \ind{canonical approach} and the \ind{path-integral approach}, although solutions to the same Schwinger-Dyson equation, seem to be \emph{different} in the case of a non-convex classical potential. So both approaches also give different results. This difference between the canonical and path-integral approach will be the main topic of this thesis. It is also this difference that might create some confusion in for example Peskin and Schroeder \cite{Peskin}. In their chapter 11 they first calculate the effective potential in the canonical approach and find it to be non-convex. Later they argue that the effective potential is always convex. They do not clearly explain how this convexity property relates to the non-convex result.

Taking the viewpoint of the canonical approach, one postulates a non-zero vacuum expectation value. This is completely self-consistent and one finds a spontaneously broken theory. One can define the effective potential as the sum of all 1PI graphs (with the appropriate factors) and one finds it to be non-convex and complex in certain regions. This does not matter however, since the proof that the effective potential is convex originates only in the path-integral approach, which is \emph{not} the same.

Taking the viewpoint of the path-integral approach one finds a convex effective potential, as can be proven on general grounds (within this approach). However, what the physics of this approach is, is unclear up to now. Also interesting is whether one can reproduce the physics as it is found in the canonical approach (with SSB and all) in this path-integral approach. Can one get the same Green's functions in this path-integral approach?

\section{Outline of this Thesis}

In the articles mentioned above several links between results from the canonical approach and results from the path-integral approach are proven to exist, although both approaches do \emph{not} give the same results in general. So a number of big questions remain: Can one somehow reproduce the canonical results from the path-integral approach? Or are both approaches fundamentally different? Can one find SSB, with all the known physics that goes with it, from the path integral? These questions will be the main topics of this thesis.

In chapter \ref{chapeffact} a short introduction to the effective action will be given. The effective action and effective potential will be defined, their meaning will be discussed and their convexity will be proven, via the path integral.

In chapter \ref{chapN1sigmamodelcan} the canonical approach to the $N=1$ linear sigma model will be discussed. We follow here the same lines as in the quantum field theory textbooks (e.g. \cite{Peskin}). The renormalized Green's functions will be computed and the counter terms will be found, so we can use them in later chapters for different approaches. Also the effective potential will be computed for several dimensions and shown to be complex where the classical potential is non-convex. Also it will be shown that it can become non-convex.

In chapter \ref{chapN1sigmamodelpath} the path-integral approach to the $N=1$ linear sigma model will be discussed. This is done in the same way as Fujimoto et al.\ \cite{Fujimoto} and Cooper et al.\ \cite{Cooper} do. We find the renormalized effective potential for several dimensions, which is indeed convex and well-defined everywhere. Also we find the renormalized Green's functions and conclude what the physics of this approach is. This physics is \emph{different} than in the canonical approach. What the Green's functions become when other particles interacting with the (Higgs) fields $\f$ are present will also be discussed.

In chapter \ref{chapN1sigmamodelfix} we will outline another path-integral approach to the $N=1$ linear sigma model. This time, hoping to reproduce the physics of the canonical approach, we will fix the paths at some time $-T$ at a specific field value over all of space. First we will show that this model is renormalizable up to 1-loop order. Then we will calculate the effective potential and the renormalized Green's functions.

In chapter \ref{chapN2sigmamodelcan} we will present the canonical approach to the $N=2$ linear sigma model. The calculations there are similar to the standard calculations done in all textbooks, like \cite{Peskin}. The renormalized Green's functions will be computed and the counter terms will be fixed, such that we can use them later throughout the thesis. The effective potential will also be calculated and shown to be complex where the classical potential is non-convex. Also it can become non-convex.

In chapter \ref{chapN2sigmamodelpathI} the path-integral approach to the $N=2$ linear sigma model will be presented. We will compute the renormalized Green's functions by naively integrating over all minima of the action. Also an approximation to the effective potential will be found in this naive way. Again, as in the $N=1$ linear sigma model, we will see that the physics of this path-integral approach is \emph{different} from the physics of the canonical approach. However, it is questionable whether the naive way of calculating here is correct.

To do the calculation from chapter \ref{chapN2sigmamodelpathI} in a better way we need the path integral in terms of polar field variables. These variables are the natural variables to describe an $O(2)$-invariant model. This complicated transformation to polar field variables will be the subject of chapter \ref{chappathintpol}. We will show what the path integral looks like in terms of polar fields, and how it should be calculated.

In chapter \ref{chapN2sigmamodelpathII} we consider again the path-integral approach to the $N=2$ linear sigma model. Now we do the calculations via the path integral in terms of polar variables, discussed in chapter \ref{chappathintpol}. We will calculate the renormalized Green's functions and the effective potential. We will compare the results obtained here with the results from chapter \ref{chapN2sigmamodelpathI}, and finally discuss the physics of the path-integral approach to the $N=2$ linear sigma model.


\chapter{The Effective Action}\label{chapeffact}

\section{Definition}

Consider a Euclidean scalar quantum field theory with any number of fields and any number of space-time points (e.g.\ infinite number). We put all field values in one single \emph{vector} $\f$. For each field value we also have a source, all these sources are put in the vector $J$. The (bare) action of this theory we denote by $S$. Then the \ind{effective action} of this theory, which is a function of a vector $\fc$, is defined by
\bq \label{defeff}
\frac{\partial\Gamma(\fc)}{\partial\fc_m} = J_m(\fc) \;,
\eq
where $J_m(\fc)$ is defined as the inverse function of
\bq
\fc_m(J) \equiv \langle \f_m \rangle(J) \equiv \frac{\int \mathcal{D}\f \; \f_m \exp\left(-{1\over\hbar}\left(S(\f)-J_m\f_m\right)\right)}{\int \mathcal{D}\f \; \exp\left(-{1\over\hbar}\left(S(\f)-J_m\f_m\right)\right)} \;.
\eq
That this inverse functional $J_m(\fc)$ always exists\index{effective action!existence} can be seen as follows. If it exists we have:
\bq \label{inv}
\fc_m(J(\fc)) = \fc_m \Rightarrow {\partial\over\partial\fc_n} \fc_m(J(\fc)) = {\partial\fc_m\over\partial J_l}{\partial J_l\over\partial\fc_n} = {1\over\hbar} \langle \f_m \f_l \rangle_{\mathrm{c}}(J){\partial J_l\over\partial\fc_n} = \delta_{mn}
\eq
This means that one requirement for $J_m(\fc)$ to exist is that the matrix $\langle \f_i \f_j \rangle_{\mathrm{c}}(J)$ has an inverse. This is true if all eigenvalues of this matrix are non-zero. That this is indeed the case follows from:
\bqa
\langle \f_i \f_j \rangle_{\mathrm{c}}(J) &=& \frac{\int\mathcal{D}\f \; \f_i \f_j P(\f)}{\int\mathcal{D}\f \; P(\f)} - \frac{\left(\int\mathcal{D}\f \; \f_i P(\f)\right)\left(\int\mathcal{D}\f \; \f_j P(\f)\right)}{\left(\int\mathcal{D}\f \; P(\f)\right)^2} \nonumber\\
&=& \frac{\int\mathcal{D}\f \mathcal{D}\f' \; \left(\f_i \f_j - \f_i \f_j'\right) P(\f) P(\f')}{\int\mathcal{D}\f \mathcal{D}\f' \; P(\f) P(\f')} \nonumber\\
&=& {1\over2} \frac{\int\mathcal{D}\f \mathcal{D}\f' \; \left(\f_i-\f_i'\right) \left(\f_j-\f_j'\right) P(\f) P(\f')}{\int\mathcal{D}\f \mathcal{D}\f' \; P(\f) P(\f')}
\eqa
with
\bq
P(\f) \equiv \exp\left(-{1\over\hbar} \left( S(\f) - J_m\f_m \right) \right) \;.
\eq
This means:
\bq \label{pos}
\rho_i \rho_j \langle \f_i \f_j \rangle_{\mathrm{c}}(J) > 0
\eq
for all $J$ and arbitrary vector $\rho$. This could not immediately be seen from $\langle (\rho\cdot\f)^2 \rangle_{\mathrm{c}}(J)$ because we are dealing with a \emph{connected} average here. Writing $\rho$ in terms of the eigenvectors $e_m$ (with eigenvalues $\lambda_m$) of the matrix $\langle \f_i \f_j \rangle_{\mathrm{c}}(J)$:
\bq
\rho_i = c_m e_{i,m} \;,
\eq
we see that also
\bq \label{poseigvalues}
\rho_i c_m e_{i,m} \lambda_m = c_n e_{i,n} c_m e_{i,m} \lambda_m = c_m^2 \lambda_m > 0 \;.
\eq
Because the $\rho$'s and thus also the $c$'s are arbitrary, all the eigenvalues $\lambda_m$ have to be positive. So we see all eigenvalues are not just non-zero, they are also strictly positive.

There is a small loop-hole here, that we have to discuss. In some special situations it can happen, for some specially chosen $\rho$, that the left-hand-side of (\ref{pos}) becomes exactly zero. We shall see in chapter \ref{chapN1sigmamodelfix} how this can happen when we introduce constraints in the path integral. In that case there is \emph{no} unique inverse. This will be discussed thoroughly in chapter \ref{chapN1sigmamodelfix}. In the rest of this chapter we shall assume that we are not in such a special situation and thus (\ref{pos}) holds.

So now we know that the matrix ${\partial J_l\over\partial\fc_n}$ exists. To find $J_l$ itself this system of partial differential equations has to be integrated. This is only possible if
\bq
{\partial^2 J_l\over\partial\fc_n\partial\fc_p} = {\partial^2 J_l\over\partial\fc_p\partial\fc_n} \;.
\eq
That this is also the case can be seen by taking another derivative in (\ref{inv}):
\bq
\frac{\partial}{\partial\fc_p}{\partial\fc_m\over\partial J_l}{\partial J_l\over\partial\fc_n} = {\partial^2\fc_m\over\partial J_q\partial J_l}{\partial J_q\over\partial\fc_p}{\partial J_l\over\partial\fc_n} + {\partial \fc_m\over\partial J_l}{\partial^2 J_l\over\partial\fc_p\partial\fc_n} = 0
\eq
Because the first term is symmetric in $p$ and $n$ the second term is too. So indeed it is possible to find the inverse $J_l(\fc)$ from (\ref{inv}).

\section{The Meaning of the Effective Action} \label{meaning}

The \ind{generating functional} $Z(J)$ of our scalar field theory is
\bq
Z(J) \equiv \int \mathcal{D}\f \; \exp\left(-{1\over\hbar}\left(S(\f)-J_m\f_m\right)\right) \;.
\eq
This functional generates the \index{Green's functions!connected}connected Green's functions:
\bq
\hbar^n\frac{\partial^n}{\partial J_{m_1} \ldots \partial J_{m_n}} \ln Z(J) = \langle \f_{m_1} \ldots \f_{m_n} \rangle_{\mathrm{c}}(J)
\eq

Now the \index{effective action!physical meaning}physical meaning of the effective action can be seen by taking derivatives and putting the source $J$ to zero in the definition (\ref{defeff}). 

First, just setting $J=0$ without taking a derivative we get:
\bq
\left. \frac{\partial\Gamma(\fc)}{\partial\fc_m} \right|_{\fc=\langle\f\rangle} = 0 \;,
\eq
where $\langle\f\rangle$ just means $\langle\f\rangle(0)$. This means the effective action has its minimum at the vacuum expectation value of the field(s).

Taking one derivative with respect to $J_n$ in (\ref{defeff}) gives:
\bqa
{\partial^2\Gamma(\fc)\over\partial\fc_l\partial\fc_m} {\partial\fc_l\over\partial J_n} &=& \delta_{mn} = {1\over\hbar} {\partial^2\Gamma(\fc)\over\partial\fc_l\partial\fc_m} \langle\f_l\f_n\rangle_{\mathrm{c}}(J) \nonumber\\
\Rightarrow {\partial^2\Gamma(\fc)\over\partial\fc_l\partial\fc_m} &=& \hbar \left(\langle\f_l\f_m\rangle_{\mathrm{c}}(J)\right)^{-1} \label{2peff}
\eqa
Now if we put $J=0$ on both sides in (\ref{2peff}) the righthand side becomes the inverse of the ordinary connected 2-point Green's function:
\bq \label{2prel}
\left. {\partial^2\Gamma(\fc)\over\partial\fc_l\partial\fc_m} \right|_{\fc=\langle\f\rangle} = \hbar\left(\langle\f_l\f_m\rangle_{\mathrm{c}}\right)^{-1}
\eq
This means that the second functional derivative of the effective action at its minimum is equal to $\hbar$ times the inverse connected 2-point Green's function. 

Taking two derivatives with respect to $J$ and putting $J=0$ in (\ref{defeff}) gives:
\bqa
{\partial^3\Gamma(\fc)\over\partial\fc_r\partial\fc_l\partial\fc_m} {\partial\fc_r\over\partial J_p} {\partial\fc_l\over\partial J_n} + {\partial^2\Gamma(\fc)\over\partial\fc_l\partial\fc_m} {\partial^2\fc_l\over\partial J_p\partial J_n} &=& 0 \nonumber\\
\Rightarrow {\partial^3\Gamma(\fc)\over\partial\fc_r\partial\fc_l\partial\fc_m} {\partial\fc_r\over\partial J_p} {\partial\fc_l\over\partial J_n} {\partial\fc_m\over\partial J_q} + {\partial^2\fc_q\over\partial J_p\partial J_n} &=& 0 \nonumber\\
\Rightarrow {\partial^3\Gamma(\fc)\over\partial\fc_r\partial\fc_l\partial\fc_m} \langle\f_r\f_p\rangle_{\mathrm{c}}(J) \langle\f_l\f_n\rangle_{\mathrm{c}}(J) \langle\f_m\f_q\rangle_{\mathrm{c}}(J) + \hbar \langle\f_q\f_p\f_n\rangle_{\mathrm{c}}(J) &=& 0 \nonumber\\
\Rightarrow \left. {\partial^3\Gamma(\fc)\over\partial\fc_r\partial\fc_l\partial\fc_m} \right|_{\fc=\langle\f\rangle} = -\hbar \langle\f_r\f_l\f_m\rangle_{\mathrm{1PI}} \label{3prel}
\eqa
This means the third functional derivative of the effective action at its minimum is equal to $-\hbar$ times the 1PI 3-point Green's function.\index{Green's functions!1PI}

By taking more derivatives with respect to $J$ and putting $J=0$ it can be shown that
\bq \label{nprel}
\left. {\partial^n\Gamma(\fc)\over\partial\fc_{m_1} \ldots \partial\fc_{m_n}} \right|_{\fc=\langle\f\rangle} = -\hbar \langle\f_{m_1} \ldots \f_{m_n} \rangle_{\mathrm{1PI}} \;.
\eq

These relations show that when we expand the effective action around its minimum, read off the propagator and the coupling constants like we would read them off from the bare action, the Feynman rules thus obtained would give the complete physical Green's functions at tree level. This is the meaning of the effective action, from it one can immediately see the physical amplitudes. 

\section{The Argument for Convexity} \label{conv}

What is `convex'? Although `convex' and `concave' are often mixed up in the world-wide literature, we shall stick to the definition that is most widely used in physics. A \ind{convex} function $V(\f)$ is a function that, for any $\f_1$ and $\f_2$ ($\f_1$ and $\f_2$ can be vectors), and any $\lambda$ in $[0,1]$ ($0\leq\lambda\leq1$), satisfies:
\bq \label{defconvex}
V(\lambda\f_1+(1-\lambda)\f_2) \leq \lambda V(\f_1) + (1-\lambda) V(\f_2) \;.
\eq
In words this means that a linear interpolation of $V(\f)$ is always larger than or equal to $V(\f)$ itself. \index{strictly convex}Strictly convex means that the linear interpolation is always larger than $V(\f)$ itself:
\bq
V(\lambda\f_1+(1-\lambda)\f_2) < \lambda V(\f_1) + (1-\lambda) V(\f_2) \;,
\eq
for all $\f_1$, $\f_2$ and $\lambda\in[0,1]$.

Now we show that the \ind{effective action} of a Euclidean quantum field theory is always \index{effective action!convexity}convex, in any dimension.

An effective action $\Gamma$ is convex if and only if
\bq \label{convex}
\left. \frac{\partial^2}{\partial\alpha^2} \Gamma\left(\fc+\alpha\rho\right) \right|_{\alpha=0} \geq 0
\eq
for all $\fc$ and arbitrary vector $\rho$. It is easy to see that this condition is equivalent to (\ref{defconvex}).

This condition is equivalent to
\bq
\rho_i \rho_j \frac{\partial^2\Gamma(\fc)}{\partial\fc_i\partial\fc_j} \geq 0 \quad \textrm{for all $\fc$} \;.
\eq
In (\ref{poseigvalues}) we showed that the eigenvalues of the matrix $\langle \f_i \f_j \rangle_{\mathrm{c}}(J)$ are all strictly positive. This means that also the eigenvalues of the inverse matrix $\left(\langle \f_i \f_j \rangle_{\mathrm{c}}(J)\right)^{-1}$ are strictly positive and that
\bq
\rho_i \rho_j \left(\langle \f_i \f_j \rangle_{\mathrm{c}}(J)\right)^{-1} > 0
\eq
for arbitrary $\rho$ and all $J$. Using (\ref{2peff}) we then see that (\ref{convex}) is indeed true, so the effective action is \index{convex}\emph{convex}.

The effective action appears to be even \emph{strictly} convex. However, this is not necessarily true. Although it is true that the eigenvalues of $\langle \f_i \f_j \rangle_{\mathrm{c}}(J)$ are all strictly positive, it can happen, in the infinite-volume limit, that one of these eigenvalues goes to infinity. Then, of course, one of the eigenvalues of $\left(\langle \f_i \f_j \rangle_{\mathrm{c}}(J)\right)^{-1}$ goes to zero in this limit. In this way the effective action can have flat directions, and it is not strictly convex, but just convex.

This completes our proof that the effective action, and with it the \ind{effective potential}, \emph{always} have to be convex for a Euclidean quantum field theory. Our argument does not depend on the dimension of space-time, nor the number of different fields in our quantum field theory. 

\section{The Effective Potential}

The \ind{effective potential} is defined as the effective action where we take all fields \emph{constant} over space-time divided by the volume of space-time. Up to now we have employed a general formalism in which we did not explicitly specify what the index of the field $\f$ meant. To obtain the effective potential we have to specify this. Let's say we have $N$ fields $\f_1,\ldots,\f_N$, all depending on space-time coordinates $x$. Here $x$ is a $d$-vector containing all space-time coordinates. 

To obtain an expression for the effective potential we first expand the effective action around its minimum. We denote the deviation of the $\f$-fields from their value $\langle\f\rangle$ by $\eta_i$, $i=1,\ldots,N$. Then the effective action can always be expanded as:
\bqa
& & \Gamma(\eta_1,\ldots,\eta_N) = \nonumber\\[5pt]
& & \qquad {1\over2!} \int d^dx_1d^dx_2 \; \left.\frac{\delta^2\Gamma}{\delta\eta_{i_1}(x_1)\delta\eta_{i_2}(x_2)}\right|_{\eta_i=0} \eta_{i_1}(x_1) \eta_{i_2}(x_2) + \nonumber\\
& & \qquad {1\over3!} \int d^dx_1d^dx_2d^dx_3 \; \left.\frac{\delta^3\Gamma}{\delta\eta_{i_1}(x_1)\delta\eta_{i_2}(x_2)\delta\eta_{i_3}(x_3)}\right|_{\eta_i=0} \eta_{i_1}(x_1) \eta_{i_2}(x_2) \eta_{i_3}(x_3) + \nonumber\\
& & \qquad \ldots
\eqa

Now we use (\ref{2prel}), (\ref{3prel}) and (\ref{nprel}) to obtain the following expression for the effective action:
\bqa
\Gamma(\eta_1,\ldots,\eta_N) &=& {1\over2!} \int d^dx_1d^dx_2 \; \hbar \left(\langle\eta_{i_1}(x_1)\eta_{i_2}(x_2)\rangle_{\mathrm{c}}\right)^{-1} \; \eta_{i_1}(x_1) \eta_{i_2}(x_2) + \nonumber\\
& & -{1\over3!} \int d^dx_1 \ldots d^dx_3 \; \hbar \langle\eta_{i_1}(x_1) \ldots \eta_{i_3}(x_3)\rangle_{\mathrm{1PI}} \; \eta_{i_1}(x_1) \ldots \eta_{i_3}(x_3) + \nonumber\\
& & -{1\over4!} \int d^dx_1 \ldots d^dx_4 \; \hbar \langle\eta_{i_1}(x_1) \ldots \eta_{i_4}(x_4)\rangle_{\mathrm{1PI}} \; \eta_{i_1}(x_1) \ldots \eta_{i_4}(x_4) + \nonumber\\
& & \ldots
\eqa
Now we can take all $\eta$-fields constant to obtain the effective \emph{potential} $V(\eta_1,\ldots,\eta_N)$:
\bqa
\left(\int d^dx\right) V(\eta_1,\ldots,\eta_N) &=& {\hbar\over2!} \eta_{i_1}\eta_{i_2} \int d^dx_1d^dx_2 \; \left(\langle\eta_{i_1}(x_1)\eta_{i_2}(x_2)\rangle_{\mathrm{c}}\right)^{-1} + \nonumber\\
& & -{\hbar\over3!} \eta_{i_1}\ldots\eta_{i_3} \int d^dx_1 \ldots d^dx_3 \; \langle\eta_{i_1}(x_1) \ldots \eta_{i_3}(x_3)\rangle_{\mathrm{1PI}} + \nonumber\\
& & -{\hbar\over4!} \eta_{i_1}\ldots\eta_{i_4} \int d^dx_1 \ldots d^dx_4 \; \langle\eta_{i_1}(x_1) \ldots \eta_{i_4}(x_4)\rangle_{\mathrm{1PI}} + \nonumber\\
& & \ldots
\eqa
Now if the fields $\eta_i$ correspond to \emph{physical} particles, then the 2-point connected Green's functions are only non-zero when the in- and outgoing lines are of the same type. Then this propagator becomes diagonal in momentum space and finding the inverse propagator is very simple, it just means literally inverting it. So, writing the Green's functions in terms of the momentum-space Green's functions, the expression for the effective potential becomes particularly simple:
\bqa
V(\eta_1,\ldots,\eta_N) &=& {\hbar\over2!}{1\over\langle\tilde{\eta}_i(0)\tilde{\eta}_i(0)\rangle_{\mathrm{c}}} \; \eta_i\eta_i - {\hbar\over3!} \langle\tilde{\eta}_{i_1}(0)\ldots\tilde{\eta}_{i_3}(0)\rangle_{\mathrm{1PI}} \; \eta_{i_1}\ldots\eta_{i_3} + \nonumber\\[5pt]
& & -{\hbar\over4!} \langle\tilde{\eta}_{i_1}(0)\ldots\tilde{\eta}_{i_4}(0)\rangle_{\mathrm{1PI}} \; \eta_{i_1}\ldots\eta_{i_4} + \ldots
\eqa
Here $0$ denotes a $d$-vector with only zeroes. Now we also know that:
\bq
\langle\tilde{\eta}_i(p)\tilde{\eta}_i(-p)\rangle_{\mathrm{c}} = \frac{\hbar}{p^2+m_i^2-\hbar\langle\tilde{\eta}_i(p)\tilde{\eta}_i(-p)\rangle_{\mathrm{1PI}}}
\eq
and with this the effective potential can be written as:
\bqa
V(\eta_1,\ldots,\eta_N) &=& \h m_i^2 \eta_i\eta_i - {\hbar\over2!} \langle\tilde{\eta}_i(0)\tilde{\eta}_i(0)\rangle_{\mathrm{1PI}} \; \eta_i\eta_i - {\hbar\over3!} \langle\tilde{\eta}_{i_1}(0)\ldots\tilde{\eta}_{i_3}(0)\rangle_{\mathrm{1PI}} \; \eta_{i_1}\ldots\eta_{i_3} + \nonumber\\[5pt]
& & -{\hbar\over4!} \langle\tilde{\eta}_{i_1}(0)\ldots\tilde{\eta}_{i_4}(0)\rangle_{\mathrm{1PI}} \; \eta_{i_1}\ldots\eta_{i_4} + \ldots
\eqa

Another convenient way to write this is:
\bq \label{vacgraphform}
V(\eta_1,\ldots,\eta_N) = \h m_i^2 \eta_i\eta_i - \hbar \underset{j_1+\ldots+j_N\geq2}{\sum_{j_1,\ldots,j_N=0}^{\infty}} {1\over j_1!\ldots j_N!} \langle\tilde{\eta}_1^{j_1}(0) \ldots \tilde{\eta}_N^{j_N}(0)\rangle_{\mathrm{1PI}} \; \eta_1^{j_1} \ldots \eta_N^{j_N}
\eq
This is the well known \ind{vacuum-graph formula}.


\chapter{The $N=1$ LSM: The Canonical Approach}\label{chapN1sigmamodelcan}

In this chapter we shall present the \index{$N=1$ linear sigma model!canonical approach}canonical approach to the $N=1$ linear sigma model. Our calculations mostly follow the well known text books on quantum field theory (e.g.\ \cite{Peskin, Itzykson}). The renormalized Green's functions and the counter terms will be computed, the latter will be used in later chapters. Also the effective potential will be calculated and shown to be complex and non-convex in general.

The Euclidean \index{action!$N=1$ LSM}linear sigma model with $N=1$ field is defined by the bare action
\bq
S = \int d^dx \; \left( \h\left(\nabla\f(x)\right)^2 - \h\mu\f^2(x) + {\lambda\over24}\f^4(x) \right) \;.
\eq
Here \ind{$x$} denotes a $d$-vector containing all space-time coordinates 
\bq
x \equiv \threev{x_1}{\vdots}{x_d} \;,
\eq
and \ind{$\nabla$} is the $d$-vector
\bq
\nabla \equiv \threev{\partial/\partial x_1}{\vdots}{\partial/\partial x_d} \;.
\eq
These notations shall be used throughout this thesis. $\mu$ Is understood to be positive, $\mu>0$, so we have a non-convex classical potential and thus SSB in the canonical approach.

\section{Green's Functions}

To compute the renormalized Green's functions we introduce the following \ind{renormalized quantities}:
\bqa
\f^{\mathrm{R}} &\equiv& {1\over\sqrt{Z}} \f \;, Z \equiv 1 + \delta_Z \nonumber\\
\mu^{\mathrm{R}} &\equiv& \mu Z - \delta_{\mu} \nonumber\\
\lambda^{\mathrm{R}} &\equiv& \lambda Z^2 - \delta_{\lambda} \label{renquant}
\eqa
From here on we shall suppress the R-superscripts, understanding that we always work with renormalized quantities from now on. Written in terms of these renormalized quantities the \index{action!$N=1$ LSM}action is
\bqa
S &=& \int d^dx \; \bigg( \h\left(\nabla\f\right)^2 - \h\mu\f^2 + {\lambda\over24}\f^4 + \nonumber\\[5pt]
& & \phantom{\int d^dx \; \bigg(} \h\delta_Z\left(\nabla\f\right)^2 - \h\delta_{\mu}\f^2 + {\delta_{\lambda}\over24}\f^4 \bigg) \;. \label{renaction}
\eqa
Now the classical action (i.e.\ the first line of (\ref{renaction})) has its \index{minima!$N=1$ LSM}minima at:
\bq \label{Z2constsols}
\f_{\mathrm{min}} = \pm v \;, v = \sqrt{{6\mu\over\lambda}} \;.
\eq
In the canonical approach it is postulated that also the quantum field has its vacuum expectation value at one of these classical minima. Which minimum does not matter for the physics, so we choose $\f=+v$\index{choosing minimum}. Therefore we express the action in terms of the field $\eta$, which indicates the deviation from this minimum:
\bq
\f \equiv v + \eta \;.
\eq
The action then becomes
\bqa
S &=& \int d^dx \; \bigg( \h\left(\nabla\eta\right)^2 + \mu\eta^2 + {\mu\over v}\eta^3 + {\mu\over4v^2}\eta^4 + \nonumber\\[5pt]
& & \phantom{\int d^dx \; \bigg(} \left(-v\delta_{\mu}+{1\over6}v^3\delta_{\lambda}\right)\eta + \h\delta_Z\left(\nabla\eta\right)^2 + \nonumber\\[5pt]
& & \phantom{\int d^dx \; \bigg(} \left(-\h\delta_{\mu}+{1\over4}v^2\delta_{\lambda}\right)\eta^2 + {1\over6}v\delta_{\lambda}\eta^3 + {1\over24}\delta_{\lambda}\eta^4 \bigg) \;.
\eqa
Now all except the first two terms are treated as perturbations. For convenience define $\mu=\h m^2$. The \index{Feynman rules!$N=1$ LSM}Feynman rules of this theory are then given by:
{\allowdisplaybreaks\bqa
\begin{picture}(100, 20)(0, 17)
\Line(20, 20)(80, 20)
\end{picture}
&\leftrightarrow& \frac{\hbar}{k^2 + m^2} \nonumber\\
\begin{picture}(100, 40)(0, 17)
\Line(20, 20)(50, 20)
\Line(50, 20)(80, 0)
\Line(50, 20)(80, 40)
\end{picture}
&\leftrightarrow& -\frac{3m^2}{\hbar v} \nonumber\\
\begin{picture}(100, 40)(0, 17)
\Line(20, 40)(80, 0)
\Line(20, 0)(80, 40)
\end{picture}
&\leftrightarrow& -\frac{3m^2}{\hbar v^2} \nonumber\\
\begin{picture}(100, 40)(0, 17)
\Line(20, 20)(50, 20)
\Vertex(50,20){3}
\end{picture}
&\leftrightarrow& {v\over\hbar}\delta_{\mu} - {1\over6}{v^3\over\hbar}\delta_{\lambda} \nonumber\\
\begin{picture}(100, 40)(0, 17)
\Line(20, 20)(80, 20)
\Vertex(50,20){3}
\end{picture}
&\leftrightarrow& -{1\over\hbar}\delta_Z k^2 + {1\over\hbar}\delta_{\mu} - \h{v^2\over\hbar}\delta_{\lambda} \nonumber\\
\begin{picture}(100, 40)(0, 17)
\Line(20, 20)(50, 20)
\Line(50, 20)(80, 0)
\Line(50, 20)(80, 40)
\Vertex(50,20){3}
\end{picture}
&\leftrightarrow& -{v\over\hbar}\delta_{\lambda} \nonumber\\
\begin{picture}(100, 40)(0, 17)
\Line(20, 40)(80, 0)
\Line(20, 0)(80, 40)
\Vertex(50,20){3}
\end{picture}
&\leftrightarrow& -{1\over\hbar}\delta_{\lambda}
\eqa}

\vspace{10pt}

Now we compute the connected momentum-space Green's functions up to one loop. In the case of the 3- and 4-point function we will calculate the 1PI-part of the connected Green's function, since it is this part that occurs in our renormalization conditions. We shall write all results in terms of the standard $d$-dimensional one-loop integral $I$\index{standard integrals}:\index{Green's functions!$N=1$ LSM}
\bqa
& & I\left(q_1,m_1,q_2,m_2,\ldots,q_n,m_n\right) \equiv \nonumber\\
& & \qquad {1\over(2\pi)^d} \int d^dk \; \frac{1}{\left(k+q_1\right)^2+m_1^2} \frac{1}{\left(k+q_2\right)^2+m_2^2} \cdots \frac{1}{\left(k+q_n\right)^2+m_n^2} \label{standint}
\eqa
\bqa
\langle \tilde{\eta} \rangle &=& \quad
\begin{picture}(50,20)(0,17)
\Line(0,20)(20,20)
\CArc(35,20)(15,-180,180)
\end{picture} \quad + \quad
\begin{picture}(50,20)(0,17)
\Line(0,20)(35,20)
\Vertex(35,20){3}
\end{picture} \nonumber\\[20pt]
&=& -{3\over2}{\hbar\over v} \; I(0,m) + {v\over m^2} \; \delta_{\mu}|_{\hbar} - {1\over6}{v^3\over m^2} \; \delta_{\lambda}|_{\hbar}
\eqa
\bqa
\langle \tilde{\eta}\left(p\right) \tilde{\eta}\left(q\right) \rangle_{\mathrm{c}} &=& \quad
\begin{picture}(70,40)(0,17)
\Line(0,20)(70,20)
\end{picture} \quad + \quad
\begin{picture}(70,40)(0,17)
\Line(0,20)(20,20)
\CArc(35,20)(15,-180,180)
\Line(50,20)(70,20)
\end{picture} \quad + \quad
\begin{picture}(70,40)(0,17)
\Line(0,5)(70,5)
\CArc(35,20)(15,-90,270)
\end{picture} \quad + \nonumber\\
& & \quad \begin{picture}(70,40)(0,17)
\Line(0,5)(70,5)
\Line(35,5)(35,15)
\GCirc(35,25){10}{0.7}
\end{picture} \quad + \quad
\begin{picture}(70,40)(0,17)
\Line(0,20)(70,20)
\Vertex(35,20){3}
\end{picture} \nonumber\\[25pt]
&=& \frac{\hbar}{p^2+m^2} + \frac{1}{\left(p^2+m^2\right)^2} \Bigg( {9\over2}{\hbar^2m^4\over v^2} \; I(0,m,p,m) + 3{\hbar^2m^2\over v^2} \; I(0,m) \nonumber\\
& & \phantom{\frac{\hbar}{p^2+m^2} + \frac{1}{\left(p^2+m^2\right)^2} \Bigg(} -\hbar p^2 \; \delta_Z|_{\hbar} - 2\hbar \; \delta_{\mu}|_{\hbar} \Bigg) \label{2pfunction}
\eqa
\bqa
\langle \tilde{\eta}\left(q_1\right) \tilde{\eta}\left(q_2\right) \tilde{\eta}\left(q_3\right) \rangle_{\mathrm{1PI}} &=& \quad
\begin{picture}(50,40)(0,17)
\Line(0,20)(25,20)
\Line(25,20)(50,0)
\Line(25,20)(50,40)
\end{picture} \quad + \quad
\begin{picture}(70,40)(0,17)
\Line(0,20)(20,20)
\CArc(35,20)(15,-180,180)
\Line(48,27)(70,40)
\Line(48,13)(70,0)
\end{picture} \quad + \quad
\begin{picture}(70,40)(0,17)
\Line(0,20)(20,20)
\CArc(35,20)(15,-180,180)
\Line(50,20)(70,40)
\Line(50,20)(70,0)
\end{picture} \quad + \nonumber\\[5pt]
& & \quad \begin{picture}(70,40)(0,17)
\Line(0,5)(70,5)
\CArc(35,20)(15,-180,180)
\Line(48,27)(70,40)
\end{picture} \quad + \quad
\begin{picture}(70,40)(0,17)
\Line(0,35)(70,35)
\CArc(35,20)(15,-180,180)
\Line(48,13)(70,0)
\end{picture} \quad + \quad
\begin{picture}(50,40)(0,17)
\Line(0,20)(25,20)
\Line(25,20)(50,0)
\Line(25,20)(50,40)
\Vertex(25,20){3}
\end{picture} \nonumber\\[25pt]
&=& -{3m^2\over\hbar v} - {27m^6\over v^3} \; I(0,m,q_1,m,-q_2,m) \nonumber\\
& & +{9m^4\over2v^3} \; \left( I(0,m,q_1,m) + I(0,m,q_2,m) + I(0,m,q_3,m) \right) \nonumber\\
& & -{v\over\hbar} \; \delta_{\lambda}|_{\hbar} \label{3p}
\eqa
\bqa
\langle \tilde{\eta}\left(q_1\right) \cdots \tilde{\eta}\left(q_4\right) \rangle_{\mathrm{1PI}} &=& \quad
\begin{picture}(40,40)(0,17)
\Line(0,0)(40,40)
\Line(0,40)(40,0)
\end{picture} \quad + \quad
\begin{picture}(70,40)(0,17)
\Line(0,40)(22,27)
\Line(0,0)(22,13)
\CArc(35,20)(15,-180,180)
\Line(48,27)(70,40)
\Line(48,13)(70,0)
\end{picture} \quad + \quad \textrm{2 permutations} \quad + \nonumber\\[5pt]
& & \quad \begin{picture}(70,40)(0,17)
\Line(0,40)(20,20)
\Line(0,0)(20,20)
\CArc(35,20)(15,-180,180)
\Line(48,27)(70,40)
\Line(48,13)(70,0)
\end{picture} \quad + \quad \textrm{5 permutations} \quad + \nonumber\\[5pt]
& & \quad \begin{picture}(70,40)(0,17)
\Line(0,40)(20,20)
\Line(0,0)(20,20)
\CArc(35,20)(15,-180,180)
\Line(50,20)(70,40)
\Line(50,20)(70,0)
\end{picture} \quad + \quad \textrm{2 permutations} \quad + \nonumber\\[5pt]
& & \quad \begin{picture}(40,40)(0,17)
\Line(0,0)(40,40)
\Line(0,40)(40,0)
\Vertex(20,20){3}
\end{picture} \nonumber\\[25pt]
&=& -{3m^2\over\hbar v^2} \nonumber\\
& & +{81m^8\over v^4} \; I(0,m,q_1,m,q_1+q_3,m,-q_2,m) + \textrm{2 permutations} \nonumber\\
& &  -{27m^6\over v^4} \; I(0,m,q_3,m,q_3+q_4,m) + \textrm{5 permutations} \nonumber\\
& & +{9m^4\over2v^4} \; I(0,m,q_3+q_4,m) + \textrm{2 permutations} \nonumber\\
& & -{1\over\hbar} \; \delta_{\lambda}|_{\hbar} \label{4p}
\eqa

Now our theory contains three free parameters, $Z$, $\mu$ and $\lambda$. To fix these we need three renormalization conditions\index{renormalization conditions!$N=1$ LSM}. We shall use the following conditions:
\bqa
\begin{picture}(50,20)(0,17)
\Line(0,20)(20,20)
\GCirc(35,20){15}{0.5}
\end{picture} \quad &=& 0 \nonumber\\
\textrm{Res} \quad
\begin{picture}(70,40)(0,17)
\Line(0,20)(70,20)
\GCirc(35,20){15}{0.5}
\end{picture} \quad &=& \hbar \nonumber\\[5pt]
\begin{picture}(70,40)(0,17)
\Line(0,40)(70,0)
\Line(0,0)(70,40)
\GCirc(35,20){15}{0.5}
\Text(35,20)[cc]{1PI}
\end{picture} \quad &=& -{\lambda\over\hbar} \quad \textrm{at $q_1=\ldots=q_4=0$} \label{rencond}
\eqa\\
The last condition states that the physical 4-point coupling at $q_1=\ldots=q_4=0$ is $\lambda$, i.e.\ we have chosen the renormalized $\lambda$ to be equal to the physical 4-point coupling constant at $q_1=\ldots=q_4=0$. The first condition states that $\langle \f \rangle=v=\sqrt{6\mu/\lambda}$ at all orders in perturbation theory. This means we have also chosen the renormalized $v$ to be equal to the physical vacuum expectation value of the $\f$-field. This condition fixes (the renormalized) $\mu$. This in turn fixes the \ind{physical mass} $m_{\mathrm{ph}}$, which can be calculated from the 2-point Green's function. Note that $m^2_{\mathrm{ph}}$ is not equal to the classical value $2\mu$. The second condition fixes the wave function renormalization $Z$. This second condition is equivalent to:
\bqa
\textrm{Res} \quad
\begin{picture}(70,40)(0,17)
\Line(0,20)(70,20)
\GCirc(35,20){15}{0.5}
\end{picture} \quad &=& \textrm{Res} \; \frac{\hbar}{p^2+m^2-\hbar A(p^2)} = \hbar \lim_{p^2=-m_{\mathrm{ph}}^2} \frac{p^2+m_{\mathrm{ph}}^2}{p^2+m^2-\hbar A(p^2)} \nonumber\\
&=& \hbar \lim_{p^2=-m_{\mathrm{ph}}^2} \frac{1}{1-\hbar {d\over dp^2} A(p^2)} = \hbar \nonumber\\
&\Rightarrow& \left. {d\over dp^2} A(p^2) \right|_{p^2=m_{\mathrm{ph}}^2} = 0
\eqa
where
\bq \label{A}
A(p^2) \equiv \quad
\begin{picture}(70,40)(0,17)
\Line(0,20)(70,20)
\GCirc(35,20){15}{0.5}
\Text(35,20)[cc]{1PI}
\end{picture} \quad.
\eq\\

With these conditions it is now easy to determine the \index{counter terms!$N=1$ LSM}counter terms:
\bqa
\delta_{\mu}|_{\hbar} &=& {3\over2}{\hbar m^2\over v^2} \; I(0,m) + {9\over4}{\hbar m^4\over v^2} \; I(0,m,0,m) - 27{\hbar m^6\over v^2} \; I(0,m,0,m,0,m) + \nonumber\\
& & {81\over2}{\hbar m^8\over v^2} \; I(0,m,0,m,0,m,0,m) \nonumber\\
\delta_{\lambda}|_{\hbar} &=& {27\over2}{\hbar m^4\over v^4} \; I(0,m,0,m) - 162{\hbar m^6\over v^4} \; I(0,m,0,m,0,m) + \nonumber\\
& & 243{\hbar m^8\over v^4} \; I(0,m,0,m,0,m,0,m) \nonumber\\
\delta_Z|_{\hbar} &=& {9\over2}{\hbar m^4\over v^2} \; \left. {d\over dp^2} I(0,m,p,m) \right|_{p^2=-m_{\mathrm{ph}}^2} \label{countt}
\eqa

$m_{\mathrm{ph}}$ Can be calculated up to order $\hbar$ from the 2-point function (\ref{2pfunction}). Substituting the counter terms obtained in (\ref{countt}) and Dyson summing\index{Dyson summation} the result (\ref{2pfunction}) gives
\bq
\frac{\hbar}{p^2+m^2-\hbar A(p^2)}
\eq
with
\bqa
A(p^2) &=& {9\over2}{m^4\over v^2} \; I(0,m,p,m) - {9\over2}{m^4\over v^2} p^2 \; \left. {d\over dp^2} I(0,m,p,m) \right|_{p^2=-m_{\mathrm{ph}}^2} - {9\over2}{m^4\over v^2} \; I(0,m,0,m) + \nonumber\\
& & 54{m^6\over v^2} \; I(0,m,0,m,0,m) - 81{m^8\over v^2} \; I(0,m,0,m,0,m,0,m) \;. \label{Aexpl}
\eqa
Now the solution $p^2$ of
\bq
p^2+m^2-\hbar A(p^2) = 0
\eq
is $-m_{\mathrm{ph}}^2$\index{physical mass}. It is easy to obtain this solution, of course up to order $\hbar$:
\bqa
m_{\mathrm{ph}}^2 &=& m^2 + {9\over2}{\hbar m^4\over v^2} \; I(0,m,0,m) - {9\over2}{\hbar m^4\over v^2} \; \left. I(0,m,p,m) \right|_{p^2=-m^2} + \nonumber\\
& & -{9\over2}{\hbar m^6\over v^2} \; \left. {d\over dp^2} I(0,m,p,m) \right|_{p^2=-m^2} - 54{\hbar m^6\over v^2} \; I(0,m,0,m,0,m) + \nonumber\\
& & 81{\hbar m^8\over v^2} I(0,m,0,m,0,m,0,m) \label{phmass}
\eqa

It is easy to see that for $d\leq4$ this $m_{\mathrm{ph}}$ is indeed finite. However for $d>4$ it is not! This shows that our $N=1$ linear sigma model, like all $\f^4$-theories, is \ind{non-renormalizable} for $d>4$.

Now it can also be shown that the 3- and 4-point 1PI Green's function (\ref{3p},\ref{4p}), and in fact all $n$-point Green's functions are finite at one-loop order, for $d\leq4$.

Now we have fixed all of our parameters in terms of the physical parameters. Our renormalized $\lambda$ is just equal to the physical 4-point coupling $\lambda_{\mathrm{ph}}$, an so is our renormalized $v=v_{\mathrm{ph}}$. This then fixes our renormalized $\mu$ (and with it $m$): $\mu=\lambda_{\mathrm{ph}}v_{\mathrm{ph}}^2/6$. Then all counter terms as given in (\ref{countt}) and the physical mass as given in (\ref{phmass}) can be expressed in terms of the two physical parameters $\lambda_{\mathrm{ph}}$ and $v_{\mathrm{ph}}$.

\section{The Effective Potential}

Now we want to find also the one-loop effective potential\index{effective potential!$N=1$ LSM} of our $N=1$ linear sigma model. We know that the effective potential $V$ is given by the vacuum-graph formula, which, in the $N=1$ case, simplifies to:
\bq
V(\eta) = \h m^2\eta^2 - \hbar \sum_{n=2}^{\infty} {1\over n!} \langle \tilde{\eta}^n(0) \rangle_{\mathrm{1PI}} \; \eta^n
\eq
This means to obtain the one-loop effective potential we just have to take the sum of all 1-loop 1PI diagrams with the appropriate factors and with zero momentum in the external legs.

Let $n_3$ and $n_4$ denote the number of 3- and 4-vertices in a diagram. Then $n_3+2n_4$ is the number of external legs. Every 3-vertex in a diagram gets a $1/3!$, every 4-vertex a $1/4!$. For a diagram with $n_3$ (identical) 3-vertices we get a $1/n_3!$, likewise we get a $1/n_4!$. To see how many ways there are to connect the legs of the vertices and the external legs we first decide which legs are going to be part of the loop. For each 3-vertex there are $3\cdot2/2$ ways to choose this, for each 4-vertex there are $4\cdot3/2$ ways to choose this. The legs that have been chosen as internal can then be connected in $(2n_3+2n_4-2)!!$ ways. The other legs can be connected to the external legs in $(n_3+2n_4)!$ ways. Then the one-loop effective potential $V_1(\eta)$ is:
\bqa
V_1(\eta) &=& {1\over(2\pi)^d} \int \; d^dk \sum_{\underset{n_3+n_4\geq1}{n_3,n_4=0}}^{\infty} \left(-{3m^2\over\hbar v}\right)^{n_3} \left(-{3m^2\over\hbar v^2}\right)^{n_4} \left({\hbar\over k^2+m^2}\right)^{n_3+n_4} \nonumber\\
& & \quad \left({1\over3!}\right)^{n_3} \left({1\over4!}\right)^{n_4} {1\over n_3!} {1\over n_4!} 3^{n_3} 6^{n_4} (2n_3+2n_4-2)!! (n_3+2n_4)! \nonumber\\
& & \quad \left(-{\hbar\over(n_3+2n_4)!} \eta^{n_3+2n_4}\right) + \nonumber\\[10pt]
& & -\hbar \;
\begin{picture}(10,10)(0,7)
\Line(0,10)(10,10)
\Vertex(10,10){2}
\end{picture} \;\; \eta - {\hbar\over2} \;
\begin{picture}(20,10)(0,7)
\Line(0,10)(20,10)
\Vertex(10,10){2}
\end{picture} \;\; \eta^2 - {\hbar\over6} \;
\begin{picture}(20,10)(0,7)
\Line(0,10)(10,10)
\Line(10,10)(20,0)
\Line(10,10)(20,20)
\Vertex(10,10){2}
\end{picture} \;\; \eta^3 - {\hbar\over24} \;
\begin{picture}(20,10)(0,7)
\Line(0,0)(20,20)
\Line(0,20)(20,0)
\Vertex(10,10){2}
\end{picture} \;\; \eta^4
\eqa
By substituting $p=n_3+n_4$ we can replace the sum over $n_4$ by a sum over $p$. The sum over $n_3$ can then easily be done with help of the binomial theorem. We find:
\bqa
V_1(\eta) &=& -{\hbar\over2}{1\over(2\pi)^d} \int \; d^dk \sum_{p=1}^{\infty} {(-1)^p\over p} {1\over\left(k^2+m^2\right)^p} \left({3m^2\over v}\eta+{3m^2\over2v^2}\eta^2\right)^p + \nonumber\\[5pt]
& & -\hbar \;
\begin{picture}(10,10)(0,7)
\Line(0,10)(10,10)
\Vertex(10,10){2}
\end{picture} \;\; \eta - {\hbar\over2} \;
\begin{picture}(20,10)(0,7)
\Line(0,10)(20,10)
\Vertex(10,10){2}
\end{picture} \;\; \eta^2 - {\hbar\over6} \;
\begin{picture}(20,10)(0,7)
\Line(0,10)(10,10)
\Line(10,10)(20,0)
\Line(10,10)(20,20)
\Vertex(10,10){2}
\end{picture} \;\; \eta^3 - {\hbar\over24} \;
\begin{picture}(20,10)(0,7)
\Line(0,0)(20,20)
\Line(0,20)(20,0)
\Vertex(10,10){2}
\end{picture} \;\; \eta^4
\eqa

For $d\leq4$ only the terms with $p=1$ and $p=2$ in the sum diverge. These divergences are supposed to be cancelled by the counter terms. It is easy to check that this indeed happens when we substitute the counter terms (\ref{countt}) that we found before! After also expressing $\eta$ in terms of the original field $\f=v+\eta$ (exact relationship with the renormalization conditions that we have chosen) we find:
\bqa
V_1(\f) &=& -{27\over4}{\hbar m^6\over v^4} \left( I(0,m,0,m,0,m) - {3m^2\over2} \; I(0,m,0,m,0,m,0,m) \right) \left(\f^2-v^2\right)^2 + \nonumber\\
& & -{\hbar\over2} \sum_{p=3}^{\infty} {(-1)^p\over p} \left({3 m^2\over2v^2}(\f^2-v^2)\right)^p {1\over(2\pi)^d} \int d^dk \; {1\over\left(k^2+m^2\right)^p}
\eqa
Now for $d\leq4$ and $p\geq3$ we have
\bq \label{1loopint}
{1\over(2\pi)^d} \int d^dk \; {1\over\left(k^2+m^2\right)^p} = {1\over(4\pi)^{d/2}} m^{d-2p} {\Gamma(p-d/2)\over\Gamma(p)} \;,
\eq
so that we get\index{effective potential!$N=1$ LSM}
\bqa
V_1(\f) &=& -{27\over8}{\hbar m^d\over v^4} {1\over(4\pi)^{d/2}} \left( \Gamma(3-d/2) - {1\over2} \Gamma(4-d/2) \right) \left(\f^2-v^2\right)^2 + \nonumber\\
& & -{\hbar\over2} {1\over(4\pi)^{d/2}} m^d \sum_{p=3}^{\infty} {(-1)^p\over p} {\Gamma(p-d/2)\over\Gamma(p)} \left({3\over2v^2}(\f^2-v^2)\right)^p \label{effpotgend}
\eqa
What function this exactly is depends strongly on the dimension $d$, so to find an explicit result we have to specify $d$.

\subsection{$d=1$}

\begin{figure}[h]
\begin{center}
\epsfig{file=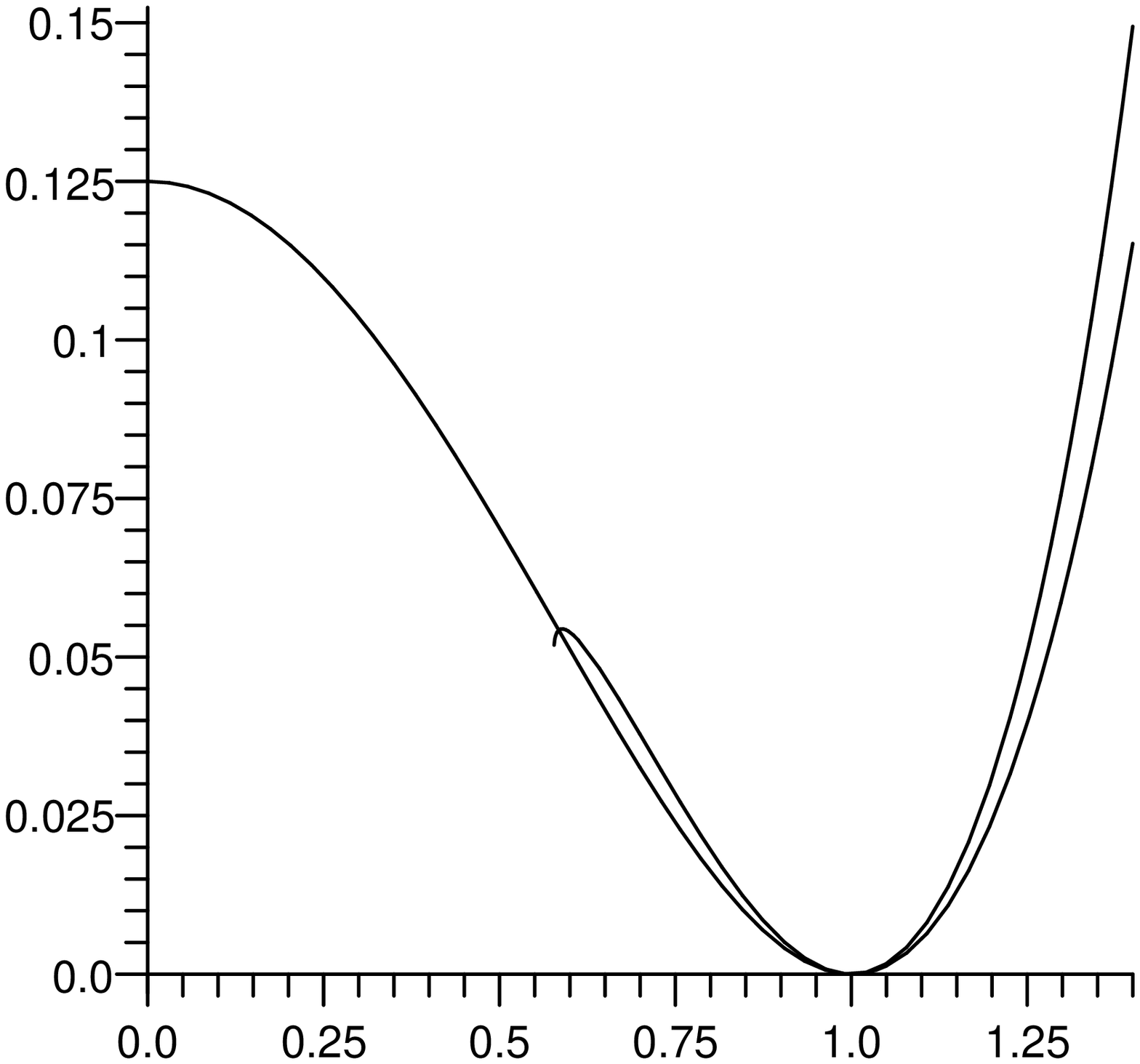,width=7cm}
\end{center}
\vspace{-1cm}
\caption{$V_0$ And $V=V_0+V_1$ as a function of $\f/v$ for $\hbar=0.1$, $m=1$, $v=1$.}
\label{effpot1mind1}
\end{figure}

For $d=1$ we find:
\bq \label{effpotd1}
V_1(\f) = -\h\hbar m + \h\hbar m \sqrt{1+{3\over2v^2}(\f^2-v^2)} - {3\over8}{\hbar m\over v^2} (\f^2-v^2) + {117\over256}{\hbar m\over v^4} (\f^2-v^2)^2
\eq
In \index{effective potential!$N=1$ LSM}figure (\ref{effpot1mind1}) $V_0={m^2\over8v^2}(\f^2-v^2)^2$ and $V=V_0+V_1$ are plotted as a function of $\f/v$ for the case $\hbar=0.1$, $m=1$, $v=1$.\index{effective potential!$N=1$ LSM}

It is easy to check that the minima of this 1-loop effective potential are still at $\f=\pm v$, like our renormaliztion conditions (\ref{rencond}) ensures. Expanding around the minimum $\f=+v$ one finds:
\bqa
V_1(\f) &=& {81\over64}{\hbar m\over v^2} (\f-v)^2 + {135\over64}{\hbar m\over v^3} (\f-v)^3 + {405\over512}{\hbar m\over v^5} (\f-v)^5 - {2943\over2048}{\hbar m\over v^6} (\f-v)^6 + \nonumber\\[5pt]
& & \mathcal{O}((\f-v)^7)
\eqa
Indeed there is no tadpole term, and also the 4-point term is absent, as imposed by our renormalization condition (\ref{rencond}). From the 2-point term one can read off a mass correction ${81\over32}{\hbar m\over v^2}$. However this is \emph{not} the correction to the physical mass $m_{\mathrm{ph}}^2$! This correction is given in (\ref{phmass}), and is in the case of $d=1$:
\bq
m_{\mathrm{ph}}^2 = m^2 + {85\over32}{\hbar m\over v^2} \;.
\eq
Here we used the standard integral results given in appendix \ref{appstandint}.

So in our renormalization scheme the 2-point part of the effective potential does \emph{not} give the \ind{physical mass}. The reason is very simple, the one-loop correction to the effective potential is given by $-\hbar A(0)$, whereas the one-loop correction to the physical mass is given by $-\hbar A(-m^2)$, with $A$ as defined in (\ref{A}). To extract the physical mass $m_{\mathrm{ph}}$ from the effective action we need the complete effective action including the dynamical part, not just the effective potential. 

The 4-point part of our effective potential does give the physical coupling constant however, simply because we put $\lambda=\lambda_{\mathrm{ph}}$ at zero incoming momentum.

Now there seems to be a \index{effective potential!complexity}problem with this 1-loop effective potential (\ref{effpotd1}). The argument of the square root in (\ref{effpotd1}) becomes negative when
\bq
\f^2 \leq {1\over3}v^2 \;.
\eq
This means the 1-loop effective potential becomes \emph{complex}\index{effective potential!complexity} where the classical effective potential becomes non-convex, i.e.\ ${d^2\over d\f^2} V_{\mathrm{cl}}(\f) \leq 0$. Even in the domain where the 1-loop effective potential is defined there is something wrong, it \emph{can} become \emph{non-convex}! In figure \ref{effpot1mind1} $V$ is indeed non-convex, however parameters can also be chosen such that it is convex where it is defined. However according to our general argument it should \emph{always} be convex. As has already been discussed in the literature it is no problem that one finds a non-convex effective potential in the canonical approach, the proof for the convexity originates \emph{only} in the path-integral approach. Also, \emph{only} in the path-integral approach one can argue that the effective potential should be real and well-defined everywhere (see chapter \ref{chapeffact}).

\subsection{$d=4$}

\begin{figure}[h]
\begin{center}
\epsfig{file=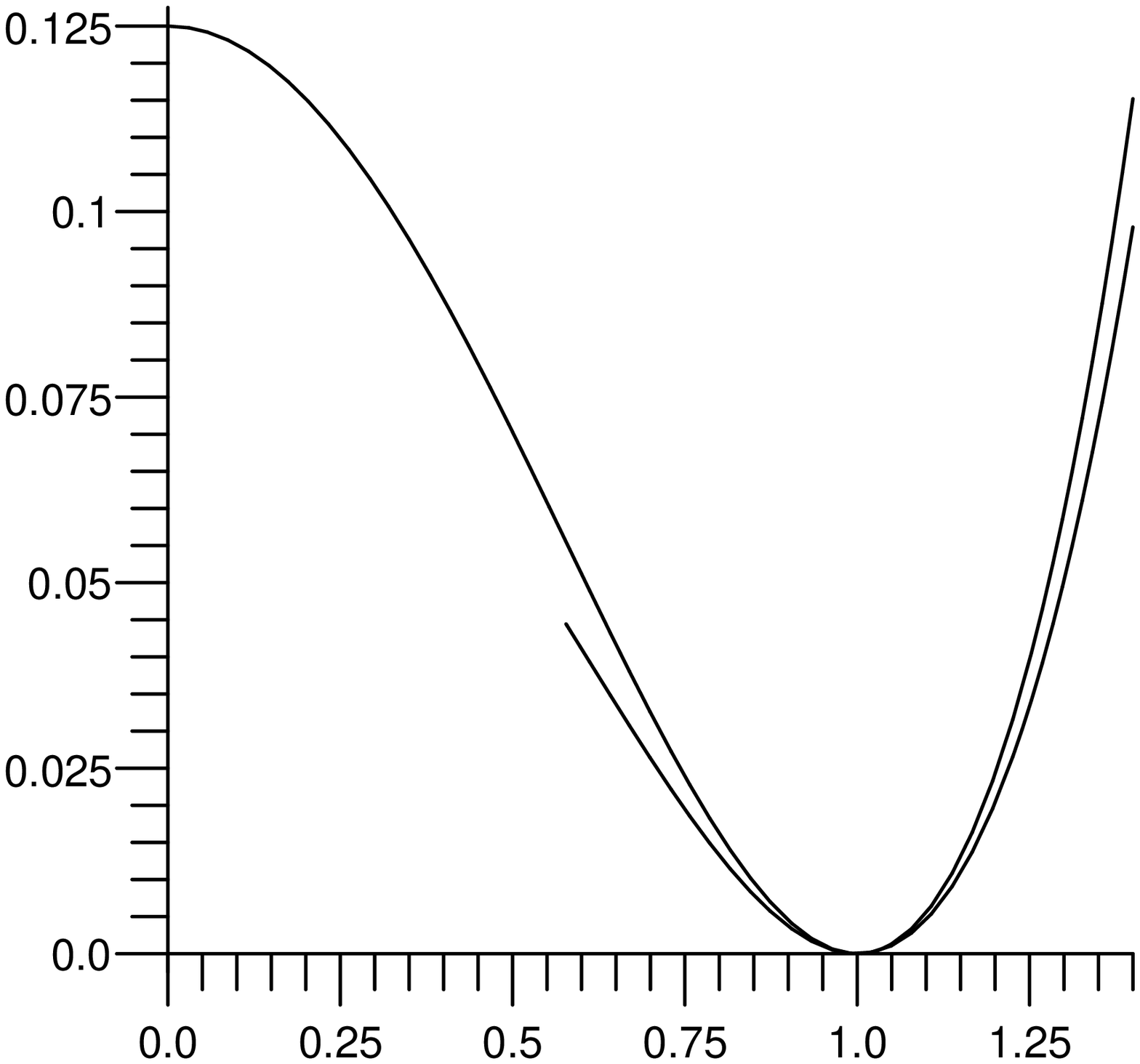,width=7cm}
\end{center}
\vspace{-1cm}
\caption{$V_0$ And $V=V_0+V_1$ as a function of $\f/v$ for $\hbar=2$, $m=1$, $v=1$.}
\label{effpot1mind4}
\end{figure}

For the case $d=4$ we find, using (\ref{effpotgend}):
\bqa
V_1(\f) &=& {1\over4}{1\over(4\pi)^2}\hbar m^4 \left(1+{3\over2}{1\over v^2}(\f^2-v^2)\right)^2 \ln\left(1+{3\over2}{1\over v^2}(\f^2-v^2)\right) \nonumber\\[5pt]
& & -{3\over8}{1\over(4\pi)^2}{\hbar m^4\over v^2} (\f^2-v^2) - {81\over32}{1\over(4\pi)^2}{\hbar m^4\over v^4} (\f^2-v^2)^2 \label{effpotd4}
\eqa
In \index{effective potential!$N=1$ LSM}figure (\ref{effpot1mind4}) $V_0$ and $V=V_0+V_1$ are plotted as a function of $\f/v$ for the case $\hbar=2$, $m=1$, $v=1$.\index{effective potential!$N=1$ LSM}

The minima are again at $\f=\pm v$. Expanding (\ref{effpotd4}) around $\f=+v$ gives:
\bqa
V_1(\f) &=& -{27\over4}{1\over(4\pi)^2}{\hbar m^4\over v^2} (\f-v)^2 - {9\over2}{1\over(4\pi)^2}{\hbar m^4\over v^3} (\f-v)^3 + {27\over80}{1\over(4\pi)^2}{\hbar m^4\over v^5} (\f-v)^5 \nonumber\\[5pt]
& & -{9\over40}{1\over(4\pi)^2}{\hbar m^4\over v^6} (\f-v)^6 + \mathcal{O}((\f-v)^7)
\eqa
Again we see that there is no tadpole and 4-point part, in accordance with our renormalization conditions (\ref{rencond}).

The \ind{physical mass} $m_{\mathrm{ph}}$ can again be calculated from (\ref{phmass}). This calculation is somewhat more involved now since we have divergences occurring at intermediate steps in the calculation. By using the standard integral results from appendix \ref{appstandint} we find:
\bq
m_{\mathrm{ph}}^2 = m^2 - {27\over16}{\hbar m^4\over v^2}{1\over\pi^2} + {5\sqrt{3}\over32}{\hbar m^4\over v^2}{1\over\pi}
\eq

Again, as in $d=1$, at first sight there is a problem with (\ref{effpotd4}). For $\f^2\leq{1\over3}v^2$ the argument of the logarithm becomes zero or negative, such that the one-loop \index{effective potential!complexity} effective potential is complex in this domain\index{effective potential!complexity}. Also it can become non-convex in this domain. Indeed the $V$ in figure \ref{effpot1mind4} is non-convex (although this might be a bit hard to see, it can be seen more clearly by plotting the derivative of $V$).

\section{Instantons}

\index{instantons}Instantons are classical solutions of the equations of motions, i.e.\ configurations that minimize the classical action, which are \emph{not} constant (like (\ref{Z2constsols})), but which go from one minimum to the other. Also these instantons should have a \emph{finite} action. For a nice book on these instantons see \cite{Rajaraman}.

So what about instantons in the $N=1$ linear sigma model? For dimension 1 it can easily be shown that there is a classical solution that takes one from one minimum to the other and that has a finite action. So for $d=1$ we should have included these instanton solutions in a complete treatment. In \cite{Das} such a treatment for the $N=1$ linear sigma model can be found. When including these instantons one will \emph{not} find a spontaneously broken theory. This is well known, in one dimension there is \emph{no} SSB, on account of tunneling.

However, it is shown in general, by Derrick \cite{Derrick}, that for $d>1$ \emph{no} instanton solutions exist. Of course there exist solutions of the classical equations of motion that go from one minimum to the other, but these solutions have infinite action. Since we are mostly interested in higher dimensions, i.e.\ $d>1$, we shall \emph{not} include instantons at all in this thesis.


\chapter{The $N=1$ LSM: The Path-Integral Approach}\label{chapN1sigmamodelpath}

In this chapter we discuss the path-integral approach to the \index{$N=1$ linear sigma model!path-integral approach}$N=1$ Euclidean linear sigma model. This means we take the path integral seriously and take into account \emph{both} minima in our calculations. These calculations generally follow Fujimoto at al.\ \cite{Fujimoto} and Cooper at al.\ \cite{Cooper}. We will calculate the effective potential, which will be convex and well-defined. Also we will calculate some Green's functions and say something about the physics resulting from this approach.

The \index{action!$N=1$ LSM}action we will use is:
\bq
S = \int d^dx \; \left( \h\left(\nabla\f\right)^2 - \h\mu\f^2 + {\lambda\over24}\f^4 - J\f\right) \;.
\eq
Here, in contrast to the previous chapter, we have included a source term in the action, to be able to compute the effective potential via this source. In this action the field depends on the space-time coordinates $x$, and so does the source $J$ in general. We however limit ourselves to the case where $J$ is constant over space-time, since we are only interested in the effective \emph{potential} and not the complete effective action.

\section{The Effective Potential}

To find the renormalized generating functional and Green's functions we have to introduce renormalized quantities, as in (\ref{renquant}). The source $J$ is renormalized\index{source renormalization} as:
\bq
J^{\mathrm{R}} = \sqrt{Z} J
\eq
By renormalizing $J$ in this way we ensure that by taking derivatives with respect to $J$ one gets the renormalized Green's functions. As in the previous chapter we will drop the superscript R from now on. The action becomes: 
\bqa
S &=& \int d^dx \; \bigg( \h\left(\nabla\f\right)^2 - \h\mu\f^2 + {\lambda\over24}\f^4 -J\f + \nonumber\\[5pt]
& & \phantom{\int d^dx \; \bigg(} \h\delta_Z\left(\nabla\f\right)^2 - \h\delta_{\mu}\f^2 + {\delta_{\lambda}\over24}\f^4 \bigg) \;. \label{renactionws}
\eqa

Now the two \index{minima!$N=1$ LSM}minima of the first line in (\ref{renactionws}) can be parameterized as:
\bqa
J &=& {2\mu v\over3\sqrt{3}} \sin(3\alpha) \;, \quad -{\pi\over6} < \alpha \leq {\pi\over6} \nonumber\\
\f_\pm &=& {2v\over\sqrt{3}} \sin\left(\alpha\pm{\pi\over3}\right) \equiv {2v\over\sqrt{3}} \sin_\pm
\eqa
with $v^2\equiv6\mu/\lambda$, as in the previous chapter. We see that in this parametrization 
\bq
-{2\mu v\over3\sqrt{3}} < J \leq {2\mu v\over3\sqrt{3}} \;.
\eq
These limits are exactly the values of $J$ where one of the minima becomes unstable. When this happens it is a good approximation to take along only \emph{one} minimum and the effective potential of the previous chapter can be used in this region.

Now the action can be expanded around one of the minima $\f_\pm$:
\bqa
S_\pm &=& \int d^dx \; \bigg( \h\left(\nabla\eta\right)^2 + \mu\left(2\sin_\pm^2-\h\right)\eta^2 + {2\over\sqrt{3}}{\mu\over v}\sin_\pm\eta^3 + {\mu\over4v^2}\eta^4 + \nonumber\\[5pt]
& & \phantom{\int d^dx \; \bigg(} {2\over3}\mu v^2\sin_\pm^2 - {4\over3}\mu v^2\sin_\pm^4 + \nonumber\\[5pt]
& & \phantom{\int d^dx \; \bigg(} \left(-{2\over\sqrt{3}}v\sin_\pm\delta_{\mu}+{4\over9\sqrt{3}}v^3\sin_\pm^3\delta_{\lambda}\right)\eta + \h\delta_Z\left(\nabla\eta\right)^2 + \nonumber\\[5pt]
& & \phantom{\int d^dx \; \bigg(} \left(-\h\delta_{\mu}+{1\over3}v^2\sin_\pm^2\delta_{\lambda}\right)\eta^2 + {1\over3\sqrt{3}}v\sin_\pm\delta_{\lambda}\eta^3 + {1\over24}\delta_{\lambda}\eta^4 + \nonumber\\[5pt]
& & \phantom{\int d^dx \; \bigg(} -{2\over3}v^2\sin_\pm^2\delta_{\mu} + {2\over27}v^4\sin_\pm^4\delta_{\lambda} \bigg) \;.
\eqa

The \ind{counter terms} cannot depend on the source $J$, so we can just as well use the counter terms as derived in (\ref{countt}). These counter terms will still make all results finite. However they will not make our theory have the same physics as in the previous chapter, also because we take into account \emph{both} minima now. For now we use (\ref{countt}) and we will see what physics we get.

The complete \ind{generating functional} $Z(J)$ is given by\index{summing over minima}
\bq
Z(J) = Z_+(J) + Z_-(J)
\eq
with
\bq
Z_\pm = \int \mathcal{D}\f \; \exp\left(-{1\over\hbar}S_\pm\right) \;.
\eq
Now we take only the \ind{saddle-point approximation} to $Z_\pm$, i.e.\ discard all interaction terms (defined as all terms in the action of higher order than $\hbar$). Below we will see that this saddle-point approximation will already produce the one-loop correction to the effective potential, which we can then compare to our previous one-loop effective potentials.
\bqa
Z_\pm(J) &=& \exp\left(-{1\over\hbar} \int d^dx \left( {1\over3}m^2v^2\sin_\pm^2 - {2\over3}m^2v^2\sin_\pm^4 - {2\over3}v^2\sin_\pm^2\delta_{\mu} + {2\over27}v^4\sin_\pm^4\delta_{\lambda} \right)\right) \cdot \nonumber\\[5pt]
& & \qquad \int \mathcal{D}\eta \; \exp\left(-{1\over\hbar} \int d^dx \left( \h\left(\nabla\eta\right)^2 + m^2\left(\sin_\pm^2-{1\over4}\right)\eta^2 \right)\right)
\eqa

This last line can be calculated as follows (We call the space-time volume $\Omega\equiv\int d^dx$.):
\bqa
A &\equiv& \int \mathcal{D}\eta \; \exp\left(-{1\over\hbar} \int d^dx \left( \h\left(\nabla\eta\right)^2 + \h M^2\eta^2 \right)\right) \nonumber\\
\Rightarrow -{2\hbar\over\Omega}\frac{\partial}{\partial M^2} \; A &=& \hbar A \; I(0,M) \nonumber\\
\Rightarrow A &\sim& \exp\left(-\h\Omega {1\over(2\pi)^d}\int d^dk \; \ln\left(k^2+M^2\right)\right) \label{Gaussianpathint}
\eqa
Next we expand the logarithm around $\sin_\pm^2={3\over4}$:
\bqa
& & {1\over(2\pi)^d}\int d^dk \; \ln\left(k^2+m^2\left(2\sin_\pm^2-\h\right)\right) = \nonumber\\
& & \qquad {1\over(2\pi)^d}\int d^dk \left( \ln(k^2+m^2) - \sum_{p=1}^\infty {(-2m^2)^p\over p} \left(\sin_\pm^2-{3\over4}\right)^p {1\over\left(k^2+m^2\right)^p} \right)
\eqa
The first term in this expansion can be discarded since it is merely a constant, not depending on $J$ and thus unimportant for physical quantities. Now the counter terms (\ref{countt}) can be inserted and (\ref{1loopint}) can be used to simplify the integrals in $Z_\pm(J)$. After some algebra one finds:
\bq
Z_\pm(J) = \exp \left( -{1\over\hbar}\Omega \left( V_0\left(\f_\pm\right) + V_1\left(\f_\pm\right) - J\f_\pm \right) \right)
\eq
where $V_0$ is the zero-loop effective potential when one takes along only \emph{one} minimum, which is just the classical potential,
\bq
V_0(\f) = -\h\mu\f^2 + {\mu\over4v^2}\f^4
\eq
and $V_1$ is the one-loop effective potential when one takes along only \emph{one} minimum, which is given in (\ref{effpotgend}). Notice that this $Z_\pm$ has a $J$ dependence through $J$ \emph{and} $\f_\pm$.

Now one can also see that if we take into account only \emph{one} minimum, e.g.\ $Z=Z_+$, we again get $V_0(\fc)+V_1(\fc)$ for the effective potential, like we got in the previous chapter. This also demonstrates that the saddle-point approximation to $Z_\pm$ was enough to get the one-loop effective potential. It also demonstrates that the canonical approach is in fact equivalent to taking into account only \emph{one} minimum in the path-integral approach.

Continuing our calculation of $Z=Z_++Z_-$ and defining:
\bqa
& & \alpha \equiv \h(\f_++\f_-) \;, \quad \beta \equiv \h(\f_+-\f_-) \;, \nonumber\\
& & A_0 \equiv \h\left(V_0(\f_+)+V_0(\f_-)\right) \;, \quad A_1 \equiv \h\left(V_1(\f_+)+V_1(\f_-)\right) \;, \nonumber\\
& & B_0 \equiv \h\left(V_0(\f_+)-V_0(\f_-)\right) \;, \quad B_1 \equiv \h\left(V_1(\f_+)-V_1(\f_-)\right) \;,
\eqa
we find
\bq
Z = 2\exp\left(-{1\over\hbar}\Omega(A_0+A_1-J\alpha)\right) \cosh\left({1\over\hbar}\Omega(B_0+B_1-J\beta)\right)
\eq
and
\bqa
\langle\f\rangle(J) &=& \hbar\frac{\delta}{\delta J} \ln Z = {\hbar\over\Omega}\frac{\partial}{\partial J} \ln Z \nonumber\\
&=& \alpha - \frac{\partial A_1}{\partial J} + \left(\beta-\frac{\partial B_1}{\partial J}\right) \tanh\left({1\over\hbar}\Omega(J\beta-B_0-B_1)\right) \;. \label{phi2min}
\eqa
Now by inverting this $\langle\f\rangle(J)$ we find $J(\fc)$, which is equal to the derivative of the effective potential $V^{\mathrm{2min}}$, where the superscript `2min' denotes that we included \emph{both} minima.

\begin{figure}[h]
\begin{center}
\epsfig{file=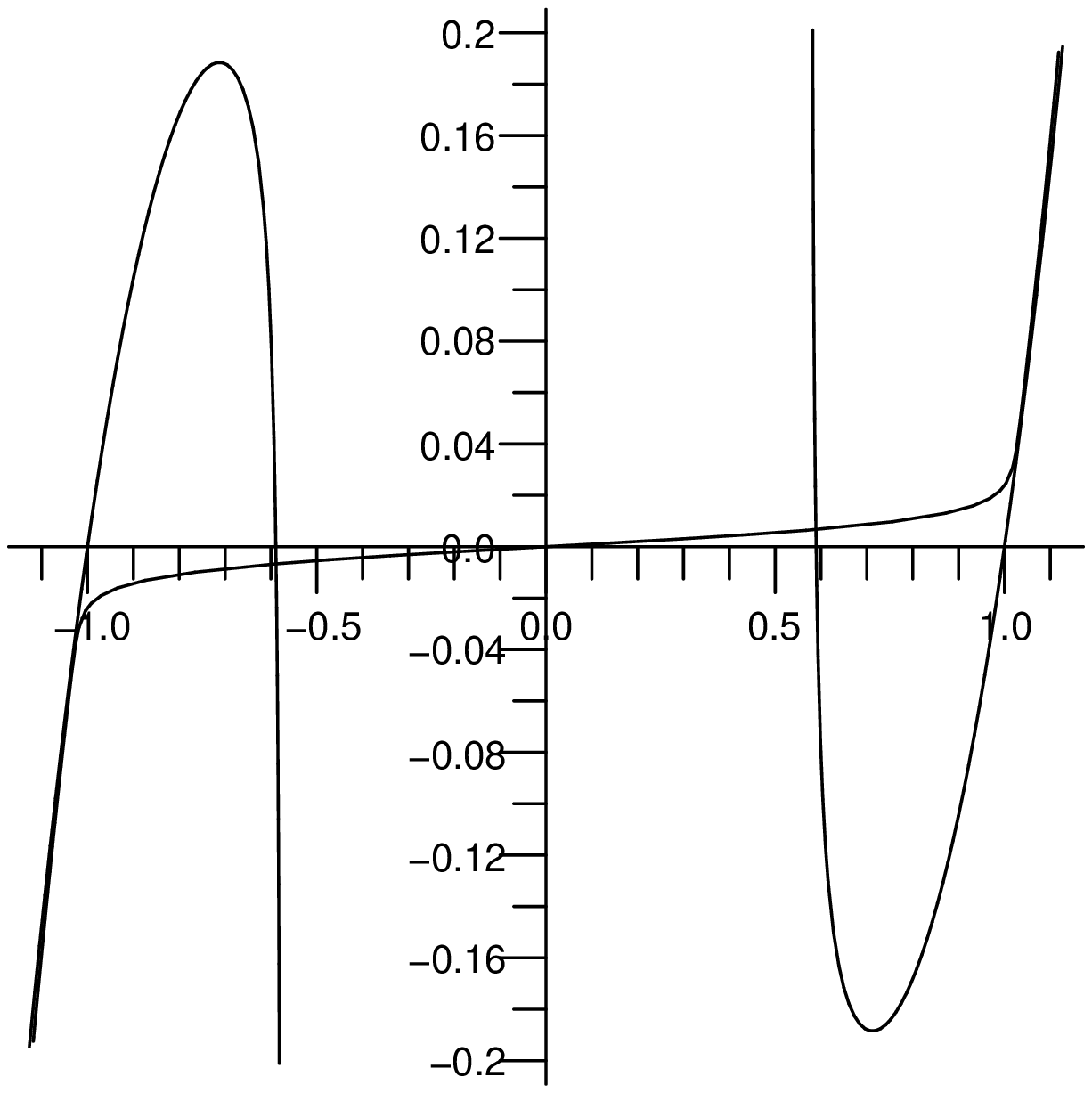,width=10cm}
\end{center}
\vspace{-0.5cm}
\caption{${\partial\over\partial\fc} V^{\mathrm{2min}}$ And ${\partial\over\partial\fc} V$ as a function of $\f/v$ for $d=1$ and $\hbar=0.1$, $m=1$, $v=1$, $\Omega=10$.}
\label{effpot2mind1}
\end{figure}

\begin{figure}[h]
\begin{center}
\epsfig{file=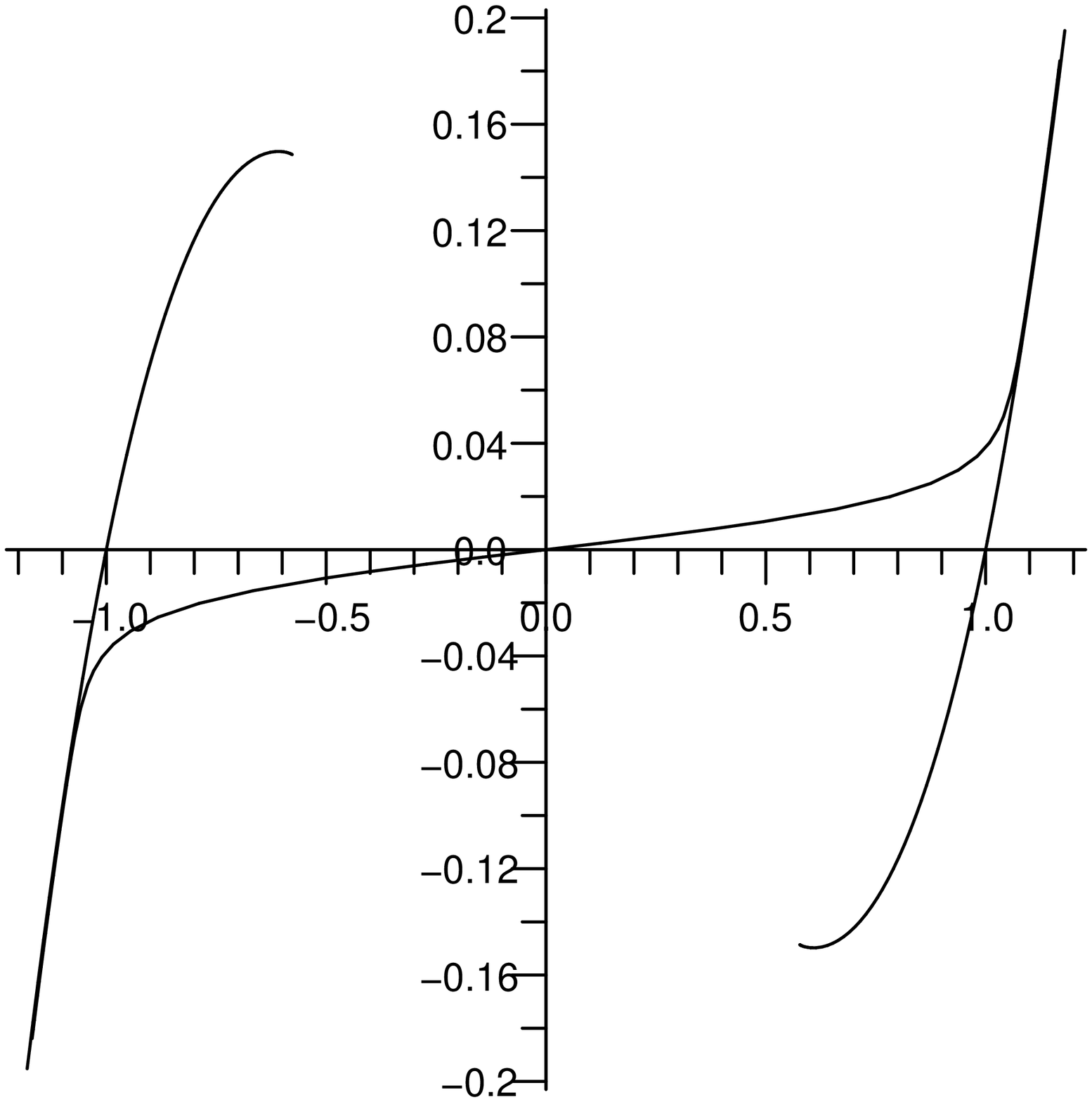,width=7cm}
\end{center}
\vspace{-0.5cm}
\caption{${\partial\over\partial\fc} V^{\mathrm{2min}}$ And ${\partial\over\partial\fc} V$ as a function of $\f/v$ for $d=4$ and $\hbar=2$, $m=1$, $v=1$, $\Omega=100$.}
\label{effpot2mind4}
\end{figure}

This derivative of $V^{\mathrm{2min}}$ is plotted in \index{effective potential!$N=1$ LSM}figure (\ref{effpot2mind1}) for the case of $d=1$ and $\hbar=0.1$, $m=1$, $v=1$, $\Omega=10$. Also the derivative of the one-minimum effective potential (\ref{effpotd1}) is plotted for comparison.
In \index{effective potential!$N=1$ LSM}figure (\ref{effpot2mind4}) the derivative of $V^{\mathrm{2min}}$ is plotted for the case of $d=4$ and $\hbar=2$, $m=1$, $v=1$, $\Omega=100$. Also the derivative of the one-minimum effective potential (\ref{effpotd4}) is plotted again.

By plotting the same case for larger and larger $\Omega$ it is easy to see that in the limit $\Omega\rightarrow\infty$ one gets an effective potential with a flat bottom. This is the \ind{Maxwell construction} of the one-minimum effective potential. That one gets this \ind{Maxwell construction} can also be seen from (\ref{phi2min}). It is easy to see that for small $J$ we have:
\bq \label{kink}
\langle\f\rangle(J) = v\tanh\left({1\over\hbar}\Omega Jv + \mathcal{O}(J^2)\right) + \mathcal{O}(J) \approx v\tanh\left({1\over\hbar}\Omega Jv \right)
\eq
For $\Omega\rightarrow\infty$ this becomes a kink.

So indeed we get a \emph{convex} and \emph{real} effective potential in this case, as dictated by our general arguments from chapter \ref{chapeffact}. For finite space-time volume $\Omega$ it is even \emph{strictly} convex.

For $\Omega\rightarrow\infty$ our effective potential is not strictly convex because in this limit we have that
\bq
\langle \f(x) \f(y) \rangle_{\mathrm{c}} = v^2 + \mathcal{O}(\hbar) \;,
\eq
as will be shown in the next section. If we consider this as a matrix, like we did in section (\ref{conv}), one can see that this matrix has eigenvalues equal to infinity. A function $f(x)$ and a value $k_f$ are eigenfunction and eigenvalue of our matrix $\langle \f(x) \f(y) \rangle_{\mathrm{c}}$ if:
\bq
\int d^dy \; \langle \f(x) \f(y) \rangle_{\mathrm{c}} \; f(y) = k_f f(x)
\eq
If our matrix is just $v^2$ (at lowest order) one finds
\bq
v^2 \int d^dy \; f(y) = k_f f(x) \Rightarrow f(x) = f \Rightarrow k_f = v^2\Omega \;,
\eq
which shows that all eigenvalues are infinite. If we include higher order terms in $\langle \f(x) \f(y) \rangle_{\mathrm{c}}$ (many of these are normalizable, i.e.\ give a finite result when integrated over) the argument changes, but one will still find at least one eigenvalue equal to infinity. This means that the eigenvalues of the inverse matrix can become zero for infinite volume and that the effective action is \emph{not} strictly convex, but of course still convex.

\section{The Green's Functions}

Now we calculate the Green's functions\index{Green's functions!$N=1$ LSM} of the theory, to discover what physics this theory gives. In section (\ref{meaning}) we showed that the functional derivatives of the effective action are related to the Green's functions of the theory. Because we find a flat bottom here, we also get different Green's functions than in the canonical approach.

First we compute the tadpole:
\bqa
\langle \f(x) \rangle &=& \frac{\int \mathcal{D}\f \; \f(x) \; e^{-{1\over\hbar}S(J=0)}}{\int \mathcal{D}\f \; e^{-{1\over\hbar}S(J=0)}} \nonumber\\
&=& \frac{\int \mathcal{D}\eta \; (\f_+(0)+\eta(x)) \; e^{-{1\over\hbar}S_+(J=0)} + \int \mathcal{D}\eta \; (\f_-(0)+\eta(x)) \; e^{-{1\over\hbar}S_-(J=0)}}{\int \mathcal{D}\eta \; e^{-{1\over\hbar}S_+(J=0)} + \int \mathcal{D}\eta \; e^{-{1\over\hbar}S_-(J=0)}} \nonumber\\
&=& \frac{(\f_+(0)+\langle\eta(x)\rangle_+(0)) Z_+(0) + (\f_-(0)+\langle\eta(x)\rangle_-(0)) Z_-(0)}{Z_+(0) + Z_-(0)}
\eqa
Here $\f_+$, $\f_-$, $S_+$, $S_-$, $Z_+$ and $Z_-$ are all defined earlier in this chapter, and $\langle\eta\rangle_\pm$ is given by:
\bq
\langle\eta\rangle_\pm = {\hbar\over\Omega}{\partial\over\partial J} \ln Z_\pm \;.
\eq
This is just the $\eta$-tadpole from the canonical approach. Now we know:
\bqa
\f_\pm(0) &=& \pm v \nonumber\\
Z_\pm(0) &=& \int \mathcal{D}\eta \; e^{-{1\over\hbar}S_\pm(J=0)} = Z_+ \nonumber\\
\langle\eta\rangle_\pm(0) &=& \int \mathcal{D}\eta \; \eta(x) \; e^{-{1\over\hbar}S_\pm(J=0)} = \pm\langle\eta\rangle_+(0)
\eqa
So we find
\bq
\langle \f(x) \rangle = 0 \;.
\eq

In the same way we can find the 2-point Green's function.
\bqa
\langle \f(x)\f(y) \rangle &=& \frac{\int \mathcal{D}\f \; \f(x)\f(y) \; e^{-{1\over\hbar}S(J=0)}}{\int \mathcal{D}\f \; e^{-{1\over\hbar}S(J=0)}} \nonumber\\
&=& \frac{\int \mathcal{D}\eta \; \left(\f_+^2(0)+\f_+(0)(\eta(x)+\eta(y))+\eta(x)\eta(y)\right) \; e^{-{1\over\hbar}S_+(J=0)} + (+\rightarrow-)}{\int \mathcal{D}\eta \; e^{-{1\over\hbar}S_+(J=0)} + (+\rightarrow-)} \nonumber\\
&=& v^2 + 2v \langle\eta\rangle_+(0) + \langle\eta(x)\eta(y)\rangle_+(0)
\eqa
Here $\langle\eta(x)\eta(y)\rangle_+(0)$ is again just the $\eta$-propagator from the canonical approach. With the counter terms that we chose in this canonical approach (and that we will use for the $N=1$ linear sigma model in the whole of this thesis) we find, up to one-loop order:
\bq
\langle \f(x)\f(y) \rangle_\mathrm{c} = v^2 + {1\over(2\pi)^d} \int d^dp \left( \frac{\hbar}{p^2+m^2} + \left(\frac{\hbar}{p^2+m^2}\right)^2 A(p^2) \right) e^{ip\cdot(x-y)} \;,
\eq
with $A$ given in (\ref{Aexpl}).

In the canonical approach we found for $\langle\f(x)\rangle$ and $\langle\f(x)\f(y)\rangle_\mathrm{c}$:
\bqa
\langle\f(x)\rangle &=& +v \nonumber\\
\langle\f(x)\f(y)\rangle_\mathrm{c} &=& {1\over(2\pi)^d} \int d^dp \left( \frac{\hbar}{p^2+m^2} + \left(\frac{\hbar}{p^2+m^2}\right)^2 A(p^2) \right) e^{ip\cdot(x-y)}
\eqa
(Remember that we chose the \emph{positive} minimum in the chapter \ref{chapN1sigmamodelcan}.)

Clearly the tadpole and the $\f$-propagator (it is questionable whether one can still call this a propagator) are \emph{different} than in the canonical approach.\index{difference canonical and path-integral approach}

\section{The Green's Functions Near Another Particle}

In the previous section we saw that the Green's functions one finds in the path-integral approach are very different from the Green's functions that one finds in the canonical approach. Now we can ask: What happens to the field $\f$ near another particle that it couples to like $\f(x) \psi^2(x)$ (like the Higgs)? Near a $\psi$-particle we have $\psi^2(x)\neq0$. This means this $\psi$-particle acts like a source-term. So to compute the Green's functions of the $\f$-field we can proceed as in the previous section, not setting the source to zero now, but setting it to $\psi^2$. It should have become clear that when $\psi^2\neq0$ immediately one of the minima is favored non-perturbatively. Then the Green's functions become what they are in the canonical approach.

This means that in this path-integral approach the $\f$-field acts the same as in the canonical approach near other particles, however it acts completely different far away from other particles. Far away from other matter there is \emph{no} SSB, and also the propagator is very different. Near particles there \emph{is} SSB, and the $\f$-field gives a mass to the $\psi$-particles. It is a very interesting, but difficult, question what this exactly means for the physics involved in the $\f$-field sector. Is this perhaps also a good mechanism to give masses to other particles? In this thesis we shall not go into this question further.\index{Green's functions!near other particle}


\chapter{The $N=1$ LSM: Fixing the Paths}\label{chapN1sigmamodelfix}

In this chapter we will again discuss the Euclidean $N=1$ linear sigma model from the path-integral viewpoint\index{$N=1$ linear sigma model!path-integral approach}, however now we shall introduce an extra constraint in the path integral. We will keep the paths fixed at a certain time $-T$ at some specific value $\bar{\f}$ over all of space. In this way we hope to construct a path-integral model which has the same physics (i.e.\ Green's functions) as the canonical approach.

The idea of fixing the paths is very much like what Fukuda et al.\ \cite{Fukuda}, O'Raifeartaigh et al.\ \cite{Wipf} and Ringwald et al.\ \cite{Ringwald} do in their papers. However they fix some space-time average of the field to some value, whereas we fix the field itself. The latter seems the more natural thing to do, when trying to induce SSB. Also, in these articles nothing has been said about the renormalizability of these models.

In the first section below we will show how to deal with a constraint in the path integral in case of a free field theory. After that we will proceed with the $N=1$ linear sigma model. We will show that this model, together with the path fixing constraint, is renormalizable at 1-loop order. We will calculate the alternative effective potential at lowest order and some Green's functions, and find that indeed we recover the physics as found also in the canonical approach.

\section{The Free Theory with Fixed Paths}

To get some feeling for what it means to have the paths in the path integral fixed we consider a \ind{free field theory}. The action of a Euclidean free field theory, including source term, is:
\bq \label{newaction}
S = \int d^dx \left( \h\left(\nabla\f(x)\right)^2 + \h\mu\f^2(x) - J(x)\f(x) \right) \;.
\eq

Now, in the path integral, we are going to keep all paths fixed at $\bar{\f}$ at time $-T$ and over all of space:\index{constraint}
\bq \label{constr}
\f(\vec{x},-T) = \bar{\f} \quad \forall \; \vec{x} \;.
\eq
By $\vec{x}$ we mean the $d-1$-vector containing all space coordinates. The generating functional $Z(J)$ is given by:
\bq
Z(J) = \underset{\f(\vec{x},-T) = \bar{\f}}{\int} \mathcal{D}\f \; e^{-{1\over\hbar} S} \;.
\eq

To find the effective action and the Green's functions we need to know the $\f$-tadpole $\langle\f(x)\rangle(J)$. To find this we can expand the action around the classical solution. To compute the classical solution we must find the minimum of the action, taking into account that we only accept solutions satisfying the constraint (\ref{constr}). This constraint can be built in by using a \ind{Lagrange multiplier} field $\lambda(\vec{x})$. Then the problem reduces to minimizing 
\bq
S + \int d^{d-1}x \; \lambda(\vec{x}) \left(\f(\vec{x},-T)-\bar{\f}\right)
\eq
with respect to $\f(x)$ and $\lambda(\vec{x})$. This means the classical solution $\f_c$ satisfies
\bq \label{eqclasssol}
-\nabla^2\f_c(\vec{x},t) + \mu\f_c(\vec{x},t) - J(\vec{x},t) + \lambda(\vec{x})\delta(t+T) = 0 \;,
\eq
and (\ref{constr}).

By passing to Fourier fields:
\bqa
\f_c(\vec{x},t) &=& {1\over(2\pi)^d} \int d^dk \; \tilde{\f}_c(k) \; e^{ik\cdot x} \nonumber\\
J(\vec{x},t) &=& {1\over(2\pi)^d} \int d^dk \; \tilde{J}(k) \; e^{ik\cdot x} \nonumber\\
\lambda(\vec{x}) &=& {1\over(2\pi)^{d-1}} \int d^{d-1}k \; \tilde{\lambda}(\vec{k}) \; e^{i\vec{k}\cdot\vec{x}}
\eqa
equation (\ref{eqclasssol}) can easily be solved:
\bq
\f_c(\vec{x},t) = {1\over(2\pi)^d} \int d^dk \; e^{ik\cdot x} \frac{\tilde{J}(k)-\tilde{\lambda}(\vec{k})e^{ik_tT}}{k^2+\mu} \;.
\eq
The Lagrange multiplier can be fixed with the constraint (\ref{constr})
\bq
\lambda(\vec{k}) = 2\sqrt{\vec{k}^2+\mu}\left( -(2\pi)^{d-1}\bar{\f} \; \delta^{d-1}(\vec{k}) + {1\over2\pi}\int dk_t \frac{\tilde{J}(k)e^{-ik_tT}}{k^2+\mu} \right) \;.
\eq
Substituting this in the classical solution we found gives:
\bqa
\f_c(\vec{x},t) &=& \bar{\f} \; e^{-\sqrt{\mu}|t+T|} + {1\over(2\pi)^d} \int d^dk \; \tilde{J}(k) \frac{e^{ik\cdot x}}{k^2+\mu} + \nonumber\\
& & -{1\over(2\pi)^d} \int d^dk \; \tilde{J}(k) \; e^{i\vec{k}\cdot\vec{x}} \; e^{-\sqrt{\vec{k}^2+\mu}|t+T|} \; e^{-ik_tT} \frac{1}{k^2+\mu} \label{classsol}
\eqa

Now the new action (\ref{newaction}) can be expanded around this classical solution,
\bq
\f(x) = \f_c(x) + \eta(x) \;,
\eq
to obtain:
\bq
S = \int d^dx \left( \h\left(\nabla\f_c(x)\right)^2 + \h\mu\f_c^2(x) - J(x)\f_c(x) \right) + \int d^dx \left( \h\left(\nabla\eta(x)\right)^2 + \h\mu\eta^2(x) \right) \;.
\eq

In this way the $\f$-tadpole $\langle \f(x) \rangle(J)$ can easily be calculated:
\bq \label{phiJ}
\langle \f(\vec{x},t) \rangle(J) = \f_c(\vec{x},t) \;.
\eq
This simple relationship is of course caused by the fact that the path integral just gives an overall constant, not depending on the source $J$. This is because we are working with a \emph{free} theory here.

\subsection{The Green's Functions}

Now the $\f$-tadpole and propagator can be calculated. To obtain the tadpole just set $J$ to zero in (\ref{phiJ}). To obtain the propagator take a functional derivative with respect to $J(y)$ and then put $J$ to zero in (\ref{phiJ}). Taking more derivatives with respect to $J$ in (\ref{phiJ}) gives zero, so higher connected Green's functions are zero, as expected in a free theory. We obtain the following results:
\bqa
\langle \f(x) \rangle &=& \bar{\f} \; e^{-\sqrt{\mu}|t+T|} \nonumber\\
\langle \f(x) \f(y) \rangle_\mathrm{c} &=& {\hbar\over(2\pi)^d} \int d^dk \; \frac{e^{ik\cdot(x-y)}}{k^2+\mu} + \nonumber\\
& & -{\hbar\over(2\pi)^{d-1}} \int d^{d-1}k \; e^{i\vec{k}\cdot(\vec{x}-\vec{y})} \; \frac{{1\over2\pi} \int dk_t \; \frac{e^{ik_t(t_x+T)}}{\vec{k}^2+k_t^2+\mu} \cdot {1\over2\pi} \int dl_t \; \frac{e^{il_t(t_y+T)}}{\vec{k}^2+l_t^2+\mu}}{{1\over2\pi} \int dl_t \; \frac{1}{\vec{k}^2+l_t^2+\mu}} \nonumber\\ \label{freefixedGreensfuncts}
\eqa

Indeed we have, as expected:
\bqa
\langle \f(\vec{x},-T) \rangle &=& \bar{\f} \nonumber\\
\langle \f(\vec{x},-T) \f(y) \rangle_\mathrm{c} &=& 0 \nonumber\\
\langle \f(\vec{x},-T) \f(y) \rangle &=& \bar{\f} \langle \f(y) \rangle \label{GrfunctsT}
\eqa

\subsection{The Effective Action}

To obtain the effective action we have to invert the relation (\ref{classsol}). This can be done by letting the operator $-\nabla^2+\mu$ work on both sides of (\ref{classsol}). One gets:
\bq \label{inveq1}
-\nabla^2\f(x) + \mu\f(x) = \left( 2\sqrt{\mu} \bar{\f} - {1\over(2\pi)^d} \int d^dk \; 2\sqrt{\vec{k}^2+\mu} \; \frac{e^{i\vec{k}\cdot\vec{x}-ik_tT} \tilde{J}(k)}{k^2+\mu} \right) \delta(t+T) + J(x) 
\eq
From this we see that $J$ has the form:
\bq \label{J}
J(x) = -\nabla^2\f(x) + \mu\f(x) + A(\vec{x})\delta(t+T) \;,
\eq
where $A(\vec{x})$ is a some functional of the field $\f(x)$. Now the $A$ has to be fixed by inserting this $J$ in (\ref{inveq1}). However, when doing this, one finds that $A$ drops out of the expression. Instead, after some algebra, one gets a condition for $\f$:
\bq
\f(\vec{x},-T) = \bar{\f} \quad \forall \; \vec{x} \;.
\eq

This means we find the inverse $J$ to be (\ref{J}), but this inverse can only be found when the field $\f$ satisfies the constraint (\ref{constr}). This is expected, because the right hand side of (\ref{classsol}) only gives a result that obeys the constraint, if someone would come up with a $\f$ that does not satisfy (\ref{constr}), there would simply be no solution $J$ that gives this $\f$.

Also $A$ remains undetermined, simply because adding a term like $A(\vec{x})\delta(t+T)$ to the source does \emph{not} change the Green's functions (i.e.\ the physics) because the field is \emph{fixed} at $t=-T$.

Finally, by integrating (\ref{J}) with respect to the field $\f$, one can find the effective action now. Clearly this effective action is \emph{not} unique. Here we find ourselves exactly in the loop-hole situation described in the first section of chapter \ref{chapeffact}. It can easily be seen that
\bq
\int d^dx \int d^dy \; \rho(x) \rho(y) \; \langle \f(x) \f(y) \rangle_\mathrm{c}(J)
\eq
is zero for the case
\bq
\rho(x) = M(x) \delta(t+T) \;,
\eq
simply because of (\ref{GrfunctsT}). This means $M(x)\delta(t+T)$ is an eigenfunction of the connected propagator with eigenvalue zero. So the connected propagator has \emph{no} unique inverse.

So clearly the effective action cannot be defined uniquely. We saw above however that there are several functionals $J$ which are a solution to (\ref{inveq1}). So in that sense several effective actions \emph{can} be defined in this case. However these are \emph{not} necessarily convex, because the argument for the convexity in chapter \ref{chapeffact} assumes that there exists a unique inverse. It is easy to construct an $A(\vec{x})$ which gives a \emph{non-convex} effective action.

Also note that, if we choose one of the possible effective actions, it is not possible to define an effective potential. Setting the field to a constant, which is what one would normally do when finding the effective potential from the effective action, is not possible in this case because a constant field does \emph{not} satisfy the constraint (\ref{constr}) in general (except $\f=\bar{\f}$). This would bring us outside the domain where the effective action is defined.

What one \emph{can} do, is define an \ind{alternative effective potential} from (\ref{J}). If we consider (\ref{J}) and the field $\f$ for large $t$, i.e.\ $t\gg-T$, then it is allowed to put the field $\f$ to a constant. If we then integrate (\ref{J}) with respect to $\f$ we have constructed an alternative effective potential. In this free field theory this alternative effective potential will be the same as the effective potential of a free field theory without the path-fixing constraint.

\subsection{Conclusions}

In the case of a model where we fix the paths in the path integral at some value $\bar{\f}$ at some time $-T$ over all of space we have the following conclusions:
\begin{itemize}
\item The effective action is defined on the domain of fields that satisfy the constraint, but this effective action is \emph{not} unique and in general \emph{not} convex.
\item The effective potential can \emph{not} be defined in the ordinary way.
\item An alternative effective potential can be defined as the anti-derivative of $J(x)$ for times $t\gg-T$.
\end{itemize}
This last definition of an effective potential we shall use whenever we are dealing with a model with fixed paths.

\newpage

\section{The $N=1$ Linear Sigma Model}

Now we proceed with the Euclidean $N=1$ linear sigma model. The action of this model, including source term, is
\bq \label{action}
S = \int d^dx \left( \h(\nabla\f(x))^2 - \h\mu\f^2(x) + {\lambda\over24}\f^4(x) - J(x)\f(x) \right) \;.
\eq
Again we take all paths in the path integral \emph{fixed} at $\bar{\f}$ at time $t=-T$ for all space-points $\vec{x}$:
\bq \label{cond}
\f(\vec{x},-T) = \bar{\f} \quad \forall \; \vec{x} \;.
\eq

We take the source $J$ in (\ref{action}) to be a constant, first of all for practical purposes, our calculations are simply too difficult when this source is also space- and time-dependent. Secondly at the end of the day we will only be interested in the alternative effective \emph{potential}, for which it is enough to consider only a constant source $J$.

In terms of renormalized quantities the action (\ref{action}) becomes:
\bqa
S &=& \int d^dx \bigg( \h(\nabla\f(x))^2 - \h\mu\f^2(x) + {\lambda\over24}\f^4(x) - J\f(x) + \nonumber\\
& & \phantom{\int d^dx \bigg(} \h\delta_Z(\nabla\f(x))^2 - \h\delta_{\mu}\f^2(x) + {\delta_{\lambda}\over24}\f^4(x) \bigg) \;. \label{renactionfix}
\eqa

First of all the minimum, or in this case minima, of the classical action (i.e.\ the first line of (\ref{renactionfix})) have to be found. Because of the boundary condition (\ref{cond}) this is far from trivial. Then the action has to be expanded around one of these minima (later we will then sum the contributions from all minima). The action will have three parts: the classical action, the quantum fluctuations and the counter terms. Because the classical solutions will be quite complicated, also calculating the classical action will not be as easy as it sounds. Also the path integral of the quantum fluctuations has to be calculated. In our treatment we shall only take the saddle-point approximation around each minimum, which means only Gaussian fluctuations are kept, all interaction terms are discarded. Even in this approximation it is very difficult to compute the path integral, as we shall see. We will only look at the divergent parts, to see whether this theory is renormalizable up to 1-loop. We will find that this is indeed the case. Then, knowing that everything is finite we can calculate the alternative effective potential at lowest order. Also we can compute the Green's functions and compare with the canonical approach.

To make all these remarks more concrete we work out mathematically what we have to do. First we have to find the minima of the classical action, i.e.\ the first line of (\ref{renactionfix}), under the condition (\ref{cond}). To implement this condition we add to the action a Lagrange multiplier term and minimize this object with respect to the field $\f$ \emph{and} the Lagrange multiplier $\lambda'$. The action plus the \ind{Lagrange multiplier} term is given by:
\bq
\int d^dx \left( \h(\nabla\f(x))^2 - \h\mu\f^2(x) + {\lambda\over24}\f^4(x) - J\f(x) \right) + \int d^{d-1}x \; \lambda'(\vec{x})\left( \f(\vec{x},-T)-\bar{\f} \right) \;.
\eq
Minimizing these two terms with respect to the field and the Lagrange multiplier we find:
\bqa
-\nabla^2 \f(\vec{x},t) - \mu\f(\vec{x},t) + {\lambda\over6}\f^3(\vec{x},t) - J + \lambda'(\vec{x})\delta(t+T) &=& 0 \nonumber\\
\f(\vec{x},-T) &=& \bar{\f}
\eqa
Now the last term in the left hand side of the first equation forces the classical solution $\f$ to be time dependent, however there is nothing that forces the solution to be dependent on the space-coordinates $\vec{x}$. The true minima of the action will have no $\vec{x}$-dependence, since this dependence will only increase the action. So we will limit ourselves to find only classical solutions which only depend on time. In this case the Lagrange multiplier necessarily has to be constant. So we should solve:
\bqa
-{d^2\over dt^2} \f(t) - \mu\f(t) + {\lambda\over6}\f^3(t) - J + \lambda'\delta(t+T) &=& 0 \nonumber\\
\f(-T) &=& \bar{\f} \label{system}
\eqa
Solving this system will be the subject of the first section below.

When we found solutions to this system the action (\ref{renactionfix}) can be expanded around such a solution. Calling the classical solution $\f_c(t)$ and the fluctuation around it $\eta(\vec{x},t)$ we find:
\bqa
S &=& \int d^dx \bigg( \h(\nabla\f_c+\nabla\eta)^2 - \h\mu(\f_c+\eta)^2 + {\lambda\over24}(\f_c+\eta)^4 - J(\f_c+\eta) + \nonumber\\
& & \phantom{\int d^dx \bigg(} \h\delta_Z(\nabla\f_c+\nabla\eta)^2 - \h\delta_{\mu}(\f_c+\eta)^2 + {\delta_{\lambda}\over24}(\f_c+\eta)^4 \nonumber\\
&=& \int d^dx \bigg( \h\left({d\over dt}\f_c\right)^2 - \h\mu\f_c^2 + {\lambda\over24}\f_c^4 - J\f_c + \nonumber\\
& & \phantom{\int d^dx \bigg(} \h(\nabla\eta)^2 + \left(-\h\mu+{\lambda\over4}\f_c^2\right)\eta^2 + {\lambda\over6}\f_c\eta^3 + {\lambda\over24}\eta^4 + \nonumber\\
& & \phantom{\int d^dx \bigg(} -\h\delta_Z\f_c{d^2\over dt^2}\f_c - \h\delta_{\mu}\f_c^2 + {\delta_{\lambda}\over24}\f_c^4 + \nonumber\\
& & \phantom{\int d^dx \bigg(} \left(-\delta_Z{d^2\over dt^2}\f_c-\delta_{\mu}\f_c+{\delta_{\lambda}\over6}\f_c^3\right)\eta + \h\delta_Z(\nabla\eta)^2 + \left(-\h\delta_{\mu}+{\delta_{\lambda}\over4}\f_c^2\right)\eta^2 + \nonumber\\
& & \phantom{\int d^dx \bigg(} {\delta_{\lambda}\over6}\f_c\eta^3 + {\delta_{\lambda}\over24}\eta^4 \bigg) \nonumber\\ \label{complaction}
\eqa
Now we will make the saddle-point approximation, so we will only keep terms up to order $\eta^2$, i.e.\ order $\hbar$. Discarding all interaction terms and recognizing that the first line in the expression above is just the classical action $S_c$ we find:
\bq
S = S_c + \int d^dx \left( \h(\nabla\eta)^2 + \left(-\h\mu+{\lambda\over4}\f_c^2\right)\eta^2 - \h\delta_Z\f_c{d^2\over dt^2}\f_c - \h\delta_{\mu}\f_c^2 + {\delta_{\lambda}\over24}\f_c^4 \right)
\eq

With this action the generating functional around \emph{one} minimum is given by:
\bqa
Z(J) &=& e^{-{1\over\hbar}S_c} \underset{\eta(\vec{x},-T)=0}{\int} \mathcal{D}\eta \; \exp\left(-{1\over\hbar}\int d^dx \left( \h(\nabla\eta)^2 + \left(-\h\mu+{\lambda\over4}\f_c^2\right)\eta^2 \right) \right) \cdot \nonumber\\
& &  \qquad \exp\left( -{1\over\hbar}\int d^dx \left( -\h\delta_Z\f_c{d^2\over dt^2}\f_c - \h\delta_{\mu}\f_c^2 + {\delta_{\lambda}\over24}\f_c^4 \right) \right) \label{Z}
\eqa

\subsection{The Classical Solutions}

\index{classical solutions}Now we solve the system (\ref{system}). Of course we need more boundary conditions than $\f(\vec{x},-T)=\bar{\f}$ to solve this differential system. What we will demand from our solutions is that, when $t\rightarrow\pm\infty$, $\f$ will converge to one of the two static minima $\f_\pm$ of the potential $-\h\mu\f^2+{\lambda\over24}\f^4-J\f$. This means:
\bq
\f(t) \rightarrow \f_\pm \;, \quad {d\over dt}\f(t) \rightarrow 0 \quad \textrm{for $t\rightarrow\pm\infty$} \;.
\eq
Of course we use these boundary conditions because we are looking for solutions giving a finite, minimal action. Below these conditions will appear to be sufficient to solve the differential system.

Before we start calculating note first that the differential system (\ref{system}) corresponds to a mechanical problem of a particle with unit mass in a potential $V(x)=\h\mu x^2-{\lambda\over24}x^4+Jx$ when $t\neq-T$. For $t\rightarrow-\infty$ the particle starts at one of the static minima $\f_\pm$, then it travels such that it is at $\bar{\f}$ at $t=-T$. Then the Lagrange multiplier term gives the particle just such a kick that it reaches one of the static minima again for $t\rightarrow+\infty$. In this way we have a nice intuitive picture that helps us to solve the differential system (\ref{system}).

We divide our time domain in two intervals, region 1 where $t<-T$ and region 2 where $t>-T$. In these regions the delta-function term is absent of course. First we consider region 1. From the first equation in (\ref{system}) we find by multiplying by $\dot{\f}\equiv{d\over dt}\f$ and integrating with respect to time:
\bq \label{eqmo}
-\h\dot{\f}^2-\h\mu\f^2+{\lambda\over24}\f^4-J\f = \alpha
\eq
For $t\rightarrow-\infty$ $\dot{\f}$ should go to zero and $\f$ should go to one of the two static minima $\f_\pm$. We denote the minimum $\f$ goes to for $t\rightarrow-\infty$ by $\f_1$. With this we can immediately fix the constant of integration $\alpha$:
\bq
\alpha = -\h\mu\f_1^2+{\lambda\over24}\f_1^4-J\f_1
\eq
So in region 1 we should find a solution to:
\bq
\dot{\f} = \pm\sqrt{-\mu(\f^2-\f_1^2)+{\lambda\over12}(\f^4-\f_1^4)-2J(\f-\f_1)}
\eq
By dividing by the square root on both sides and integrating over time again we find:
\bq
\int_0^{\f(t)} d\f \frac{\pm1}{\sqrt{-\mu(\f^2-\f_1^2)+{\lambda\over12}(\f^4-\f_1^4)-2J(\f-\f_1)}} = t+\beta
\eq
The constant $\beta$ can be fixed with the second equation in (\ref{system}). One finds:
\bq
\beta = T + \int_0^{\bar{\f}} d\f \frac{\pm1}{\sqrt{-\mu(\f^2-\f_1^2)+{\lambda\over12}(\f^4-\f_1^4)-2J(\f-\f_1)}}
\eq
And finally the solution in region 1 becomes:
\bq \label{sol}
\int_{\bar{\f}}^{\f(t)} d\f \frac{\pm1}{\sqrt{-\mu(\f^2-\f_1^2)+{\lambda\over12}(\f^4-\f_1^4)-2J(\f-\f_1)}} = t+T
\eq

Now one should worry a little about the roots of the argument of the square root. In the corresponding mechanical problem this argument gives (twice) the energy the particle has at time $-\infty$ minus the potential energy at position $\f$. This is just the kinetic energy of the particle at position $\f$. Clearly the regions in $\f$ where this kinetic energy becomes negative are forbidden. Also the roots of the argument can only be reached for $t\rightarrow\pm\infty$, as can easily be seen in (\ref{sol}). This means a solution $\f(t)$ always stays between two roots.

The solution (\ref{sol}) can be simplified by passing to a different variable:
\bq \label{deff}
f(t) \equiv {\f(t)-\f_1\over v}
\eq
If we also define the new dimensionless quantities
\bq
m_1 \equiv {\f_1\over v} \;, \quad \bar{f} \equiv {\bar{\f}-\f_1\over v} \;, \quad x \equiv {J\over\mu v}
\eq
then the integral equation (\ref{sol}) can be written as
\bq \label{reg1}
\int_{\bar{f}}^{f(t)} df \frac{\pm1}{\sqrt{(6m_1^2-2)f^2+4m_1f^3+f^4}} = \sqrt{{\mu\over2}}(t+T) \;.
\eq
The linear term in $f$ in the square root has dropped out because $m_1$ satisfies $m_1^3-m_1-x=0$.

Of course the same steps can be done in region 2, where $t>-T$. There we obtain:
\bq \label{reg2}
\int_{\bar{g}}^{g(t)} dg \frac{\pm1}{\sqrt{(6m_2^2-2)g^2+4m_2g^3+g^4}} = \sqrt{{\mu\over2}}(t+T) \;,
\eq
with
\bq \label{defg}
g(t) \equiv {\f(t)-\f_2\over v} \;, \quad m_2 \equiv {\f_2\over v} \;, \quad \bar{g} \equiv {\bar{\f}-\f_2\over v} \;.
\eq

Now we can explicitly solve (\ref{reg1}) and (\ref{reg2}). We will demonstrate the procedure for region 1, of course things go completely similar in region 2. First remember that $f$ can never pass a root of the argument in the square root. This means that for one solution, $f$ always stays between two roots. For this reason we can write (\ref{reg1}) as
\bq \label{reg12}
\int_{\bar{f}}^{f(t)} df \frac{\pm1}{f\sqrt{(6m_1^2-2)+4m_1f+f^2}} = \sqrt{{\mu\over2}}(t+T) \;.
\eq
Of course the $\pm$ in (\ref{reg12}) can be a different $\pm$ than in (\ref{reg1}). Next we switch to a new variable $z$:
\bq \label{subst1}
z = f-\sqrt{(6m_1^2-2)+4m_1f+f^2}
\eq
Note that for $J=0$, so $m_1=\pm1$, the argument of the square root combines to $(f+2m_1)^2$ and for $f>2m_1$ the $f$'s in (\ref{subst1}) cancel, such that the variable substitution becomes nonsense. This just means that later on we have to be a bit careful in setting $J=0$. After this variable substitution we do another one,
\bq
z = \sqrt{6m_1^2-2} \; {1\over\cos\theta} \;,
\eq
after which the integral in (\ref{reg12}) becomes
\bq
-{2\over\sqrt{6m_1^2-2}} \int d\theta {1\over\sin\theta} = -{1\over\sqrt{6m_1^2-2}} \left[ \ln{1-\cos\theta\over1+\cos\theta} \right] \;.
\eq
Writing this expression in terms of the $f$ variables and substituting the appropriate boundaries gives us
\bq \label{reg13}
\frac{f-\sqrt{6m_1^2-2+4m_1f+f^2}-\sqrt{6m_1^2-2}}{f-\sqrt{6m_1^2-2+4m_1f+f^2}+\sqrt{6m_1^2-2}} = \Omega_1
\eq
with
\bq
\Omega_1 = \frac{\bar{f}-\sqrt{6m_1^2-2+4m_1\bar{f}+\bar{f}^2}-\sqrt{6m_1^2-2}}{\bar{f}-\sqrt{6m_1^2-2+4m_1\bar{f}+\bar{f}^2}+\sqrt{6m_1^2-2}} \; \exp\left(\pm\sqrt{\mu(3m_1^2-1)}(t+T)\right) \;.
\eq

Now we know that for $t\rightarrow-\infty$ $\f$ has to go to $\f_1$, so $f$ has to go to zero. From (\ref{reg13}) we see that $\Omega_1$ has to go to $\pm\infty$ in this limit. So we have to choose the minus-sign in the exponential.

Finally solving (\ref{reg13}) for $f$ gives\index{classical solutions}
\bq \label{fsol}
f(t) = \frac{4(6m_1^2-2)\Omega_1}{4m_1(1-\Omega_1)^2+2\sqrt{6m_1^2-2}(1-\Omega_1^2)} \;,
\eq
with
\bq
\Omega_1 = \frac{\bar{f}-\sqrt{6m_1^2-2+4m_1\bar{f}+\bar{f}^2}-\sqrt{6m_1^2-2}}{\bar{f}-\sqrt{6m_1^2-2+4m_1\bar{f}+\bar{f}^2}+\sqrt{6m_1^2-2}} \; \exp\left(-\sqrt{\mu(3m_1^2-1)}(t+T)\right) \;.
\eq

In region 2 we find likewise
\bq
g(t) = \frac{4(6m_2^2-2)\Omega_2}{4m_2(1-\Omega_2)^2+2\sqrt{6m_2^2-2}(1-\Omega_2^2)} \;,
\eq
with
\bq
\Omega_2 = \frac{\bar{g}-\sqrt{6m_2^2-2+4m_2\bar{g}+\bar{g}^2}-\sqrt{6m_2^2-2}}{\bar{g}-\sqrt{6m_2^2-2+4m_2\bar{g}+\bar{g}^2}+\sqrt{6m_2^2-2}} \; \exp\left(\sqrt{\mu(3m_2^2-1)}(t+T)\right) \;.
\eq
Note that here we had to choose the plus sign in the exponential in $\Omega_2$.

\begin{figure}[h]
\begin{center}
\epsfig{file=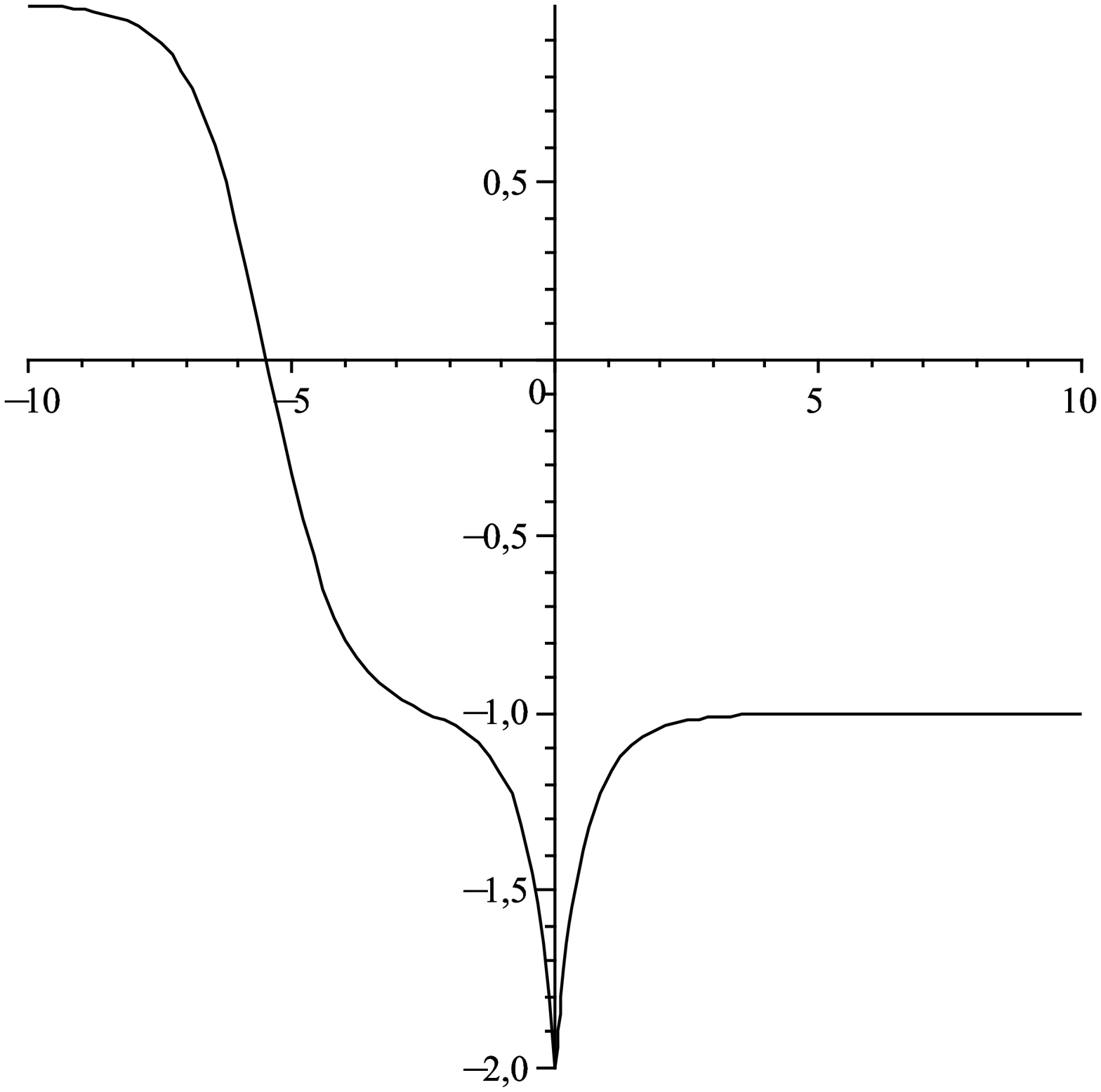,width=7cm}
\end{center}
\vspace{-0.5cm}
\caption{$\f(t)$ as a function time $t$ for $\f_1=\f_+$, $\f_2=\f_-$, $\bar{\f}=-2$, $T=0$, $\mu=1$, $v=1$ and $\alpha=0.001$.}
\label{solfix1}
\end{figure}

\begin{figure}[h]
\begin{center}
\epsfig{file=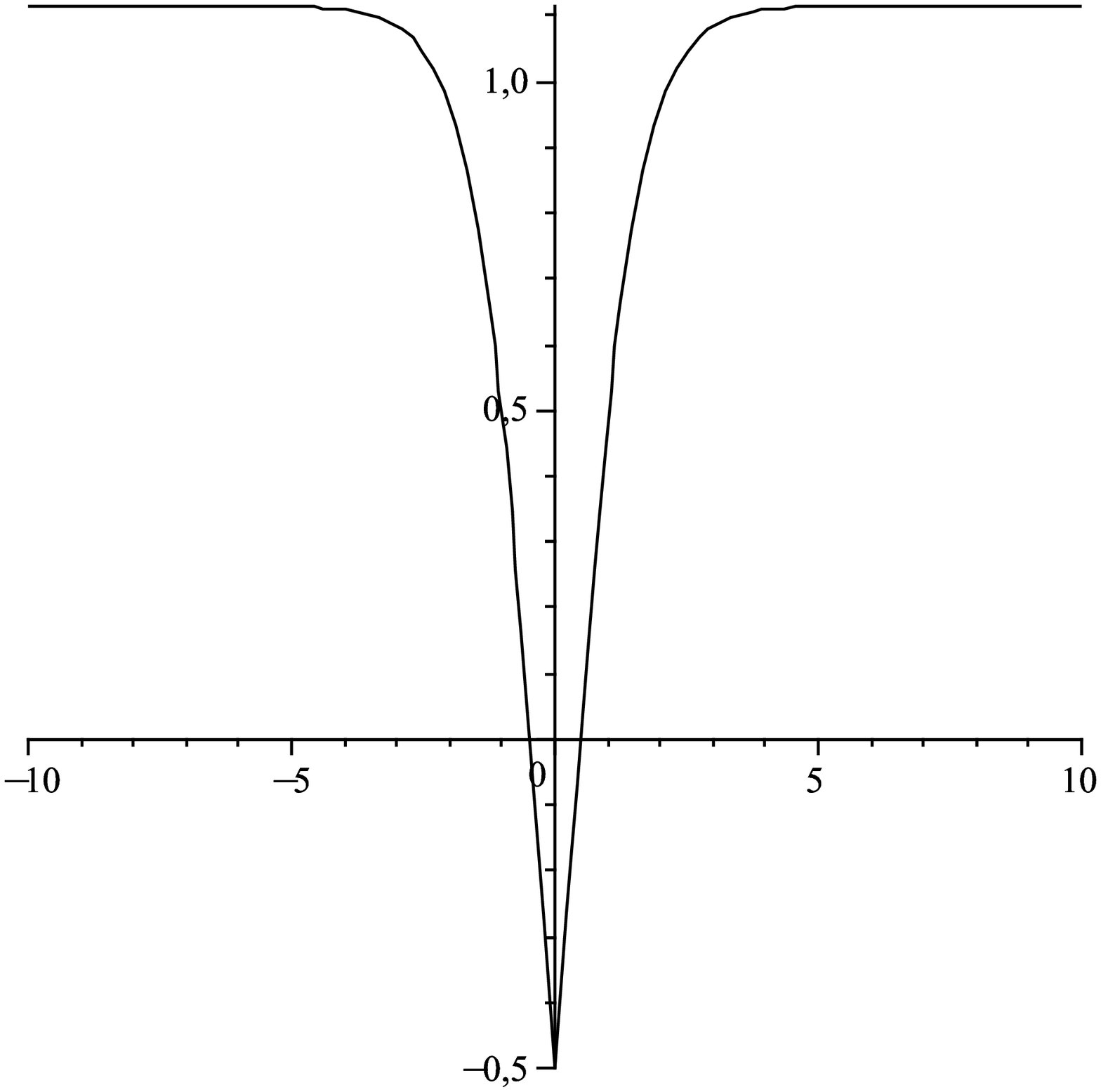,width=7cm}
\end{center}
\vspace{-0.5cm}
\caption{$\f(t)$ as a function time $t$ for $\f_1=\f_+$, $\f_2=\f_+$, $\bar{\f}=-0.5$, $T=0$, $\mu=1$, $v=1$ and $\alpha={\pi\over12}$.}
\label{solfix2}
\end{figure}

In figures \ref{solfix1} and \ref{solfix2} one can see the solution plotted for two nice cases.

Notice that when $m_1$ and $m_2$ are not equal we obtain an instanton-like solution, which takes the field from one static minimum to another. For dimensions greater than one we know that \ind{instantons} do \emph{not} contribute in the path integral (see Derrick \cite{Derrick}). For this reason we shall only consider classical solutions for which $m_1=m_2$. Then there are in general \emph{two} solutions, both symmetric around $t=-T$. However for some values of $J$ and $\bar{\f}$ it can happen that one of these two solutions does \emph{not} exist, which can easily be seen from the corresponding mechanical problem.

\subsection{The Classical Action}

The \ind{classical action} can be written as:
\bqa
S_c &=& \int d^dx \left( \h\left({d\over dt}\f_c\right)^2 - \h\mu\f_c^2 + {\lambda\over24}\f_c^4 - J\f_c \right) \nonumber\\
&=& \int d^{d-1}x \Bigg( \int_{-L/2}^{-T} dt \Bigg( \h\left({d\over dt}\f_c\right)^2 - \h\mu(\f_c^2-\f_1^2) + \nonumber\\
& & \phantom{\int d^{d-1}x \Bigg( \int_{-L/2}^{-T} dt \Bigg(} {\lambda\over24}(\f_c^4-\f_1^4) - J(\f_c-\f_1) \Bigg) + \nonumber\\
& & \phantom{\int d^{d-1}x \Bigg(} \int_{-T}^{L/2} dt \Bigg( \h\left({d\over dt}\f_c\right)^2 - \h\mu(\f_c^2-\f_2^2) + \nonumber\\
& & \phantom{\int d^{d-1}x \Bigg( \int_{-L/2}^{-T} dt \Bigg(} {\lambda\over24}(\f_c^4-\f_2^4) - J(\f_c-\f_2) \Bigg) \Bigg) + \nonumber\\
& & L^{d-1}\left(-\h\mu\f_1^2+{\lambda\over24}\f_1^4-J\f_1\right)\left({L\over2}-T\right) + \nonumber\\
& & L^{d-1}\left(-\h\mu\f_2^2+{\lambda\over24}\f_2^4-J\f_2\right)\left({L\over2}+T\right) \nonumber\\
&=& \int d^{d-1}x \left( \int_{-L/2}^{-T} dt \left({d\over dt}\f_c\right)^2 + \int_{-T}^{L/2} dt \left({d\over dt}\f_c\right)^2 \right) + \nonumber\\
& & L^{d-1}\left(-\h\mu\f_1^2+{\lambda\over24}\f_1^4-J\f_1\right)\left({L\over2}-T\right) + \nonumber\\
& & L^{d-1}\left(-\h\mu\f_2^2+{\lambda\over24}\f_2^4-J\f_2\right)\left({L\over2}+T\right) \nonumber\\
\eqa
Here we have used (\ref{eqmo}) in the last step.

Now for the solutions we are interested in, with $m_1=m_2$, this simplifies to:
\bq
S_c = 2v^2L^{d-1} \int_{-L/2}^{-T} dt \left({d\over dt}f\right)^2 + L^d\left(-\h\mu\f_1^2+{\lambda\over24}\f_1^4-J\f_1\right)
\eq
The last term is the classical action of the $N=1$ linear sigma model where one does not fix the paths. This term is
\bq
L^d\mu v^2 \left({2\over3}\sin_{\pm}^2-{4\over3}\sin_{\pm}^4\right) \;,
\eq
with
\bq
\sin_{\pm} = \sin\left(\alpha\pm{\pi\over3}\right) \;, \quad J = {2\mu v\over3\sqrt{3}} \sin(3\alpha) \;.
\eq

The first term is a little harder to calculate. Just inserting the solution for $f$ (\ref{fsol}) gives a very hard integral. However by using a clever trick things become doable. Remember
\bq
\dot{\f} = \pm\sqrt{-\mu(\f^2-\f_1^2)+{\lambda\over12}(\f^4-\f_1^4)-2J(\f-\f_1)} \;,
\eq
for $t<-T$, where the plus sign has to be taken when $\bar{\f}>\f_1$ and the minus sign when $\bar{\f}<\f_1$. This means
\bq
df = \pm\sqrt{{\mu\over2}}\sqrt{(6m_1^2-2)f^2+4m_1f^3+f^4} \; dt \;.
\eq
With this variable substitution the integral can be written as:
\bqa
\int_{-\infty}^{-T} dt \left({d\over dt}f\right)^2 &=& \pm \int_0^{\bar{f}} df \sqrt{{\mu\over2}}\sqrt{(6m_1^2-2)f^2+4m_1f^3+f^4} \nonumber\\
&=& \sqrt{{\mu\over2}} \int_0^{\bar{f}} df \; f \sqrt{6m_1^2-2+4m_1f+f^2}
\eqa
After some hard work this gives:
\bqa
& & \int_{-\infty}^{-T} dt \left({d\over dt}f\right)^2 = \nonumber\\
& & \hspace{50pt} {1\over6}\sqrt{2\mu} \left( \left(\bar{f}^2+m_1\bar{f}-2\right) \sqrt{6m_1^2-2+4m_1\bar{f}+\bar{f}^2} + 2\sqrt{6m_1^2-2} \right) + \nonumber\\
& & \hspace{50pt} -\sqrt{2\mu} \; m_1\left(m_1^2-1\right) \ln\left(\frac{2m_1+\bar{f}+\sqrt{6m_1^2-2+4m_1\bar{f}+\bar{f}^2}}{2m_1+\sqrt{6m_1^2-2}}\right)
\eqa

So finally we obtain for the \ind{classical action} around the $\pm$-minimum (realizing that $m_1$ can be $m_\pm\equiv\f_\pm/v$ and $\bar{f}$ can be $\bar{f}_\pm\equiv(\bar{\f}-\f_\pm)/v$):
\bqa
S_c^\pm &=& L^d\mu v^2 \left({2\over3}\sin_{\pm}^2-{4\over3}\sin_{\pm}^4\right) + \nonumber\\
& & {1\over3}v^2L^{d-1} \sqrt{2\mu} \left( \left(\bar{f}_\pm^2+m_\pm\bar{f}_\pm-2\right) \sqrt{6m_\pm^2-2+4m_\pm\bar{f}_\pm+\bar{f}_\pm^2} + 2\sqrt{6m_\pm^2-2} \right) + \nonumber\\
& & -2v^2L^{d-1} \sqrt{2\mu} \; m_\pm\left(m_\pm^2-1\right) \ln\left(\frac{2m_\pm+\bar{f}_\pm+\sqrt{6m_\pm^2-2+4m_\pm\bar{f}_\pm+\bar{f}_\pm^2}}{2m_\pm+\sqrt{6m_\pm^2-2}}\right) \nonumber\\
\label{classactionfixed}
\eqa
For $\bar{f}_\pm=0$ the last two line vanish, as expected, because then the classical solution just becomes a constant ($\f_\pm$).

\subsection{The Path Integral}

Now we wish to calculate the path integral
\bq \label{pathintegralA}
A \equiv \underset{\eta(\vec{x},-T)=0}{\int} \mathcal{D}\eta \; \exp\left(-{1\over\hbar}\int d^dx \left( \h(\nabla\eta(x))^2 + \h M(t)\eta^2(x) \right) \right) \;,
\eq
with
\bq
M(t) = -\mu+{\lambda\over2}\f_c(t)^2 \;.
\eq
Here $\f_c$ can be any of the two classical solutions we found (starting and ending at $\f_{\pm} = {2v\over\sqrt{3}}\sin_{\pm}$).

To calculate this path integral we first perform a \ind{Fourier transform} in space. To do this neatly we take all coordinates in the domain $[-L/2,L/2]$. Then $\eta$ can be written as
\bq
\eta(\vec{x},t) = \sum_{k_1,\ldots,k_{d-1}=-\infty}^{\infty} \tilde{\eta}_{k_1,\ldots,k_{d-1}}(t) \; e^{-{2\pi i\over L}(k_1x_1+\ldots+k_{d-1}x_{d-1})}
\eq
with the condition
\bq
\tilde{\eta}_{k_1,\ldots,k_{d-1}} = \tilde{\eta}^*_{-k_1,\ldots,-k_{d-1}}
\eq
to make $\eta(\vec{x},t)$ real. From this condition one can see which variables are independent. We shall choose the following set of independent variables:
\bqa
& & \textrm{Re} \; \tilde{\eta}_{k_1,\ldots,k_{d-1}} \;, \quad k_1+\ldots+k_{d-1} \geq 0 \nonumber\\
& & \textrm{Im} \; \tilde{\eta}_{k_1,\ldots,k_{d-1}} \;, \quad k_1+\ldots+k_{d-1} > 0
\eqa
Now we can write:
{\allowdisplaybreaks\bqa
& & \int_{-L/2}^{L/2} d^dx \left( \h(\nabla\eta)^2 + \h M(t)\eta^2 \right) \nonumber\\
& & \qquad = \h L^{d-1} \int_{-L/2}^{L/2} dt \sum_{k_1,\ldots,k_{d-1}=-\infty}^{\infty} \tilde{\eta}_{k_1,\ldots,k_{d-1}}(t) \cdot \nonumber\\
& & \qquad \qquad \left( -{d^2\over dt^2} + \left({2\pi k_1\over L}\right)^2 + \ldots + \left({2\pi k_{d-1}\over L}\right)^2 + M(t) \right) \tilde{\eta}^*_{k_1,\ldots,k_{d-1}}(t) \nonumber\\
& & \qquad = L^{d-1} \int_{-L/2}^{L/2} dt \underset{k_1+\ldots+k_{d-1}>0}{\sum_{k_1,\ldots,k_{d-1}=-\infty}^{\infty}} \Bigg[ \textrm{Re} \; \tilde{\eta}_{k_1,\ldots,k_{d-1}}(t) \cdot \nonumber\\
& & \qquad \qquad \left( -{d^2\over dt^2} + \left({2\pi k_1\over L}\right)^2 + \ldots + \left({2\pi k_{d-1}\over L}\right)^2 + M(t) \right) \textrm{Re} \; \tilde{\eta}_{k_1,\ldots,k_{d-1}}(t) + \nonumber\\[5pt]
& & \qquad \qquad \textrm{Im} \; \tilde{\eta}_{k_1,\ldots,k_{d-1}}(t) \cdot \nonumber\\[5pt]
& & \qquad \qquad \left( -{d^2\over dt^2} + \left({2\pi k_1\over L}\right)^2 + \ldots + \left({2\pi k_{d-1}\over L}\right)^2 + M(t) \right) \textrm{Im} \; \tilde{\eta}_{k_1,\ldots,k_{d-1}}(t) \Bigg] + \nonumber\\
& & \qquad \phantom{=} \h L^{d-1} \int_{-L/2}^{L/2} dt \underset{k_1+\ldots+k_{d-1}=0}{\sum_{k_1,\ldots,k_{d-1}=-\infty}^{\infty}} \textrm{Re} \; \tilde{\eta}_{k_1,\ldots,k_{d-1}}(t) \cdot \nonumber\\
& & \qquad \qquad \left( -{d^2\over dt^2} + \left({2\pi k_1\over L}\right)^2 + \ldots + \left({2\pi k_{d-1}\over L}\right)^2 + M(t) \right) \textrm{Re} \; \tilde{\eta}_{k_1,\ldots,k_{d-1}}(t) \nonumber\\
& & \qquad = L^{d-1} \int_{-L/2}^{L/2} dt \sum_{k_1,\ldots,k_{d-1}=-\infty}^{\infty} f_{k_1,\ldots,k_{d-1}}(t) \cdot \nonumber\\
& & \qquad \qquad \left( -{d^2\over dt^2} + \left({2\pi k_1\over L}\right)^2 + \ldots + \left({2\pi k_{d-1}\over L}\right)^2 + M(t) \right) f_{k_1,\ldots,k_{d-1}}(t)
\eqa}
In the last step we defined $f$ as:
\bq
f_{k_1,\ldots,k_{d-1}}(t) = \left\{ \begin{array}{ll}
\textrm{Re} \; \tilde{\eta}_{k_1,\ldots,k_{d-1}}(t) & \textrm{for } k_1+\ldots+k_{d-1}>0 \\
{1\over\sqrt{2}} \textrm{Re} \; \tilde{\eta}_{k_1,\ldots,k_{d-1}}(t) & \textrm{for } k_1+\ldots+k_{d-1}=0 \\
\textrm{Im} \; \tilde{\eta}_{-k_1,\ldots,-k_{d-1}}(t) & \textrm{for } k_1+\ldots+k_{d-1}<0
\end{array} \right.
\eq
Finally the path integral $A$ becomes:
\bqa
A &\sim& \underset{f(-T)=0}{\int} \mathcal{D}f \; \exp\Bigg( -{1\over\hbar}L^{d-1} \int_{-L/2}^{L/2} dt \sum_{k_1,\ldots,k_{d-1}=-\infty}^{\infty} f_{k_1,\ldots,k_{d-1}}(t) \cdot \nonumber\\
& & \qquad \qquad \left( -{d^2\over dt^2} + \left({2\pi k_1\over L}\right)^2 + \ldots + \left({2\pi k_{d-1}\over L}\right)^2 + M(t) \right) f_{k_1,\ldots,k_{d-1}}(t) \Bigg) \nonumber\\
&=& \prod_{k_1,\ldots,k_{d-1}=-\infty}^{\infty} \underset{f(-T)=0}{\int} \mathcal{D}f(t) \; \exp\Bigg( -{1\over\hbar}L^{d-1} \int_{-L/2}^{L/2} dt \; f(t) \cdot \nonumber\\
& & \qquad \qquad \left( -{d^2\over dt^2} + \left({2\pi k_1\over L}\right)^2 + \ldots + \left({2\pi k_{d-1}\over L}\right)^2 + M(t) \right) f(t) \Bigg)
\eqa

Note that we have reduced the problem of calculating the $d$-dimensional path integral $A$ to the problem of calculating a 1-dimensional path integral.

Now focus on this 1-dimensional path integral, which we call $B$:
\bq
B \equiv \underset{f(-T)=0}{\int} \mathcal{D}f(t) \; \exp\left( -{1\over\hbar}L^{d-1} \int_{-L/2}^{L/2} dt \; f(t) \left( -{d^2\over dt^2} + \beta(t) \right) f(t) \right)
\eq
where $\beta(t)$, in our case, is:
\bq
\beta(t) = \left({2\pi k_1\over L}\right)^2 + \ldots + \left({2\pi k_{d-1}\over L}\right)^2 + M(t) \;.
\eq

We will try to calculate this $B$ with the following formula for \ind{Gaussian integrals}:
\bq \label{Gaussian}
\int dx_0 dx_1 \ldots dx_{N-1} \; \exp\left(-\h\vec{x}\cdot A\vec{x}\right) = (2\pi)^{N/2} \frac{1}{\sqrt{\det{A}}} \;,
\eq
where $A$ is an $N\times N$ real symmetric matrix.

If we make the time interval $[-L/2,L/2]$ discrete by defining
\bq
t = \Delta i - L/2 \; \quad i = 0,\ldots,N-1 \;,
\eq
and if we define the discrete index $j$ as the index belonging to the time $-T$:
\bq
-T = \Delta j -L/2 \;,
\eq
then $B$ can be written as:
\bq
B = \left({L^{d-1}\over{\pi\hbar\Delta}}\right)^{N/2} \int df_0 \ldots df_{j-1} df_{j+1} \ldots df_{N-1} \; \exp\left(-{L^{d-1}\over\hbar\Delta} \vec{f} \cdot E_{N-1} \vec{f}\right) \;,
\eq
where $\vec{f}$ is a vector of length $N-1$:
\bq
\vec{f} = \left(\begin{array}{c} f_0 \\ \vdots \\ f_{j-1} \\ f_{j+1} \\ \vdots \\ f_{N-1} \end{array}\right) \;,
\eq
and $E_{N-1}$ is the $N-1\times N-1$ matrix:
\bq
E_{N-1} = \scriptsize \left(\begin{array}{cccccccc} 
2+\Delta^2\beta_0 & -1 & & & \ldots & & & -1 \\[10pt]
-1 & 2+\Delta^2\beta_1 & -1 & & & & & \\[10pt]
& -1 & \ddots & -1 & & & \emptyset & \\[10pt]
& & -1 & 2+\Delta^2\beta_{j-1} & 0 & & & \\[10pt]
& & & 0 & 2+\Delta^2\beta_{j+1} & -1 & & \\[10pt]
& \emptyset & & & -1 & \ddots & -1 & \\[10pt]
& & & & & -1 & 2+\Delta^2\beta_{N-2} & -1 \\[10pt]
-1 & & & \ldots & & & -1 & 2+\Delta^2\beta_{N-1}
\end{array} \right) \normalsize \;.
\eq
Note that the $-1$ in the lower left and upper right corner mean that we have periodic boundary conditions in our path integral.

Now using (\ref{Gaussian}) we find for $B$:
\bq
B = \sqrt{{L^{d-1}\over\pi\hbar\Delta}} \frac{1}{\sqrt{\det{E_{N-1}}}} \;.
\eq
Now it is easy to see that:
\bq
\det{E_{N-1}} = \det{F_{N-1}} \;,
\eq
with
\bq
F_{N-1} = \scriptsize \left(\begin{array}{cccccccc} 
2+\Delta^2\beta_{j+1} & -1 & & & & & & \\[10pt]
-1 & 2+\Delta^2\beta_{j+2} & -1 & & & & & \\[10pt]
& -1 & \ddots & -1 & & & \emptyset & \\[10pt]
& & -1 & 2+\Delta^2\beta_{N-1} & -1 & & & \\[10pt]
& & & -1 & 2+\Delta^2\beta_0 & -1 & & \\[10pt]
& \emptyset & & & -1 & \ddots & -1 & \\[10pt]
& & & & & -1 & 2+\Delta^2\beta_{j-2} & -1 \\[10pt]
& & & & & & -1 & 2+\Delta^2\beta_{j-1}
\end{array} \right) \normalsize \;.
\eq

So now we are interested in $\Delta\det{F_{N-1}}$ in the limit of $\Delta\rightarrow 0$. This quantity can be found as follows. For $\det{F_{N-1}}$ one can easily derive the \ind{recursion relation}:
\bq \label{recur}
\det{F_i} = \left(2+\Delta^2\beta_{\m{i+j-N}{N}}\right) \det{F_{i-1}} - \det{F_{i-2}} \;
\eq
where $F_i$ is a $i\times i$ matrix with always $2+\Delta^2\beta_{j+1}$ in the upper left corner, and:
\bq
\det{F_0} = 1 \;, \quad \det{F_1} = 2+\Delta^2\beta_{j+1} \;.
\eq

Now define
\bq
\alpha(t) \equiv \Delta\det{F_i} \qquad (t=\Delta i - L/2) \;.
\eq
Then, in the continuum limit the recursion relation (\ref{recur}) becomes a differential equation:
\bqa
\frac{d^2\alpha(t)}{dt^2} - \beta(t-T+L/2) \; \alpha(t) &=& 0 \quad \textrm{for $-L/2<t<T$} \nonumber\\
\frac{d^2\alpha(t)}{dt^2} - \beta(t-T-L/2) \; \alpha(t) &=& 0 \quad \textrm{for $T<t<L/2$} \;, \label{diffsystalpha}
\eqa
with boundary conditions
\bq \label{boundcondalpha}
\alpha(-L/2) = 0 \;, \quad \frac{d\alpha}{dt}(-L/2) = 1 \;.
\eq

The 1-dimensional path integral $B$ is now (in the continuum) given by:
\bq
B \sim \frac{1}{\sqrt{\alpha(L/2)}} \;.
\eq

In essence we have now proven the theorem (7.40) in Das' book \cite{Das}. There he states that the determinant of an operator $\hat{O}$, with boundary conditions $\eta(-L/2)=\eta(L/2)=0$, is proportional to $\psi(L/2)$, where $\psi(t)$ is a solution to the differential equation
\bq
\hat{O} \psi(t) = 0
\eq
with boundary conditions $\psi(-L/2)=0, \; {d\psi\over dt}(-L/2) = 1$. In our case the time is just shifted because we do not have $\eta(-L/2)=\eta(L/2)=0$, but $\eta(-T)=\eta(-T+L)=0$.

Now the big question is: How do we find a solution to the differential system (\ref{diffsystalpha}) \& (\ref{boundcondalpha})?

\subsubsection{Trying to Solve (\ref{diffsystalpha})}

So the differential equation we wish to solve has the form
\bq
\frac{d^2\alpha(t)}{dt^2} - b(t) \; \alpha(t) = 0
\eq
with
\bq
b(t) \equiv \left\{\begin{array}{ll}
\beta(t-T+L/2) & \textrm{for $-L/2<t<T$} \\
\beta(t-T-L/2) & \textrm{for $T<t<L/2$}
\end{array}\right.
\eq
Our differential equation can be transformed to a \ind{Riccati equation} by the transformation:
\bq
\gamma(t) \equiv {d\over dt}\ln \alpha(t) \;.
\eq
Then we get:
\bq
{d\gamma(t)\over dt} + \gamma^2(t) = b(t) \;.
\eq

It is known that these Riccati type differential equations are very hard to solve, so there is little hope to solve our differential system.

One last thing that can be done here is write $\beta(t)$ in a more convenient way, it appears this $\beta(t)$ can also be written as:
\bq \label{beta}
\beta(t) = \left({2\pi k_1\over L}\right)^2 + \ldots + \left({2\pi k_{d-1}\over L}\right)^2 + \frac{\dddot{f}(t)}{\dot{f}(t)} \;,
\eq
for $-L/2<t<-T$, where $f$ is the solution in region 1, as defined in (\ref{deff}). For $-T<t<L/2$ this $f$ should of course be replaced by the $g$ from (\ref{defg}).

However, even with this simplification it is very hard to solve our differential equation. If we would only have had the first $d-1$ terms in (\ref{beta}), which are just a constant, the equation can easily be solved. Also, if we would only have the last term in (\ref{beta}), then the equation can also be solved for general $f$, because of the specific form of a triple derivative divided by a first derivative. However, when we have both terms, as in our case, it is very hard to solve the differential equation.

\subsection{The Divergences}

As shown in the previous section it is very hard to compute our path integral (\ref{pathintegralA}) exactly. One important thing we \emph{can} do however, is find the divergences\index{divergences!$N=1$ LSM} hidden in this path integral and see whether they can be cancelled by the counter terms, such that (the physical part of) the generating functional (\ref{Z}) becomes finite. If we can indeed cancel all divergences with the counter terms, then we know at least that the corrections from the path integral to all physical quantities are higher order, i.e.\ small. Therefore the classical approximation to these physical quantities is already a good approximation.

So let's see whether the divergences cancel. To this end we will compare the infinite parts of
\bq \label{ctdiv}
-\hbar\frac{\partial}{\partial J} \left( -{1\over\hbar}\int d^dx \left( -\h\delta_Z\f_c{d^2\over dt^2}\f_c - \h\delta_{\mu}\f_c^2 + {\delta_{\lambda}\over24}\f_c^4 \right) \right)
\eq
and
\bq \label{pathintdiv}
-\hbar\frac{\partial}{\partial J} \; \ln \left( \underset{\eta(\vec{x},-T)=0}{\int} \mathcal{D}\eta \; \exp\left(-{1\over\hbar}\int d^dx \left( \h(\vec{\nabla}\eta)^2 + \left(-\h\mu+{\lambda\over4}\f_c^2\right)\eta^2 \right) \right) \right) \;.
\eq
Here we have taken a derivative with respect to the source $J$, since only the $J$-dependent part of the generating functional is important for physical quantities, so only in this part it is necessary that all divergences cancel.

First consider the infinite part of the counter terms (\ref{ctdiv}). We use the counter terms that we found in the canonical approach (\ref{countt}). One does not expect the counter terms to change just because one fixes the paths. If we use these counter terms the infinite part of (\ref{ctdiv}) becomes:
\bqa
& & L^{d-1} \int dt \; \f_c(t) \frac{\partial\f_c(t)}{\partial J} \bigg( -{3\over2}{\hbar m^2\over v^2} \; I(0,m) - {9\over4}{\hbar m^4\over v^2} \; I(0,m,0,m) \nonumber\\
& & \phantom{L^{d-1} \int dt \; \f_c(t) \frac{\partial\f_c(t)}{\partial J} \bigg(} +{9\over4}{\hbar m^4\over v^4} \f_c^2(t) \; I(0,m,0,m) \bigg) \label{ctdiv2}
\eqa
Notice that it is far from obvious that divergences in here are going to cancel against divergences from the path integral because of the time dependent factors multiplying the divergences, i.e.\ the structure of the divergent terms is very different here than in models without path-fixing.

Now consider (\ref{pathintdiv}). Using what we found in the previous section this (\ref{pathintdiv}) can be written as:
\bqa
& & -\hbar \frac{\partial}{\partial J} \; \ln \left( \prod_{k_1,\ldots,k_{d-1}=-\infty}^{\infty} \frac{1}{\sqrt{\alpha(L/2)}} \right) \nonumber\\
&=& {\hbar\over2} \sum_{k_1,\ldots,k_{d-1}=-\infty}^{\infty} \frac{\partial}{\partial J} \ln\left(\alpha(L/2)\right) \nonumber\\
&=& {\hbar\over2} \sum_{k_1,\ldots,k_{d-1}=-\infty}^{\infty} \int dt \; \frac{\partial\f_c^2(t)}{\partial J} \frac{\delta}{\delta\f_c^2(t)} \ln\left(\alpha(L/2)\right) \nonumber\\
&=& \hbar \int dt \; \f_c(t) \frac{\partial\f_c(t)}{\partial J} \sum_{k_1,\ldots,k_{d-1}=-\infty}^{\infty} \frac{\delta}{\delta\f_c^2(t)} \ln\left(\alpha(L/2)\right) \label{pathintdiv2}
\eqa

Now, to find the divergent parts in this expression we can write down the first few terms of the Taylor expansion of
\bq
\frac{\delta}{\delta\f_c^2(t)} \ln\left(\alpha(L/2)\right)
\eq
around $\f_c^2(t) = v^2$:
\bqa
\frac{\delta}{\delta\f_c^2(t)} \ln\left(\alpha(L/2)\right) &=& \left. \frac{\frac{\delta\alpha(L/2)}{\delta\f_c^2(t)}}{\alpha(L/2)} \right|_{\f_c^2(t) = v^2} + \nonumber\\
& & \int dt' \; \left(\f_c^2(t')-v^2\right) \left.\left( \frac{\frac{\delta^2\alpha(L/2)}{\delta\f_c^2(t')\delta\f_c^2(t)}}{\alpha(L/2)} - \frac{\frac{\delta\alpha(L/2)}{\delta\f_c^2(t')} \frac{\delta\alpha(L/2)}{\delta\f_c^2(t)}}{\alpha(L/2)^2} \right)\right|_{\f_c^2(t) = v^2} + \nonumber\\[10pt]
& & \ldots \label{Taylorexp}
\eqa
So, to find these first two terms of the expansion we have to know:
\bq
\left.\alpha(L/2)\right|_{\f_c^2(t) = v^2} \quad, \quad \left.\frac{\delta\alpha(L/2)}{\delta\f_c^2(t)}\right|_{\f_c^2(t) = v^2} \quad \textrm{and} \quad \left.\frac{\delta^2\alpha(L/2)}{\delta\f_c^2(t')\delta\f_c^2(t)}\right|_{\f_c^2(t) = v^2}
\eq
These can all be found from the differential equation (\ref{diffsystalpha}). 

$\alpha(t)|_{\f_c^2(t) = v^2}$ Can of course be found from (\ref{diffsystalpha}) by just setting $\f_c^2(t) = v^2$ everywhere in the differential equation. If we define the \emph{constant} $\beta$ as:
\bq
\beta \equiv \beta(t)|_{\f_c^2(t) = v^2} = \left({2\pi k_1\over L}\right)^2 + \ldots + \left({2\pi k_{d-1}\over L}\right)^2 + 2\mu \;,
\eq
then the differential equation is
\bq
{d^2\over dt^2}\alpha(t)|_{\f_c^2(t) = v^2} - \beta \; \alpha(t)|_{\f_c^2(t) = v^2} = 0 \;,
\eq
and the solution that satisfies the boundary conditions (\ref{boundcondalpha}) can easily be found to be
\bq
\alpha(t)|_{\f_c^2(t) = v^2} = \frac{\sinh\left(\sqrt{\beta}(t+L/2)\right)}{\sqrt{\beta}} \;.
\eq

Now $\left.\frac{\delta\alpha(t)}{\delta\f_c^2(t')}\right|_{\f_c^2(t) = v^2}$ Can be found by first taking a functional derivative with respect to $\f_c^2(t')$ in the differential equation (\ref{diffsystalpha}) and then setting $\f_c^2(t) = v^2$ everywhere. We find the differential equation:
\bqa
\frac{d^2}{dt^2} \left.\frac{\delta\alpha(t)}{\delta\f_c^2(t')}\right|_{\f_c^2(t) = v^2} &=& {\lambda\over2} \; \delta(t-T+L/2-t') \; \alpha(t)|_{\f_c^2(t) = v^2} + \beta \; \left.\frac{\delta\alpha(t)}{\delta\f_c^2(t')}\right|_{\f_c^2(t) = v^2} \nonumber\\[5pt]
& & \hspace{200pt} \textrm{for $-L/2<t<T$} \nonumber\\[10pt]
\frac{d^2}{dt^2} \left.\frac{\delta\alpha(t)}{\delta\f_c^2(t')}\right|_{\f_c^2(t) = v^2} &=& {\lambda\over2} \; \delta(t-T-L/2-t') \; \alpha(t)|_{\f_c^2(t) = v^2} + \beta \; \left.\frac{\delta\alpha(t)}{\delta\f_c^2(t')}\right|_{\f_c^2(t) = v^2} \nonumber\\[5pt]
& & \hspace{200pt} \textrm{for $T<t<L/2$} \nonumber\\
\eqa
with boundary conditions:
\bq
\left.\frac{\delta\alpha(-L/2)}{\delta\f_c^2(t')}\right|_{\f_c^2(t) = v^2} = 0 \quad \;, \quad {d\over dt} \left.\frac{\delta\alpha(L/2)}{\delta\f_c^2(t')}\right|_{\f_c^2(t) = v^2} = 0 \;.
\eq
This differential system can also be solved easily, we have to distinguish the cases $-L/2<t'<-T$ and $-T<t'<L/2$ however. In the case $-L/2<t'<-T$ we find:
\bqa
& & \left.\frac{\delta\alpha(t)}{\delta\f_c^2(t')}\right|_{\f_c^2(t) = v^2} = \nonumber\\[2pt]
& & \hspace{25pt} \left\{\begin{array}{ll}
0 & \textrm{for $t<t'+T+L/2$} \\
{\lambda\over2\beta} \sinh\left(\sqrt{\beta}(t'+T+L)\right) \sinh\left(\sqrt{\beta}(t-t'-T-L/2)\right) & \textrm{for $t>t'+T+L/2$}
\end{array}\right. \nonumber\\
\eqa
In the case $-T<t'<L/2$ we find:
\bqa
& & \left.\frac{\delta\alpha(t)}{\delta\f_c^2(t')}\right|_{\f_c^2(t) = v^2} = \nonumber\\[2pt]
& & \hspace{25pt} \left\{\begin{array}{ll}
0 & \textrm{for $t<t'+T-L/2$} \\
{\lambda\over2\beta} \sinh\left(\sqrt{\beta}(t'+T)\right) \sinh\left(\sqrt{\beta}(t-t'-T+L/2)\right) & \textrm{for $t>t'+T-L/2$}
\end{array}\right. \nonumber\\
\eqa
At $t=L/2$ we can write the solution for all cases as:
\bq
\left.\frac{\delta\alpha(L/2)}{\delta\f_c^2(t')}\right|_{\f_c^2(t) = v^2} = -{\lambda\over2\beta} \sinh\left(\sqrt{\beta}|t'+T|\right) \sinh\left(\sqrt{\beta}(|t'+T|-L)\right)
\eq

Finally $\left.\frac{\delta^2\alpha(t)}{\delta\f_c^2(t')\delta\f_c^2(t'')}\right|_{\f_c^2(t) = v^2}$ can be found from the differential equation (\ref{diffsystalpha}) by taking two functional derivatives, with respect to $\f_c^2(t')$ and $\f_c^2(t'')$, and then putting $\f_c^2(t) = v^2$ everywhere. The differential equation one gets then is:
\bqa
\frac{d^2}{dt^2} \left.\frac{\delta^2\alpha(t)}{\delta\f_c^2(t')\delta\f_c^2(t'')}\right|_{\f_c^2(t) = v^2} &=& {\lambda\over2} \; \delta(t-T+L/2-t') \; \left.\frac{\delta\alpha(t)}{\delta\f_c^2(t'')}\right|_{\f_c^2(t) = v^2} + \nonumber\\
& & {\lambda\over2} \; \delta(t-T+L/2-t'') \; \left.\frac{\delta\alpha(t)}{\delta\f_c^2(t')}\right|_{\f_c^2(t) = v^2} + \nonumber\\
& & \beta \; \left.\frac{\delta^2\alpha(t)}{\delta\f_c^2(t')\delta\f_c^2(t'')}\right|_{\f_c^2(t) = v^2} \hspace{50pt} \textrm{for $-L/2<t<T$} \nonumber\\
\frac{d^2}{dt^2} \left.\frac{\delta^2\alpha(t)}{\delta\f_c^2(t')\delta\f_c^2(t'')}\right|_{\f_c^2(t) = v^2} &=& {\lambda\over2} \; \delta(t-T-L/2-t') \; \left.\frac{\delta\alpha(t)}{\delta\f_c^2(t'')}\right|_{\f_c^2(t) = v^2} + \nonumber\\
& & {\lambda\over2} \; \delta(t-T-L/2-t'') \; \left.\frac{\delta\alpha(t)}{\delta\f_c^2(t')}\right|_{\f_c^2(t) = v^2} + \nonumber\\
& & \beta \; \left.\frac{\delta^2\alpha(t)}{\delta\f_c^2(t')\delta\f_c^2(t'')}\right|_{\f_c^2(t) = v^2} \hspace{50pt} \textrm{for $T<t<L/2$} \nonumber\\
\eqa
Also this differential equation can be solved, now we have to distinguish 6 different cases however. After a lot of algebra one finds for the solution at $t=L/2$:
\bqa
\left.\frac{\delta^2\alpha(L/2)}{\delta\f_c^2(t')\delta\f_c^2(t'')}\right|_{\f_c^2(t) = v^2} &=& {\lambda^2\over4\beta^{3/2}} \sinh\left(\sqrt{\beta}\left(\textrm{max}(t',t'',-T)-\textrm{min}(t',t'',-T)-L\right)\right) \cdot \nonumber\\[-4pt]
& & \phantom{{\lambda^2\over4\beta^{3/2}}} \sinh\left(\sqrt{\beta}\left(\textrm{mid}(t',t'',-T)-\textrm{min}(t',t'',-T)\right)\right) \cdot \nonumber\\
& & \phantom{{\lambda^2\over4\beta^{3/2}}} \sinh\left(\sqrt{\beta}\left(\textrm{mid}(t',t'',-T)-\textrm{max}(t',t'',-T)\right)\right)
\eqa
where $\textrm{min}(t',t'',-T)$, $\textrm{mid}(t',t'',-T)$ and $\textrm{max}(t',t'',-T)$ give respectively the smallest, middle and largest variable in the set $\{t',t'',-T\}$.

Now all the expressions we found should be inserted in the Taylor expansion (\ref{Taylorexp}). Also remember that we are only interested in the divergent part of (\ref{pathintdiv2}). Divergences in (\ref{pathintdiv2}) are of course caused by the sum over the $d-1$ $k$'s, when the summand does not drop to zero fast enough for large $k$'s. So, to find only the divergent terms in (\ref{pathintdiv2}), we should only keep the terms that do not go to zero very fast in the Taylor expansion (\ref{Taylorexp}).

Now the first term in (\ref{Taylorexp}) is:
\bqa
\left.\frac{\frac{\delta\alpha(L/2)}{\delta\f_c^2(t)}}{\alpha(L/2)}\right|_{\f_c^2(t) = v^2} &=& -{\lambda\over2\sqrt{\beta}} \frac{\sinh\left(\sqrt{\beta}|t+T|\right) \sinh\left(\sqrt{\beta}(|t+T|-L)\right)}{\sinh(\sqrt{\beta}L)} \nonumber\\
&\overset{L\rightarrow\infty}{\longrightarrow}& {\lambda\over4\sqrt{\beta}} \left( 1 - e^{-2\sqrt{\beta}|t+T|} \right)
\eqa
The second term will not give a divergence under the sum, since large $k$'s are exponentially damped. (For $t=-T$ this exponential is of course not there, however for $t=-T$ the whole expressions (\ref{ctdiv2}) and (\ref{pathintdiv2}) become zero because ${\partial\f_c(-T)\over\partial J}=0$.) The first term will give a divergence, and we shall only keep this term. When we insert this term in (\ref{pathintdiv2}) we get:
\bqa
& & \hbar \int dt \; \f_c(t) \frac{\partial\f_c(t)}{\partial J} \sum_{k_1,\ldots,k_{d-1}=-\infty}^{\infty} {\lambda\over4} \frac{1}{\sqrt{\left({2\pi k_1\over L}\right)^2 + \ldots + \left({2\pi k_{d-1}\over L}\right)^2 + 2\mu}} \nonumber\\
&\overset{L\rightarrow\infty}{\longrightarrow}& {\lambda\over4}\hbar L^{d-1} \int dt \; \f_c(t) \frac{\partial\f_c(t)}{\partial J} \; {1\over(2\pi)^{d-1}} \int d^{d-1}k \frac{1}{\sqrt{\vec{k}^2+m^2}} \nonumber\\
&=& {\lambda\over2}\hbar L^{d-1} \int dt \; \f_c(t) \frac{\partial\f_c(t)}{\partial J} \; I(0,m) \nonumber\\
&=& {3\over2}{\hbar m^2\over v^2} L^{d-1} \int dt \; \f_c(t) \frac{\partial\f_c(t)}{\partial J} \; I(0,m)
\eqa
This cancels exactly the first term in the counter term part (\ref{ctdiv2}) of the generating functional!

Now consider the second order term in the Taylor expansion (\ref{Taylorexp}).
\bqa
& & \left.\left( \frac{\frac{\delta^2\alpha(L/2)}{\delta\f_c^2(t')\delta\f_c^2(t)}}{\alpha(L/2)} - \frac{\frac{\delta\alpha(L/2)}{\delta\f_c^2(t')} \frac{\delta\alpha(L/2)}{\delta\f_c^2(t)}}{\alpha(L/2)^2} \right)\right|_{\f_c^2(t) = v^2} = {\lambda^2\over4\beta} \frac{1}{\sinh^2(\sqrt{\beta}L)} \cdot \nonumber\\
& & \hspace{100pt} \bigg[ \sinh\left(\sqrt{\beta}L\right) \sinh\left(\sqrt{\beta}\left(\textrm{max}-\textrm{min}-L\right)\right) \cdot \nonumber\\
& & \hspace{100pt} \phantom{\bigg[} \sinh\left(\sqrt{\beta}\left(\textrm{mid}-\textrm{min}\right)\right) \sinh\left(\sqrt{\beta}\left(\textrm{mid}-\textrm{max}\right)\right) \nonumber\\
& & \hspace{100pt} \phantom{\bigg[} -\sinh\left(\sqrt{\beta}|t+T|\right) \sinh\left(\sqrt{\beta}(L-|t+T|)\right) \cdot \nonumber\\
& & \hspace{100pt} \phantom{\bigg[} \sinh\left(\sqrt{\beta}|t'+T|\right) \sinh\left(\sqrt{\beta}(L-|t'+T|)\right) \bigg] \nonumber\\
\eqa
After some hard work this can be written to:
\bqa
& & \left.\left( \frac{\frac{\delta^2\alpha(L/2)}{\delta\f_c^2(t')\delta\f_c^2(t)}}{\alpha(L/2)} - \frac{\frac{\delta\alpha(L/2)}{\delta\f_c^2(t')} \frac{\delta\alpha(L/2)}{\delta\f_c^2(t)}}{\alpha(L/2)^2} \right)\right|_{\f_c^2(t) = v^2} = -{\lambda^2\over4\beta} \frac{1}{\sinh^2(\sqrt{\beta}L)} \cdot \nonumber\\
& & \hspace{50pt} \bigg[ \cosh\left(\sqrt{\beta}(|t-t'|-L/2)\right) \cosh\left({\sqrt{\beta}L\over2}\right) + \nonumber\\
& & \hspace{50pt} \phantom{\bigg(} -\cosh\left(\sqrt{\beta}(|t+T|-L/2)\right) \cosh\left(\sqrt{\beta}(|t'+T|-L/2)\right) \bigg]^2
\eqa
Now, all the terms, when worked out for large $L$, will still contain exponentials which dampen the large $k$ values in the sum. The square of the first term between the straight brackets will also contain such an exponential, however the argument of this exponential ($\sim |t-t'|$) can become zero without $t$ or $t'$ becoming $-T$. So in the case of $t=t'$ this first term will give a divergence. The other terms in the expression above (when the square is worked out) will always contain an exponential with arguments proportional to $|t+T|$ or $|t'+T|$, such that these will only give divergences at $t=-T$ or $t'=-T$, where the whole expression (\ref{pathintdiv2}) becomes zero.

So let's only keep the first term between the straight brackets and work out what we get for large $L$. 
\bq
-{\lambda^2\over4\beta} \frac{\cosh^2\left(\sqrt{\beta}(|t-t'|-L/2)\right) \cosh^2\left({\sqrt{\beta}L\over2}\right)}{\sinh^2(\sqrt{\beta}L)} \overset{L\rightarrow\infty}{\longrightarrow} -{\lambda^2\over16\beta} e^{-2\sqrt{\beta}|t-t'|}
\eq
Substituting this in the Taylor expansion (\ref{Taylorexp}) and then in (\ref{pathintdiv2}) we find:
\bq
-{\lambda^2\hbar\over16} \int dt \int dt' \; \f_c(t) \frac{\partial\f_c(t)}{\partial J} \left(\f_c^2(t')-v^2\right) \sum_{k_1,\ldots,k_{d-1}=-\infty}^{\infty} {1\over\beta} e^{-2\sqrt{\beta}|t-t'|}
\eq
As argued above, a divergence can only arise when $t=t'$, so when we are only interested in this part we can safely set $\f_c^2(t')=\f_c^2(t)$. Then we can do the $t'$ integral and we find:
\bqa
& & -{\lambda^2\hbar\over16} \int dt \; \f_c(t) \left(\f_c^2(t)-v^2\right) \frac{\partial\f_c(t)}{\partial J} \sum_{k_1,\ldots,k_{d-1}=-\infty}^{\infty} {1\over\beta^{3/2}} \nonumber\\
&\overset{L\rightarrow\infty}{\longrightarrow}& -{\lambda^2\hbar\over16} L^{d-1} \int dt \; \f_c(t) \left(\f_c^2(t)-v^2\right) \frac{\partial\f_c(t)}{\partial J} \; {1\over(2\pi)^{d-1}} \int d^{d-1}k \frac{1}{\left(\vec{k}^2+m^2\right)^{3/2}} \nonumber\\
&=& -{\lambda^2\hbar\over4} L^{d-1} \int dt \; \f_c(t) \left(\f_c^2(t)-v^2\right) \frac{\partial\f_c(t)}{\partial J} \; I(0,m,0,m) \nonumber\\
&=& -{9\over4}{\hbar m^4\over v^4} L^{d-1} \int dt \; \f_c(t) \left(\f_c^2(t)-v^2\right) \frac{\partial\f_c(t)}{\partial J} \; I(0,m,0,m)
\eqa
And this cancels exactly the second and third term in the counter term part (\ref{ctdiv2})!

So finally we have proven that the physical part of the 1-loop generating functional can be made finite with the same counter terms as in the canonical approach to the $N=1$ linear sigma model. Of course also all 1-loop Green's functions are finite then.

Notice that the cancellation of the divergences does \emph{not} depend on the specific form of $\f_c(t)$. We have proven here that the divergences cancel in all physical quantities independent of what the classical solution looks like explicitly.\index{renormalizable}

\subsection{The Alternative Effective Potential}

Because we know now that the physical part of the generating functional is finite a good approximation to this $Z(J)$ is the classical approximation:\index{summing over minima}
\bq
Z(J) = e^{-{1\over\hbar}S_c^+} + e^{-{1\over\hbar}S_c^-} \equiv Z_+(J) + Z_-(J)
\eq
Actually this is the only approximation we \emph{can} do to find $Z(J)$, since the 1-loop, or saddle-point, approximation would already involve the difficult path integral (\ref{pathintegralA}). Defining $A$ and $B$ as
\bq
A \equiv \h(S_c^++S_c^-) \;, \quad B \equiv \h(S_c^+-S_c^-) \;,
\eq
we have
\bq
Z(J) = 2 \; e^{-{1\over\hbar}A} \cosh\left({1\over\hbar}B\right) \;.
\eq
Then we also have:
\bq
\hbar{\partial\over\partial J} \ln Z = \int d^dx \; \langle\f(x)\rangle(J) = -\frac{\partial A}{\partial J} + \frac{\partial B}{\partial J} \tanh\left({1\over\hbar}B\right)
\eq

\begin{figure}[h]
\begin{center}
\epsfig{file=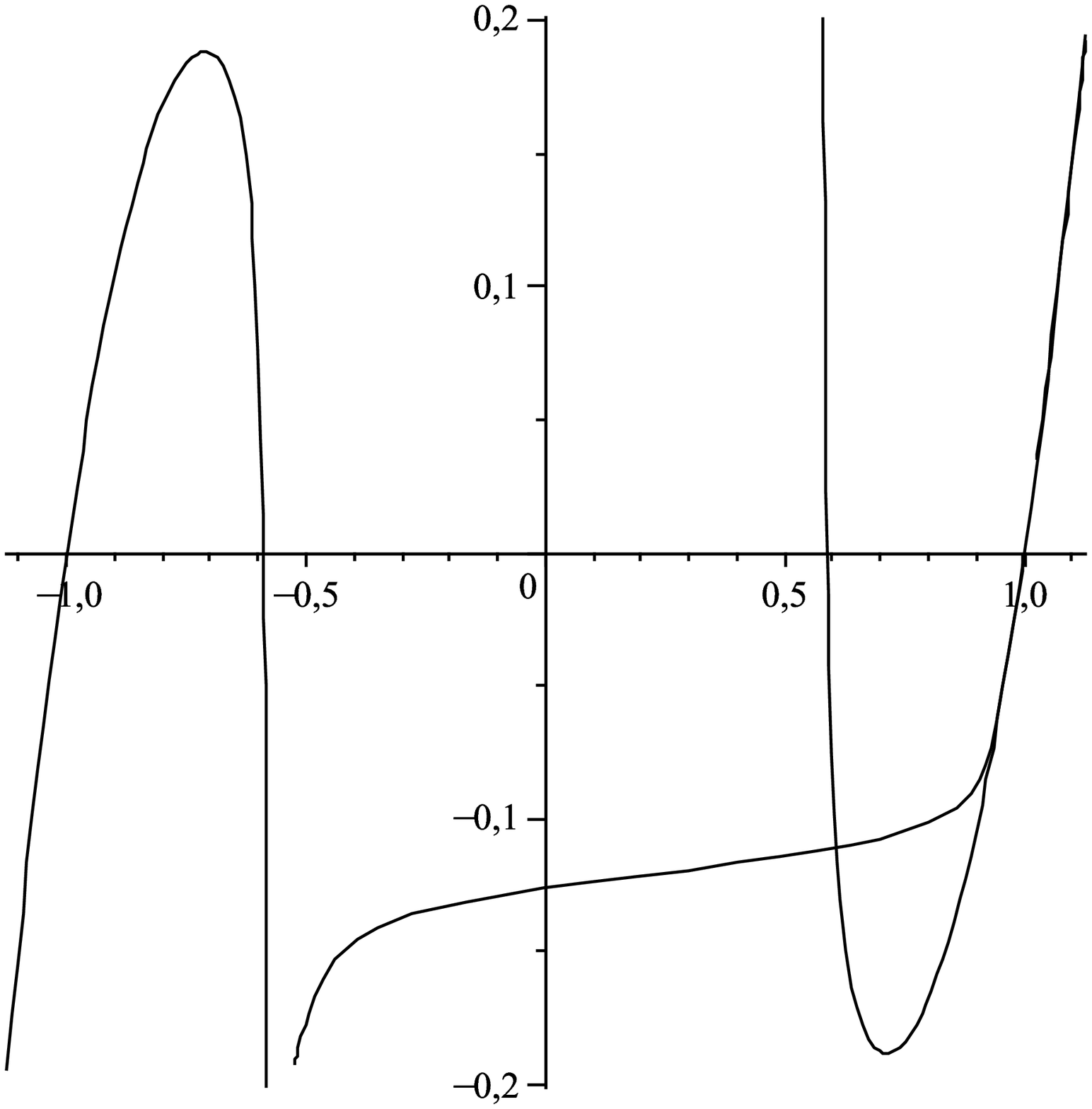,width=7cm}
\epsfig{file=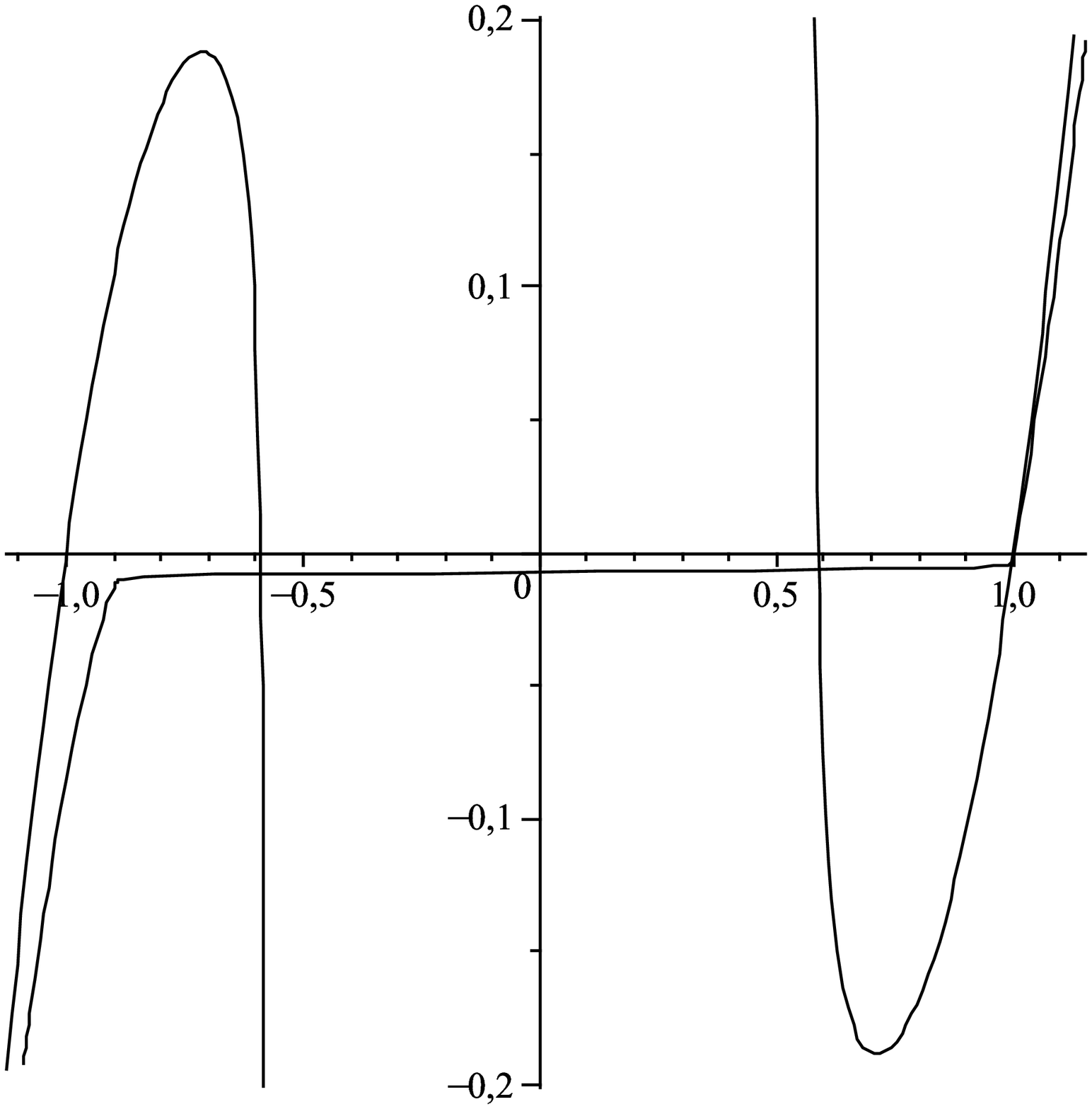,width=7cm}
\end{center}
\vspace{-0.5cm}
\caption{The derivative of the alternative effective potential for $L=10$ and $L=100$ for the case $d=1$, $\hbar=0.1$, $\mu=1$, $v=1$, $\bar{\f}=1$ and $T=0$.}
\label{effpotfixd1}
\end{figure}

\begin{figure}[h]
\begin{center}
\epsfig{file=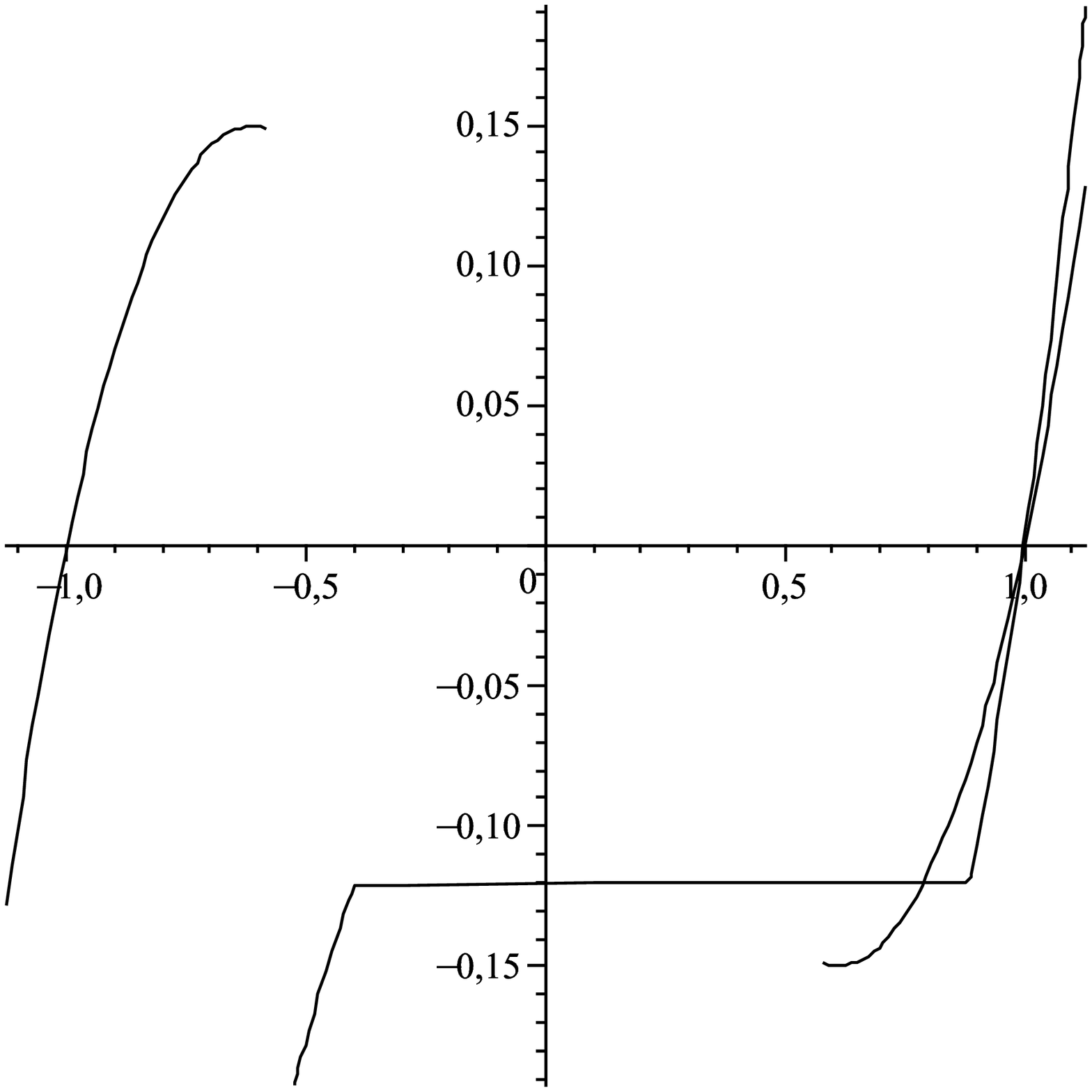,width=7cm}
\epsfig{file=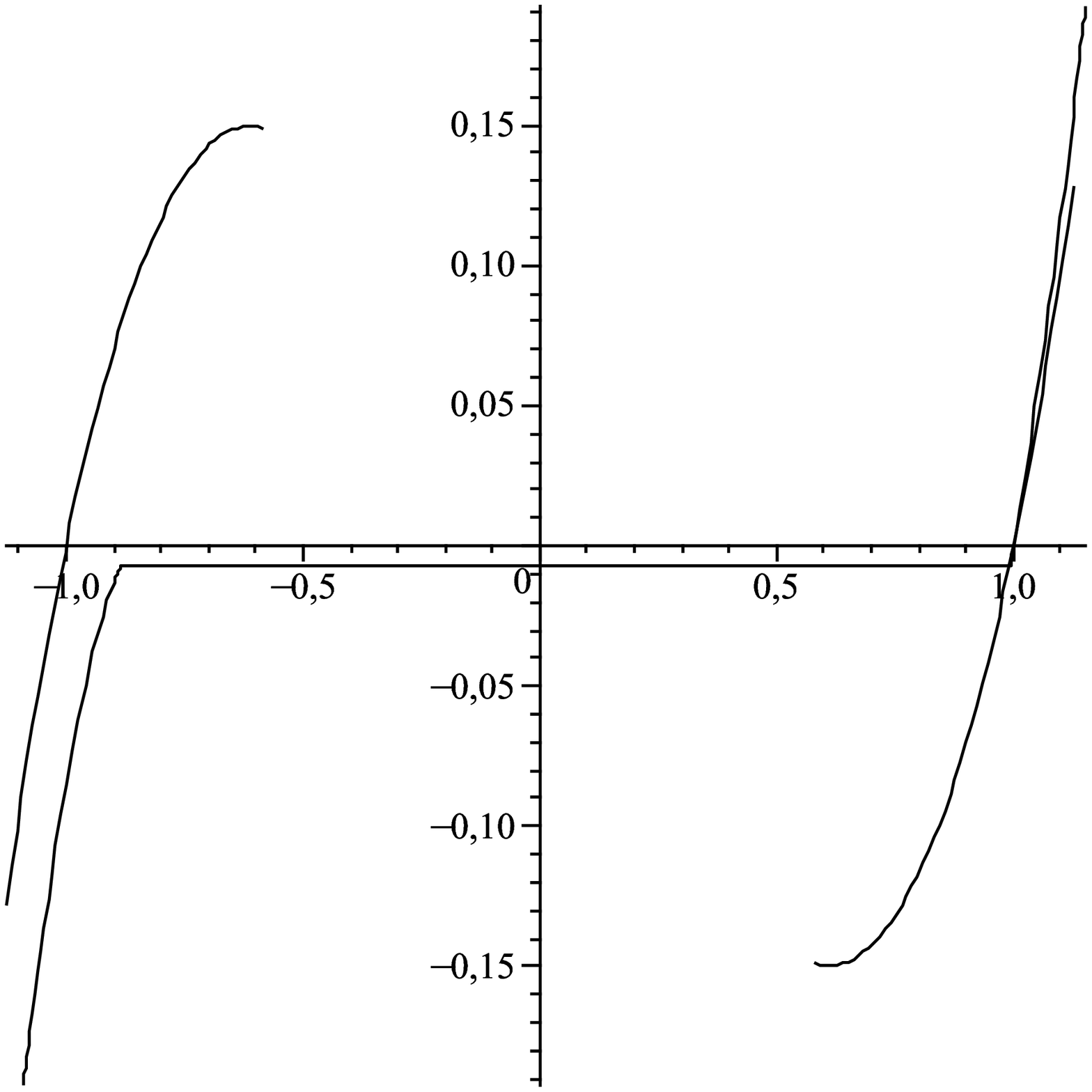,width=7cm}
\end{center}
\vspace{-0.5cm}
\caption{The derivative of the alternative effective potential for $L=10$ and $L=100$ for the case $d=4$, $\hbar=2$, $\mu=1$, $v=1$, $\bar{\f}=1$ and $T=0$.}
\label{effpotfixd4}
\end{figure}

To find the \ind{alternative effective potential} we have to know $J(x)$ as a functional of $\langle\f(x)\rangle$ for large $t$, i.e.\ $t\gg-T$. So we first have to know $\langle\f(x)\rangle$ as a function of $J(x)$ for large $t$. For large $t$ this tadpole goes to a constant, which can be read off from the formula above:
\bq
\Omega \langle\f\rangle(J) = -\frac{\partial A}{\partial J} + \frac{\partial B}{\partial J} \tanh\left({1\over\hbar}B\right)
\eq
So indeed we see, to obtain the tadpole for large times it was only necessary to do our calculations for constant source $J$. From this formula one can now find the inverse and construct the alternative effective potential. In figure \ref{effpotfixd1} the derivative of this alternative effective potential is plotted for $L=10$ and $L=100$, for the case $d=1$, $\hbar=0.1$, $\mu=1$, $v=1$, $\bar{\f}=1$ and $T=0$. For comparison we have also plotted the derivative of the effective potential from the canonical approach. In figure \ref{effpotfixd4} the derivative of the alternative effective potential is plotted for $L=10$ and $L=100$, for the case $d=4$, $\hbar=2$, $\mu=1$, $v=1$, $\bar{\f}=1$ and $T=0$, also together with the canonical result.\index{alternative effective potential!$N=1$ LSM}

One can see that these alternative effective potentials come out to be \emph{convex}, which is not obvious beforehand. Also they converge to the effective potentials from the path-integral approach without fixing in the limit $L\rightarrow\infty$.

\subsection{The Green's Functions}

Now we compute the Green's functions\index{Green's functions!$N=1$ LSM} in this approach. Again, as in the previous chapter, we have:
\bqa
\langle \f(x) \rangle &=& \frac{\underset{\f(\vec{x},-T)=\bar{\f}}{\int} \mathcal{D}\f \; \f(x) \; e^{-{1\over\hbar}S(J=0)}}{\underset{\f(\vec{x},-T)=\bar{\f}}{\int} \mathcal{D}\f \; e^{-{1\over\hbar}S(J=0)}} \nonumber\\
&=& \frac{(\f_+(t)(J=0)+\langle\eta(x)\rangle_+) Z_+(0) + (\f_-(t)(J=0)+\langle\eta(x)\rangle_-) Z_-(0)}{Z_+(0) + Z_-(0)} \;, \nonumber\\
\eqa
where
\bq \label{etapathfixed}
\langle \eta(x) \rangle_\pm = \frac{\underset{\eta(\vec{x},-T)=0}{\int} \mathcal{D}\eta \; \eta(x) \; e^{-{1\over\hbar}S(\f_c=\f_\pm(J=0))}}{\underset{\eta(\vec{x},-T)=0}{\int} \mathcal{D}\f \; e^{-{1\over\hbar}S(\f_c=\f_\pm(J=0))}}
\eq
where $S$ is the action in (\ref{complaction}) and $\f_\pm(t)$ is the classical solution $\f_c(t)$ around the $\pm$-minumum.

Now we do not have $Z_+(0)=Z_-(0)$, as in the previous chapter. Looking only at the classical actions around each minimum we have (substituting $J=0$ in \ref{classactionfixed}):
\bq
S_c^\pm(J=0) = -{1\over4}L^d \mu v^2 + {1\over3}mv^2L^{d-1} \left( \left({\bar{\f}\over v}\mp2\right) \left({\bar{\f}\over v}\pm1\right) \left|{\bar{\f}\over v}\pm1\right| + 4 \right)
\eq
The first term is the same for both minima, and thus unimportant. The second term is not the same for both minima, and thus this term determines which of the minima is non-perturbatively favored in the infinite-volume limit. It is easy to see that:
\bqa
S_c^+(J=0) &<& S_c^-(J=0) \quad \textrm{for} \; \bar{\f}>0 \nonumber\\
S_c^+(J=0) &>& S_c^-(J=0) \quad \textrm{for} \; \bar{\f}<0
\eqa
So for $\bar{\f}>0$ the negative minimum is non-perturbatively suppressed, whereas for $\bar{\f}<0$ the positive minimum is suppressed. So the minimum closest to $\bar{\f}$, i.e.\ the point where the paths are fixed, survives in the infinite-volume limit, the other minimum is suppressed non-perturbatively. We do \emph{not} even have to know the quantum part of $Z_\pm$ to say this, since this quantum part is small with respect to the classical part one can already see from the classical part which minimum wins.

Now we choose $\bar{\f}>0$. Then the tadpole becomes:
\bq
\langle \f(x) \rangle = \f_+(t)(J=0) + \langle\eta(x)\rangle_+
\eq
Notice that this is an \emph{exact} expression for $L\rightarrow\infty$.

Now we will assume that for large $t$, i.e.\ $t\gg-T$, the tadpole $\langle \eta(x) \rangle_\pm$ does \emph{not} notice the path-fixing and the time-dependent mass and coupling constants anymore. Remember that, in the free field theory considered at the beginning of this chapter, the tadpole and propagator did indeed \emph{not} notice the path-fixing anymore for large times. See (\ref{freefixedGreensfuncts}). This makes the current assumption plausible.

Under this assumption we find for large $t$:
\bq
\langle \f(x) \rangle = v + \langle\eta\rangle \;,
\eq
where $\langle\eta\rangle$ is defined as in (\ref{etapathfixed}), but without the path-fixing constraint and with $\f_c=+v$. This is the canonical result.

Now consider the $\f$-propagator. This propagator can be written as:
\bqa
& & \langle \f(x) \f(y) \rangle = \frac{1}{Z_+(0) + Z_-(0)} \cdot \nonumber\\
& & \quad \bigg( \left(\f_+^2(t)(J=0) + \f_+(t)(J=0)\left(\langle\eta(x)\rangle_++\langle\eta(y)\rangle_+\right) + \langle\eta(x)\eta(y)\rangle_+\right)Z_+(0) + \nonumber\\
& & \quad \phantom{\bigg(} \left(\f_-^2(t)(J=0) + \f_-(t)(J=0)\left(\langle\eta(x)\rangle_-+\langle\eta(y)\rangle_-\right) + \langle\eta(x)\eta(y)\rangle_-\right)Z_-(0) \bigg) \nonumber\\
\eqa
with
\bq \label{etaetapathfixed}
\langle \eta(x) \eta(y) \rangle_\pm = \frac{\underset{\eta(\vec{x},-T)=0}{\int} \mathcal{D}\eta \; \eta(x) \eta(y) \; e^{-{1\over\hbar}S(\f_c=\f_\pm(J=0))}}{\underset{\eta(\vec{x},-T)=0}{\int} \mathcal{D}\f \; e^{-{1\over\hbar}S(\f_c=\f_\pm(J=0))}} \;.
\eq

Again the minimum closest to the fixing point $\bar{\f}$ dominates completely for $L\rightarrow\infty$. Since we took $\bar{\f}>0$ this propagator becomes:
\bqa
\langle \f(x) \f(y) \rangle &=& \f_+^2(t) + \f_+(t)\left(\langle\eta(x)\rangle_++\langle\eta(y)\rangle_+\right) + \langle\eta(x)\eta(y)\rangle_+ \nonumber\\
\langle \f(x) \f(y) \rangle_\mathrm{c} &=& \langle\eta(x)\eta(y)\rangle_{+,c}
\eqa
Also this is an \emph{exact} expression in the limit $L\rightarrow\infty$.

For large times $t$, $t\gg-T$, again assuming that the $\eta$-propagator does \emph{not} notice the path-fixing and the explicit time dependence in the action, the connected propagator becomes:
\bq
\langle \f(x) \f(y) \rangle_\mathrm{c} = \langle\eta(x)\eta(y)\rangle_\mathrm{c} \;,
\eq
where $\langle\eta(x)\eta(y)\rangle$ is defined as in (\ref{etaetapathfixed}) without the path fixing constraint. This is the canonical propagator.

In this way all Green's functions can be shown to be equal to the canonical Green's functions for times $t\gg-T$.

So, finally we have shown that in the path-integral approach with fixed paths we get the same Green's functions as in the canonical approach, including SSB. Now we get a \emph{convex} alternative effective potential\index{alternative effective potential!convexity} however.


\chapter{The $N=2$ LSM: The Canonical Approach}\label{chapN2sigmamodelcan}

In this chapter we will calculate the same things as in chapter \ref{chapN1sigmamodelcan}, now for the $N=2$ linear sigma model\index{$N=2$ linear sigma model!canonical approach}. This Euclidean linear sigma model with $N=2$ fields is defined by the bare action\index{action!$N=2$ LSM}
\bq
S = \int d^dx \; \left( \h\left(\nabla\f_1\right)^2 + \h\left(\nabla\f_2\right)^2 - \h\mu\left(\f_1^2+\f_2^2\right) + {\lambda\over24}\left(\f_1^2+\f_2^2\right)^2 \right) \;.
\eq
We will outline the \index{$N=2$ linear sigma model!canonical approach}canonical treatment of this model for the case $\mu>0$, i.e.\ for the case with \ind{spontaneous symmetry breaking}. Our calculations will be much like those found in most textbooks, like e.g.\ Peskin and Schroeder \cite{Peskin}.

\section{Green's Functions}

To compute the renormalized Green's functions\index{Green's functions!renormalized} of this theory we introduce renormalized quantities as in (\ref{renquant}). The action in terms of these renormalized quantities is (again we suppress the $R$ superscripts from now on):
\bqa
S &=& \int d^dx \; \bigg( \h\left(\nabla\f_1\right)^2 + \h\left(\nabla\f_2\right)^2 - \h\mu\left(\f_1^2+\f_2^2\right) + {\lambda\over24}\left(\f_1^2+\f_2^2\right)^2 + \nonumber\\[5pt]
& & \phantom{\int d^dx \; \bigg(} \h\delta_Z\left(\nabla\f_1\right)^2 + \h\delta_Z\left(\nabla\f_2\right)^2 - \h\delta_{\mu}\left(\f_1^2+\f_2^2\right) + {\delta_{\lambda}\over24}\left(\f_1^2+\f_2^2\right)^2 \bigg) \;.
\eqa
Now the classical action, i.e.\ the first line has its minima\index{minima!$N=2$ LSM} on the circle
\bq
\f_1^2+\f_2^2 = v^2 \;, \quad v \equiv \sqrt{{6\mu\over\lambda}} \;.
\eq
We choose\index{choosing minimum}
\bq
\f_1=v \;, \quad \f_2=0
\eq
as the true minimum in this canonical approach. Then define
\bq
\f_1 \equiv v+\eta_1 \;, \quad \f_2 \equiv \eta_2 \;.
\eq
In terms of these $\eta$ fields the action becomes:
\bqa
S &=& \int d^dx \; \bigg( \h\left(\nabla\eta_1\right)^2 + \h\left(\nabla\eta_2\right)^2 + \mu\eta_1^2 + {\mu\over v}\eta_1^3 + {\mu\over v}\eta_1\eta_2^2 + \nonumber\\[5pt]
& & \phantom{\int d^dx \; \bigg(} {\mu\over4v^2}\eta^4 + {\mu\over4v^2}\eta_2^2 + {\mu\over2v^2}\eta_1^2\eta_2^2 + \nonumber\\[5pt]
& & \phantom{\int d^dx \; \bigg(} \left(-v\delta_{\mu}+{1\over6}v^3\delta_{\lambda}\right)\eta_1 + \h\delta_Z\left(\nabla\eta_1\right)^2 + \left(-\h\delta_{\mu}+{1\over4}v^2\delta_{\lambda}\right)\eta_1^2 + \nonumber\\[5pt]
& & \phantom{\int d^dx \; \bigg(} \h\delta_Z\left(\nabla\eta_2\right)^2 + \left(-\h\delta_{\mu}+{1\over12}v^2\delta_{\lambda}\right)\eta_2^2 + {1\over6}v\delta_{\lambda}\eta_1^3 + {1\over6}v\delta_{\lambda}\eta_1\eta_2^2 + \nonumber\\[5pt]
& & \phantom{\int d^dx \; \bigg(} {1\over24}\delta_{\lambda}\eta_1^4 + {1\over24}\delta_{\lambda}\eta_2^4 + {1\over12}\delta_{\lambda}\eta_1^2\eta_2^2 \bigg) \;. \label{N2etaaction}
\eqa

Define again $\mu\equiv\h m^2$. Then the The \index{Feynman rules!$N=2$ LSM}Feynman rules are:
{\allowdisplaybreaks\bqa
\begin{picture}(100, 20)(0, 17)
\Line(20, 20)(80, 20)
\end{picture}
&\leftrightarrow& \frac{\hbar}{k^2 + m^2} \nonumber\\
\begin{picture}(100, 40)(0, 17)
\DashLine(20, 20)(80, 20){3}
\end{picture}
&\leftrightarrow& \frac{\hbar}{k^2} \nonumber\\
\begin{picture}(100, 40)(0, 17)
\Line(20, 20)(50, 20)
\Line(50, 20)(80, 0)
\Line(50, 20)(80, 40)
\end{picture}
&\leftrightarrow& -\frac{3m^2}{\hbar v} \nonumber\\
\begin{picture}(100, 40)(0, 17)
\Line(20, 20)(50, 20)
\DashLine(50, 20)(80, 0){3}
\DashLine(50, 20)(80, 40){3}
\end{picture}
&\leftrightarrow& -\frac{m^2}{\hbar v} \nonumber\\
\begin{picture}(100, 40)(0, 17)
\Line(20, 40)(80, 0)
\Line(20, 0)(80, 40)
\end{picture}
&\leftrightarrow& -\frac{3m^2}{\hbar v^2} \nonumber\\
\begin{picture}(100, 40)(0, 17)
\DashLine(20, 40)(80, 0){3}
\DashLine(20, 0)(80, 40){3}
\end{picture}
&\leftrightarrow& -\frac{3m^2}{\hbar v^2} \nonumber\\
\begin{picture}(100, 40)(0, 17)
\Line(20, 40)(80, 0)
\DashLine(20, 0)(80, 40){3}
\end{picture}
&\leftrightarrow& -\frac{m^2}{\hbar v^2} \nonumber\\
\begin{picture}(100, 40)(0, 17)
\Line(20, 20)(50, 20)
\Vertex(50,20){3}
\end{picture}
&\leftrightarrow& {v\over\hbar}\delta_{\mu} - {1\over6}{v^3\over\hbar}\delta_{\lambda} \nonumber\\
\begin{picture}(100, 40)(0, 17)
\Line(20, 20)(80, 20)
\Vertex(50,20){3}
\end{picture}
&\leftrightarrow& -{1\over\hbar}\delta_Z k^2 + {1\over\hbar}\delta_{\mu} - \h{v^2\over\hbar}\delta_{\lambda} \nonumber\\
\begin{picture}(100, 40)(0, 17)
\DashLine(20, 20)(80, 20){3}
\Vertex(50,20){3}
\end{picture}
&\leftrightarrow& -{1\over\hbar}\delta_Z k^2 + {1\over\hbar}\delta_{\mu} - {1\over6}{v^2\over\hbar}\delta_{\lambda} \nonumber\\
\begin{picture}(100, 40)(0, 17)
\Line(20, 20)(50, 20)
\Line(50, 20)(80, 0)
\Line(50, 20)(80, 40)
\Vertex(50,20){3}
\end{picture}
&\leftrightarrow& -{v\over\hbar}\delta_{\lambda} \nonumber\\
\begin{picture}(100, 40)(0, 17)
\Line(20, 20)(50, 20)
\DashLine(50, 20)(80, 0){3}
\DashLine(50, 20)(80, 40){3}
\Vertex(50,20){3}
\end{picture}
&\leftrightarrow& -{1\over3}{v\over\hbar}\delta_{\lambda} \nonumber\\
\begin{picture}(100, 40)(0, 17)
\Line(20, 40)(80, 0)
\Line(20, 0)(80, 40)
\Vertex(50,20){3}
\end{picture}
&\leftrightarrow& -{1\over\hbar}\delta_{\lambda} \nonumber\\
\begin{picture}(100, 40)(0, 17)
\DashLine(20, 40)(80, 0){3}
\DashLine(20, 0)(80, 40){3}
\Vertex(50,20){3}
\end{picture}
&\leftrightarrow& -{1\over\hbar}\delta_{\lambda} \nonumber\\
\begin{picture}(100, 40)(0, 17)
\Line(20, 40)(80, 0)
\DashLine(20, 0)(80, 40){3}
\Vertex(50,20){3}
\end{picture}
&\leftrightarrow& -{1\over3}{1\over\hbar}\delta_{\lambda}
\eqa}

\vspace{10pt}

Now let's calculate the momentum space Green's functions\index{Green's functions!$N=2$ LSM} of this theory, up to one loop. Again we shall write the results in terms of the standard integrals (\ref{standint}).

Notice that standard integrals like $I(0,0,\ldots,0)$ are zero in the \ind{dimensional regularization scheme}. We shall \emph{not} specify the \ind{regularization scheme} in our calculations however, and keep everything general.

\bqa
\langle \tilde{\eta}_1 \rangle &=& \quad
\begin{picture}(50,20)(0,17)
\Line(0,20)(20,20)
\CArc(35,20)(15,-180,180)
\end{picture} \quad + \quad
\begin{picture}(50,20)(0,17)
\Line(0,20)(20,20)
\DashCArc(35,20)(15,-180,180){3}
\end{picture} \quad + \quad
\begin{picture}(50,20)(0,17)
\Line(0,20)(35,20)
\Vertex(35,20){3}
\end{picture} \nonumber\\[20pt]
&=& -{3\over2}{\hbar\over v} \; I(0,m) - \h{\hbar\over v} \; I(0,0) + {v\over m^2} \; \delta_{\mu}|_{\hbar} - {1\over6}{v^3\over m^2} \; \delta_{\lambda}|_{\hbar}
\eqa
\bq
\langle \tilde{\eta}_2 \rangle = 0
\eq
\bqa
\langle \tilde{\eta}_1\left(p\right) \tilde{\eta}_1\left(q\right) \rangle_{\mathrm{c}} &=& \quad
\begin{picture}(70,40)(0,17)
\Line(0,20)(70,20)
\end{picture} \quad + \quad
\begin{picture}(70,40)(0,17)
\Line(0,20)(20,20)
\CArc(35,20)(15,-180,180)
\Line(50,20)(70,20)
\end{picture} \quad + \quad
\begin{picture}(70,40)(0,17)
\Line(0,20)(20,20)
\DashCArc(35,20)(15,-180,180){3}
\Line(50,20)(70,20)
\end{picture} \quad + \nonumber\\
& & \quad \begin{picture}(70,40)(0,17)
\Line(0,5)(70,5)
\CArc(35,20)(15,-90,270)
\end{picture} \quad + \quad
\begin{picture}(70,40)(0,17)
\Line(0,5)(70,5)
\DashCArc(35,20)(15,-90,270){3}
\end{picture} \quad + \quad
\begin{picture}(70,40)(0,17)
\Line(0,5)(70,5)
\Line(35,5)(35,15)
\GCirc(35,25){10}{0.7}
\end{picture} \quad + \nonumber\\
& & \quad \begin{picture}(70,40)(0,17)
\Line(0,20)(70,20)
\Vertex(35,20){3}
\end{picture} \nonumber\\[25pt]
&=& \frac{\hbar}{p^2+m^2} + \frac{1}{\left(p^2+m^2\right)^2} \Bigg( {9\over2}{\hbar^2m^4\over v^2} \; I(0,m,p,m) + \h{\hbar^2m^4\over v^2} \; I(0,0,p,0) + \nonumber\\
& & \phantom{\frac{\hbar}{p^2+m^2} + \frac{1}{\left(p^2+m^2\right)^2} \Bigg(} 3{\hbar^2m^2\over v^2} \; I(0,m) + {\hbar^2m^2\over v^2} \; I(0,0) + \nonumber\\
& & \phantom{\frac{\hbar}{p^2+m^2} + \frac{1}{\left(p^2+m^2\right)^2} \Bigg(} -\hbar p^2 \; \delta_Z|_{\hbar} - 2\hbar \; \delta_{\mu}|_{\hbar} \Bigg) \label{eta1propcan}
\eqa
\bqa
\langle \tilde{\eta}_2\left(p\right) \tilde{\eta}_2\left(q\right) \rangle_{\mathrm{c}} &=& \quad
\begin{picture}(70,40)(0,17)
\DashLine(0,20)(70,20){3}
\end{picture} \quad + \quad
\begin{picture}(70,40)(0,17)
\DashLine(0,20)(20,20){3}
\CArc(35,20)(15,-180,0)
\DashCArc(35,20)(15,0,180){3}
\DashLine(50,20)(70,20){3}
\end{picture} \quad + \quad
\begin{picture}(70,40)(0,17)
\DashLine(0,5)(70,5){3}
\CArc(35,20)(15,-90,270)
\end{picture} \quad + \nonumber\\
& & \quad \begin{picture}(70,40)(0,17)
\DashLine(0,5)(70,5){3}
\DashCArc(35,20)(15,-90,270){3}
\end{picture} \quad + \quad
\begin{picture}(70,40)(0,17)
\DashLine(0,5)(70,5){3}
\Line(35,5)(35,15)
\GCirc(35,25){10}{0.7}
\end{picture} \quad + \quad
\begin{picture}(70,40)(0,17)
\DashLine(0,20)(70,20){3}
\Vertex(35,20){3}
\end{picture} \nonumber\\[25pt]
&=& \frac{\hbar}{p^2} + \frac{1}{\left(p^2\right)^2} \Bigg( {\hbar^2m^4\over v^2} \; I(0,0,p,m) + {\hbar^2m^2\over v^2} \; I(0,m) - {\hbar^2m^2\over v^2} \; I(0,0) + \nonumber\\
& & \phantom{\frac{\hbar}{p^2} + \frac{1}{\left(p^2\right)^2} \Bigg(} -\hbar p^2 \; \delta_Z|_{\hbar} \Bigg)
\eqa
\bq
\langle \tilde{\eta}_1\left(p\right) \tilde{\eta}_2\left(q\right) \rangle_{\mathrm{c}} = 0
\eq
{\allowdisplaybreaks\bqa
\langle \tilde{\eta}_1\left(q_1\right) \tilde{\eta}_1\left(q_2\right) \tilde{\eta}_1\left(q_3\right) \rangle_{\mathrm{1PI}} &=& \quad
\begin{picture}(50,40)(0,17)
\Line(0,20)(25,20)
\Line(25,20)(50,0)
\Line(25,20)(50,40)
\end{picture} \quad + \nonumber\\[5pt]
& & \quad \begin{picture}(70,40)(0,17)
\Line(0,20)(20,20)
\CArc(35,20)(15,-180,180)
\Line(48,27)(70,40)
\Line(48,13)(70,0)
\end{picture} \quad + \quad
\begin{picture}(70,40)(0,17)
\Line(0,20)(20,20)
\DashCArc(35,20)(15,-180,180){3}
\Line(48,27)(70,40)
\Line(48,13)(70,0)
\end{picture} \quad + \nonumber\\[5pt]
& & \quad \begin{picture}(70,40)(0,17)
\Line(0,20)(20,20)
\CArc(35,20)(15,-180,180)
\Line(50,20)(70,40)
\Line(50,20)(70,0)
\end{picture} \quad + \quad
\begin{picture}(70,40)(0,17)
\Line(0,20)(20,20)
\DashCArc(35,20)(15,-180,180){3}
\Line(50,20)(70,40)
\Line(50,20)(70,0)
\end{picture} \quad + \nonumber\\[5pt]
& & \quad \begin{picture}(70,40)(0,17)
\Line(0,5)(70,5)
\CArc(35,20)(15,-180,180)
\Line(48,27)(70,40)
\end{picture} \quad + \quad
\begin{picture}(70,40)(0,17)
\Line(0,5)(70,5)
\DashCArc(35,20)(15,-180,180){3}
\Line(48,27)(70,40)
\end{picture} \quad + \nonumber\\[5pt]
& & \quad \begin{picture}(70,40)(0,17)
\Line(0,35)(70,35)
\CArc(35,20)(15,-180,180)
\Line(48,13)(70,0)
\end{picture} \quad + \quad 
\begin{picture}(70,40)(0,17)
\Line(0,35)(70,35)
\DashCArc(35,20)(15,-180,180){3}
\Line(48,13)(70,0)
\end{picture} \quad + \nonumber\\[5pt]
& & \quad \begin{picture}(50,40)(0,17)
\Line(0,20)(25,20)
\Line(25,20)(50,0)
\Line(25,20)(50,40)
\Vertex(25,20){3}
\end{picture} \nonumber\\[25pt]
&=& -{3m^2\over\hbar v} - {27m^6\over v^3} \; I(0,m,q_1,m,-q_3,m) \nonumber\\
& & -{m^6\over v^3} \; I(0,0,q_1,0,-q_3,0) \nonumber\\
& & +{9m^4\over2v^3} \; \left( I(0,m,q_1,m) + I(0,m,q_2,m) + I(0,m,q_3,m) \right) \nonumber\\
& & +{m^4\over2v^3} \; \left( I(0,0,q_1,0) + I(0,0,q_2,0) + I(0,0,q_3,0) \right) \nonumber\\
& & -{v\over\hbar} \; \delta_{\lambda}|_{\hbar}
\eqa
\bq
\langle \tilde{\eta}_1\left(q_1\right) \tilde{\eta}_1\left(q_2\right) \tilde{\eta}_2\left(q_3\right) \rangle_{\mathrm{1PI}} = 0
\eq
\bqa
\langle \tilde{\eta}_1\left(q_1\right) \tilde{\eta}_2\left(q_2\right) \tilde{\eta}_2\left(q_3\right) \rangle_{\mathrm{1PI}} &=& \quad
\begin{picture}(50,40)(0,17)
\Line(0,20)(25,20)
\DashLine(25,20)(50,0){3}
\DashLine(25,20)(50,40){3}
\end{picture} \quad + \nonumber\\[5pt]
& & \quad \begin{picture}(70,40)(0,17)
\Line(0,20)(20,20)
\DashCArc(35,20)(15,-30,30){3}
\CArc(35,20)(15,30,-30)
\DashLine(48,27)(70,40){3}
\DashLine(48,13)(70,0){3}
\end{picture} \quad + \quad
\begin{picture}(70,40)(0,17)
\Line(0,20)(20,20)
\DashCArc(35,20)(15,30,-30){3}
\CArc(35,20)(15,-30,30)
\DashLine(48,27)(70,40){3}
\DashLine(48,13)(70,0){3}
\end{picture} \quad + \nonumber\\[5pt]
& & \quad \begin{picture}(70,40)(0,17)
\Line(0,20)(20,20)
\CArc(35,20)(15,-180,180)
\DashLine(50,20)(70,40){3}
\DashLine(50,20)(70,0){3}
\end{picture} \quad + \quad
\begin{picture}(70,40)(0,17)
\Line(0,20)(20,20)
\DashCArc(35,20)(15,-180,180){3}
\DashLine(50,20)(70,40){3}
\DashLine(50,20)(70,0){3}
\end{picture} \quad + \nonumber\\[5pt]
& & \quad \begin{picture}(70,40)(0,17)
\Line(0,5)(35,5)
\DashLine(35,5)(70,5){3}
\CArc(35,20)(15,30,270)
\DashCArc(35,20)(15,-90,30){3}
\DashLine(48,27)(70,40){3}
\end{picture} \quad + \quad
\begin{picture}(70,40)(0,17)
\Line(0,35)(35,35)
\DashLine(35,35)(70,35){3}
\CArc(35,20)(15,90,-30)
\DashCArc(35,20)(15,-30,90){3}
\DashLine(48,13)(70,0){3}
\end{picture} \quad + \nonumber\\[5pt]
& & \quad \begin{picture}(50,40)(0,17)
\Line(0,20)(25,20)
\DashLine(25,20)(50,0){3}
\DashLine(25,20)(50,40){3}
\Vertex(25,20){3}
\end{picture} \nonumber\\[25pt]
&=& -{m^2\over\hbar v} \nonumber\\
& & -{3m^6\over v^3} \; I(0,m,q_1,m,-q_3,0) - {m^6\over v^3} \; I(0,0,q_1,0,-q_3,m) \nonumber\\
& & +{3m^4\over2v^3} \; I(0,m,q_1,m) + {3m^4\over2v^3}\; I(0,0,q_1,0) \nonumber\\
& & +{m^4\over v^3} \; I(0,m,q_2,0) + {m^4\over v^3} \; I(0,m,q_3,0) \nonumber\\
& & -{v\over3\hbar} \; \delta_{\lambda}|_{\hbar}
\eqa
\bq
\langle \tilde{\eta}_2\left(q_1\right) \tilde{\eta}_2\left(q_2\right) \tilde{\eta}_2\left(q_3\right) \rangle_{\mathrm{1PI}} = 0
\eq
\bqa
\langle \tilde{\eta}_1\left(q_1\right) \cdots \tilde{\eta}_1\left(q_4\right) \rangle_{\mathrm{1PI}} &=& \quad
\begin{picture}(40,40)(0,17)
\Line(0,0)(40,40)
\Line(0,40)(40,0)
\end{picture} \quad + \quad
\begin{picture}(70,40)(0,17)
\Line(0,40)(22,27)
\Line(0,0)(22,13)
\CArc(35,20)(15,-180,180)
\Line(48,27)(70,40)
\Line(48,13)(70,0)
\end{picture} \quad + \quad \textrm{2 perm's} \quad + \nonumber\\[5pt]
& & \quad \begin{picture}(70,40)(0,17)
\Line(0,40)(22,27)
\Line(0,0)(22,13)
\DashCArc(35,20)(15,-180,180){3}
\Line(48,27)(70,40)
\Line(48,13)(70,0)
\end{picture} \quad + \quad \textrm{2 perm's} \quad + \nonumber\\[5pt]
& & \quad \begin{picture}(70,40)(0,17)
\Line(0,40)(20,20)
\Line(0,0)(20,20)
\CArc(35,20)(15,-180,180)
\Line(48,27)(70,40)
\Line(48,13)(70,0)
\end{picture} \quad + \quad \textrm{5 perm's} \quad + \nonumber\\[5pt]
& & \quad \begin{picture}(70,40)(0,17)
\Line(0,40)(20,20)
\Line(0,0)(20,20)
\DashCArc(35,20)(15,-180,180){3}
\Line(48,27)(70,40)
\Line(48,13)(70,0)
\end{picture} \quad + \quad \textrm{5 perm's} \quad + \nonumber\\[5pt]
& & \quad \begin{picture}(70,40)(0,17)
\Line(0,40)(20,20)
\Line(0,0)(20,20)
\CArc(35,20)(15,-180,180)
\Line(50,20)(70,40)
\Line(50,20)(70,0)
\end{picture} \quad + \quad \textrm{2 perm's} \quad + \nonumber\\[5pt]
& & \quad \begin{picture}(70,40)(0,17)
\Line(0,40)(20,20)
\Line(0,0)(20,20)
\DashCArc(35,20)(15,-180,180){3}
\Line(50,20)(70,40)
\Line(50,20)(70,0)
\end{picture} \quad + \quad \textrm{2 perm's} \quad + \quad
\begin{picture}(40,40)(0,17)
\Line(0,0)(40,40)
\Line(0,40)(40,0)
\Vertex(20,20){3}
\end{picture} \nonumber\\[25pt]
&=& -{3m^2\over\hbar v^2} \nonumber\\
& & +{81m^8\over v^4} \; I(0,m,q_1,m,q_1+q_3,m,-q_2,m) + \textrm{2 perm's} \nonumber\\
& & +{m^8\over v^4} \; I(0,0,q_1,0,q_1+q_3,0,-q_2,0) + \textrm{2 perm's} \nonumber\\
& & -{27m^6\over v^4} \; I(0,m,q_3,m,q_3+q_4,m) + \textrm{5 perm's} \nonumber\\
& & -{m^6\over v^4} \; I(0,0,q_3,0,q_3+q_4,0) + \textrm{5 perm's} \nonumber\\
& & +{9m^4\over2v^4} \; I(0,m,q_3+q_4,m) + \textrm{2 perm's} \nonumber\\
& & +{m^4\over2v^4} \; I(0,0,q_3+q_4,0) + \textrm{2 perm's} \nonumber\\
& & -{1\over\hbar} \; \delta_{\lambda}|_{\hbar}
\eqa
\bq
\langle \tilde{\eta}_1\left(q_1\right) \tilde{\eta}_1\left(q_2\right) \tilde{\eta}_1\left(q_3\right) \tilde{\eta}_2\left(q_4\right) \rangle_{\mathrm{1PI}} = 0
\eq
\bqa
\langle \tilde{\eta_1}\left(q_1\right) \tilde{\eta}_1\left(q_2\right) \tilde{\eta}_2\left(q_3\right) \tilde{\eta}_2\left(q_4\right) \rangle_{\mathrm{1PI}} &=& \quad
\begin{picture}(40,40)(0,17)
\Line(0,0)(20,20)
\Line(0,40)(20,20)
\DashLine(20,20)(40,40){3}
\DashLine(20,20)(40,0){3}
\end{picture} \quad + \quad
\begin{picture}(70,40)(0,17)
\Line(0,40)(22,27)
\Line(0,0)(22,13)
\CArc(35,20)(15,30,-30)
\DashCArc(35,20)(15,-30,30){3}
\DashLine(48,27)(70,40){3}
\DashLine(48,13)(70,0){3}
\end{picture} \quad + \quad \textrm{1 perm} \quad + \nonumber\\[5pt]
& & \quad \begin{picture}(70,40)(0,17)
\Line(0,40)(22,27)
\Line(0,0)(22,13)
\CArc(35,20)(15,-30,30)
\DashCArc(35,20)(15,30,-30){3}
\DashLine(48,27)(70,40){3}
\DashLine(48,13)(70,0){3}
\end{picture} \quad + \quad \textrm{1 perm} \quad + \nonumber\\[5pt]
& & \quad \begin{picture}(70,40)(0,17)
\Line(0,40)(20,20)
\Line(0,0)(20,20)
\CArc(35,20)(15,30,-30)
\DashCArc(35,20)(15,-30,30){3}
\DashLine(48,27)(70,40){3}
\DashLine(48,13)(70,0){3}
\end{picture} \quad + \quad
\begin{picture}(70,40)(0,17)
\Line(0,40)(20,20)
\Line(0,0)(20,20)
\CArc(35,20)(15,-30,30)
\DashCArc(35,20)(15,30,-30){3}
\DashLine(48,27)(70,40){3}
\DashLine(48,13)(70,0){3}
\end{picture} \quad + \nonumber\\[5pt]
& & \quad \begin{picture}(70,40)(0,17)
\Line(0,40)(22,27)
\Line(0,0)(22,13)
\CArc(35,20)(15,-180,180)
\DashLine(50,20)(70,40){3}
\DashLine(50,20)(70,0){3}
\end{picture} \quad + \quad
\begin{picture}(70,40)(0,17)
\Line(0,40)(22,27)
\Line(0,0)(22,13)
\DashCArc(35,20)(15,-180,180){3}
\DashLine(50,20)(70,40){3}
\DashLine(50,20)(70,0){3}
\end{picture} \quad + \nonumber\\[5pt]
& & \quad \begin{picture}(70,40)(0,17)
\Line(0,40)(35,35)
\DashLine(35,35)(70,40){3}
\CArc(35,20)(15,90,-30)
\DashCArc(35,20)(15,-30,90){3}
\Line(0,0)(22,13)
\DashLine(48,13)(70,0){3}
\end{picture} \quad + \quad \textrm{3 perm's} \quad + \nonumber\\[5pt]
& & \quad \begin{picture}(70,40)(0,17)
\Line(0,40)(35,35)
\DashLine(35,35)(70,40){3}
\CArc(35,20)(15,-30,90)
\DashCArc(35,20)(15,90,-30){3}
\Line(0,0)(22,13)
\DashLine(48,13)(70,0){3}
\end{picture} \quad + \quad \textrm{3 perm's} \quad + \nonumber\\[5pt]
& & \quad \begin{picture}(70,40)(0,17)
\Line(0,40)(20,20)
\Line(0,0)(20,20)
\CArc(35,20)(15,-180,180)
\DashLine(50,20)(70,40){3}
\DashLine(50,20)(70,0){3}
\end{picture} \quad + \quad
\begin{picture}(70,40)(0,17)
\Line(0,40)(20,20)
\Line(0,0)(20,20)
\DashCArc(35,20)(15,-180,180){3}
\DashLine(50,20)(70,40){3}
\DashLine(50,20)(70,0){3}
\end{picture} \quad + \nonumber\\[5pt]
& & \quad \begin{picture}(70,40)(0,17)
\Line(0,40)(35,35)
\DashLine(35,35)(70,40){3}
\CArc(35,20)(15,90,-90)
\DashCArc(35,20)(15,-90,90){3}
\Line(0,0)(35,5)
\DashLine(35,5)(70,0){3}
\end{picture} \quad + \textrm{1 perm} \quad + \nonumber\\[5pt]
& & \quad \begin{picture}(40,40)(0,17)
\Line(0,0)(20,20)
\Line(0,40)(20,20)
\DashLine(20,20)(40,40){3}
\DashLine(20,20)(40,0){3}
\Vertex(20,20){3}
\end{picture} \nonumber\\[25pt]
&=& -{m^2\over\hbar v^2} \nonumber\\
& & +{9m^8\over v^4} \; I(0,m,q_1,m,q_1+q_3,0,-q_2,m) + \textrm{1 perm} \nonumber\\
& & +{m^8\over v^4} \; I(0,0,q_1,0,q_1+q_3,m,-q_2,0) + \textrm{1 perm} \nonumber\\
& & -{9m^6\over v^4} \; I(0,m,q_3,0,q_3+q_4,m) \nonumber\\
& & -{m^6\over v^4} \; I(0,0,q_3,m,q_3+q_4,0) \nonumber\\
& & -{9m^6\over v^4} \; I(0,m,q_1,m,-q_2,m) \nonumber\\
& & -{3m^6\over v^4} \; I(0,0,q_1,0,-q_2,0) \nonumber\\
& & -{3m^6\over v^4} \; I(0,0,q_4,m,q_2+q_4,m) + \textrm{3 perm's} \nonumber\\
& & -{m^6\over v^4} \; I(0,m,q_4,0,q_2+q_4,0) + \textrm{3 perm's} \nonumber\\
& & +{3m^4\over2v^4} \; I(0,m,q_3+q_4,m) \nonumber\\
& & +{3m^4\over2v^4} \; I(0,0,q_3+q_4,0) \nonumber\\
& & +{m^4\over v^4} \; I(0,0,q_2+q_4,m) + \textrm{1 perm} \nonumber\\
& & -{1\over3\hbar} \; \delta_{\lambda}|_{\hbar}
\eqa
\bq
\langle \tilde{\eta}_1\left(q_1\right) \tilde{\eta}_2\left(q_2\right) \ldots \tilde{\eta}_2\left(q_4\right) \rangle_{\mathrm{1PI}} = 0
\eq
\bqa
\langle \tilde{\eta}_2\left(q_1\right) \ldots \tilde{\eta}_2\left(q_4\right) \rangle_{\mathrm{1PI}} &=& \quad
\begin{picture}(40,40)(0,17)
\DashLine(0,0)(40,40){3}
\DashLine(0,40)(40,0){3}
\end{picture} \quad + \quad
\begin{picture}(70,40)(0,17)
\DashLine(0,40)(22,27){3}
\DashLine(0,0)(22,13){3}
\CArc(35,20)(15,30,150)
\DashCArc(35,20)(15,150,210){3}
\CArc(35,20)(15,210,-30)
\DashCArc(35,20)(15,-30,30){3}
\DashLine(48,27)(70,40){3}
\DashLine(48,13)(70,0){3}
\end{picture} \quad + \quad \textrm{2 perm's} \quad + \nonumber\\[5pt]
& & \quad \begin{picture}(70,40)(0,17)
\DashLine(0,40)(22,27){3}
\DashLine(0,0)(22,13){3}
\CArc(35,20)(15,-30,30)
\DashCArc(35,20)(15,30,150){3}
\CArc(35,20)(15,150,210)
\DashCArc(35,20)(15,210,-30){3}
\DashLine(48,27)(70,40){3}
\DashLine(48,13)(70,0){3}
\end{picture} \quad + \quad \textrm{2 perm's} \quad + \nonumber\\[5pt]
& & \quad \begin{picture}(70,40)(0,17)
\DashLine(0,40)(20,20){3}
\DashLine(0,0)(20,20){3}
\CArc(35,20)(15,30,-30)
\DashCArc(35,20)(15,-30,30){3}
\DashLine(48,27)(70,40){3}
\DashLine(48,13)(70,0){3}
\end{picture} \quad + \quad \textrm{5 perm's} \quad + \nonumber\\[5pt]
& & \quad \begin{picture}(70,40)(0,17)
\DashLine(0,40)(20,20){3}
\DashLine(0,0)(20,20){3}
\CArc(35,20)(15,-30,30)
\DashCArc(35,20)(15,30,-30){3}
\DashLine(48,27)(70,40){3}
\DashLine(48,13)(70,0){3}
\end{picture} \quad + \quad \textrm{5 perm's} \quad + \nonumber\\[5pt]
& & \quad \begin{picture}(70,40)(0,17)
\DashLine(0,40)(20,20){3}
\DashLine(0,0)(20,20){3}
\CArc(35,20)(15,-180,180)
\DashLine(50,20)(70,40){3}
\DashLine(50,20)(70,0){3}
\end{picture} \quad + \quad \textrm{2 perm's} \quad + \nonumber\\[5pt]
& & \quad \begin{picture}(70,40)(0,17)
\DashLine(0,40)(20,20){3}
\DashLine(0,0)(20,20){3}
\DashCArc(35,20)(15,-180,180){3}
\DashLine(50,20)(70,40){3}
\DashLine(50,20)(70,0){3}
\end{picture} \quad + \quad \textrm{2 perm's} \quad + \nonumber\\[5pt]
& & \quad \begin{picture}(40,40)(0,17)
\Line(0,0)(20,20)
\Line(0,40)(20,20)
\DashLine(20,20)(40,40){3}
\DashLine(20,20)(40,0){3}
\Vertex(20,20){3}
\end{picture} \nonumber\\[25pt]
&=& -{3m^2\over\hbar v^2} \nonumber\\
& & +{m^8\over v^4} \; I(0,0,q_1,m,q_1+q_3,0,-q_2,m) + \textrm{2 perm's} \nonumber\\
& & +{m^8\over v^4} \; I(0,m,q_1,0,q_1+q_3,m,-q_2,0) + \textrm{2 perm's} \nonumber\\
& & -{m^6\over v^4} \; I(0,m,q_3,0,q_3+q_4,m) + \textrm{5 perm's} \nonumber\\
& & -{3m^6\over v^4} \; I(0,0,q_3,m,q_3+q_4,0) + \textrm{5 perm's} \nonumber\\
& & +{m^4\over2v^4} \; I(0,m,q_3+q_4,m) + \textrm{2 perm's} \nonumber\\
& & +{9m^4\over2v^4} \; I(0,0,q_3+q_4,0) + \textrm{2 perm's} \nonumber\\
& & -{1\over\hbar} \; \delta_{\lambda}|_{\hbar}
\eqa}

Now, as in the case of the $N=1$ linear sigma model our theory contains three free parameters, $Z$, $\mu$ and $\lambda$, which have to be fixed by three renormalization conditions. We could try to use the same renormalization conditions\index{renormalization conditions!$N=2$ LSM} as in chapter \ref{chapN1sigmamodelcan} (\ref{rencond}) (with the $\eta$-lines replaced by $\eta_1$-lines):
\bqa
\begin{picture}(50,20)(0,17)
\Line(0,20)(20,20)
\GCirc(35,20){15}{0.5}
\end{picture} \quad &=& 0 \nonumber\\
\textrm{Res} \quad
\begin{picture}(70,40)(0,17)
\Line(0,20)(70,20)
\GCirc(35,20){15}{0.5}
\end{picture} \quad &=& \hbar \nonumber\\[5pt]
\begin{picture}(70,40)(0,17)
\Line(0,40)(70,0)
\Line(0,0)(70,40)
\GCirc(35,20){15}{0.5}
\Text(35,20)[cc]{1PI}
\end{picture} \quad &=& -{\lambda\over\hbar} \quad \textrm{at $q_1=\ldots=q_4=0$} \label{rencond2}
\eqa\\
This is \emph{not} a good idea however, because some of our amplitudes are singular at zero incoming momentum. These singularities are of course caused by loops with the \ind{Goldstone boson}. If we would set the 4-point 1PI amplitude to $-\lambda/\hbar$ at zero external momenta we would be absorbing \ind{infrared divergences}, which only occur for very specific external momenta, in the counter terms. The most straightforward thing to do now is change the renormalization point. However, this will complicate the calculations greatly.

What we shall do is just remove these infrared divergences from our counter term $\delta_{\lambda}$ by hand.

This is somewhat similar to what is done in Peskin and Schroeder \cite{Peskin}, there they work in the dimensional regularization scheme, in which the infrared divergences are invisible anyway. (Their renormalization point is at $s=4m^2, \; u=t=0$ however, but also at this point there occur IR divergences.)

If we strictly use the conditions (\ref{rencond2}) the counter terms become, up to order $\hbar$:
\bqa
\delta_{\mu}|_{\hbar} &=& {3\over2}{\hbar m^2\over v^2} \; I(0,m) + \h{\hbar m^2\over v^2} \; I(0,0) + {1\over6}v^2 \; \delta_{\lambda}|_{\hbar} \nonumber\\
\delta_{\lambda}|_{\hbar} &=& {27\over2}{\hbar m^4\over v^4} \; I(0,m,0,m) + {3\over2}{\hbar m^4\over v^4} \; I(0,0,0,0) + \nonumber\\
& & -162{\hbar m^6\over v^4} \; I(0,m,0,m,0,m) - 6{\hbar m^6\over v^4} \; I(0,0,0,0,0,0) + \nonumber\\
& & 243{\hbar m^8\over v^4} \; I(0,m,0,m,0,m,0,m) + 3{\hbar m^8\over v^4} \; I(0,0,0,0,0,0,0,0) \nonumber\\
\delta_Z|_{\hbar} &=& {9\over2}{\hbar m^4\over v^2} \; \left. {d\over dp^2} I(0,m,p,m) \right|_{p^2=-m_{\mathrm{ph,1}}^2} + \h{\hbar m^4\over v^2} \; \left. {d\over dp^2} I(0,0,p,0) \right|_{p^2=-m_{\mathrm{ph,1}}^2}
\eqa
Now we see that the second, fourth and sixth term in $\delta_{\lambda}|_{\hbar}$ contain infrared divergences, which we should \emph{not} include. ($I(0,0)$ Is \emph{not} IR divergent for dimensions greater than 2.) The easiest thing to do is introduce a mass in these terms, such that the infrared divergences are regularized. We shall just use $m$ for this mass, to keep the calculation as simple as possible. After this manual procedure the counter terms\index{counter terms!$N=2$ LSM} are:
\bqa
\delta_{\mu}|_{\hbar} &=& {3\over2}{\hbar m^2\over v^2} \; I(0,m) + \h{\hbar m^2\over v^2} \; I(0,0) + {5\over2}{\hbar m^4\over v^2} \; I(0,m,0,m) + \nonumber\\
& & -28{\hbar m^6\over v^2} \; I(0,m,0,m,0,m) + 41{\hbar m^8\over v^2} \; I(0,m,0,m,0,m,0,m) \nonumber\\
\delta_{\lambda}|_{\hbar} &=& 15{\hbar m^4\over v^4} \; I(0,m,0,m) - 168{\hbar m^6\over v^4} \; I(0,m,0,m,0,m) + \nonumber\\
& & 246{\hbar m^8\over v^4} \; I(0,m,0,m,0,m,0,m) \nonumber\\
\delta_Z|_{\hbar} &=& {9\over2}{\hbar m^4\over v^2} \; \left. {d\over dp^2} I(0,m,p,m) \right|_{p^2=-m_{\mathrm{ph,1}}^2} + \h{\hbar m^4\over v^2} \; \left. {d\over dp^2} I(0,0,p,0) \right|_{p^2=-m_{\mathrm{ph,1}}^2} \label{counttermsN2}
\eqa

The physical masses\index{physical mass} of the $\eta_1$- and $\eta_2$-particle, $m_{\mathrm{ph,1}}$ and $m_{\mathrm{ph,2}}$, can now be calculated from the Dyson summed propagators. The Dyson summed $\eta_1$ propagator is\index{Dyson summation}
\bq \label{eta1prop}
\frac{\hbar}{p^2+m^2-\hbar A_1(p^2)}
\eq
with
\bqa
A_1(p^2) &=& {9\over2}{m^4\over v^2} \; I(0,m,p,m) - {9\over2}{m^4\over v^2} p^2 \; \left. {d\over dp^2} I(0,m,p,m) \right|_{p^2=-m_{\mathrm{ph,1}}^2} + \nonumber\\
& & \h{m^4\over v^2} \; I(0,0,p,0) - \h{m^4\over v^2} p^2 \; \left. {d\over dp^2} I(0,0,p,0) \right|_{p^2=-m_{\mathrm{ph,1}}^2} + \nonumber\\
& & -5{m^4\over v^2} \; I(0,m,0,m) + 56{m^6\over v^2} \; I(0,m,0,m,0,m) + \nonumber\\
& & -82{m^8\over v^2} \; I(0,m,0,m,0,m,0,m) \;. \label{A1canon}
\eqa
The location of the pole of (\ref{eta1prop}) gives $-m_{\mathrm{ph,1}}^2$. Up to order $\hbar$ we can easily find this pole:
\bqa
m_{\mathrm{ph,1}}^2 &=& m^2 + 5{\hbar m^4\over v^2} \; I(0,m,0,m) + \nonumber\\
& & -{9\over2}{\hbar m^4\over v^2} \; \left. I(0,m,p,m) \right|_{p^2=-m^2} - \h{\hbar m^4\over v^2} \; \left. I(0,0,p,0) \right|_{p^2=-m^2} + \nonumber\\
& & -{9\over2}{\hbar m^6\over v^2} \; \left. {d\over dp^2} I(0,m,p,m) \right|_{p^2=-m^2} - \h{\hbar m^6\over v^2} \; \left. {d\over dp^2} I(0,0,p,0) \right|_{p^2=-m^2} + \nonumber\\
& & -56{\hbar m^6\over v^2} \; I(0,m,0,m,0,m) + 82{\hbar m^8\over v^2} \; I(0,m,0,m,0,m,0,m) \;.
\eqa
For $d\leq4$ this $m_{\mathrm{ph,1}}$ is finite, for $d>4$ it is not, which shows that the linear sigma model is non-renormalizable for $d>4$.

Likewise we can obtain the physical mass of the $\eta_2$ particle $m_{\mathrm{ph,2}}$. The Dyson summed $\eta_2$ propagator is:
\bq \label{eta2prop}
\frac{\hbar}{p^2-\hbar A_2(p^2)}
\eq
with
\bqa
A_2(p^2) &=& {m^4\over v^2} \; I(0,0,p,m) + {m^2\over v^2} \; I(0,m) - {m^2\over v^2} \; I(0,0) + \nonumber\\
& & -{9\over2}{m^4\over v^2} p^2 \; \left. {d\over dp^2} I(0,m,p,m) \right|_{p^2=-m_{\mathrm{ph,1}}^2} - \h{m^4\over v^2} p^2 \; \left. {d\over dp^2} I(0,0,p,0) \right|_{p^2=-m_{\mathrm{ph,1}}^2} \;. \label{A2canon}
\eqa
Again the pole of (\ref{eta2prop}) is easily found up to order $\hbar$:
\bq \label{mph2}
m_{\mathrm{ph,2}}^2 = -{\hbar m^4\over v^2} \; I(0,0,0,m) - {\hbar m^2\over v^2} \; I(0,m) + {\hbar m^2\over v^2} \; I(0,0) = 0
\eq
This is an illustration of the \ind{Goldstone theorem}, which states that in the case of spontaneous symmetry breaking the mass of the Goldstone boson remains zero at \emph{all} orders.

Actually things are a bit trickier than they look here. The above result for $m_{\mathrm{ph,2}}$ seems to hold for all dimensions below 5, where the theory is \ind{renormalizable}. This would mean that also for $d=1$, where we know that \emph{no} \ind{spontaneous symmetry breaking} can occur, $m_{\mathrm{ph,2}}$ would remain zero up to order $\hbar$. This is \emph{not} true in general. In general (\ref{mph2}) is wrong for $d=1$ because we put the momentum $p$, flowing through the propagator, to zero \emph{before} we have done the loop integral. Actually we have to compute the integral and only then put $p$ to zero. The two operations do not commute. In case of the $N=2$ linear sigma model it happens to be that (\ref{mph2}) is correct after all, the problem with setting $p$ to zero before doing the loop integrals only shows up in 2-loop integrals. One can explicitly verify that at 2-loop order $m_{\mathrm{ph,2}}$ is no longer zero for $d=1$. For $d>2$ (\ref{mph2}) is \emph{always} correct however, which is in complete agreement with the Goldstone theorem. (Remember that $d=2$ is a special case, see Coleman \cite{Coleman} and Coleman et al.\ \cite{Jackiw}.

\section{The Effective Potential}

Now we want to calculate the effective potential\index{effective potential!$N=2$ LSM}. We will again use the vacuum-graph formula (\ref{vacgraphform}), as we did in the $N=1$ linear sigma model. However our calculation will be much more involved now because we now have \emph{two} types of lines, which complicates how we connect the lines inside the loop.

To deal with this complication we consider the same vertex, of which a different set of legs is going to be part of the loop, as \emph{different}. In this way each 1-loop diagram is characterized by 8 numbers, each denoting the number of a certain type of vertices in the diagram. These numbers are defined as follows:
\vspace{-17pt}
\bq
\begin{array}{lll}
\begin{picture}(100, 40)(0, 17)
\Line(20, 20)(50, 20)
\Line(50, 20)(80, 0)
\Line(50, 20)(80, 40)
\end{picture}
\leftrightarrow n_3 &
\begin{picture}(100, 40)(0, 17)
\Line(20, 20)(50, 20)
\DashLine(50, 20)(80, 0){3}
\DashLine(50, 20)(80, 40){3}
\end{picture}
\leftrightarrow m_3 &
\begin{picture}(100, 40)(0, 17)
\DashLine(20, 20)(50, 20){3}
\DashLine(50, 20)(80, 0){3}
\Line(50, 20)(80, 40)
\end{picture}
\leftrightarrow q_3 \\[10pt]
\begin{picture}(100, 40)(0, 17)
\Line(20, 0)(50, 20)
\Line(20,40)(50,20)
\Line(50, 20)(80, 0)
\Line(50, 20)(80, 40)
\end{picture}
\leftrightarrow n_4 &
\begin{picture}(100, 40)(0, 17)
\Line(20, 0)(50, 20)
\Line(20,40)(50,20)
\DashLine(50, 20)(80, 0){3}
\DashLine(50, 20)(80, 40){3}
\end{picture}
\leftrightarrow m_4 &
\begin{picture}(100, 40)(0, 17)
\DashLine(20, 0)(50, 20){3}
\Line(20,40)(50,20)
\Line(50, 20)(80, 0)
\DashLine(50, 20)(80, 40){3}
\end{picture}
\leftrightarrow q_4 \\[10pt]
\begin{picture}(100, 40)(0, 17)
\DashLine(20, 0)(50, 20){3}
\DashLine(20,40)(50,20){3}
\DashLine(50, 20)(80, 0){3}
\DashLine(50, 20)(80, 40){3}
\end{picture}
\leftrightarrow p_4 & &
\begin{picture}(100, 40)(0, 17)
\DashLine(20, 0)(50, 20){3}
\DashLine(20,40)(50,20){3}
\Line(50, 20)(80, 0)
\Line(50, 20)(80, 40)
\end{picture}
\leftrightarrow r_4
\end{array}
\eq\\
Here it is understood that the legs pointing to the right are going to be part of the loop. Before we can write down the expression for the 1-loop effective potential, i.e.\ the sum of all 1-loop 1PI diagrams weighed with the appropriate factors, we have to know in how many ways we can connect the internal legs. If we denote the number of vertices that give two solid lines to go into the loop as $p$, the number of vertices that give two dashed lines as $q$ and the number of vertices that give one solid and one dashed line as $r$, then the number of ways to connect these vertices to give a loop is:
\bq
2^{p+q} \frac{\left({r\over2}+p-1\right)!}{\left({r\over2}-1\right)!} \frac{\left({r\over2}+q-1\right)!}{\left({r\over2}-1\right)!} (r-1)! 
\eq

In our case we have of course $p=n_3+n_4+r_4, \; q=m_3+m_4+p_4, \; r=q_3+q_4$. The 1-loop effective potential $V_1$ is now given by:
\bqa
V_1(\eta_1,\eta_2) &=& {1\over(2\pi)^d} \int \; d^dk \sum_{\underset{n_3+n_4+p_4+m_3+m_4+q_3+q_4+r_4\geq1}{n_3,n_4,p_4,m_3,m_4,q_3,q_4,r_4=0}}^{\infty} 3^{n_3} 6^{n_4} 6^{p_4} 2^{q_3} 4^{q_4} \nonumber\\
& & \left(-{3m^2\over\hbar v}\right)^{n_3} \left(-{3m^2\over\hbar v^2}\right)^{n_4} \left(-{3m^2\over\hbar v^2}\right)^{p_4} \left(-{m^2\over\hbar v}\right)^{m_3+q_3} \left(-{m^2\over\hbar v^2}\right)^{m_4+q_4+r_4} \nonumber\\
& & \left({1\over3!}\right)^{n_3} \left({1\over4!}\right)^{n_4} \left({1\over4!}\right)^{p_4} \left({1\over2!}\right)^{m_3+q_3} \left({1\over2!2!}\right)^{m_4+q_4+r_4} \nonumber\\
& & {1\over n_3!} {1\over n_4!} {1\over p_4!} {1\over m_3!} {1\over q_3!} {1\over m_4!} {1\over q_4!} {1\over r_4!} \sum_{n=0}^\infty \delta_{2n,q_3+q_4} \nonumber\\
& & \left(\frac{\hbar}{k^2}\right)^{m_3+m_4+p_4+\h q_3+\h q_4} \left(\frac{\hbar}{k^2+m^2}\right)^{n_3+n_4+r_4+\h q_3+\h q_4} \nonumber\\
& & 2^{n_3+n_4+r_4+m_3+m_4+p_4} \frac{(q_3+q_4-1)!}{(\h q_3+\h q_4-1)!^2} \nonumber\\
& & \left(\h q_3+\h q_4+n_3+n_4+r_4-1\right)! \left(\h q_3+\h q_4+m_3+m_4+p_4-1\right)! \nonumber\\
& & (2p_4+q_3+q_4+2r_4)! (n_3+2n_4+m_3+2m_4+q_4)! \nonumber\\
& & \bigg(-\hbar{1\over(2p_4+q_3+q_4+2r_4)!} {1\over(n_3+2n_4+m_3+2m_4+q_4)!} \cdot \nonumber\\
& & \phantom{\bigg(} \eta_1^{n_3+2n_4+m_3+2m_4+q_4} \eta_2^{2p_4+q_3+q_4+2r_4} \bigg) + \nonumber\\[10pt]
& & -\hbar \;
\begin{picture}(10,10)(0,7)
\Line(0,10)(10,10)
\Vertex(10,10){2}
\end{picture} \;\; \eta_1 - {\hbar\over2} \;
\begin{picture}(20,10)(0,7)
\Line(0,10)(20,10)
\Vertex(10,10){2}
\end{picture} \;\; \eta_1^2 - {\hbar\over2} \;
\begin{picture}(20,10)(0,7)
\DashLine(0,10)(20,10){3}
\Vertex(10,10){2}
\end{picture} \;\; \eta_2^2 - {\hbar\over6} \;
\begin{picture}(20,10)(0,7)
\Line(0,10)(10,10)
\Line(10,10)(20,0)
\Line(10,10)(20,20)
\Vertex(10,10){2}
\end{picture} \;\; \eta_1^3 - {\hbar\over2} \;
\begin{picture}(20,10)(0,7)
\Line(0,10)(10,10)
\DashLine(10,10)(20,0){3}
\DashLine(10,10)(20,20){3}
\Vertex(10,10){2}
\end{picture} \;\; \eta_1\eta_2^2 + \nonumber\\[5pt]
& & -{\hbar\over24} \;
\begin{picture}(20,10)(0,7)
\Line(0,0)(20,20)
\Line(0,20)(20,0)
\Vertex(10,10){2}
\end{picture} \;\; \eta_1^4 - {\hbar\over24} \;
\begin{picture}(20,10)(0,7)
\DashLine(0,0)(20,20){3}
\DashLine(0,20)(20,0){3}
\Vertex(10,10){2}
\end{picture} \;\; \eta_2^4 - {\hbar\over4} \;
\begin{picture}(20,10)(0,7)
\Line(0,0)(20,20)
\DashLine(0,20)(20,0){3}
\Vertex(10,10){2}
\end{picture} \;\; \eta_1^2\eta_2^2
\eqa

After a long calculation this can be written to:
\bqa
V_1(\f_1,\f_2) &=& {\hbar\over2} {1\over(2\pi)^d} \int \; d^dk \Bigg( \ln\left( 1 + {1\over k^2} \h{m^2\over v^2}(\f_1^2+\f_2^2-v^2) \right) + \nonumber\\
& & \phantom{{\hbar\over2} {1\over(2\pi)^d} \int \; d^dk \Bigg(} \ln\left( 1 + {1\over k^2+m^2} {3\over2}{m^2\over v^2}(\f_1^2+\f_2^2-v^2) \right) \Bigg) + \nonumber\\
& & -\h\left( \delta_\mu|_{\hbar}-{1\over6}v^2 \; \delta_\lambda|_{\hbar} \right) (\f_1^2+\f_2^2-v^2) + {1\over24} \; \delta_\lambda|_{\hbar} \; (\f_1^2+\f_2^2-v^2)^2 \nonumber\\ \label{effpotN2}
\eqa
This result is identical to what Peskin and Schroeder \cite{Peskin} find in their formula (11.74), of course taking into account differences in definitions of coupling constants and counter terms.

\subsection{Zero Dimensions}

There is a much quicker, though less straightforward, way to obtain the 1-loop effective potential (\ref{effpotN2}). In zero dimensions it is very easy to find the 1-loop effective action through the \ind{Schwinger-Dyson equations}. Of course in zero dimensions this effective action is equal to the effective potential. The diagrammatic structure of this 1-loop effective potential in zero dimensions is exactly the same as in $d$ dimensions, only the mathematical expressions corresponding to the diagrams is different. For the 1-loop case however the difference in mathematical expression is not so big: the propagators in the loop, which are $1/m^2$ in zero dimensions just become $1/(k^2+m^2)$ in $d$ dimensions. So if we are able to find the zero-dimensional 1-loop effective potential we can do this replacement to obtain the $d$-dimensional effective potential.

So we first have to calculate the zero-dimensional 1-loop effective potential through the Schwinger-Dyson equations. We write the zero-dimensional action of our $N=2$ linear sigma model generically as:
\bq
S = \h m_1\eta_1^2 + \h m_2\eta_2^2 + {1\over3!}g_1\eta_1^3 + {1\over2!}g_2\eta_1\eta_2^2 + {1\over4!}\lambda_1\eta_1^4 + {1\over4!}\lambda_2\eta_2^4 + {1\over2!2!}\lambda_3\eta_1^2\eta_2^2
\eq
Notice that we have included a mass $\sqrt{m_2}$ for the $\eta_2$-particle now, to be able to do the replacement $m_2\rightarrow k^2$ later. (Also for $m_2=0$ the propagator would not even exist in zero dimensions.)

In diagrammatic form the \ind{Schwinger-Dyson equations} read:
\bqa
\begin{picture}(20,20)(0,7)
\Line(0,10)(15,10)
\GCirc(15,10){5}{0.5}
\end{picture} &=&
\begin{picture}(20,20)(0,7)
\Line(0,10)(15,10)
\Line(13,8)(17,12)
\Line(13,12)(17,8)
\end{picture} + \h
\begin{picture}(30,20)(0,7)
\Line(0,10)(15,10)
\Line(15,10)(25,20)
\Line(15,10)(25,0)
\GCirc(25,20){5}{0.5}
\GCirc(25,0){5}{0.5}
\end{picture} + \h
\begin{picture}(45,20)(0,7)
\Line(0,10)(15,10)
\CArc(25,10)(10,0,360)
\GCirc(35,10){8}{0.5}
\end{picture} + \h
\begin{picture}(30,20)(0,7)
\Line(0,10)(15,10)
\DashLine(15,10)(25,20){2}
\DashLine(15,10)(25,0){2}
\GCirc(25,20){5}{0.5}
\GCirc(25,0){5}{0.5}
\end{picture} + \h
\begin{picture}(45,20)(0,7)
\Line(0,10)(15,10)
\DashCArc(25,10)(10,0,360){2}
\GCirc(35,10){8}{0.5}
\end{picture} + {1\over3!}
\begin{picture}(40,20)(0,7)
\Line(0,10)(30,10)
\Line(15,10)(25,20)
\Line(15,10)(25,0)
\GCirc(25,20){5}{0.5}
\GCirc(25,0){5}{0.5}
\GCirc(30,10){5}{0.5}
\end{picture} + \nonumber\\
& & \h \begin{picture}(35,20)(0,7)
\Line(0,10)(15,10)
\Line(15,10)(25,20)
\CArc(20,5)(7,0,360)
\GCirc(25,20){5}{0.5}
\GCirc(25,0){5}{0.5}
\end{picture} + {1\over3!}
\begin{picture}(45,20)(0,7)
\Line(0,10)(35,10)
\CArc(25,10)(10,0,360)
\GCirc(35,10){8}{0.5}
\end{picture} + \h
\begin{picture}(40,20)(0,7)
\Line(0,10)(30,10)
\DashLine(15,10)(25,20){2}
\DashLine(15,10)(25,0){2}
\GCirc(25,20){5}{0.5}
\GCirc(25,0){5}{0.5}
\GCirc(30,10){5}{0.5}
\end{picture} + \h
\begin{picture}(35,20)(0,7)
\Line(0,10)(15,10)
\Line(15,10)(25,20)
\DashCArc(20,5)(7,0,360){2}
\GCirc(25,20){5}{0.5}
\GCirc(25,0){5}{0.5}
\end{picture} +
\begin{picture}(35,20)(0,7)
\Line(0,10)(15,10)
\DashLine(15,10)(25,20){2}
\CArc(20,5)(7,0,142)
\DashCArc(20,5)(7,142,360){2}
\GCirc(25,20){5}{0.5}
\GCirc(25,0){5}{0.5}
\end{picture} + \h
\begin{picture}(45,20)(0,7)
\Line(0,10)(35,10)
\DashCArc(25,10)(10,0,360){2}
\GCirc(35,10){8}{0.5}
\end{picture} \nonumber\\[10pt]
\begin{picture}(20,20)(0,7)
\DashLine(0,10)(15,10){2}
\GCirc(15,10){5}{0.5}
\end{picture} &=&
\begin{picture}(20,20)(0,7)
\DashLine(0,10)(15,10){2}
\Line(13,8)(17,12)
\Line(13,12)(17,8)
\end{picture} +
\begin{picture}(30,20)(0,7)
\DashLine(0,10)(15,10){2}
\DashLine(15,10)(25,20){2}
\Line(15,10)(25,0)
\GCirc(25,20){5}{0.5}
\GCirc(25,0){5}{0.5}
\end{picture} +
\begin{picture}(45,20)(0,7)
\DashLine(0,10)(15,10){2}
\CArc(25,10)(10,180,360)
\DashCArc(25,10)(10,0,180){2}
\GCirc(35,10){8}{0.5}
\end{picture} + \h
\begin{picture}(40,20)(0,7)
\DashLine(0,10)(15,10){2}
\DashLine(15,10)(25,20){2}
\Line(15,10)(35,10)
\Line(15,10)(25,0)
\GCirc(25,20){5}{0.5}
\GCirc(25,0){5}{0.5}
\GCirc(30,10){5}{0.5}
\end{picture} + \h
\begin{picture}(35,20)(0,7)
\DashLine(0,10)(15,10){2}
\DashLine(15,10)(25,20){2}
\CArc(20,5)(7,0,360)
\GCirc(25,20){5}{0.5}
\GCirc(25,0){5}{0.5}
\end{picture} + 
\begin{picture}(35,20)(0,7)
\DashLine(0,10)(15,10){2}
\Line(15,10)(25,20)
\CArc(20,5)(7,0,142)
\DashCArc(20,5)(7,142,360){2}
\GCirc(25,20){5}{0.5}
\GCirc(25,0){5}{0.5}
\end{picture} + \nonumber\\
& & \h \begin{picture}(45,20)(0,7)
\DashLine(0,10)(35,10){2}
\CArc(25,10)(10,0,360)
\GCirc(35,10){8}{0.5}
\end{picture} + {1\over3!}
\begin{picture}(40,20)(0,7)
\DashLine(0,10)(30,10){2}
\DashLine(15,10)(25,20){2}
\DashLine(15,10)(25,0){2}
\GCirc(25,20){5}{0.5}
\GCirc(25,0){5}{0.5}
\GCirc(30,10){5}{0.5}
\end{picture} + \h
\begin{picture}(35,20)(0,7)
\DashLine(0,10)(15,10){2}
\DashLine(15,10)(25,20){2}
\DashCArc(20,5)(7,0,360){2}
\GCirc(25,20){5}{0.5}
\GCirc(25,0){5}{0.5}
\end{picture} + {1\over3!}
\begin{picture}(45,20)(0,7)
\DashLine(0,10)(35,10){2}
\DashCArc(25,10)(10,0,360){2}
\GCirc(35,10){8}{0.5}
\end{picture}
\eqa
Here the little crosses indicate the vertices from the sources, respectively $J_1/\hbar$ and $J_2/\hbar$. If we denote the tadpoles by $\phi(J_1,J_2)$ and $\psi(J_1,J_2)$:
\bq
\begin{picture}(20,20)(0,7)
\Line(0,10)(15,10)
\GCirc(15,10){5}{0.5}
\end{picture} \equiv \phi(J_1,J_2) \;, \quad
\begin{picture}(20,20)(0,7)
\DashLine(0,10)(15,10){2}
\GCirc(15,10){5}{0.5}
\end{picture} \equiv \psi(J_1,J_2) \;,
\eq
and their derivatives by
\bqa
\phi_{i_1i_2\ldots i_n} &\equiv& \frac{\partial^n}{\partial J_{i_1} \partial J_{i_2} \ldots \partial J_{i_n}} \; \phi \nonumber\\
\psi_{i_1i_2\ldots i_n} &\equiv& \frac{\partial^n}{\partial J_{i_1} \partial J_{i_2} \ldots \partial J_{i_n}} \; \psi \;,
\eqa
then the Schwinger-Dyson equations read
\bqa
m_1\phi &=& J_1 - \h g_1(\phi^2+\hbar\phi_1) - \h g_2(\psi^2+\hbar\psi_2) - {1\over6}\lambda_1(\phi^3+3\hbar\phi\phi_1+\hbar^2\phi_{11}) + \nonumber\\
& & -\h\lambda_3(\phi\psi^2+\hbar\phi\psi_2+2\hbar\psi\phi_2+\hbar^2\phi_{22}) \nonumber\\
m_2\psi &=& J_2 - g_2(\phi\psi+\hbar\phi_2) - \h\lambda_3(\phi^2\psi+\hbar\phi_1\psi+2\hbar\phi\phi_2+\hbar^2\phi_{12}) + \nonumber\\
& & -{1\over6}\lambda_2(\psi^3+3\hbar\psi\psi_2+\hbar^2\psi_{22})
\eqa

Now the definition of the effective action is
\bqa
\frac{\partial\Gamma}{\partial\phi}(\phi,\psi) &=& J_1(\phi,\psi) \nonumber\\
\frac{\partial\Gamma}{\partial\psi}(\phi,\psi) &=& J_2(\phi,\psi) \;,
\eqa
from which one can derive
\bqa
\phi_1 &=& \frac{-\frac{\partial^2\Gamma}{\partial\psi^2}}{\left(\left(\frac{\partial^2\Gamma}{\partial\phi\partial\psi}\right)^2 - \frac{\partial^2\Gamma}{\partial\phi^2}\frac{\partial^2\Gamma}{\partial\psi^2} \right)} \;, \nonumber\\
\psi_1 &=& \frac{\frac{\partial^2\Gamma}{\partial\phi\partial\psi}}{\left(\left(\frac{\partial^2\Gamma}{\partial\phi\partial\psi}\right)^2 - \frac{\partial^2\Gamma}{\partial\phi^2}\frac{\partial^2\Gamma}{\partial\psi^2} \right)} \;, \nonumber\\
\phi_2 &=& \frac{\frac{\partial^2\Gamma}{\partial\phi\partial\psi}}{\left(\left(\frac{\partial^2\Gamma}{\partial\phi\partial\psi}\right)^2 - \frac{\partial^2\Gamma}{\partial\phi^2}\frac{\partial^2\Gamma}{\partial\psi^2} \right)} \;, \nonumber\\
\psi_2 &=& \frac{-\frac{\partial^2\Gamma}{\partial\phi^2}}{\left(\left(\frac{\partial^2\Gamma}{\partial\phi\partial\psi}\right)^2 - \frac{\partial^2\Gamma}{\partial\phi^2}\frac{\partial^2\Gamma}{\partial\psi^2} \right)} \;.
\eqa
Through these relations we can write the Schwinger-Dyson equations in terms of (partial derivatives of) the effective action and the tadpole. Then it appears one can solve these partial differential equations iteratively up to some order to express the effective action in terms of the tadpole. Assuming that the effective action starts with a term of order $\hbar^0$, which is characteristic of the canonical approach, and writing
\bq
\Gamma(\phi,\psi) = A(\phi,\psi) + \hbar B(\phi,\psi) + \ldots\;
\eq
we find:
\bqa
A(\phi,\psi) &=& \h m_1\phi^2 + \h m_2\psi^2 + {1\over6}g_1\phi^3 + {1\over2}g_2\phi\psi^2 + {1\over24}\lambda_1\phi^4 + {1\over24}\lambda_2\psi^4 + {1\over4}\lambda_3\phi^2\psi^2 \nonumber\\
B(\phi,\psi) &=& \h \ln\left( \frac{\partial^2A}{\partial\phi^2} \frac{\partial^2A}{\partial\psi^2} - \left(\frac{\partial^2A}{\partial\phi\partial\psi}\right)^2 \right) + C \label{effactAB}
\eqa
Here the constant $C$ is just a constant of integration, which is unimportant for the physics. It is convenient to fix it however by demanding that for a free theory $B=0$, which gives $C=-\h\ln m_1m_2$.

Now to obtain the 1-loop effective potential in $d$ dimensions (excluding counter terms) we have to make the replacements
\bq
m_1 \rightarrow k^2+m^2 \;, \quad m_2 \rightarrow k^2 \;,
\eq
in $\hbar B(\phi,\psi)$ and add the integration ${1\over(2\pi)^d}\int d^dk$. 

If we do this, specify all the masses and coupling constants $m$, $g$ and $\lambda$ to the masses and coupling constants we have in the $N=2$ linear sigma model, and write the $\eta_1$- and $\eta_2$-field in terms of the $\f_1$- and $\f_2$-field again we find exactly (\ref{effpotN2}), of course excluding the counter terms.

Also notice that (\ref{effactAB}) shows in general (for a 2-field theory) that the effective potential becomes complex when the classical potential $A$ becomes non-convex. Inside the logarithm in $B$ in (\ref{effactAB}) is the Hessian of the function $A$, which is negative where the function $A$ is non-convex.

\subsection{Calculating The Effective Potential}

To proceed calculating (\ref{effpotN2}) we have to expand the logarithms again to let any divergent parts cancel the divergences in the counter terms. Of course when we expand the logarithm with the $1/k^2$ a lot of infrared divergences are going to appear. These divergences should later sum up to something finite again, but for the moment we have to regularize them, which we do by introducing a mass $\varepsilon$ for the $\eta_2$-particle. The 1-loop effective potential becomes:
\bqa
V_1 &=& \bigg( {1\over16}{\hbar m^4\over v^4} \left( I(0,m,0,m) - I(0,\varepsilon,0,\varepsilon) \right) - 7{\hbar m^6\over v^4} \; I(0,m,0,m,0,m) + \nonumber\\
& & \phantom{\bigg(} {41\over4}{\hbar m^8\over v^4} \; I(0,m,0,m,0,m,0,m) \bigg) (\f_1^2+\f_2^2-v^2)^2 + \nonumber\\
& & -{\hbar\over2} \sum_{n=3}^\infty {1\over n} \left( -\h{m^2\over v^2}(\f_1^2+\f_2^2-v^2) \right)^n {1\over(2\pi)^d} \int d^dk \; \frac{1}{(k^2+\varepsilon^2)^n} + \nonumber\\
& & -{\hbar\over2} \sum_{n=3}^\infty {1\over n} \left( -{3\over2}{m^2\over v^2}(\f_1^2+\f_2^2-v^2) \right)^n {1\over(2\pi)^d} \int d^dk \; \frac{1}{(k^2+m^2)^n}
\eqa
In the first term the ultraviolet divergences cancel, we can write this term as:
\bqa
& & I(0,m,0,m) - I(0,\varepsilon,0,\varepsilon) = \nonumber\\
& & \qquad 2(\varepsilon^2-m^2) \; I(0,m,0,m,0,\varepsilon) - (\varepsilon^2-m^2)^2 \; I(0,m,0,m,0,\varepsilon,0,\varepsilon)
\eqa

Now using that for $d\leq4$ and $n\geq3$ we have
\bq
{1\over(2\pi)^d} \int d^dk \; {1\over\left(k^2+m^2\right)^n} = {1\over(4\pi)^{d/2}} m^{d-2n} {\Gamma(n-d/2)\over\Gamma(n)} \;,
\eq
and for $d\leq4$ and $n+p\geq3$ we have
\bqa
& & {1\over(2\pi)^d} \int d^dk \; {1\over\left(k^2+m^2\right)^n} {1\over\left(k^2+\varepsilon^2\right)^p} = \nonumber\\
& & \qquad {1\over(4\pi)^{d/2}} \frac{\Gamma(n+p-d/2)}{\Gamma(n)\Gamma(p)} \int_0^1 dx \; x^{n-1} (1-x)^{p-1} (xm^2+(1-x)\varepsilon^2)^{d/2-n-p} \;, \nonumber\\
\eqa
we find for the 1-loop effective potential
\bqa
V_1 &=& \bigg( {1\over16}{\hbar m^4\over v^4} \left(m^{d-4}-\varepsilon^{d-4}\right) {1\over(4\pi)^{d/2}} \Gamma(2-d/2) - {7\over2}{\hbar m^d\over v^4} {1\over(4\pi)^{d/2}} \Gamma(3-d/2) + \nonumber\\
& & \phantom{\bigg(} {41\over24}{\hbar m^d\over v^4} {1\over(4\pi)^{d/2}} \Gamma(4-d/2) \bigg) (\f_1^2+\f_2^2-v^2)^2 + \nonumber\\
& & -{1\over(4\pi)^{d/2}} \varepsilon^d {\hbar\over2} \sum_{n=3}^\infty {1\over n!} \left( -\h{m^2\over v^2\varepsilon^2}(\f_1^2+\f_2^2-v^2) \right)^n \Gamma(n-d/2) + \nonumber\\
& & -{1\over(4\pi)^{d/2}} m^d {\hbar\over2} \sum_{n=3}^\infty {1\over n!} \left( -{3\over2}{1\over v^2}(\f_1^2+\f_2^2-v^2) \right)^n \Gamma(n-d/2) \;. \label{effpotN2ddim}
\eqa

We see that for $d\leq4$ all ultraviolet divergences cancel, which shows again that the theory is \ind{renormalizable} for $d\leq4$.

To find a more explicit expression for $V_1$ we have to specify the dimension $d$.

\subsection{$d=1$ And $d=2$}

If one substitutes $d=1$ in (\ref{effpotN2ddim}), performs the sums and works everything out one finds that the divergences for $\varepsilon\rightarrow0$ do \emph{not} cancel. The same happens for $d=2$. This is generally known, in one and two dimensions there is \emph{no} SSB\index{no spontaneous symmetry breaking}, which is manifested by the remaining \ind{infrared divergences}. See for example Coleman \cite{Coleman} and Coleman, Jackiw and Politzer \cite{Jackiw}.

\subsection{$d=4$}

In $d=4$ the infrared divergences do cancel and one finds:
\bqa
V_1 &=& -{3\over8}{\hbar m^4\over v^2}{1\over16\pi^2} (\f_1^2+\f_2^2-v^2) - {131\over48}{\hbar m^4\over v^4}{1\over16\pi^2} (\f_1^2+\f_2^2-v^2)^2 + \nonumber\\
& & {1\over16}{\hbar m^4\over v^4}{1\over16\pi^2} (\f_1^2+\f_2^2-v^2)^2 \ln\left( \h{1\over v^2}(\f_1^2+\f_2^2-v^2) \right) + \nonumber\\
& & {1\over4}\hbar m^4{1\over16\pi^2} \left( 1+{3\over2}{1\over v^2}(\f_1^2+\f_2^2-v^2) \right)^2 \ln\left( 1+{3\over2}{1\over v^2}(\f_1^2+\f_2^2-v^2) \right) \label{effpotN2d4}
\eqa
This result does not correspond to (11.79) in Peskin and Schroeder \cite{Peskin}, simply because we use \emph{different} renormalization conditions.

\begin{figure}[h]
\begin{center}
\epsfig{file=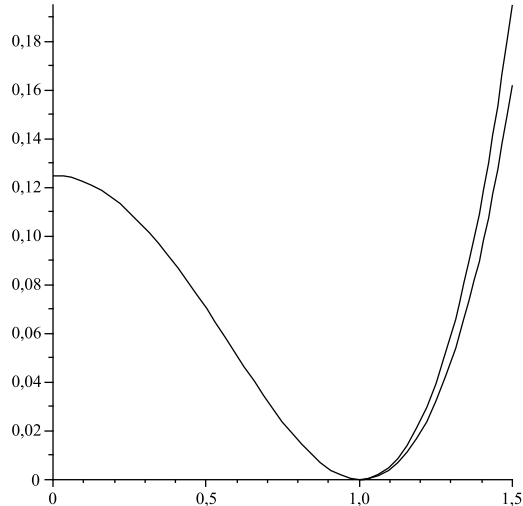,width=7cm}
\end{center}
\vspace{-1cm}
\caption{$V_0$ And $V=V_0+V_1$ as a function of $\sqrt{\f_1^2+\f_2^2}/v$ for $\hbar=2$, $m=1$, $v=1$.}
\label{effpotN21mind4}
\end{figure}

In figure \ref{effpotN21mind4} the complete effective potential\index{effective potential!$N=2$ LSM} (up to one loop) and the classical potential are plotted for the case $\hbar=2$, $m=1$, $v=1$.

The minimum of the effective potential is at $\f_1^2+\f_2^2=v^2$, as our tadpole renormalization condition in (\ref{rencond2}) ensures. Also exactly at this point the effective potential becomes complex because of the first logarithm in (\ref{effpotN2d4}). This shows again that the effective potential in the canonical approach becomes complex\index{effective potential!complexity} where the classical potential becomes non-convex. Because the effective potential becomes complex exactly at the location of the minima we \emph{cannot} compute the $n$-points Green's functions from it. This is related to the fact that all these $n$-points Green's functions suffer from \ind{infrared divergences} at zero incoming momentum.

Note that the effective potential we have computed here is \emph{convex} where it is defined. This is \emph{not} always the case. We could easily have chosen other renormalization conditions such that the minima of the effective potential occur for $\f_1^2+\f_2^2>v^2$ (by for example adding a constant term to $\delta_\mu|_{\hbar}$). Then there is a non-convex region between these minima and the circle $\f_1^2+\f_2^2=v^2$, where the effective potential becomes complex again. In fact in Peskin and Schroeder \cite{Peskin} such an effective potential is found in (11.79). They use the $\overline{MS}$ renormalization scheme. Their remark that fortunately the minima of the effective potential occur outside the region where it becomes complex is somewhat inappropriate, since we have shown here that this is \emph{not} always the case.

So also for the $N=2$ linear sigma model there is an apparent \ind{convexity problem}. Again, as has been thoroughly discussed in the $N=1$-case this problem is resolved by realizing that the canonical and path-integral approach are \emph{not} the same in the case of a non-convex classical action.


\chapter{The $N=2$ LSM: The Path-Integral Approach I}\label{chapN2sigmamodelpathI}

In this chapter we will discuss the path-integral approach to the Euclidean $N=2$ linear sigma model\index{$N=2$ linear sigma model!path-integral approach}. This means we want to calculate the path integral of this model in some approximation. For the $N=1$ linear sigma model we calculated the path integral (in chapter \ref{chapN1sigmamodelpath}) with a saddle-point approximation. This means we expand the generating functional around each minimum and add all these generating functionals to obtain the complete generating functional. In the $N=1$-case this is a good approximation because the minima lie far away from each other. In the $N=2$-case we can also use such a saddle-point approximation, however now the minima form a continuous set and do \emph{not} lie far apart. So it is questionable whether expanding around each minimum and then summing\index{summing over minima}, or rather integrating, the contributions from each minimum gives a reasonable approximation to the path integral.

Another questionable point is the \ind{perturbative expansion} around each minimum. When making this expansion one has replaced the, in principle damped, $\eta_2$-direction (i.e.\ tangential direction) by a non-damped straight line. There is an $\eta_2^4$-term that damps oscillations in the $\eta_2$-direction in principle, however in perturbation theory the exponential of this term is expanded, and \emph{not} all terms are kept. In this way we loose the damping effect in the tangential direction, which \emph{is} actually there.

In this chapter we shall just use this naive \ind{saddle-point approximation}, even though the arguments above advise strongly against it. There is also an argument in favor of this naive approach. We know that expanding around \emph{one} minimum (i.e.\ the canonical approach) gives a self-consistent theory and the Green's functions calculated in this way satisfy the \ind{Schwinger-Dyson equations}. Also the generating functional calculated by including only \emph{one} minimum satisfies the Schwinger-Symanzik equations. Because the Schwinger-Dyson and Schwinger-Symanzik equations are linear (in the \emph{full} Green's functions or generating functional) also the sum of several \emph{full} Green's functions or generating functionals around different minima are solutions to these equations. So we know at least that the full Green's functions and generating functional obtained by summing or integrating over \emph{all} minima are solutions to the Schwinger-Dyson and Schwinger-Symanzik equations. 

\section{Green's Functions}

The renormalized action of the $d$-dimensional $N=2$ linear sigma model is:
\bqa
S &=& \int d^dx \; \bigg( \h\left(\nabla\f_1\right)^2 + \h\left(\nabla\f_2\right)^2 - \h\mu\left(\f_1^2+\f_2^2\right) + {\lambda\over24}\left(\f_1^2+\f_2^2\right)^2 + \nonumber\\
& & \phantom{\int d^dx \; \bigg(} \h\delta_Z\left(\nabla\f_1\right)^2 + \h\delta_Z\left(\nabla\f_2\right)^2 - \h\delta_{\mu}\left(\f_1^2+\f_2^2\right) + {\delta_{\lambda}\over24}\left(\f_1^2+\f_2^2\right)^2 \bigg) \;.
\eqa
The minima\index{minima!$N=2$ LSM} of the first line are given by
\bqa
\f_1 &=& v\cos\delta \equiv v_1(\delta) \nonumber\\
\f_2 &=& v\sin\delta \equiv v_2(\delta) \;,
\eqa
with $v=\sqrt{6\mu/\lambda}$ again. Now we expand the action around \emph{one} of these minima:
\bq
\f_1 = v_1 + \eta_1 \;, \quad \f_2 = v_2 + \eta_2
\eq
When writing the action in terms of these $\eta$-fields the Gaussian part becomes non-diagonal in $\eta_1$ and $\eta_2$. To make this part diagonal again we introduce the $\psi$-fields:
\bqa
\psi_1 &=& {1\over v}(v_1\eta_1+v_2\eta_2) \nonumber\\
\psi_2 &=& {1\over v}(v_2\eta_1-v_1\eta_2)
\eqa
\bqa
\eta_1 &=& {1\over v}(v_1\psi_1+v_2\psi_2) \nonumber\\
\eta_2 &=& {1\over v}(v_2\psi_1-v_1\psi_2)
\eqa
In terms of these $\psi$-fields the action reads (again defining $\mu\equiv\h m^2$):
\bqa
S &=& \int d^dx \; \bigg( \h\left(\nabla\psi_1\right)^2 + \h\left(\nabla\psi_2\right)^2 + \h m^2\psi_1^2 + \nonumber\\
& & \phantom{\int d^dx \; \bigg(} \h{m^2\over v}\psi_1^3 + \h{m^2\over v}\psi_1\psi_2^2 + {1\over8}{m^2\over v^2}\psi_1^4 + {1\over8}{m^2\over v^2}\psi_2^4 + {1\over4}{m^2\over v^2}\psi_1^2\psi_2^2 + \nonumber\\
& & \phantom{\int d^dx \; \bigg(} \h\delta_Z\left(\nabla\psi_1\right)^2 + \h\delta_Z\left(\nabla\psi_2\right)^2 + \left(-v\delta_\mu+{1\over6}v^3\delta_\lambda\right)\psi_1 + \nonumber\\
& & \phantom{\int d^dx \; \bigg(} \left(-\h\delta_\mu+{1\over4}v^2\delta_\lambda\right)\psi_1^2 + \left(-\h\delta_\mu+{1\over12}v^2\delta_\lambda\right)\psi_2^2 + \nonumber\\
& & \phantom{\int d^dx \; \bigg(} {1\over6}v\delta_\lambda\psi_1^3 + {1\over6}v\delta_\lambda\psi_1\psi_2^2 + {\delta_\lambda\over24}\psi_1^4 + {\delta_\lambda\over24}\psi_2^4 + {\delta_\lambda\over12}\psi_1^2\psi_2^2 \bigg)
\eqa

Notice that this action does \emph{not} depend on $\delta$ anymore, as is expected from the $O(2)$-invariance of this model. Also notice that the action for the $\psi$-fields is exactly the same as the action for the $\eta$-fields in the canonical approach (\ref{N2etaaction}). This means the $\psi$-Green's functions are also identical to the $\eta$-Green's functions in the canonical approach, and for these Green's functions we can use the results from the previous chapter.

Now we wish to obtain the $\f$-Green's functions. As stated in the introduction we are going to calculate these by just integrating over the contributions from all minima, i.e.\ integrate over $\delta$. One should keep in mind here that the $\psi$-Green's functions do \emph{not} depend on $\delta$ anymore.\index{summing over minima}
\bqa
\langle \f_1(x) \rangle &=& {1\over2\pi} \int_{-\pi}^\pi d\delta \; \left( v_1 + \langle \eta_1(x) \rangle \right) = {1\over2\pi} \int_{-\pi}^\pi d\delta \; \left( v_1 + {v_1\over v}\langle \psi_1(x) \rangle + {v_2\over v} \langle \psi_2(x) \rangle \right) = 0 \nonumber\\
\langle \f_2(x) \rangle &=& {1\over2\pi} \int_{-\pi}^\pi d\delta \; \left( v_2 + \langle \eta_2(x) \rangle \right) = {1\over2\pi} \int_{-\pi}^\pi d\delta \; \left( v_2 + {v_2\over v}\langle \psi_1(x) \rangle - {v_1\over v} \langle \psi_2(x) \rangle \right) = 0 \nonumber\\
\eqa
\bqa
\langle \f_1(x)\f_1(y) \rangle &=& {1\over2\pi} \int_{-\pi}^\pi d\delta \; \langle \left(v_1 + \eta_1(x)\right)\left(v_1 + \eta_1(y)\right) \rangle \nonumber\\
&=& \h v^2 + \h v \langle\psi_1(x)\rangle + \h v \langle\psi_1(y)\rangle + \h \langle\psi_1(x)\psi_1(y)\rangle + \h \langle\psi_2(x)\psi_2(y)\rangle \nonumber\\
\langle \f_2(x)\f_2(y) \rangle &=& \h v^2 + \h v \langle\psi_1(x)\rangle + \h v \langle\psi_1(y)\rangle + \h \langle\psi_1(x)\psi_1(y)\rangle + \h \langle\psi_2(x)\psi_2(y)\rangle \nonumber\\
\langle \f_1(x)\f_2(y) \rangle &=& 0 \label{Cartavintermspsi}
\eqa
In this last line we also used $\langle\psi_2(x)\rangle = \langle\psi_1(x)\psi_2(y)\rangle = 0$.

With the results of the previous chapter it is now easy to obtain the $\f_1$- and $\f_2$-propagator up to 1-loop order. If we use the same counter terms as in the canonical approach we have, up to 1-loop order (using the tadpole renormalization condition (\ref{rencond}) and the Dyson summed propagators (\ref{eta1prop}) and (\ref{eta2prop})):
\bqa
\langle\psi_1(x)\rangle &=& 0 \nonumber\\
\langle\psi_1(x)\psi_1(y)\rangle &=& {1\over(2\pi)^d} \int d^dp \; e^{ip\cdot(x-y)} \; \frac{\hbar}{p^2+m^2-\hbar A_1(p^2)} \nonumber\\
\langle\psi_2(x)\psi_2(y)\rangle &=& {1\over(2\pi)^d} \int d^dp \; e^{ip\cdot(x-y)} \; \frac{\hbar}{p^2-\hbar A_2(p^2)}
\eqa
with $A_1$ and $A_2$ given in (\ref{A1canon}) and (\ref{A2canon}). Finally we find:\index{Green's functions!$N=2$ LSM}
\bqa
\langle \f_1(x)\f_1(y) \rangle = \langle \f_2(x)\f_2(y) \rangle &=& \h v^2 + \nonumber\\
& & \h {1\over(2\pi)^d} \int d^dp \; e^{ip\cdot(x-y)} \; \frac{\hbar}{p^2+m^2-\hbar A_1(p^2)} + \nonumber\\
& & \h {1\over(2\pi)^d} \int d^dp \; e^{ip\cdot(x-y)} \; \frac{\hbar}{p^2-\hbar A_2(p^2)} \;.
\eqa

With the formulas (\ref{Cartavintermspsi}) it is also easy to calculate the $\f_1$- and $\f_2$-propagator up to order $\hbar^2$. All we need more is $\langle\psi_1(x)\rangle$ at order $\hbar^2$. This quantity can easily be calculated with the Feynam rules from chapter \ref{chapN2sigmamodelcan}. In this case we shall not specify the counter terms, but keep them general. This will later be convenient when comparing the upcoming result for the $\f_1$- and $\f_2$-propagator to the result obtained from a calculation via the path integral in terms of polar field variables. $\langle\psi_1(x)\rangle$ At order $\hbar^2$ is now:
{\allowdisplaybreaks\bqa
\langle \tilde{\psi}_1 \rangle|_{\hbar^2} &=& \quad
\begin{picture}(70,40)(0,18)
\Line(0,20)(20,20)
\GCirc(55,20){15}{0.7}
\DashCArc(37,20)(17,50,180){3}
\DashCArc(37,20)(17,180,310){3}
\end{picture} \quad + \quad
\begin{picture}(70,40)(0,18)
\Line(0,20)(20,20)
\GCirc(55,20){15}{0.7}
\CArc(37,20)(17,50,180)
\CArc(37,20)(17,180,310)
\end{picture} \quad + \quad
\quad \begin{picture}(70,40)(0,18)
\Line(0,20)(40,20)
\GCirc(55,20){15}{0.7}
\CArc(20,30)(10,-90,270)
\end{picture} \quad + \nonumber\\[10pt]
& & \quad \begin{picture}(70,40)(0,18)
\Line(0,20)(40,20)
\GCirc(55,20){15}{0.7}
\DashCArc(20,30)(10,-90,270){3}
\end{picture} \quad + \quad
\quad \begin{picture}(70,40)(0,18)
\Line(0,20)(25,20)
\Line(25,20)(60,33)
\Line(25,20)(60,7)
\GCirc(60,33){10}{0.7}
\GCirc(60,7){10}{0.7}
\end{picture} \quad + \quad
\begin{picture}(70,40)(0,18)
\Line(0,20)(70,20)
\DashCArc(50,20)(20,-180,180){3}
\end{picture} \quad + \nonumber\\[10pt]
& & \quad \begin{picture}(70,40)(0,18)
\Line(0,20)(70,20)
\CArc(50,20)(20,-180,180)
\end{picture} \quad + \quad
\begin{picture}(70,40)(0,18)
\Line(0,20)(50,20)
\Vertex(50,20){3}
\end{picture} \quad + \quad
\begin{picture}(70,40)(0,18)
\Line(0,20)(40,20)
\GCirc(55,20){15}{0.7}
\Vertex(20,20){3}
\end{picture} \quad + \nonumber\\[10pt]
& & \quad \begin{picture}(70,40)(0,18)
\Line(0,20)(40,20)
\CArc(55,20)(15,0,360)
\Vertex(40,20){3}
\end{picture} \quad + \quad
\begin{picture}(70,40)(0,18)
\Line(0,20)(40,20)
\DashCArc(55,20)(15,-180,180){3}
\Vertex(40,20){3}
\end{picture} \nonumber\\[30pt]
&=& -{1\over8} {\hbar^2\over v^3} \; I(0,0)^2 - {3\over4} {\hbar^2\over v^3} \; I(0,0) I(0,m) - {9\over8} {\hbar^2\over v^3} \; I(0,m)^2 \nonumber\\
& & -{3\over2} {\hbar^2m^2\over v^3} \; I(0,0) I(0,m,0,m) - {9\over2} {\hbar^2m^2\over v^3} \; I(0,m) I(0,m,0,m) \nonumber\\
& & -\h {\hbar^2m^2\over v^3} \; I(0,m) I(0,0,0,0) + \h {\hbar^2m^2\over v^3} \; I(0,0) I(0,0,0,0) \nonumber\\
& & +\h{\hbar^2m^2\over v^3} \; D_{00m} + {3\over2} {\hbar^2m^2\over v^3} \; D_{mmm} - {3\over4} {\hbar^2m^4\over v^3} \; B_{m00} - {27\over4} {\hbar^2m^4\over v^3} \; B_{mmm} \nonumber\\
& & -\h {\hbar^2m^4\over v^3} \; B_{00m} \nonumber\\
& & +{3\over2} {\hbar\over v} \; I(0,m) \; \delta_Z|_{\hbar} + {3\over2} {\hbar\over m^2v} \; I(0,m) \; \delta_\mu|_{\hbar} - {1\over4} {\hbar v\over m^2} \; I(0,m) \; \delta_{\lambda}|_{\hbar} \nonumber\\
& & +\h {\hbar\over v} \; I(0,0) \; \delta_Z|_{\hbar} + \h {\hbar\over m^2v} \; I(0,0) \; \delta_\mu|_{\hbar} - {1\over12} {\hbar v\over m^2} \; I(0,0) \; \delta_{\lambda}|_{\hbar} \nonumber\\
& & -{3\over2} {\hbar m^2\over v} \; I(0,m,0,m) \; \delta_Z|_{\hbar} + 3 {\hbar\over v} \; I(0,m,0,m) \; \delta_\mu|_{\hbar} + \nonumber\\
& & -\h {v\over m^4} \; \left(\delta_\mu|_{\hbar}\right)^2 + {1\over24} {v^5\over m^4} \; \left(\delta_{\lambda}|_{\hbar}\right)^2 - {1\over6} {v^3\over m^4} \; \delta_\mu|_{\hbar} \delta_{\lambda}|_{\hbar} + {v\over m^2} \; \delta_\mu|_{\hbar^2} - {1\over6} {v^3\over m^2} \; \delta_{\lambda}|_{\hbar^2} \nonumber\\
\eqa}

Here we have defined the following new two-loop \ind{standard integrals}:
\bqa
D_{m_1m_2m_3} &\equiv& {1\over(2\pi)^{2d}} \int d^dk \; d^dl \; \frac{1}{k^2+m_1^2} \; \frac{1}{l^2+m_2^2} \; \frac{1}{(k-l)^2+m_3^2} \nonumber\\
B_{m_1m_2m_3} &\equiv& {1\over(2\pi)^{2d}} \int d^dk \; d^dl \; \frac{1}{(k^2+m_1^2)^2} \; \frac{1}{l^2+m_2^2} \; \frac{1}{(k-l)^2+m_3^2}
\eqa

Substituting this and the already obtained $\psi_1$- and $\psi_2$-propagator in (\ref{Cartavintermspsi}) gives:
{\allowdisplaybreaks\bqa
\langle \varphi_1(0) \varphi_1(x) \rangle_{\mathrm{c}} &=& +\h v^2 \nonumber\\
& & -\h \hbar \; I(0,0) + \h \hbar \; A_0(x) - {3\over2} \hbar \; I(0,m) + \h \hbar \; A_m(x) \nonumber\\
& & +{v^2\over m^2} \; \delta_\mu|_{\hbar} - {1\over6} {v^4\over m^2} \; \delta_{\lambda}|_{\hbar} \nonumber\\
& & +\h {\hbar^2m^2\over v^2} \; I(0,0) I(0,0,0,0) - \h {\hbar^2m^2\over v^2} \; I(0,m) I(0,0,0,0) \nonumber\\
& & -{3\over2} {\hbar^2m^2\over v^2} \; I(0,0) I(0,m,0,m) - {9\over2} {\hbar^2m^2\over v^2} \; I(0,m) I(0,m,0,m) \nonumber\\
& & +\h {\hbar^2m^2\over v^2} \; D_{00m} + {3\over2} {\hbar^2m^2\over v^2} \; D_{mmm} - \h {\hbar^2m^4\over v^2} \; B_{00m} - {3\over4} {\hbar^2m^4\over v^2} \; B_{m00} \nonumber\\
& & -{27\over4} {\hbar^2m^4\over v^2} \; B_{mmm} \nonumber\\
& & -\h {\hbar^2m^2\over v^2} \; I(0,0) C_{00}(x) + \h {\hbar^2m^2\over v^2} \; I(0,m) C_{00}(x) \nonumber\\
& & +\h {\hbar^2m^2\over v^2} \; I(0,0) C_{mm}(x) + {3\over2} {\hbar^2m^2\over v^2} \; I(0,m) C_{mm}(x) \nonumber\\
& & +\h {\hbar^2m^4\over v^2} \; B_{00m}(x) + {1\over4} {\hbar^2m^4\over v^2} \; B_{m00}(x) + {9\over4} {\hbar^2m^4\over v^2} \; B_{mmm}(x) \nonumber\\
& & +\h \hbar \; I(0,0) \; \delta_Z|_{\hbar} - \h \hbar \; A_0(x) \; \delta_Z|_{\hbar} + {3\over2} \hbar \; I(0,m) \; \delta_Z|_{\hbar} - \h \hbar \; A_m(x) \; \delta_Z|_{\hbar} \nonumber\\
& & -{3\over2} \hbar m^2 \; I(0,m,0,m) \; \delta_Z|_{\hbar} + 3 \hbar \; I(0,m,0,m) \; \delta_\mu|_{\hbar} \nonumber\\
& & +\h \hbar m^2 \; C_{mm}(x) \; \delta_Z|_{\hbar} - \hbar \; C_{mm}(x) \; \delta_\mu|_{\hbar} \nonumber\\
& & +{1\over18}{v^6\over m^4} \; \left(\delta_{\lambda}|_{\hbar}\right)^2 - {1\over3}{v^4\over m^4} \; \delta_\mu|_{\hbar} \; \delta_{\lambda}|_{\hbar} + {v^2\over m^2} \; \delta_\mu|_{\hbar^2} - {1\over6}{v^4\over m^2} \; \delta_{\lambda}|_{\hbar^2} \nonumber\\[5pt]
& & +\mathcal{O}\left(\hbar^3\right) \label{phi1propchap7}
\eqa}

Here we have defined another four \ind{standard integrals}:
\bqa
A_m(x) &\equiv& {1\over(2\pi)^d} \int d^dk \; e^{ik\cdot x} \; \frac{1}{k^2+m^2} \nonumber\\
C_{m_1m_2}(x) &\equiv& {1\over(2\pi)^d} \int d^dk \; e^{ik\cdot x} \; \frac{1}{k^2+m_1^2} \; \frac{1}{k^2+m_2^2} \nonumber\\
D_{m_1m_2m_3}(x) &\equiv& {1\over(2\pi)^{2d}} \int d^dk \; d^dl \; e^{ik\cdot x} \; \frac{1}{k^2+m_1^2} \; \frac{1}{l^2+m_2^2} \; \frac{1}{(k-l)^2+m_3^2} \nonumber\\
B_{m_1m_2m_3}(x) &\equiv& {1\over(2\pi)^{2d}} \int d^dk \; d^dl \; e^{ik\cdot x} \; \frac{1}{(k^2+m_1^2)^2} \; \frac{1}{l^2+m_2^2} \; \frac{1}{(k-l)^2+m_3^2}
\eqa

\section{The Effective Potential}

Now we will try to find the effective potential\index{effective potential!$N=2$ LSM} of the $N=2$ linear sigma model. To this end we introduce source terms in the action again. Because we are only interested in the effective \emph{potential} we shall take the sources to be constant over space time. Including these source terms the action is:
\bqa
S &=& \int d^dx \; \bigg( \h\left(\nabla\f_1\right)^2 + \h\left(\nabla\f_2\right)^2 - \h\mu\left(\f_1^2+\f_2^2\right) + {\lambda\over24}\left(\f_1^2+\f_2^2\right)^2 - J_1\f_1 - J_2\f_2 + \nonumber\\
& & \phantom{\int d^dx \; \bigg(} \h\delta_Z\left(\nabla\f_1\right)^2 + \h\delta_Z\left(\nabla\f_2\right)^2 - \h\delta_{\mu}\left(\f_1^2+\f_2^2\right) + {\delta_{\lambda}\over24}\left(\f_1^2+\f_2^2\right)^2 \bigg) \;.
\eqa

Now we have to find the minima of the first line again. Only for the case $J_1=J_2=0$ we have a ring of minima, as found in the previous section. For one of the sources non-zero however there is only \emph{one} minimum (and \emph{one} saddle point). This means that when both sources are of order $\hbar^0$ taking into account \emph{one} minimum is a good approximation. Below we shall show that taking into account \emph{one} minimum is equivalent to the canonical approach, outlined in the previous chapter. However, when the sources become of order $\hbar$ the minimum becomes so unstable that quantum fluctuations along the ring become important. Clearly in this regime it is a bad approximation to take into account only this single minimum, although it is the \emph{only} true minimum (for $J\neq0$). In this regime we have to take notice of all the points in the ring. What all the points in the ring have in common is that they are minima in $r=\sqrt{\f_1^2+\f_2^2}$. So to find these points, also for non-zero sources, we have to minimize the classical action with respect to $r$. Writing the classical field as:
\bqa
\f_1 &=& r\cos\delta \nonumber\\
\f_2 &=& r\sin\delta \;, \label{classpoints}
\eqa
we find the equation
\bq
-\mu r + {\lambda\over6}r^3 - J_1\cos\delta - J_2\sin\delta = 0 \;.
\eq
Writing
\bqa
J_1 &=& J\cos\beta \nonumber\\
J_2 &=& J\sin\beta
\eqa
and parameterizing $J$ as
\bq
J = {2\mu v\over3\sqrt{3}} \frac{\sin(3\alpha)}{\cos(\delta-\beta)} \;, \quad \alpha = 0,\ldots,{\pi\over6}
\eq
we find the solution
\bq \label{r}
r = {2v\over\sqrt{3}} \sin\left(\alpha+{\pi\over3}\right) \;.
\eq
So for each angle $\delta$ we have a point on the ring given by (\ref{r}).

Again we should expand the action around the classical points (\ref{classpoints}). To make the action diagonal we have to introduce the $\psi$-fields again:
\bqa
\psi_1 &=& \cos\delta \; \eta_1 + \sin\delta \; \eta_2 \nonumber\\
\psi_2 &=& \sin\delta \; \eta_1 - \cos\delta \; \eta_2
\eqa
The action becomes:
{\allowdisplaybreaks\bqa
S &=& \int d^dx \; \bigg( \h\mu r^2 - {3\over4}{\mu\over v^2}r^4 - \h\delta_\mu r^2 + {\delta_\lambda\over24}r^4 + \nonumber\\
& & \phantom{\int d^dx \; \bigg(} \h\left(\nabla\psi_1\right)^2 + \h\left(\nabla\psi_2\right)^2 + \left( -\h\mu+{3\over2}{\mu\over v^2}r^2 \right)\psi_1^2 + \nonumber\\
& & \phantom{\int d^dx \; \bigg(} -J\sin(\delta-\beta)\psi_2 + \left( -\h\mu+\h{\mu\over v^2}r^2 \right)\psi_2^2 + \nonumber\\
& & \phantom{\int d^dx \; \bigg(} {\mu\over v^2}r\psi_1^3 + {\mu\over v^2}r\psi_1\psi_2^2 + {1\over4}{\mu\over v^2}\psi_1^4 + {1\over4}{\mu\over v^2}\psi_2^4 + \h{\mu\over v^2}\psi_1^2\psi_2^2 + \nonumber\\
& & \phantom{\int d^dx \; \bigg(} \h\delta_Z\left(\nabla\psi_1\right)^2 + \h\delta_Z\left(\nabla\psi_2\right)^2 + \left(-\delta_\mu r+{1\over6}\delta_\lambda r^3\right) \psi_1 + \nonumber\\
& & \phantom{\int d^dx \; \bigg(} \left(-\h\delta_\mu+{1\over4}\delta_\lambda r^2\right) \psi_1^2 + \left(-\h\delta_\mu+{1\over12}\delta_\lambda r^2\right) \psi_2^2 + \nonumber\\
& & \phantom{\int d^dx \; \bigg(} {1\over6}\delta_\lambda r \psi_1^3 + {1\over6}\delta_\lambda r \psi_1\psi_2^2 + {1\over24}\delta_\lambda \psi_1^4 + {1\over24}\delta_\lambda \psi_2^4 + {1\over12}\delta_\lambda \psi_1^2\psi_2^2 \bigg)
\eqa}

Now we shall take the magnitude of the source $J=\sqrt{J_1^2+J_2^2}$ to be of order $\hbar$. To proceed further with the calculation one has to make an approximation. The most straightforward option is to treat all terms of order higher than $\hbar$ in the action as a perturbation. This means we should also expand $r$ in $\hbar$:
\bqa
r(\delta) &=& {2v\over\sqrt{3}} \sin\left( {1\over3}\arcsin\left({3\sqrt{3}\over2\mu v}J\cos(\delta-\beta)\right) + {\pi\over3} \right) \nonumber\\
&=& v + {1\over2\mu}J\cos(\delta-\beta) - {3\over8}{1\over v\mu^2}J^2\cos^2(\delta-\beta) + \mathcal{O}(\hbar^3) \label{r2}
\eqa
Then one can read off the Feynman rules from the action and calculate the generating functional and the $\psi_1$- and $\psi_2$-tadpole with Feynman diagrams. This is all straightforward, but at the end one finds an infrared-divergent expression. One might have expected this from the results of the previous chapter. There we saw that, in $d=3$ and $d=4$, the infrared divergences only sum up to something finite if we include \emph{all} 1-loop graphs. Because we take $J$ of order $\hbar$ here it means effectively that we \emph{cannot} calculate \emph{any} $n$-points Green's functions from our generating functional. For this one would need to know the exact $J$-dependence. This in turn means we are \emph{not} including all 1-loop graphs and we cannot expect the \ind{infrared divergences} to disappear.

Another thing one can do, which is less straightforward, but gives results without remaining infrared divergences, is ignore the term
\bq \label{discpsi2term}
-J\sin(\delta-\beta)\psi_2
\eq
in the action, because $J$ is small anyway. Then $\psi_1^2$ and $\psi_2^2$ are of order $\hbar$ and we shall only keep the Gaussian terms. Doing this we find for the generating functional:
\bqa
Z_{\delta} &=& \exp\left(-{1\over\hbar}\Omega\left( \h\mu r^2 - {3\over4}{\mu\over v^2}r^4 - \h\delta_\mu r^2 + {\delta_\lambda\over24}r^4 \right)\right) \cdot \nonumber\\
& & \int \mathcal{D}\psi_1 \mathcal{D}\psi_2 \; \exp\Bigg(-{1\over\hbar}\Bigg[ \h\left(\nabla\psi_1\right)^2 + \h\left(\nabla\psi_2\right)^2 + \nonumber\\
& & \phantom{\int \mathcal{D}\psi_1 \mathcal{D}\psi_2 \; \exp\Bigg(-{1\over\hbar}\Bigg(} \left( -\h\mu+{3\over2}{\mu\over v^2}r^2 \right)\psi_1^2 + \left( -\h\mu+\h{\mu\over v^2}r^2 \right)\psi_2^2 \Bigg]\Bigg) \nonumber\\
\eqa
Notice that this generating functional depends on $\delta$, as well as on the sources $J_1$ and $J_2$. With formula (\ref{Gaussianpathint}) one can compute $Z_\delta$ further. After some algebra one finds:
\bqa
Z_{\delta} &\sim& \exp\Bigg(-{1\over\hbar}\Omega\bigg[ V_0\big(r(\delta)\cos\delta, r(\delta)\sin\delta\big) + V_1\big(r(\delta)\cos\delta, r(\delta)\sin\delta\big) \nonumber\\
& & \phantom{\exp\Bigg(-{1\over\hbar}\Omega\bigg[} -J_1r(\delta)\cos\delta - J_2r(\delta)\sin\delta \bigg]\Bigg) \;,
\eqa
with $V_0$ the classical potential
\bq
V_0 = -\h\mu\left(\f_1^2+\f_2^2\right) + {\lambda\over24}\left(\f_1^2+\f_2^2\right)^2
\eq
and $V_1$ the 1-loop effective potential found in the canonical approach, given in (\ref{effpotN2}).

\subsection{Including One Minimum}

Now we can see what happens if, for some reason, we would only include the single minimum. For non-zero source this minimum is at $\delta=\beta$ and $r$ given by (\ref{r}). Notice that in this case it is correct to discard the term (\ref{discpsi2term}), because $\delta=\beta$. So in this case the generating functional is given by $Z_\beta$ and the $\f_1$- and $\f_2$-tadpole can be calculated as follows.
\bqa
\langle \f_1 \rangle(J_1,J_2) &=& {\hbar\over\Omega}\frac{\partial}{\partial J_1} \ln Z_\beta \nonumber\\
&=& \left(-\frac{\partial V_0}{\partial\f_1}-\frac{\partial V_1}{\partial\f_1}+J_1\right) \frac{\partial\f_1}{\partial J_1} + \left(-\frac{\partial V_0}{\partial\f_2}-\frac{\partial V_1}{\partial\f_2}+J_2\right) \frac{\partial\f_2}{\partial J_1} + \f_1 \nonumber\\
\langle \f_2 \rangle(J_1,J_2) &=& {\hbar\over\Omega}\frac{\partial}{\partial J_2} \ln Z_\beta \nonumber\\
&=& \left(-\frac{\partial V_0}{\partial\f_1}-\frac{\partial V_1}{\partial\f_1}+J_1\right) \frac{\partial\f_1}{\partial J_2} + \left(-\frac{\partial V_0}{\partial\f_2}-\frac{\partial V_1}{\partial\f_2}+J_2\right) \frac{\partial\f_2}{\partial J_2} + \f_2 \nonumber\\
\eqa
Using
\bqa
\frac{\partial V_{0/1}}{\partial\f_1} &=& \frac{\partial V_{0/1}}{\partial r}\cos\beta \nonumber\\
\frac{\partial V_{0/1}}{\partial\f_2} &=& \frac{\partial V_{0/1}}{\partial r}\sin\beta
\eqa
and
\bqa
\frac{\partial\f_1}{\partial J_1} &=& \frac{\partial r}{\partial J_1}\cos\beta + {r\over J}\sin^2\beta \nonumber\\
\frac{\partial\f_2}{\partial J_1} &=& \frac{\partial r}{\partial J_1}\sin\beta - {r\over J}\cos\beta\sin\beta \nonumber\\
\frac{\partial\f_1}{\partial J_2} &=& \frac{\partial r}{\partial J_2}\cos\beta - {r\over J}\cos\beta\sin\beta \nonumber\\
\frac{\partial\f_2}{\partial J_2} &=& \frac{\partial r}{\partial J_2}\sin\beta + {r\over J}\cos^2\beta
\eqa
one finds
\bqa
\langle \f_1 \rangle(J_1,J_2) &=& \f_1 - \frac{\partial V_1}{\partial J_1} \nonumber\\
\langle \f_2 \rangle(J_1,J_2) &=& \f_2 - \frac{\partial V_1}{\partial J_2} \;.
\eqa

These equations can easily be inverted, up to order $\hbar$, to obtain $J_1$ and $J_2$ as a function of $\langle \f_1 \rangle$ and $\langle \f_2 \rangle$. One finds:
\bqa
J_1(\f_1,\f_2) &=& \frac{\partial V_0}{\partial\f_1}(\f_1,\f_2) + \frac{\partial V_1}{\partial\f_1}(\f_1,\f_2) \nonumber\\
J_2(\f_1,\f_2) &=& \frac{\partial V_0}{\partial\f_2}(\f_1,\f_2) + \frac{\partial V_1}{\partial\f_2}(\f_1,\f_2)
\eqa
This can be integrated to give for the effective potential\index{effective potential!$N=2$ LSM}, up to order $\hbar$:
\bq
V(\f_1,\f_2) = V_0(\f_1,\f_2) + V_1(\f_1,\f_2) \;.
\eq

Indeed we see that including \emph{one} minimum in the path integral gives the canonical effective potential.

\subsection{Including All Minima}

Including \emph{all} minima, i.e.\ all points on the ring, means:\index{summing over minima}
\bq
Z = {1\over2\pi} \int_{-\pi}^\pi d\delta \; Z_\delta \;.
\eq
This generating functional can be calculated further. If we define the function $\mathcal{R}$ as
\bq
\mathcal{R}\left(J\cos(\delta-\beta)\right) = r(\delta) \;,
\eq
with $r$ defined in (\ref{r2}), the generating functional $Z$ can be written as
\bqa
Z &\sim& \int_{-\pi}^\pi d\delta \; \exp\bigg( -{1\over\hbar}\Omega\big( V_0\left(\mathcal{R}(J\cos(\delta-\beta))\right) + V_1\left(\mathcal{R}(J\cos(\delta-\beta))\right) + \nonumber\\
& & \phantom{\int_{-\pi}^\pi d\delta \; \exp\bigg( -{1\over\hbar}\Omega\big(} -\mathcal{R}(J\cos(\delta-\beta))J\cos(\delta-\beta) \big) \bigg) \nonumber\\
&\sim& \int_0^\pi d\delta \; \exp\bigg( -{1\over\hbar}\Omega\big( V_0\left(\mathcal{R}(J\cos\delta)\right) + V_1\left(\mathcal{R}(J\cos\delta)\right) + \nonumber\\
& & \phantom{\int_{-\pi}^\pi d\delta \; \exp\bigg( -{1\over\hbar}\Omega\big(} -\mathcal{R}(J\cos\delta)J\cos\delta \big) \bigg) \label{ZCartallmin}
\eqa

For the tadpoles we find
\bqa
\langle \f_1 \rangle(J_1,J_2) &=& {\hbar\over\Omega}\frac{\partial}{\partial J_1} \ln Z = {\hbar\over\Omega}\left(\frac{\partial J}{\partial J_1}\frac{\partial}{\partial J}+\frac{\partial\beta}{\partial J_1}\frac{\partial}{\partial\beta}\right) \ln Z = {\hbar\over\Omega} \cos\beta \frac{\partial}{\partial J} \ln Z \nonumber\\
\langle \f_2 \rangle(J_1,J_2) &=& {\hbar\over\Omega}\frac{\partial}{\partial J_2} \ln Z = {\hbar\over\Omega}\left(\frac{\partial J}{\partial J_2}\frac{\partial}{\partial J}+\frac{\partial\beta}{\partial J_2}\frac{\partial}{\partial\beta}\right) \ln Z = {\hbar\over\Omega} \sin\beta \frac{\partial}{\partial J} \ln Z \nonumber\\
\eqa
\bqa
& & \sqrt{\langle \f_1 \rangle^2(J_1,J_2)+\langle \f_2 \rangle^2(J_1,J_2)} = {\hbar\over\Omega}\left|\frac{\partial}{\partial J} \ln Z\right| = \nonumber\\
& & \hspace{70pt} \frac{\int_0^\pi d\delta \left[ \mathcal{R}(J\cos\delta) - \frac{\partial V_1}{\partial r}\left(\mathcal{R}(J\cos\delta)\right) \mathcal{R}'(J\cos\delta) \right]\cos\delta \;  \exp\left( \ldots \right)}{\int_0^\pi d\delta \; \exp\left( \ldots \right)} \nonumber\\ \label{phiCartallmin}
\eqa
In this last line the argument of the exponent is the same as in (\ref{ZCartallmin}).

We see that the magnitude of the $\f$-field only depends on the magnitude of the sources $J$, as expected because of the $O(2)$-symmetry.

Now this last expression is only valid for small $J$, because we discarded the term (\ref{discpsi2term}). So we will expand our result (\ref{phiCartallmin}) also in $J$ and keep all terms up to order $\hbar$. (Remember that $J$ is also of order $\hbar$.) We find:
\bq
\sqrt{\langle \f_1 \rangle^2+\langle \f_2 \rangle^2} = \frac{\int_0^\pi d\delta \left[ v\cos\delta + {J\over m^2}\cos^2\delta + \h{\Omega\over\hbar}{vJ^2\over m^2}\cos^3\delta \right] \exp\left( {\Omega vJ\over\hbar}\cos\delta \right)}{\int_0^\pi d\delta \left[ 1 + \h{\Omega\over\hbar}{J^2\over m^2}\cos^2\delta \right] \exp\left( {\Omega vJ\over\hbar}\cos\delta \right)}
\eq
This can be calculated analytically:
\bq
\sqrt{\langle \f_1 \rangle^2+\langle \f_2 \rangle^2} = \frac{vI_1\left({\Omega vJ\over\hbar}\right) + \h{J\over m^2}\left(I_0\left({\Omega vJ\over\hbar}\right)+I_2\left({\Omega vJ\over\hbar}\right)\right) + {1\over8}{\Omega\over\hbar}{vJ^2\over m^2}\left(3I_1\left({\Omega vJ\over\hbar}\right)+I_3\left({\Omega vJ\over\hbar}\right)\right)}{I_0\left({\Omega vJ\over\hbar}\right) + {1\over4}{\Omega\over\hbar}{J^2\over m^2}\left(I_0\left({\Omega vJ\over\hbar}\right)+I_2\left({\Omega vJ\over\hbar}\right)\right)}
\eq
Here $I_n(x)$ is the modified Bessel function of the first kind.

\begin{figure}[h]
\begin{center}
\epsfig{file=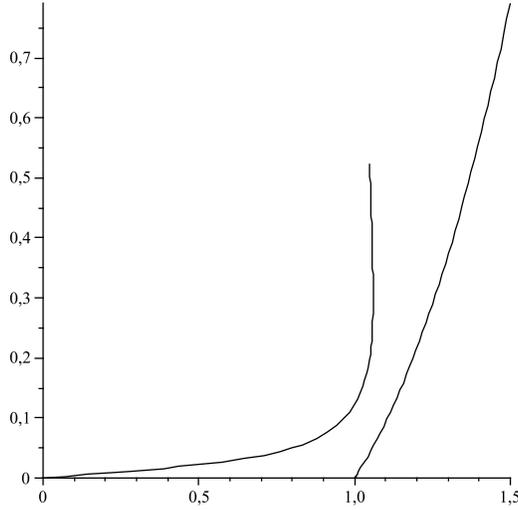,width=7cm}
\end{center}
\vspace{-1cm}
\caption{The derivative of the effective potential as a function of $\sqrt{\f_1^2+\f_2^2}/v$ for $\hbar=2$, $m=1$, $v=1$, $\Omega=100$. The left curve is only valid for small $J$, i.e. small $\sqrt{\f_1^2+\f_2^2}/v$, the right curve is only valid for large $J$.}
\label{effpotN2allmind4}
\end{figure}

This result is plotted in figure \ref{effpotN2allmind4}. The left curve is $J$, so the derivative of the effective potential, as a function of $\sqrt{\langle \f_1 \rangle^2+\langle \f_2 \rangle^2}$. The right curve is the derivative of the canonical effective potential as a function of $\sqrt{\f_1^2+\f_2^2}$. Both curves do not join at some point, the left curve is only valid for very small $J$, whereas the right curve is only valid for large $J$.\index{effective potential!$N=2$ LSM}

Apparently the way we calculate here, simply integrating over the ring of minima (in $r$), is \emph{not} a good way to cover the whole range of $J$, from small $J$ of order $\hbar$, to $J$ of order 1. However we \emph{do} find that the effective potential has a flat bottom in the limit $\Omega\rightarrow\infty$.


\chapter{Path Integrals in Polar Variables}\label{chappathintpol}

\section{Introduction}

In the next chapter we shall calculate the Green's functions in the $N=2$ linear sigma model via the path integral in terms of \ind{polar field variables}. These polar variables are the natural variables to describe an $O(2)$-invariant model. Before we can do these calculations however, we have to know how to transform to polar variables in a $d$-dimensional path integral.

When dealing with a normal integral a common technique to solve it is to transform to a different integration variable. Since the path integral is merely an infinite-dimensional integral, one should also be able to transform to different integration variables in this case. In particular, in the case of a path integral over two fields, $\f_1$ and $\f_2$, one can transform to polar field variables, $r$ and $\theta$, defined as:\index{transformation to polar fields}
\bqa
\f_1(x) &=& r(x)\cos\theta(x) \nonumber\\
\f_2(x) &=& r(x)\sin\theta(x) \;.
\eqa
It is this transformation to polar field variables that we shall study in this chapter. We wish to emphasize here that \emph{not} the space-time coordinates, but the quantum fields $\f_1$ and $\f_2$ are transformed to polar fields $r$ and $\theta$.

The infinite dimensionality of the path integral makes such a variable transformation very complicated. Several difficult questions immediately pop up:
\begin{enumerate}
\item In principle the whole path integral is only defined on a \ind{lattice}, so the transformation should also be done with the path integral in discrete form. This means one \emph{cannot} simply let the transformation work on the continuous action. Instead one has to write out the action in discrete form and only then let the transformation work. After this one gets a complicated action, with also terms proportional to the lattice spacing $\Delta$. These terms \emph{cannot} be discarded, since one has to perform the path integration first and then take the continuum limit $\Delta\rightarrow\infty$. It is not obvious that the terms proportional to $\Delta$ will not give a finite contribution in this continuum limit. In fact, in this chapter we will see that they \emph{do} give finite contributions to Green's functions.
\item After the transformation the domain of integration is not $\langle-\infty,\infty\rangle$. For the $r$-variables it is $[0,\infty\rangle$, whereas for the $\theta$-variables it is $[-\pi,\pi]$. How does one evaluate such a path integral, especially because we can only compute path integrals with perturbation theory? To do perturbation theory we have to be able to identify a Gaussian part, and the fields in such a Gaussian part are always integrated from $-\infty$ to $\infty$.
\item After the transformation one gets a \ind{Jacobian}, how does one deal with this? Since we can only do perturbation theory we also have to identify the Gaussian part and the perturbative part of this Jacobian.
\end{enumerate}

From these questions it is clear that transforming to polar variables in a $d$-dimensional path integral is very complicated. For 1-dimensional systems, i.e.\ quantum-mechanical systems, there is quite some literature on the transformation to polar variables.

In his textbook \cite{Lee} Lee derives the quantum-mechanical path integral in curvilinear coordinates in chapter 19. The result (19.49) is a path integral with a new action $L_{\textrm{eff}}$, which is \emph{not} equal to the action one would find by transforming to polar coordinates in the \emph{continuum} action (in Cartesian coordinates).

Edwards et al.\ \cite{Edwards} and Peak et al.\ \cite{Peak} also transform to polar coordinates in the discrete quantum-mechanical path integral. They find that terms of order $\Delta$ or higher, which arise when transforming to polar coordinates in the discrete action (in Cartesian coordinates), \emph{cannot} all be neglected.

\section{A Conjecture}

From the above it should be clear that transforming a path integral in terms of the normal (i.e.\ Cartesian) fields $\f_1$ and $\f_2$ to a path integral in terms of polar fields is far from trivial. To be able to do computations at all we will present a conjecture in this section. In the next sections we will then try to make this conjecture plausible by considering certain toy models where we can see that the conjecture actually works.

The generic form of a \ind{$d$-dimensional Euclidean path integral} $P$ in two fields is:
\bqa
& & P = \int \mathcal{D}\f_1\mathcal{D}\f_2 \; \f_1(x_1)\cdots\f_1(x_n) \; \f_2(y_1)\cdots\f_2(y_m) \cdot \nonumber\\
& & \hspace{70pt} \exp\left(-{1\over\hbar}\int d^dx\left( \h\left(\nabla\f_1\right)^2 + \h\left(\nabla\f_2\right)^2 + V(\f_1,\f_2) \right)\right) \label{genericpathint}
\eqa
Very naively, one could hope that the transformation to polar variables\index{transformation to polar fields} works as:
\bqa
\int_{-\infty}^\infty \mathcal{D}\f_1 \int_{-\infty}^\infty \mathcal{D}\f_2 &\rightarrow& \int_{-\infty}^\infty \mathcal{D}r \; \prod_x r(x) \int_{-\infty}^\infty \; \mathcal{D}\theta \nonumber\\
\f_1(x) &\rightarrow& r(x)\cos\theta(x) \nonumber\\
\f_2(x) &\rightarrow& r(x)\sin\theta(x) \nonumber\\
\int d^dx\left( \h\left(\nabla\f_1\right)^2 + \h\left(\nabla\f_2\right)^2 \right) &\rightarrow& \int d^dx\left( \h\left(\nabla r\right)^2 + \h r^2\left(\nabla\theta\right)^2 \right) \nonumber\\
\int d^dx \;  V(\f_1,\f_2) &\rightarrow& \int d^dx \; V(r\cos\theta,r\sin\theta)
\eqa
Here one has just extended the integration domains for $r$ and $\theta$ to $\langle-\infty,\infty\rangle$. In the fourth line one has just transformed the continuum action to an action in terms of polar fields, disregarding the fact that one should do this on the lattice, where the path integral is defined.

To make all these expressions completely continuous we still have to do something about the \ind{Jacobian} factor
\bq
\prod_x r(x) \;,
\eq
since this is still a lattice expression. We shall write this Jacobian as:
\bqa
\prod_x r(x) &=& \prod_x \exp\left(-{1\over\hbar}\left(-\hbar\ln r(x)\right)\right) \nonumber\\
&=& \exp\left(-{1\over\hbar} \sum_x \left(-\hbar\ln r(x)\right)\right) \nonumber\\
&\rightarrow& \exp\left(-{1\over\hbar}{1\over\Delta^d} \int d^dx \left(-\hbar\ln r(x)\right)\right) \nonumber\\
&\rightarrow& \exp\left(-{1\over\hbar} \left[{1\over(2\pi)^d}\int d^dk\right] \int d^dx \left(-\hbar\ln r(x)\right)\right)
\eqa
This is a continuum expression, which looks a lot like the exponential of an action. 

So, we might hope that the continuum form of a path integral in polar field variables is given by making the substitutions
\bqa
\int_{-\infty}^\infty \mathcal{D}\f_1 \int_{-\infty}^\infty \mathcal{D}\f_2 &\rightarrow& \int_{-\infty}^\infty \mathcal{D}r \; \int_{-\infty}^\infty \; \mathcal{D}\theta \cdot \nonumber\\
& & \quad \exp\left(-{1\over\hbar} \left[{1\over(2\pi)^d}\int d^dk\right] \int d^dx \left(-\hbar\ln r(x)\right)\right) \nonumber\\
\f_1(x) &\rightarrow& r(x)\cos\theta(x) \nonumber\\
\f_2(x) &\rightarrow& r(x)\sin\theta(x) \nonumber\\
\int d^dx\left( \h\left(\nabla\f_1\right)^2 + \h\left(\nabla\f_2\right)^2 \right) &\rightarrow& \int d^dx\left( \h\left(\nabla r\right)^2 + \h r^2\left(\nabla\theta\right)^2 \right) \nonumber\\
\int d^dx \;  V(\f_1,\f_2) &\rightarrow& \int d^dx \; V(r\cos\theta,r\sin\theta) \label{conjecture}
\eqa
in the path integral in Cartesian form.

Now the \ind{conjecture} we are going to make is:\\[15pt]
\fbox{\parbox{15.64cm}{\textbf{Conjecture} \;\; \emph{It is correct to transform to polar variables naively, as in (\ref{conjecture}), provided one does the calculation in a $d$-dimensional way.}}}\\[15pt]
This seems a very strange conjecture, and with all the remarks we made in the introduction it is hard to imagine how it can work. However in the next two sections we shall demonstrate that indeed this conjecture is true for two toy models. Also there we will demonstrate what is meant exactly by `calculating in a $d$-dimensional way'.\index{calculating in a $d$-dimensional way}

These two toy models are selected with the following two criteria:
\begin{enumerate}
\item The model should have \emph{one} minimum.
\item This minimum should \emph{not} be at $\f_1=0,\;\f_2=0$.
\end{enumerate}
The reason for the first criterion is that we want a toy model where we can calculate Green's functions both through the path integral in normal (Cartesian) fields \emph{and} through the path integral in polar fields. Only then can we check whether the conjecture, used in the calculation through polar fields, works. Notice that the $N=2$ linear sigma model does not satisfy the first criterion, there we have an infinite set of minima. In that case it is not clear whether a calculation through the Cartesian path integral (as done in the previous chapter) is correct. The reason for the second criterion is that at $\f_1=0,\;\f_2=0$, i.e.\ $r=0$, the transformation to polar fields becomes problematic. To be on the safe side we simply avoid these difficulties by only considering toy models which have their minimum away from $r=0$, such that we can do perturbation theory around a point where the transformation is well-defined.

\section{The Shifted Toy Model}

In this section we shall calculate several Green's functions in the so-called \ind{shifted toy model}. This model has an action\index{action!shifted toy model}
\bq
S = \int d^dx\left( \h\left(\nabla\f_1\right)^2 + \h\left(\nabla\f_2\right)^2 + \h m^2(\f_1-v)^2 + \h m^2\f_2^2 \right) \;.
\eq
This is just the action of a free model with the $\f_1$-field shifted, hence the name. We shifted this field such that the minimum of the action is at $\f_1=v,\;\f_2=0$.

To prove that the conjecture indeed works in the case of this model we shall now calculate several Green's functions through the normal, Cartesian, path integral \emph{and} through the path integral in terms of polar fields. Then we can compare results.

\subsection{Cartesian Results}

Because the shifted toy model is just a free theory with one field shifted it is very easy to obtain the \emph{exact full} Green's functions. They are:\index{Green's functions!shifted toy model}
\bqa
\langle \f_1(x) \rangle &=& v \nonumber\\
\langle \f_2(x) \rangle &=& 0 \nonumber\\
\langle \f_1(x) \f_1(y) \rangle &=& v^2 + {\hbar\over(2\pi)^d} \int d^dk \; \frac{e^{ik \cdot (x-y)}}{k^2+m^2} \nonumber\\
&=& v^2 + \hbar \; A_m(x-y) \nonumber\\
\langle \f_2(x) \f_2(y) \rangle &=& {\hbar\over(2\pi)^d} \int d^dk \; \frac{e^{ik \cdot (x-y)}}{k^2+m^2} \nonumber\\
&=& \hbar \; A_m(x-y) \nonumber\\
\langle \f_1(x) \f_2(y) \rangle &=& 0
\eqa

\subsection{Polar Results}

Now we perform the transformation to polar fields in the path integral:
\bqa
\f_1(x) &=& r(x)\cos\left({w(x)\over v}\right) \nonumber\\
\f_2(x) &=& r(x)\sin\left({w(x)\over v}\right) \;.
\eqa
Here we have used $w(x)/v$ instead of $\theta(x)$ to have a new angular field-variable with also the dimensions of a field. This is purely a matter of convenience.

To calculate the Green's functions through the path integral in polar fields now we will use the conjecture. According to this conjecture, the new action we have to work with is:
\bq
S = \int d^dx \left( {1\over2} \left( \nabla r(x) \right)^2 + {1\over2v^2} r^2(x) \left( \nabla w(x) \right)^2 + {1\over2}m^2 r^2(x) - m^2v r(x) \cos(w(x)/v) \right).
\eq
The minimum of the action is at $r(x)=v$ and $w(x)=0$. Expanding around $r(x)=v$ and writing
\bq
r(x) \equiv v + \eta(x)
\eq
we get
\bqa
S &=& \int d^dx \bigg( {1\over2} \left( \nabla \eta(x) \right)^2 + {1\over2} \left( \nabla w(x) \right)^2 + {1\over v} \eta(x) \left( \nabla w(x) \right)^2 + {1\over2v^2} \eta^2(x) \left( \nabla w(x) \right)^2 + \nonumber\\
& & \phantom{\int d^dx \bigg(} m^2v \eta(x) + {1\over2}m^2 \eta^2(x) - m^2v^2 \cos(w(x)/v) - m^2v \eta(x) \cos(w(x)/v) \bigg) \;. \nonumber\\
\eqa
Expanding also around $w(x)=0$, i.e.\ expanding the cosine gives:
\bqa
S &=& \int d^dx \bigg( {1\over2} \left( \nabla \eta(x) \right)^2 + {1\over2} \left( \nabla w(x) \right)^2 + {1\over v} \eta(x) \left( \nabla w(x) \right)^2 + {1\over2v^2} \eta^2(x) \left( \nabla w(x) \right)^2 + \nonumber\\
& & \phantom{\int d^dx \bigg(} {1\over2}m^2 \eta^2(x) + {1\over2}m^2 w^2(x) - {m^2\over24v^2} w^4(x) + {m^2\over2v} \eta(x) w^2(x) - {m^2\over24v^3} \eta(x) w^4(x) + \nonumber\\
& & \phantom{\int d^dx \bigg(} \mathcal{O}(\hbar^3) \bigg).
\eqa

The Jacobian gives, according to the conjecture:
\bqa
& & \exp\left(-{1\over\hbar} \left[{1\over(2\pi)^d}\int d^dk\right] \int d^dx \left(-\hbar\ln r(x)\right)\right) \sim \nonumber\\
& & \exp\left(-{1\over\hbar} \left[{1\over(2\pi)^d}\int d^dk\right] \int d^dx \left(-\hbar\ln\left(1+{\eta(x)\over v}\right)\right)\right) = \nonumber\\
& & \exp\left(-{1\over\hbar} \left[{1\over(2\pi)^d}\int d^dk\right] \int d^dx \left(-\hbar \sum_{n=1}^\infty \frac{(-1)^{n+1}}{n} \left({\eta(x)\over v}\right)^n \right)\right)
\eqa

It is convenient to define also the standard integral $I$:\ind{standard integrals}
\bq \label{standintI}
I \equiv {1\over(2\pi)^d}\int d^dk \;.
\eq

Now can read off the Feynman rules from the action and the Jacobian, and the conjecture states that we can calculate everything in the continuum, provided we do a $d$-dimensional calculation. The Feynman rules are (up to order $\hbar^{5/2}$):
{\allowdisplaybreaks\bqa
\begin{picture}(100, 20)(0, 17)
\Line(20, 20)(80, 20)
\end{picture}
&\leftrightarrow& \frac{\hbar}{k^2+m^2} \nonumber\\
\begin{picture}(100, 20)(0, 17)
\DashLine(20, 20)(80, 20){5}
\end{picture}
&\leftrightarrow& \frac{\hbar}{k^2+m^2} \nonumber\\
\begin{picture}(100, 40)(0, 17)
\Line(20, 20)(50, 20)
\DashLine(50, 20)(80, 0){5}
\DashLine(50, 20)(80, 40){5}
\Text(80,9)[c]{$k_1$}
\Text(80,31)[c]{$k_2$}
\end{picture}
&\leftrightarrow& {2\over\hbar v} k_1 \cdot k_2 - {m^2\over\hbar v} \nonumber\\
\begin{picture}(100, 40)(0, 17)
\Line(20,40)(50,20)
\Line(20,0)(50,20)
\DashLine(80,0)(50,20){5}
\DashLine(80,40)(50,20){5}
\Text(80,9)[c]{$k_1$}
\Text(80,31)[c]{$k_2$}
\end{picture}
&\leftrightarrow& {2\over\hbar v^2} k_1 \cdot k_2 \nonumber\\
\begin{picture}(100, 50)(0, 17)
\DashLine(20, 0)(80, 40){5}
\DashLine(20, 40)(80, 0){5}
\end{picture}
&\leftrightarrow& {m^2\over\hbar v^2} \nonumber\\
\begin{picture}(100, 40)(0, 17)
\Line(20, 20)(50, 20)
\DashLine(20, 0)(80, 40){5}
\DashLine(20, 40)(80, 0){5}
\end{picture}
&\leftrightarrow& {m^2\over\hbar v^3} \nonumber\\
\begin{picture}(100, 40)(0, 17)
\Line(20, 20)(50, 20)
\Vertex(50, 20){3}
\end{picture}
&\leftrightarrow& {1\over v} \; I \nonumber\\
\begin{picture}(100, 40)(0, 17)
\Line(20, 20)(80, 20)
\Vertex(50, 20){3}
\end{picture}
&\leftrightarrow& -{1\over v^2} \; I \nonumber\\
\begin{picture}(100, 40)(0, 17)
\Line(20, 20)(50, 20)
\Line(50, 20)(80, 40)
\Line(50, 20)(80, 0)
\Vertex(50, 20){3}
\end{picture}
&\leftrightarrow& {2\over v^3} \; I
\eqa}\\[5pt]
Here all momenta are counted as going into the vertex. With these rules we can now compute Green's functions up to order $\hbar^2$.

First we will demonstrate however what we mean exactly by `calculating in a $d$-dimensional way'.\index{calculating in a $d$-dimensional way} What we mean can most easily be seen in the following `$d$-dimensional calculation' of a tadpole diagram.
\vspace{-15pt}
{\allowdisplaybreaks\bqa
\begin{picture}(100,50)(0,25)
\Line(0,25)(50,25)
\DashCArc(75,25)(25,0,360){3}
\Line(75,0)(75,50)
\end{picture} &=& \h{\hbar\over m^2} {1\over(2\pi)^{2d}}\int d^dk \; d^dl \; \left(-{2\over\hbar v}k^2-{m^2\over\hbar v}\right) \left(-{2\over\hbar v}k\cdot l-{m^2\over\hbar v}\right)^2 \nonumber\\
& & \phantom{\h{\hbar\over m^2} {1\over(2\pi)^{2d}}\int d^dk \; d^dl \;} \left(\frac{\hbar}{k^2+m^2}\right)^2 \frac{\hbar}{l^2+m^2} \frac{\hbar}{(k-l)^2+m^2} \nonumber\\
&=& \h{\hbar^2\over m^2v^3} {1\over(2\pi)^{2d}}\int d^dk \; d^dl \; (-2k^2-m^2)(-2k\cdot l-m^2)^2 \nonumber\\
& & \phantom{\h{\hbar^2\over m^2v^3} {1\over(2\pi)^{2d}}\int d^dk \; d^dl \;} \left(\frac{1}{k^2+m^2}\right)^2 \frac{1}{l^2+m^2} \frac{1}{(k-l)^2+m^2} \nonumber\\
&=& -{\hbar^2\over m^2v^3} {1\over(2\pi)^{2d}}\int d^dk \; d^dl \; (-2k\cdot l-m^2)^2 \cdot \nonumber\\
& & \phantom{-{\hbar^2\over m^2v^3} {1\over(2\pi)^{2d}}\int d^dk \; d^dl \;} \frac{1}{k^2+m^2} \frac{1}{l^2+m^2} \frac{1}{(k-l)^2+m^2} + \nonumber\\
& & \h{\hbar^2\over v^3} {1\over(2\pi)^{2d}}\int d^dk \; d^dl \; (-2k\cdot l-m^2)^2 \cdot \nonumber\\
& & \phantom{\h{\hbar^2\over v^3} {1\over(2\pi)^{2d}}\int d^dk \; d^dl \;} \left(\frac{1}{k^2+m^2}\right)^2 \frac{1}{l^2+m^2} \frac{1}{(k-l)^2+m^2} \nonumber\\
\eqa}
Now use
\bq
-2k\cdot l-m^2 = \left[(k-l)^2+m^2\right] - \left[k^2+m^2\right] - \left[l^2+m^2\right] \;.
\eq
Then the tadpole diagram becomes:
\bqa
& & -{\hbar^2\over m^2v^3} {1\over(2\pi)^{2d}}\int d^dk \; d^dl \; (-2k\cdot l-m^2) \nonumber\\
& & \phantom{-{\hbar^2\over m^2v^3} {1\over(2\pi)^{2d}}\int d^dk \; d^dl \;} \left( \frac{1}{k^2+m^2} \frac{1}{l^2+m^2} - 2\frac{1}{k^2+m^2} \frac{1}{(k-l)^2+m^2} \right) + \nonumber\\
& & \h{\hbar^2\over v^3} {1\over(2\pi)^{2d}}\int d^dk \; d^dl \; (-2k\cdot l-m^2) \nonumber\\
& & \phantom{\h{\hbar^2\over v^3} {1\over(2\pi)^{2d}}\int d^dk \; d^dl \;} \bigg( \frac{1}{(k^2+m^2)^2} \frac{1}{l^2+m^2} - \frac{1}{k^2+m^2} \frac{1}{l^2+m^2} \frac{1}{(k-l)^2+m^2} + \nonumber\\
& & \phantom{\h{\hbar^2\over v^3} {1\over(2\pi)^{2d}}\int d^dk \; d^dl \; \bigg(} -\frac{1}{(k^2+m^2)^2} \frac{1}{(k-l)^2+m^2} \bigg) \nonumber\\
&=& {\hbar^2\over v^3} \; I(0,m)^2 + 2{\hbar^2\over m^2v^3} {1\over(2\pi)^{2d}}\int d^dk \; d^dl \; (-2k\cdot l-2k^2-m^2) \frac{1}{k^2+m^2} \frac{1}{l^2+m^2} + \nonumber\\
& & -\h{\hbar^2m^2\over v^3} \; I(0,m)I(0,m,0,m) + \nonumber\\
& & -\h{\hbar^2\over v^3} {1\over(2\pi)^{2d}}\int d^dk \; d^dl \; \left( \frac{1}{k^2+m^2} \frac{1}{l^2+m^2} - 2\frac{1}{k^2+m^2} \frac{1}{(k-l)^2+m^2} \right) + \nonumber\\
& & -\h{\hbar^2\over v^3} {1\over(2\pi)^{2d}}\int d^dk \; d^dl \; (-2k\cdot l-2k^2-m^2) \frac{1}{(k^2+m^2)^2} \frac{1}{l^2+m^2} \nonumber\\
&=& {9\over2}{\hbar^2\over v^3} \; I(0,m)^2 - 4{\hbar^2\over m^2v^3} \; I(0,m) I - {\hbar^2m^2\over v^3} \; I(0,m)I(0,m,0,m)
\eqa
In the above steps, and in all $d$-dimensional calculations, one essentially uses three rules: One writes dot-products from the vertices in terms of the denominators of the propagators, to let them cancel as much as possible, one can shift all loop momenta and one can set
\bq
\int d^dk \; \frac{k_i}{k^2+m^2} = 0 \;.
\eq
Using these three rules is what we mean by a `$d$-dimensional calculation'.

Notice that, for example in dimension 1, where $k\cdot l$ becomes a simple product, we could also have combined momenta coming from different vertices and let them cancel denominators. This simplifies the calculation of the tadpole diagram considerably, however this is \emph{not} what we mean by a `$d$-dimensional calculation'. The result in terms of standard integrals is also \emph{different}. Even the numerical result is \emph{different} because the diagram contains a divergence, also in $d=1$. (The divergence comes from $I$.) In order for the conjecture to work we \emph{have} to perform a $d$-dimensional calculation.

Now we compute $\langle\f_1(x)\rangle$, $\langle\f_2(x)\rangle$, $\langle\f_1(x)\f_1(y)\rangle$, $\langle\f_2(x)\f_2(y)\rangle$ and $\langle\f_1(x)\f_2(y)\rangle$ via the polar fields, up to order $\hbar^2$. To do this we first have to express these Green's functions in terms of the $\eta$- and $w$-Green's-functions.
{\allowdisplaybreaks\bqa
\langle\f_1(x)\rangle &=& \langle(v+\eta(x))\cos(w(x)/v)\rangle \nonumber\\
&=& v + \langle\eta(x)\rangle - {1\over2v}\langle w^2(x)\rangle - {1\over2v^2}\langle\eta(x)w^2(x)\rangle + {1\over24v^3}\langle w^4(x)\rangle + \mathcal{O}(\hbar^3) \nonumber\\
\langle\f_2(x)\rangle &=& \langle(v+\eta(x))\sin(w(x)/v)\rangle \nonumber\\
&=& \langle w(x)\rangle + {1\over v}\langle\eta(x)w(x)\rangle - {1\over6v^2}\langle w^3(x)\rangle - {1\over6v^3}\langle\eta(x)w^3(x)\rangle + \mathcal{O}(\hbar^3) \nonumber\\
\langle\f_1(x)\f_1(y)\rangle &=& \langle(v+\eta(x))(v+\eta(y))\cos(w(x)/v)\cos(w(y)/v)\rangle \nonumber\\
&=& v^2 + 2v\langle\eta(x)\rangle + \langle\eta(x)\eta(y)\rangle - \langle w^2(x)\rangle - {1\over v}\langle\eta(x)w^2(x)\rangle + \nonumber\\
& & -{1\over2v}\langle\eta(x)w^2(y)\rangle - {1\over2v}\langle\eta(y)w^2(x)\rangle + {1\over12v^2}\langle w^4(x)\rangle + \nonumber\\
& & {1\over4v^2}\langle w^2(x)w^2(y)\rangle - {1\over2v^2}\langle\eta(x)\eta(y)w^2(x)\rangle - {1\over2v^2}\langle\eta(x)\eta(y)w^2(y)\rangle + \nonumber\\[5pt]
& & \mathcal{O}(\hbar^3) \nonumber\\[5pt]
\langle\f_2(x)\f_2(y)\rangle &=& \langle(v+\eta(x))(v+\eta(y))\sin(w(x)/v)\sin(w(y)/v)\rangle \nonumber\\
&=& \langle w(x)w(y)\rangle + {1\over v}\langle\eta(x)w(x)w(y)\rangle + {1\over v}\langle\eta(y)w(x)w(y)\rangle + \nonumber\\
& & {1\over v^2}\langle\eta(x)\eta(y)w(x)w(y)\rangle - {1\over6v^2}\langle w^3(x)w(y)\rangle - {1\over6v^2}\langle w(x)w^3(y)\rangle + \nonumber\\[5pt]
& & \mathcal{O}(\hbar^3) \nonumber\\[5pt]
\langle\f_1(x)\f_2(y)\rangle &=& \langle(v+\eta(x))(v+\eta(y))\cos(w(x)/v)\sin(w(y)/v)\rangle \nonumber\\
&=& v\langle w(y)\rangle + \langle\eta(x)w(y)\rangle + \langle\eta(y)w(y)\rangle - {1\over6v}\langle w^3(y)\rangle + \nonumber\\
& & -{1\over2v}\langle w^2(x)w(y)\rangle + {1\over v}\langle\eta(x)\eta(y)w(y)\rangle - {1\over6v^2}\langle\eta(x)w^3(y)\rangle + \nonumber\\
& & -{1\over6v^2}\langle\eta(y)w^3(y)\rangle - {1\over2v^2}\langle\eta(x)w^2(x)w(y)\rangle - {1\over2v^2}\langle\eta(y)w^2(x)w(y)\rangle + \nonumber\\[5pt]
& & \mathcal{O}(\hbar^3) \label{Cartaverages}
\eqa}
Notice that in these formulas only the \emph{full} $\eta$- and $w$-Green's-functions occur.

Now we compute all the $\eta$- and $w$-Green's-functions that we need, up to order $\hbar^2$. All results will be expressed in the standard integrals again. One-loop standard integrals have already been defined in (\ref{standint}). One would expect that we also need the two-loop standard integrals here, since two-loop diagrams will occur in the $\eta$-tadpole at order $\hbar^2$. However it will appear that no two-loop integrals occur in the final expressions for the tadpole diagrams, all expressions can be written in terms of one-loop standard integrals, probably because of the simple action of the shifted toy model. From the Feynman rules we can also immediately see that any Green's function with an odd number of $w$'s is zero, we will not list them explicitly below.

\subsubsection{The $\eta$-Tadpole}

Below we list all the diagrams contributing to $\langle\tilde{\eta}\rangle=\langle\eta\rangle$ up to order $\hbar^2$. 
{\allowdisplaybreaks\bqa
\begin{picture}(100,50)(0,25)
\Line(0,25)(50,25)
\DashCArc(75,25)(25,0,360){3}
\end{picture} \qquad &=& \qquad -{\hbar\over m^2v} \; I + {1\over2}{\hbar\over v} \; I(0,m) \nonumber\\
\begin{picture}(100,50)(0,25)
\Line(0,25)(50,25)
\Vertex(50,25){4}
\end{picture} \qquad &=& \qquad {\hbar\over m^2v} \; I \nonumber\\
\begin{picture}(100,50)(0,25)
\Line(0,25)(25,25)
\GCirc(37.5,25){12.5}{0.5}
\Line(50,25)(75,25)
\GCirc(87.5,25){12.5}{0.5}
\end{picture} \qquad &=& \qquad -{1\over2}\frac{\hbar^2}{v^3} \; I(0,m)^2 + {1\over4}\frac{\hbar^2m^2}{v^3} \; I(0,m,0,m) \; I(0,m) \nonumber\\
\begin{picture}(100,50)(0,25)
\Line(0,25)(50,25)
\DashCArc(75,25)(25,0,360){3}
\Line(75,0)(75,50)
\end{picture} \qquad &=& \qquad -4{\hbar^2\over m^2v^3} \; I(0,m) \; I + {9\over2}\frac{\hbar^2}{v^3} \; I(0,m)^2 + \nonumber\\
& & \qquad -\frac{\hbar^2m^2}{v^3} \; I(0,m,0,m) \; I(0,m) \nonumber\\
\begin{picture}(100,50)(0,25)
\Line(0,25)(100,25)
\DashCArc(75,25)(25,0,360){3}
\end{picture} \qquad &=& \qquad 2{\hbar^2\over m^2v^3} \; I(0,m) \; I - 2\frac{\hbar^2}{v^3} \; I(0,m)^2 \nonumber\\
\begin{picture}(100,50)(0,25)
\Line(0,25)(20,25)
\DashCArc(40,25)(20,0,360){3}
\DashCArc(80,25)(20,0,360){3}
\end{picture} \qquad &=& \qquad -{1\over2}\frac{\hbar^2}{v^3} \; I(0,m)^2 + {1\over4}\frac{\hbar^2m^2}{v^3} \; I(0,m,0,m) \; I(0,m) \nonumber\\
\begin{picture}(100,50)(0,25)
\Line(0,25)(20,25)
\DashCArc(40,25)(20,0,360){3}
\CArc(80,25)(20,0,360)
\end{picture} \qquad &=& \qquad {\hbar^2\over m^2v^3} \; I(0,m) \; I - {3\over2}\frac{\hbar^2}{v^3} \; I(0,m)^2 + \nonumber\\
& & \qquad \h\frac{\hbar^2m^2}{v^3} \; I(0,m,0,m) \; I(0,m) \nonumber\\
\begin{picture}(100,50)(0,25)
\Line(0,25)(75,25)
\DashCArc(75,37.5)(12.5,0,360){3}
\DashCArc(75,12.5)(12.5,0,360){3}
\end{picture} \qquad &=& \qquad {1\over8} \frac{\hbar^2}{v^3} \; I(0,m)^2 \nonumber\\
\begin{picture}(100,50)(0,25)
\Line(0,25)(50,25)
\CArc(75,25)(25,0,360)
\Vertex(50,25){4}
\end{picture} \qquad &=& \qquad {\hbar^2\over m^2v^3} \; I(0,m) \; I
\eqa}\\

The complete result for the $\eta$-tadpole is finally:
\bq
\langle\tilde{\eta}\rangle = \h{\hbar\over v} \; I(0,m) + {1\over8}{\hbar^2\over v^3} \; I(0,m)^2.
\eq
Notice that all the (infinite) $I$-integrals from the Jacobian have nicely cancelled against identical terms from $w$-loops.

\subsubsection{The $\eta$-Propagator}

Below we list the diagrams contributing to the \emph{connected} momentum-space $\eta$-propagator $\langle\tilde{\eta}(p)\tilde{\eta}(-p)\rangle_{\mathrm{c}}$. 
\bqa
\begin{picture}(70,40)(0,18)
\Line(0,20)(70,20)
\end{picture} \quad &=& \frac{\hbar}{p^2+m^2} \nonumber\\
\begin{picture}(70,40)(0,18)
\Line(0,20)(20,20)
\DashCArc(35,20)(15,-180,180){3}
\Line(50,20)(70,20)
\end{picture} \quad &=& 2{\hbar^2\over v^2} \frac{1}{(p^2+m^2)^2} \; I - {\hbar^2\over v^2}\frac{p^2+2m^2}{(p^2+m^2)^2} \; I(0,m) + \h{\hbar^2\over v^2}\;  I(0,m,p,m) \nonumber\\
\begin{picture}(70,40)(0,18)
\Line(0,5)(70,5)
\DashCArc(35,20)(15,-90,270){3}
\end{picture} \quad &=& -{\hbar^2\over v^2} \frac{1}{(p^2+m^2)^2} \; I + {\hbar^2m^2\over v^2}\frac{1}{(p^2+m^2)^2} \; I(0,m) \nonumber\\
\begin{picture}(70,40)(0,18)
\Line(0,20)(70,20)
\Vertex(35,20){4}
\end{picture} \quad &=& -{\hbar^2\over v^2} \frac{1}{(p^2+m^2)^2} \; I
\eqa\\
For the connected $\eta$-propagator we get:
\bq
\langle \tilde{\eta}(p)\tilde{\eta}(-p) \rangle_\mathrm{c} = \frac{\hbar}{p^2+m^2} - {\hbar^2\over v^2}\frac{1}{p^2+m^2} \; I(0,m) + \h{\hbar^2\over v^2} I(0,m,p,m) 
\eq

\subsubsection{The $w$-Propagator}

Below we list the diagrams contributing to the \emph{connected} $w$-propagator $\langle\tilde{w}(p)\tilde{w}(-p)\rangle_\mathrm{c}$. 
{\allowdisplaybreaks\bqa
\begin{picture}(70,40)(0,18)
\DashLine(0,20)(70,20){3}
\end{picture} \quad &=& \frac{\hbar}{p^2+m^2} \nonumber\\
\begin{picture}(70,40)(0,18)
\DashLine(0,20)(20,20){3}
\DashCArc(35,20)(15,-180,0){3}
\CArc(35,20)(15,0,180)
\DashLine(50,20)(70,20){3}
\end{picture} \quad &=& {\hbar^2\over v^2} \; I(0,m,p,m) + {\hbar^2\over v^2}\frac{2p^2}{(p^2+m^2)^2} \; I(0,m) \nonumber\\
\begin{picture}(70,40)(0,18)
\DashLine(0,5)(70,5){3}
\Line(35,5)(35,25)
\GCirc(35,25){10}{0.5}
\end{picture} \quad &=& -{\hbar^2\over v^2}\frac{p^2+\h m^2}{(p^2+m^2)^2} \; I(0,m) \nonumber\\
\begin{picture}(70,40)(0,18)
\DashLine(0,5)(70,5){3}
\DashCArc(35,20)(15,-90,270){3}
\end{picture} \quad &=& \h{\hbar^2m^2\over v^2}\frac{1}{(p^2+m^2)^2} \; I(0,m) \nonumber\\
\begin{picture}(70,40)(0,18)
\DashLine(0,5)(70,5){3}
\CArc(35,20)(15,-90,270)
\end{picture} \quad &=& -{\hbar^2\over v^2}\frac{p^2}{(p^2+m^2)^2} \; I(0,m) \nonumber\\
\eqa}
\bq
\langle\tilde{w}(p)\tilde{w}(-p)\rangle_\mathrm{c} = \frac{\hbar}{p^2+m^2} + {\hbar^2\over v^2} \; I(0,m,p,m)
\eq

\subsubsection{Configuration-Space Green's Functions}

Knowing these $\eta$- and $w$-Green's-functions we can write down the configuration space Green's functions needed in (\ref{Cartaverages}), up to order $\hbar^2$.
{\allowdisplaybreaks\bqa
\langle\eta(x)\rangle &=& \h{\hbar\over v} \; I(0,m) + {1\over8}{\hbar^2\over v^3} \; I(0,m)^2 \nonumber\\
\langle w(x)\rangle &=& 0 \nonumber\\
\langle\eta(x)\eta(y)\rangle &=& {1\over(2\pi)^d}\int d^dp \; e^{ip\cdot(x-y)} \; \langle\tilde{\eta}(p)\tilde{\eta}(-p)\rangle_{\mathrm{c}} + \langle\eta(x)\rangle^2 \nonumber\\
&=& {1\over4}{\hbar^2\over v^2} \; I(0,m)^2 + {\hbar\over(2\pi)^d}\int d^dp \; \frac{e^{ip\cdot(x-y)}}{p^2+m^2} + \nonumber\\
& & -{\hbar^2\over v^2} \; I(0,m) \; {1\over(2\pi)^d}\int d^dp \; \frac{e^{ip\cdot(x-y)}}{p^2+m^2} + \nonumber\\
& & \h{\hbar^2\over v^2} {1\over(2\pi)^d}\int d^dp \; e^{ip\cdot(x-y)} \; I(0,m,p,m) \nonumber\\
&=& {1\over4}{\hbar^2\over v^2} \; I(0,m)^2 + \hbar \; A_m(x-y) - {\hbar^2\over v^2} \; I(0,m) A_m(x-y) + \nonumber\\
& & \h{\hbar^2\over v^2} \; A_m(x-y)^2 \nonumber\\
\langle w(x)w(y)\rangle &=& {1\over(2\pi)^d}\int d^dp \; e^{ip\cdot(x-y)} \; \langle\tilde{w}(p)\tilde{w}(-p)\rangle_{\mathrm{c}} \nonumber\\
&=& {\hbar\over(2\pi)^d}\int d^dp \; \frac{e^{ip\cdot(x-y)}}{p^2+m^2} + \nonumber\\
& & {\hbar^2\over v^2} {1\over(2\pi)^d}\int d^dp \; e^{ip\cdot(x-y)} \; I(0,m,p,m) \nonumber\\
&=& \hbar \; A_m(x-y) + {\hbar^2\over v^2} \; A_m(x-y)^2 \nonumber\\
\langle\eta(x_1)\eta(x_2)\eta(x_3)\rangle &=& \langle\eta\rangle^3 + \nonumber\\
& & \langle\eta\rangle\left( \langle\eta(x_1)\eta(x_2)\rangle_{\mathrm{c}} + \langle\eta(x_1)\eta(x_3)\rangle_{\mathrm{c}} + \langle\eta(x_2)\eta(x_3)\rangle_{\mathrm{c}} \right) + \nonumber\\
& & \langle\eta(x_1)\eta(x_2)\eta(x_3)\rangle_{\mathrm{c}} \nonumber\\
&=& \h{\hbar^2\over v} \; I(0,m) \left( A_m(x_1-x_2) + A_m(x_1-x_3) + A_m(x_2-x_3) \right) \nonumber\\
\langle\eta(x_1)\eta(x_2)w(x_3)\rangle &=& 0 \nonumber\\
\langle\eta(x_1)w(x_2)w(x_3)\rangle &=& \langle\eta\rangle \langle w(x_2)w(x_3)\rangle_{\mathrm{c}} + \langle\eta(x_1)w(x_2)w(x_3)\rangle_{\mathrm{c}} \nonumber\\
&=& \h{\hbar^2\over v} \; I(0,m) \; {1\over(2\pi)^d}\int d^dp \; \frac{e^{ip\cdot(x_2-x_3)}}{p^2+m^2} + \nonumber\\
& & {1\over(2\pi)^d}\int d^dk_1 \cdots {1\over(2\pi)^d}\int d^dk_3 \; e^{i(k_1\cdot x_1+\ldots+k_3\cdot x_3)} \cdot \nonumber\\
& & \hspace{10pt} {2k_2\cdot k_3-m^2\over\hbar v} \frac{\hbar}{k_1^2+m^2} \cdots \frac{\hbar}{k_3^2+m^2} (2\pi)^d\delta^d(k_1+\ldots+k_3) \nonumber\\
&=& \h{\hbar^2\over v} \; I(0,m) A_m(x_2-x_3) + \nonumber\\
& & {\hbar^2\over v} \; A_m(x_1-x_2) A_m(x_1-x_3) + \nonumber\\
& & -{\hbar^2\over v} \; A_m(x_1-x_2) A_m(x_2-x_3) + \nonumber\\
& & -{\hbar^2\over v} \; A_m(x_1-x_3) A_m(x_2-x_3) \nonumber\\
\langle w(x_1)w(x_2)w(x_3)\rangle &=& 0 \nonumber\\
\langle\eta(x_1)\cdots\eta(x_4)\rangle &=& \langle\eta\rangle^4 + \nonumber\\
& & \langle\eta\rangle^2\big( \langle\eta(x_1)\eta(x_2)\rangle_{\mathrm{c}} + \langle\eta(x_1)\eta(x_3)\rangle_{\mathrm{c}} + \langle\eta(x_1)\eta(x_4)\rangle_{\mathrm{c}} + \nonumber\\
& & \phantom{\langle\eta\rangle^2\big(} \langle\eta(x_2)\eta(x_3)\rangle_{\mathrm{c}} + \langle\eta(x_2)\eta(x_4)\rangle_{\mathrm{c}} + \langle\eta(x_3)\eta(x_4)\rangle_{\mathrm{c}} \big) + \nonumber\\
& & \langle\eta\rangle \big( \langle\eta(x_1)\eta(x_2)\eta(x_3)\rangle_{\mathrm{c}} + \langle\eta(x_1)\eta(x_2)\eta(x_4)\rangle_{\mathrm{c}} + \nonumber\\
& & \phantom{\langle\eta\rangle \big(} \langle\eta(x_1)\eta(x_3)\eta(x_4)\rangle_{\mathrm{c}} + \langle\eta(x_2)\eta(x_3)\eta(x_4)\rangle_{\mathrm{c}} \big) + \nonumber\\
& & \langle\eta(x_1)\eta(x_2)\rangle_{\mathrm{c}} \langle\eta(x_3)\eta(x_4)\rangle_{\mathrm{c}} + \nonumber\\
& & \langle\eta(x_1)\eta(x_3)\rangle_{\mathrm{c}} \langle\eta(x_2)\eta(x_4)\rangle_{\mathrm{c}} + \nonumber\\
& & \langle\eta(x_1)\eta(x_4)\rangle_{\mathrm{c}} \langle\eta(x_2)\eta(x_3)\rangle_{\mathrm{c}} + \nonumber\\
& & \langle\eta(x_1)\eta(x_2)\eta(x_3)\eta(x_4)\rangle_{\mathrm{c}} \nonumber\\
&=& \hbar^2 \; A_m(x_1-x_2) A_m(x_3-x_4) + \nonumber\\
& & \hbar^2 \; A_m(x_1-x_3) A_m(x_2-x_4) + \nonumber\\
& & \hbar^2 \; A_m(x_1-x_4) A_m(x_2-x_3) \nonumber\\
\langle\eta(x_1)\cdots\eta(x_3)w(x_4)\rangle &=& 0 \nonumber\\
\langle\eta(x_1)\eta(x_2)w(x_3)w(x_4)\rangle &=& \hbar^2 \; A_m(x_1-x_2) A_m(x_3-x_4) \nonumber\\
\langle\eta(x_1)w(x_2)\cdots w(x_4)\rangle &=& 0 \nonumber\\
\langle w(x_1)\cdots w(x_4)\rangle &=& \hbar^2 \; A_m(x_1-x_2) A_m(x_3-x_4) + \nonumber\\
& & \hbar^2 \; A_m(x_1-x_3) A_m(x_2-x_4) + \nonumber\\
& & \hbar^2 \; A_m(x_1-x_4) A_m(x_2-x_3) \nonumber\\
\eqa}

Now, substituting all these results in (\ref{Cartaverages}) gives us finally:\index{Green's functions!shifted toy model}
\bqa
\langle \f_1(x) \rangle &=& v \nonumber\\
\langle \f_2(x) \rangle &=& 0 \nonumber\\
\langle \f_1(x) \f_1(y) \rangle &=& v^2 + \hbar \; A_m(x-y) \nonumber\\
\langle \f_2(x) \f_2(y) \rangle &=& \hbar \; A_m(x-y) \nonumber\\
\langle \f_1(x) \f_2(y) \rangle &=& 0
\eqa
These are indeed the correct results for the Green's functions. So the conjecture is verified for several Green's functions in the shifted toy model up to order $\hbar^2$.

\newpage

\section{The Arctangent Toy Model}

As another illustration of the conjecture we now consider the \ind{arctangent toy model}. The action of this model is:\index{action!arctangent toy model}
\bq \label{Cartactionarctan}
S = \int dx \left( \h(\nabla\f_1)^2 + \h(\nabla\f_2)^2 + \frac{m^2}{4v^2}(\f_1^2+\f_2^2-v^2)^2 + \frac{m^2v^4}{2(\f_1^2+\f_2^2)} \arctan^2 \left( \frac{\f_2}{\f_1} \right) \right)
\eq
This action has a single minimum at
\bq \label{minarctantoy}
\f_1 = v \;, \quad \f_2 = 0 \;.
\eq

We want to stress that this model is not at all a physical model. The action has an infinite number of vertices (by expanding the arctangent), which means this model is \emph{not} renormalizable. We just want to use this model as a toy model to test the conjecture. Especially because it is not renormalizable, so \emph{no} big cancellations can be expected to occur, the arctangent toy model is a very good test of the conjecture.

\subsection{Cartesian Results} \label{arctanCartsect}

To find the Cartesian Green's functions we expand the action around the minimum (\ref{minarctantoy}):
\bq
\f_1(x) = v + \eta_1(x) \;, \quad \f_2(x) = \eta_2(x) \;.
\eq
Notice that also the arctangent in the action has to be expanded, this term will give an infinite number of vertices. Up to order $\hbar^{5/2}$ the Feynman rules are:
{\allowdisplaybreaks\bqa
\begin{picture}(100, 20)(0, 17)
\Line(20, 20)(80, 20)
\end{picture}
&\leftrightarrow& \frac{\hbar}{k^2+\mu^2} \nonumber\\
\begin{picture}(100, 20)(0, 17)
\DashLine(20, 20)(80, 20){5}
\end{picture}
&\leftrightarrow& \frac{\hbar}{k^2+m^2} \nonumber\\
\begin{picture}(100, 40)(0, 17)
\Line(20, 20)(50, 20)
\Line(50, 20)(80, 0)
\Line(50, 20)(80, 40)
\end{picture}
&\leftrightarrow& -\frac{1}{\hbar}\frac{6m^2}{v} \nonumber\\
\begin{picture}(100, 40)(0, 17)
\Line(20, 20)(50, 20)
\DashLine(50, 20)(80, 0){5}
\DashLine(50, 20)(80, 40){5}
\end{picture}
&\leftrightarrow& \frac{1}{\hbar}\frac{2m^2}{v} \nonumber\\
\begin{picture}(100, 40)(0, 17)
\Line(20, 0)(80, 40)
\Line(20, 40)(80, 0)
\end{picture}
&\leftrightarrow& -\frac{1}{\hbar}\frac{6m^2}{v^2} \nonumber\\
\begin{picture}(100, 40)(0, 17)
\DashLine(20, 0)(80, 40){5}
\Line(20, 40)(80, 0)
\end{picture}
&\leftrightarrow& -\frac{1}{\hbar}\frac{22m^2}{v^2} \nonumber\\
\begin{picture}(100, 50)(0, 17)
\DashLine(20, 0)(80, 40){5}
\DashLine(20, 40)(80, 0){5}
\end{picture}
&\leftrightarrow& \frac{1}{\hbar}\frac{14m^2}{v^2} \nonumber\\
\begin{picture}(100, 40)(0, 17)
\Line(20, 0)(50, 20)
\Line(20, 20)(50, 20)
\Line(20, 40)(50, 20)
\DashLine(50, 20)(80, 40){5}
\DashLine(50, 20)(80, 0){5}
\end{picture}
&\leftrightarrow& \frac{1}{\hbar}\frac{120m^2}{v^3} \nonumber\\
\begin{picture}(100, 40)(0, 17)
\Line(20, 20)(50, 20)
\DashLine(20, 0)(80, 40){5}
\DashLine(20, 40)(80, 0){5}
\end{picture}
&\leftrightarrow& -\frac{1}{\hbar}\frac{120m^2}{v^3}
\eqa}\\[10pt]
Here the solid lines indicate the $\eta_1$-particle, the dashed lines indicate the $\eta_2$-particle and $\mu$ is given by $\mu = \sqrt{2}m$. (Notice that this is a different definition of $\mu$ than in the rest of this thesis, this definition will only be used in calculations in the arctangent toy model.) With these Feynman rules we can now compute some Green's functions up to order $\hbar^2$. We shall not present all diagrams here, since this Cartesian calculation is straightforward and quite lengthy.

This time we will also get two-loop integrals in our expressions for the Green's functions. They don't drop out in this case, as in they did in the shifted toy model, because this model has a more complicated action.

Now the Green's functions are, up to order $\hbar^2$:\index{Green's functions!arctangent toy model}
{\allowdisplaybreaks\bqa
\langle\f_1(x)\rangle &=& v + \langle\eta_1(x)\rangle \nonumber\\
&=& v - {3\over2}{\hbar\over v} \; I(0,\mu) + \h{\hbar\over v} \; I(0,m) + \nonumber\\
& & -{9\over8}{\hbar^2\over v^3} \; I(0,\mu)^2 - {85\over8}{\hbar^2\over v^3} \; I(0,m)^2 + {99\over4}{\hbar^2\over v^3} \; I(0,\mu)I(0,m) + \nonumber\\
& & -9{\hbar^2m^2\over v^3} \; I(0,\mu,0,\mu)I(0,\mu) + 4{\hbar^2m^2\over v^3} \; I(0,m,0,m)I(0,m) + \nonumber\\
& & 21{\hbar^2m^2\over v^3} \; I(0,\mu,0,\mu)I(0,m) - 7{\hbar^2m^2\over v^3} \; I(0,m,0,m)I(0,\mu) + \nonumber\\
& & -27{\hbar^2m^4\over v^3} \; B_{\mu\mu\mu} - 3{\hbar^2m^4\over v^3} \; B_{\mu mm} + 2{\hbar^2m^4\over v^3} \; B_{m\mu m} + \nonumber\\
& & 3{\hbar^2m^2\over v^3} \; D_{\mu\mu\mu} - 11{\hbar^2m^2\over v^3} \; D_{mm\mu} \nonumber\\[5pt]
\langle\f_2(x)\rangle &=& \langle\eta_2(x)\rangle = 0 \nonumber\\[5pt]
\langle\f_1(x)\f_1(y)\rangle &=& v^2 + 2v\langle\eta_1(x)\rangle + \langle\eta_1(x)\eta_1(y)\rangle \nonumber\\
&=& v^2 - 3\hbar \; I(0,\mu) + \hbar \; I(0,m) + \hbar \; A_\mu(x-y) + \nonumber\\
& & -21{\hbar^2\over v^2} \; I(0,m)^2 + 48{\hbar^2\over v^2} \; I(0,\mu)I(0,m) -18{\hbar^2m^2\over v^2} \; I(0,\mu,0,\mu)I(0,\mu) + \nonumber\\
& & 8{\hbar^2m^2\over v^2} \; I(0,m,0,m)I(0,m) + 42{\hbar^2m^2\over v^2} \; I(0,\mu,0,\mu)I(0,m) + \nonumber\\
& & -14{\hbar^2m^2\over v^2} \; I(0,m,0,m)I(0,\mu) - 54{\hbar^2m^4\over v^2} \; B_{\mu\mu\mu} - 6{\hbar^2m^4\over v^2} \; B_{\mu mm} + \nonumber\\
& & 4{\hbar^2m^4\over v^2} \; B_{m\mu m} + 6{\hbar^2m^2\over v^2} \; D_{\mu\mu\mu} - 22{\hbar^2m^2\over v^2} \; D_{mm\mu} + \nonumber\\
& & 6{\hbar^2m^2\over v^2} \; I(0,\mu) C_{\mu\mu}(x-y) - 14{\hbar^2m^2\over v^2} \; I(0,m) C_{\mu\mu}(x-y) + \nonumber\\
& & 18{\hbar^2m^4\over v^2} \; B_{\mu\mu\mu}(x-y) + 2{\hbar^2m^4\over v^2} \; B_{\mu mm}(x-y) \nonumber\\
\langle\f_2(x)\f_2(y)\rangle &=& \langle\eta_2(x)\eta_2(y)\rangle \nonumber\\
&=& \hbar \; A_m(x-y) + \nonumber\\
& & -14{\hbar^2m^2\over v^2} \; I(0,\mu) C_{mm}(x-y) + 8{\hbar^2m^2\over v^2} \; I(0,m) C_{mm}(x-y) + \nonumber\\
& & 4{\hbar^2m^4\over v^2} \; B_{mm\mu}(x-y) \nonumber\\[5pt]
\langle\f_1(x)\f_2(y)\rangle &=& v\langle\eta_2(y)\rangle + \langle\eta_1(x)\eta_2(y)\rangle = 0 \label{Cartarctan}
\eqa}

\subsection{Polar Results}

According to the conjecture we can just transform the continuum action (\ref{Cartactionarctan}) to an action in terms of the polar field variables to obtain the Feynman rules for the polar calculation. So the action becomes:
\bq
S = \int d^dx \left( \h\left(\nabla r\right)^2 + \h{r^2\over v^2}\left(\nabla w\right)^2 + {m^2\over4v^2}(r^2-v^2)^2 + \h m^2v^2{w^2\over r^2} \right)
\eq
This can be expanded around $r=v$ again. Defining
\bq
r(x) \equiv v + \eta(x) \;,
\eq
we find the following Feynman rules for the $\eta$- and $w$-field (up to order $\hbar^{5/2}$). The vertices from the Jacobian are exactly the same as in the shifted toy model.
{\allowdisplaybreaks\bqa
\begin{picture}(100, 20)(0, 17)
\Line(20, 20)(80, 20)
\end{picture}
&\leftrightarrow& \frac{\hbar}{k^2+\mu^2} \nonumber\\
\begin{picture}(100, 20)(0, 17)
\DashLine(20, 20)(80, 20){5}
\end{picture}
&\leftrightarrow& \frac{\hbar}{k^2+m^2} \nonumber\\
\begin{picture}(100, 40)(0, 17)
\Line(20, 20)(50, 20)
\Line(50, 20)(80, 0)
\Line(50, 20)(80, 40)
\end{picture}
&\leftrightarrow& -\frac{1}{\hbar}\frac{6m^2}{v} \nonumber\\
\begin{picture}(100, 40)(0, 17)
\Line(20, 20)(50, 20)
\DashLine(50, 20)(80, 0){5}
\DashLine(50, 20)(80, 40){5}
\Text(80,30)[c]{$k_1$}
\Text(80,10)[c]{$k_2$}
\end{picture}
&\leftrightarrow& \frac{2}{\hbar v} k_1\cdot k_2 + \frac{1}{\hbar}\frac{2m^2}{v} \nonumber\\
\begin{picture}(100, 40)(0, 17)
\Line(20, 0)(80, 40)
\Line(20, 40)(80, 0)
\end{picture}
&\leftrightarrow& -\frac{1}{\hbar}\frac{6m^2}{v^2} \nonumber\\
\begin{picture}(100, 40)(0, 17)
\Line(20,0)(50,20)
\Line(20,40)(50,20)
\DashLine(50,20)(80,40){5}
\DashLine(50,20)(80,0){5}
\Text(80,30)[c]{$k_1$}
\Text(80,10)[c]{$k_2$}
\end{picture}
&\leftrightarrow& \frac{2}{\hbar v^2} k_1\cdot k_2 - \frac{1}{\hbar}\frac{6m^2}{v^2} \nonumber\\
\begin{picture}(100, 40)(0, 17)
\Line(20, 0)(50, 20)
\Line(20, 20)(50, 20)
\Line(20, 40)(50, 20)
\DashLine(50, 20)(80, 40){5}
\DashLine(50, 20)(80, 0){5}
\end{picture}
&\leftrightarrow& \frac{1}{\hbar}\frac{24m^2}{v^3} \nonumber\\
\begin{picture}(100, 40)(0, 17)
\Line(20, 20)(50, 20)
\Vertex(50, 20){3}
\end{picture}
&\leftrightarrow& {1\over v} \; I \nonumber\\
\begin{picture}(100, 40)(0, 17)
\Line(20, 20)(80, 20)
\Vertex(50, 20){3}
\end{picture}
&\leftrightarrow& -{1\over v^2} \; I \nonumber\\
\begin{picture}(100, 40)(0, 17)
\Line(20, 20)(50, 20)
\Line(50, 20)(80, 40)
\Line(50, 20)(80, 0)
\Vertex(50, 20){3}
\end{picture}
&\leftrightarrow& {2\over v^3} \; I
\eqa}\\
Here the solid lines denote the $\eta$-field, the dashed lines denote the $w$-field and all momenta are counted into the vertex. Also we have defined $\mu=\sqrt{2}m$ again, as in the Cartesian calculation. (Note that this is a different definition of $\mu$ than in the rest of this thesis.)

Now we can again compute the $\eta$- and $w$-Green's-functions up to order $\hbar^2$.

\subsubsection{The $\eta$-Tadpole}

Below we list the diagrams contributing to the $\eta$-tadpole up to order $\hbar^2$.
{\allowdisplaybreaks\bqa
\begin{picture}(100,50)(0,25)
\Line(0,25)(50,25)
\CArc(75,25)(25,0,360)
\end{picture} \qquad &=& \quad -{3\over2}{\hbar\over v} \; I(0,\mu) \nonumber\\
\begin{picture}(100,50)(0,25)
\Line(0,25)(50,25)
\DashCArc(75,25)(25,0,360){3}
\end{picture} \qquad &=& \quad -\h{\hbar\over m^2v} \; I + {\hbar\over v} \; I(0,m) \nonumber\\
\begin{picture}(100,50)(0,25)
\Line(0,25)(75,25)
\Vertex(75,25){4}
\end{picture} \qquad &=& \quad \h{\hbar\over m^2v} \; I \nonumber\\
\begin{picture}(100,50)(0,25)
\Line(0,25)(35,25)
\CArc(60,25)(25,0,360)
\GCirc(85,25){15}{0.5}
\end{picture} \qquad &=& \quad -{3\over4}{\hbar^2\over v^3} \; I(0,m)^2 + {3\over2}{\hbar^2\over v^3} \; I(0,\mu) I(0,m) + \nonumber\\
& & \quad -9{\hbar^2m^2\over v^3} \; I(0,\mu,0,\mu) I(0,\mu) + \nonumber\\
& & \quad 21{\hbar^2m^2\over v^3} \; I(0,\mu,0,\mu) I(0,m) + \nonumber\\
& & \quad -27{\hbar^2m^4\over v^3} \; B_{\mu\mu\mu} - 3{\hbar^2m^4\over v^3} \; B_{\mu mm} - 3{\hbar^2m^2\over v^3} \; D_{mm\mu} \nonumber\\
\begin{picture}(100,50)(0,25)
\Line(0,25)(35,25)
\DashCArc(60,25)(25,0,360){3}
\GCirc(85,25){15}{0.5}
\end{picture} \qquad &=& \quad {\hbar^2\over m^2v^3} \; I(0,m) \; I - 3{\hbar^2\over m^2v^3} \; I(0,\mu) \; I - 7{\hbar^2\over v^3} \; I(0,m)^2 + \nonumber\\
& & \quad 16{\hbar^2\over v^3} \; I(0,\mu) I(0,m) + 8{\hbar^2m^2\over v^3} \; I(0,m,0,m) I(0,m) + \nonumber\\
& & \quad -14{\hbar^2m^2\over v^3} \; I(0,m,0,m) I(0,\mu) + \nonumber\\
& & \quad 4{\hbar^2m^4\over v^3} \; B_{m\mu m} - 6{\hbar^2m^2\over v^3} \; D_{mm\mu} \nonumber\\
\begin{picture}(100,50)(0,25)
\Line(0,25)(60,25)
\CArc(30,40)(15,0,360)
\GCirc(80,25){20}{0.5}
\end{picture} \qquad &=& \quad {9\over4}\frac{\hbar^2}{v^3} \; I(0,\mu)^2 - {3\over2}\frac{\hbar^2}{v^3} \; I(0,\mu) I(0,m) \nonumber\\
\begin{picture}(100,50)(0,25)
\Line(0,25)(60,25)
\DashCArc(30,40)(15,0,360){3}
\GCirc(80,25){20}{0.5}
\end{picture} \qquad &=& \quad -\h{\hbar^2\over m^2v^3} \; I(0,m) \; I + {3\over4}{\hbar^2\over m^2v^3} \; I(0,\mu) \; I + \nonumber\\
& & \quad -\frac{\hbar^2}{v^3} \; I(0,m)^2 + {3\over2}\frac{\hbar^2}{v^3} \; I(0,\mu) I(0,m) \nonumber\\
\begin{picture}(100,50)(0,25)
\Line(0,25)(60,25)
\Vertex(30,25){4}
\GCirc(80,25){20}{0.5}
\end{picture} \qquad &=& \quad -\h{\hbar^2\over m^2v^3} \; I(0,m) \; I + {3\over4}{\hbar^2\over m^2v^3} \; I(0,\mu) \; I \nonumber\\
\begin{picture}(100,50)(0,25)
\Line(0,25)(50,25)
\Line(50,25)(85,42)
\Line(50,25)(85,8)
\GCirc(85,42){15}{0.5}
\GCirc(85,8){15}{0.5}
\end{picture} \qquad &=& \quad -{27\over8}\frac{\hbar^2}{v^3} \; I(0,\mu)^2 - {3\over2}\frac{\hbar^2}{v^3} \; I(0,m)^2 + {9\over2}\frac{\hbar^2}{v^3} \; I(0,\mu) I(0,m) \nonumber\\
\begin{picture}(100,50)(0,25)
\Line(0,25)(100,25)
\CArc(75,25)(25,0,360)
\end{picture} \qquad &=& \quad 3\frac{\hbar^2m^2}{v^3} \; D_{\mu\mu\mu} \nonumber\\
\begin{picture}(100,50)(0,25)
\Line(0,25)(100,25)
\DashCArc(75,25)(25,0,360){3}
\end{picture} \qquad &=& \quad {\hbar^2\over m^2v^3} \; I(0,\mu) \; I - {\hbar^2\over v^3} \; I(0,m)^2 + {\hbar^2\over v^3} \; I(0,\mu) I(0,m) + \nonumber\\
& & \quad -3\frac{\hbar^2m^2}{v^3} \; D_{mm\mu} \nonumber\\
\begin{picture}(100,50)(0,25)
\Line(0,25)(75,25)
\CArc(75,37.5)(12.5,0,360)
\DashCArc(75,12.5)(12.5,0,360){3}
\end{picture} \qquad &=& \quad 3\frac{\hbar^2}{v^3} \; I(0,\mu) I(0,m) \nonumber\\
\begin{picture}(100,50)(0,25)
\Line(0,25)(50,25)
\CArc(75,25)(25,0,360)
\Vertex(50,25){4}
\end{picture} \qquad &=& \quad \h{\hbar^2\over m^2v^3} \; I(0,\mu) \; I
\eqa}

\vspace{15pt}

The complete result for the $\eta$-tadpole is finally:
\bqa
\langle\tilde{\eta}\rangle &=& -{3\over2}{\hbar\over v} \; I(0,\mu) + {\hbar\over v} \; I(0,m) + \nonumber\\
& & -{9\over8}{\hbar^2\over v^3} \; I(0,\mu)^2 - {45\over4}{\hbar^2\over v^3} \; I(0,m)^2 + 26{\hbar^2\over v^3} \; I(0,\mu) I(0,m) + \nonumber\\
& & -9{\hbar^2m^2\over v^3} \; I(0,\mu,0,\mu) I(0,\mu) + 8{\hbar^2m^2\over v^3} \; I(0,m,0,m) I(0,m) + \nonumber\\
& & 21{\hbar^2m^2\over v^3} \; I(0,\mu,0,\mu) I(0,m) - 14{\hbar^2m^2\over v^3} \; I(0,m,0,m) I(0,\mu) + \nonumber\\
& & -27{\hbar^2m^4\over v^3} \; B_{\mu\mu\mu} - 3{\hbar^2m^4\over v^3} \; B_{\mu mm} + 4{\hbar^2m^4\over v^3} \; B_{m\mu m} + \nonumber\\
& & 3{\hbar^2m^2\over v^3} \; D_{\mu\mu\mu} - 12{\hbar^2m^2\over v^3} \; D_{mm\mu}
\eqa

\subsubsection{The $\eta$-Propagator}

The diagrams for the momentum-space $\eta$-propagator are:
{\allowdisplaybreaks\bqa
\begin{picture}(70,40)(0,18)
\Line(0,20)(70,20)
\end{picture} \quad &=& \frac{\hbar}{p^2+\mu^2} \nonumber\\
\begin{picture}(70,40)(0,18)
\Line(0,20)(20,20)
\CArc(35,20)(15,-180,180)
\Line(50,20)(70,20)
\end{picture} \quad &=& 18{\hbar^2m^4\over v^2}\frac{1}{(p^2+\mu^2)^2} \; I(0,\mu,p,\mu) \nonumber\\
\begin{picture}(70,40)(0,18)
\Line(0,20)(20,20)
\DashCArc(35,20)(15,-180,180){3}
\Line(50,20)(70,20)
\end{picture} \quad &=& 2{\hbar^2\over v^2}\frac{1}{(p^2+\mu^2)^2} \; I - 6{\hbar^2m^2\over v^2}\frac{1}{(p^2+\mu^2)^2} \; I(0,m) + \nonumber\\
& & -{\hbar^2\over v^2}\frac{1}{p^2+\mu^2} \; I(0,m) + 2{\hbar^2m^2\over v^2}\frac{1}{p^2+\mu^2} \; I(0,m,p,m) + \nonumber\\
& & \h{\hbar^2\over v^2} \; I(0,m,p,m) + 2{\hbar^2m^4\over v^2}\frac{1}{(p^2+\mu^2)^2} \; I(0,m,p,m) \nonumber\\
\begin{picture}(70,40)(0,18)
\Line(0,5)(70,5)
\CArc(35,20)(15,-90,270)
\end{picture} \quad &=& -3{\hbar^2m^2\over v^2}\frac{1}{(p^2+\mu^2)^2} \; I(0,\mu) \nonumber\\
\begin{picture}(70,40)(0,18)
\Line(0,5)(70,5)
\DashCArc(35,20)(15,-90,270){3}
\end{picture} \quad &=& -{\hbar^2\over v^2}\frac{1}{(p^2+\mu^2)^2} \; I - 2{\hbar^2m^2\over v^2}\frac{1}{(p^2+\mu^2)^2} \; I(0,m) \nonumber\\
\begin{picture}(70,40)(0,18)
\Line(0,5)(70,5)
\Line(35,5)(35,25)
\GCirc(35,25){10}{0.5}
\end{picture} \quad &=& 9{\hbar^2m^2\over v^2}\frac{1}{(p^2+\mu^2)^2} \; I(0,\mu) - 6{\hbar^2m^2\over v^2}\frac{1}{(p^2+\mu^2)^2} \; I(0,m) \nonumber\\
\begin{picture}(70,40)(0,18)
\Line(0,20)(70,20)
\Vertex(35,20){4}
\end{picture} \quad &=& -{\hbar^2\over v^2}\frac{1}{(p^2+\mu^2)^2} \; I
\eqa}\\
For the connected $\eta$-propagator we get:
\bqa
\langle \tilde{\eta}(p)\tilde{\eta}(-p) \rangle_{\mathrm{c}} &=& \frac{\hbar}{p^2+\mu^2} + 6{\hbar^2m^2\over v^2}\frac{1}{(p^2+\mu^2)^2} \; I(0,\mu) - 14{\hbar^2m^2\over v^2}\frac{1}{(p^2+\mu^2)^2} \; I(0,m) + \nonumber\\
& & -{\hbar^2\over v^2}\frac{1}{p^2+\mu^2} \; I(0,m) + 18{\hbar^2m^4\over v^2}\frac{1}{(p^2+\mu^2)^2} \; I(0,\mu,p,\mu) + \nonumber\\
& & 2{\hbar^2m^4\over v^2}\frac{1}{(p^2+\mu^2)^2} \; I(0,m,p,m) + 2{\hbar^2m^2\over v^2}\frac{1}{p^2+\mu^2} \; I(0,m,p,m) + \nonumber\\
& & \h{\hbar^2\over v^2} \; I(0,m,p,m) \nonumber\\
\eqa

\subsubsection{The $w$-Propagator}

The diagrams contributing to the $w$-propagator up to order $\hbar^2$ are:
{\allowdisplaybreaks\bqa
\begin{picture}(70,40)(0,18)
\DashLine(0,20)(70,20){3}
\end{picture} \quad &=& \frac{\hbar}{p^2+m^2} \nonumber\\
\begin{picture}(70,40)(0,18)
\DashLine(0,20)(20,20){3}
\DashCArc(35,20)(15,-180,0){3}
\CArc(35,20)(15,0,180)
\DashLine(50,20)(70,20){3}
\end{picture} \quad &=& -{\hbar^2\over v^2}\frac{1}{p^2+m^2} \; I(0,m) + 4{\hbar^2m^2\over v^2}\frac{1}{(p^2+m^2)^2} \; I(0,m) + \nonumber\\
& & 3{\hbar^2\over v^2}\frac{1}{p^2+m^2} \; I(0,\mu) - 6{\hbar^2m^2\over v^2}\frac{1}{(p^2+m^2)^2} \; I(0,\mu) + \nonumber\\
& & 4{\hbar^2m^4\over v^2}\frac{1}{(p^2+m^2)^2} \; I(0,\mu,p,m) - 4{\hbar^2m^2\over v^2}\frac{1}{p^2+m^2} \; I(0,\mu,p,m) + \nonumber\\
& & {\hbar^2\over v^2} \; I(0,\mu,p,m)  \nonumber\\
\begin{picture}(70,40)(0,18)
\DashLine(0,5)(70,5){3}
\Line(35,5)(35,25)
\GCirc(35,25){10}{0.5}
\end{picture} \quad &=& -2{\hbar^2\over v^2}\frac{1}{p^2+m^2} \; I(0,m) + 4{\hbar^2m^2\over v^2}\frac{1}{(p^2+m^2)^2} \; I(0,m) + \nonumber\\
& & 3{\hbar^2\over v^2}\frac{1}{p^2+m^2} \; I(0,\mu) - 6{\hbar^2m^2\over v^2}\frac{1}{(p^2+m^2)^2} \; I(0,\mu) \nonumber\\
\begin{picture}(70,40)(0,18)
\DashLine(0,5)(70,5){3}
\CArc(35,20)(15,-90,270)
\end{picture} \quad &=& -{\hbar^2\over v^2}\frac{1}{p^2+m^2} \; I(0,\mu) -2{\hbar^2m^2\over v^2}\frac{1}{(p^2+m^2)^2} \; I(0,\mu) \nonumber\\
\eqa}
\bqa
\langle\tilde{w}(p)\tilde{w}(-p)\rangle_\mathrm{c} &=& \frac{\hbar}{p^2+m^2} + 8{\hbar^2m^2\over v^2}\frac{1}{(p^2+m^2)^2} \; I(0,m) - 3{\hbar^2\over v^2}\frac{1}{p^2+m^2} \; I(0,m) \nonumber\\
& & -14{\hbar^2m^2\over v^2}\frac{1}{(p^2+m^2)^2} \; I(0,\mu) + 5{\hbar^2\over v^2}\frac{1}{p^2+m^2} \; I(0,\mu) + {\hbar^2\over v^2} \; I(0,\mu,p,m) \nonumber\\
& & -4{\hbar^2m^2\over v^2}\frac{1}{p^2+m^2} \; I(0,\mu,p,m) + 4{\hbar^2m^4\over v^2}\frac{1}{(p^2+m^2)^2} \; I(0,\mu,p,m) \nonumber\\
\eqa

\subsubsection{Configuration-Space Green's Functions}

To compute the Cartesian $\f_1$- and $\f_2$-Green's-functions we can again use the expansions (\ref{Cartaverages}), since these expansions are model independent. First we have to find the configuration-space $\eta$- and $w$-Green's functions however.
{\allowdisplaybreaks\bqa
\langle\eta(x)\rangle &=& -{3\over2}{\hbar\over v} \; I(0,\mu) + {\hbar\over v} \; I(0,m) + \nonumber\\
& & -{9\over8}{\hbar^2\over v^3} \; I(0,\mu)^2 - {45\over4}{\hbar^2\over v^3} \; I(0,m)^2 + 26{\hbar^2\over v^3} \; I(0,\mu) I(0,m) + \nonumber\\
& & -9{\hbar^2m^2\over v^3} \; I(0,\mu,0,\mu) I(0,\mu) + \nonumber\\
& & 8{\hbar^2m^2\over v^3} \; I(0,m,0,m) I(0,m) + \nonumber\\
& & 21{\hbar^2m^2\over v^3} \; I(0,\mu,0,\mu) I(0,m) + \nonumber\\
& & -14{\hbar^2m^2\over v^3} \; I(0,m,0,m) I(0,\mu) + \nonumber\\
& & -27{\hbar^2m^4\over v^3} \; B_{\mu\mu\mu} - 3{\hbar^2m^4\over v^3} \; B_{\mu mm} + 4{\hbar^2m^4\over v^3} \; B_{m\mu m} + \nonumber\\
& & 3{\hbar^2m^2\over v^3} \; D_{\mu\mu\mu} - 12{\hbar^2m^2\over v^3} \; D_{mm\mu} \nonumber\\
\langle w(x)\rangle &=& 0 \nonumber\\
\langle\eta(x)\eta(y)\rangle &=& {1\over(2\pi)^d}\int d^dp \; e^{ip\cdot(x-y)} \; \langle\tilde{\eta}(p)\tilde{\eta}(-p)\rangle_{\mathrm{c}} + \langle\eta(x)\rangle^2 \nonumber\\
&=& {9\over4}{\hbar^2\over v^2} \; I(0,\mu)^2 + {\hbar^2\over v^2} \; I(0,m)^2 - 3{\hbar^2\over v^2} \; I(0,\mu) I(0,m) + \nonumber\\
& & \hbar \; A_m(x-y) + \nonumber\\
& & -{\hbar^2\over v^2} \; I(0,m) A_\mu(x-y) - 14{\hbar^2m^2\over v^2} \; I(0,m) C_{\mu\mu}(x-y) + \nonumber\\
& & 6{\hbar^2m^2\over v^2} \; I(0,\mu) C_{\mu\mu}(x-y) + \h{\hbar^2\over v^2} \; A_m(x-y)^2 + \nonumber\\
& & 2{\hbar^2m^2\over v^2} \; D_{\mu mm}(x-y) + 2{\hbar^2m^4\over v^2} \; B_{\mu mm}(x-y) + \nonumber\\
& & 18{\hbar^2m^4\over v^2} \; B_{\mu\mu\mu}(x-y) \nonumber\\
\langle w(x)w(y)\rangle &=& {1\over(2\pi)^d}\int d^dp \; e^{ip\cdot(x-y)} \; \langle\tilde{w}(p)\tilde{w}(-p)\rangle_{\mathrm{c}} \nonumber\\
&=& \hbar \; A_m(x-y) + \nonumber\\
& & -3{\hbar^2\over v^2} \; I(0,m) A_m(x-y) + 5{\hbar^2\over v^2} \; I(0,\mu) A_m(x-y) + \nonumber\\
& & 8{\hbar^2m^2\over v^2} \; I(0,m) C_{mm}(x-y) + \nonumber\\
& & -14{\hbar^2m^2\over v^2} \; I(0,\mu) C_{mm}(x-y) + \nonumber\\
& & {\hbar^2\over v^2} \; A_\mu(x-y) A_m(x-y) - 4{\hbar^2m^2\over v^2} \; D_{m\mu m}(x-y) + \nonumber\\
& & 4{\hbar^2m^4\over v^2} \; B_{m\mu m}(x-y) \nonumber\\
\langle\eta(x_1)\eta(x_2)w(x_3)\rangle &=& 0 \nonumber\\
\langle\eta(x_1)w(x_2)w(x_3)\rangle &=& \langle\eta\rangle \langle w(x_2)w(x_3)\rangle_{\mathrm{c}} + \langle\eta(x_1)w(x_2)w(x_3)\rangle_{\mathrm{c}} \nonumber\\
&=& {\hbar^2\over v} \; I(0,m) \; {1\over(2\pi)^d}\int d^dp \; \frac{e^{ip\cdot(x_2-x_3)}}{p^2+m^2} + \nonumber\\
& & -{3\over2}{\hbar^2\over v} \; I(0,\mu) \; {1\over(2\pi)^d}\int d^dp \; \frac{e^{ip\cdot(x_2-x_3)}}{p^2+m^2} + \nonumber\\
& & {1\over(2\pi)^d}\int d^dk_1 \cdots {1\over(2\pi)^d}\int d^dk_3 \; e^{i(k_1\cdot x_1+\ldots+k_3\cdot x_3)} \cdot \nonumber\\
& & \hspace{30pt} {2k_2\cdot k_3+2m^2\over\hbar v} \frac{\hbar}{k_1^2+\mu^2} \frac{\hbar}{k_2^2+m^2} \frac{\hbar}{k_3^2+m^2} \cdot \nonumber\\
& & \hspace{30pt} (2\pi)^d\delta^d(k_1+\ldots+k_3) \nonumber\\
&=& {\hbar^2\over v} \; I(0,m) A_m(x_2-x_3) - {3\over2}{\hbar^2\over v} \; I(0,\mu) A_m(x_2-x_3) + \nonumber\\
& & {\hbar^2\over v} \; A_m(x_1-x_2) A_m(x_1-x_3) + \nonumber\\
& & -{\hbar^2\over v} \; A_m(x_1-x_2) A_m(x_2-x_3) + \nonumber\\
& & -{\hbar^2\over v} \; A_m(x_1-x_3) A_m(x_2-x_3) + \nonumber\\
& & 2{\hbar^2m^2\over v} {1\over(2\pi)^{2d}}\int d^dk d^dl \; \frac{e^{ik\cdot(x_1-x_2)}}{k^2+m^2} \frac{e^{il\cdot(x_1-x_3)}}{l^2+m^2} \frac{1}{(k+l)^2+\mu^2} \nonumber\\
\langle w(x_1)w(x_2)w(x_3)\rangle &=& 0 \nonumber\\
\langle\eta(x_1)\eta(x_2)w(x_3)w(x_4)\rangle &=& \hbar^2 \; A_\mu(x_1-x_2) A_m(x_3-x_4) \nonumber\\
\langle\eta(x_1)w(x_2)w(x_3)w(x_4)\rangle &=& 0 \nonumber\\
\langle w(x_1)w(x_2)w(x_3)w(x_4)\rangle &=& \hbar^2 \; A_m(x_1-x_2) A_m(x_3-x_4) + \nonumber\\
& & \hbar^2 \; A_m(x_1-x_3) A_m(x_2-x_4) + \nonumber\\
& & \hbar^2 \; A_m(x_1-x_4) A_m(x_2-x_3) \nonumber\\
\eqa}

Now these results can be substituted in (\ref{Cartaverages}). Doing this one finds again the results (\ref{Cartarctan}). So also in case of the arctangent toy model the conjecture is verified for several Green's functions up to order $\hbar^2$.\index{Green's functions!arctangent toy model}

\newpage

\section{Proof of the Conjecture}\label{proofconj}

In the last two sections evidence for the truth of the \ind{conjecture} has accumulated.\index{proof conjecture} In this section we shall prove this conjecture for a general model in $d$ space-time dimensions. A general action (in terms of normal, Cartesian fields) for a $d$-dimensional model with two fields is given by:
\bq \label{generalaction}
S = \int d^dx \left( \h\left(\nabla\f_1\right)^2 + \h\left(\nabla\f_2\right)^2 + V(\f_1,\f_2) \right) \;.
\eq

Our proof will be based on the fact that the transformation to polar field variables actually has to be performed in the path integral on the \ind{lattice}, i.e.\ in the path integral formulated in a discrete way. After transforming to polar fields one gets a path integral in terms of polar fields formulated on a lattice. This path integral gives a (complicated) set of Feynman rules, and diagrams actually have to be calculated with space-time still discrete. Only in the end result for the Green's function one should then take the \ind{continuum limit}, i.e.\ $\Delta\rightarrow0$.

Now a $d$-dimensional \ind{continuum calculation} is correct if one can see that all the steps one performs there to calculate a diagram correspond to a similar step in a discrete calculation. In a continuum calculation one performs the following three steps when calculating any diagram:
\begin{enumerate}
\item One writes momentum-dependent factors from the vertices in terms of the denominators of propagators, such that one can let them cancel. For example:
\bq
k\cdot l = \h((k+l)^2+m^2) - \h(k^2+m^2) - \h(l^2+m^2) + \h m^2 \;.
\eq
\item One shifts momenta, for example:
\bq
\int d^dk \frac{k^2+m^2}{(k+l)^2+m^2} = \int d^dk \frac{(k-l)^2+m^2}{k^2+m^2} \;.
\eq
\item When there is momentum dependence left in the numerator, which cannot cancel anything in the denominator anymore, one uses
\bq
\int d^dk \; \frac{k_i}{k^2+m^2} = 0 \;.
\eq
For example:
\bqa
& & \int d^dk \int d^dl \; \frac{(k-l)^2+m^2}{(k^2+m^2)(l^2+m^2)} \nonumber\\
& & = \int d^dk \int d^dl \; \frac{(k^2+m^2)+(l^2+m^2)-m^2}{(k^2+m^2)(l^2+m^2)} \nonumber\\
& & = \int d^dk \int d^dl \; \left( \frac{1}{l^2+m^2} + \frac{1}{k^2+m^2} - \frac{m^2}{(k^2+m^2)(l^2+m^2)} \right)
\eqa
\end{enumerate}
If we can somehow see that these steps are also valid in a \ind{discrete calculation}, then we have proven the conjecture. For then we know that every operation one performs in the $d$-dimensional continuum calculation corresponds to a valid operation in a discrete calculation, even though one writes down these steps in a continuum formalism.

Because we need the path integral on a lattice we define \ind{discrete space-time variables} $i$ as:
\bq
x \equiv (-L/2+\Delta i_1,-L/2+\Delta i_2,\ldots,-L/2+\Delta i_d) \;, \quad i=0,\ldots,N-1 \;, \Delta N = L
\eq
Notice that we have limited the space-time domain (each direction goes from $-L/2$ to $L/2$) \emph{and} we have made this domain discrete (in every direction there are $N$ lattice sites). The fields on the discrete lattice are denoted by:
\bqa
\f_{1 \;\; i_1,\ldots,i_d} &\equiv& \f_1(x) \nonumber\\
\f_{2 \;\; i_1,\ldots,i_d} &\equiv& \f_2(x)
\eqa
For these fields we shall assume periodic boundary conditions in all directions. For the 1-direction this means:
\bqa
\f_{1 \;\; N,i_2,\ldots,i_d} &=& \f_{1 \;\; 0,i_2,\ldots,i_d} \nonumber\\
\f_{2 \;\; N,i_2,\ldots,i_d} &=& \f_{2 \;\; 0,i_2,\ldots,i_d}
\eqa

The action (\ref{generalaction}), formulated on this lattice, is:\index{action!on lattice}
\bqa
S &=& \Delta^d \sum_{i_1,\ldots,i_d=0}^{N-1} \Bigg( \h{1\over\Delta^2}\left(\f_{1 \;\; i_1+1,i_2,\ldots,i_d}-\f_{1 \;\; i_1,i_2,\ldots,i_d}\right)^2 + \ldots + \nonumber\\
& & \phantom{\Delta^d \sum_{i_1,\ldots,i_d=0}^{N-1} \Bigg(} \h{1\over\Delta^2}\left(\f_{1 \;\; i_1,\ldots,i_{d-1},i_d+1}-\f_{1 \;\; i_1,\ldots,i_{d-1},i_d}\right)^2 + \nonumber\\
& & \phantom{\Delta^d \sum_{i_1,\ldots,i_d=0}^{N-1} \Bigg(} \h{1\over\Delta^2}\left(\f_{2 \;\; i_1+1,i_2,\ldots,i_d}-\f_{2 \;\; i_1,i_2,\ldots,i_d}\right)^2 + \ldots + \nonumber\\
& & \phantom{\Delta^d \sum_{i_1,\ldots,i_d=0}^{N-1} \Bigg(} \h{1\over\Delta^2}\left(\f_{2 \;\; i_1,\ldots,i_{d-1},i_d+1}-\f_{2 \;\; i_1,\ldots,i_{d-1},i_d}\right)^2 + \nonumber\\
& & \phantom{\Delta^d \sum_{i_1,\ldots,i_d=0}^{N-1} \Bigg(} V\left(\f_{1 \;\; i_1,\ldots,i_d},\f_{2 \;\; i_1,\ldots,i_d}\right) \Bigg)
\eqa

Now we may transform to polar field variables, since now the path integral is properly defined, it has become merely a $2N^d$-dimensional integral. The transformation goes as follows:\index{transformation to polar fields}
\bqa
\f_{1 \;\; i_1,\ldots,i_d} &=& r_{i_1,\ldots,i_d} \cos\left(\frac{w_{i_1,\ldots,i_d}}{v}\right) \nonumber\\
\f_{2 \;\; i_1,\ldots,i_d} &=& r_{i_1,\ldots,i_d} \sin\left(\frac{w_{i_1,\ldots,i_d}}{v}\right) \label{discrtransf}
\eqa
This can be substituted in the action above. To keep things readable we define the shorthand notations:
\bqa
i &\equiv& i_1,\ldots,i_d \nonumber\\
\nabla r_{i_1,\ldots,i_d} &\equiv& \left( {1\over\Delta}\left(r_{i_1+1,i_2,\ldots,i_d}-r_{i_1,i_2,\ldots,i_d}\right), \ldots, {1\over\Delta}\left(r_{i_1,\ldots,i_{d-1},i_d+1}-r_{i_1,\ldots,i_{d-1},i_d}\right) \right) \nonumber\\ \label{shorthand}
\eqa
The action becomes:
\bqa
S &=& \Delta^d \sum_{i_1,\ldots,i_d=0}^{N-1} \Bigg( \h\left(\nabla r_i\right)^2 + {1\over\Delta^2}r_i^2\left( \left(1-\cos\frac{\Delta\nabla_1w_i}{v}\right) + \ldots + \left(1-\cos\frac{\Delta\nabla_dw_i}{v}\right) \right) + \nonumber\\
& & \phantom{\Delta^d \sum_{i_1,\ldots,i_d=0}^{N-1} \Bigg(} {1\over\Delta}r_i\nabla_1r_i \left(1-\cos\frac{\Delta\nabla_1w_i}{v}\right) + \ldots + {1\over\Delta}r_i\nabla_dr_i \left(1-\cos\frac{\Delta\nabla_dw_i}{v}\right) + \nonumber\\
& & \phantom{\Delta^d \sum_{i_1,\ldots,i_d=0}^{N-1} \Bigg(} V\left(r_i\cos{w_i\over v}, r_i\sin{w_i\over v}\right) \Bigg)
\eqa

Now we will expand the cosines in the first two lines. Also we will assume that the potential is such that the minimum of the complete action is at $v$, where $v$ is some nonzero constant (the same $v$ that divides $w$ in the cosine). This assumption is necessary to avoid difficulties with the singularity at $r=0$ in the transformation (\ref{discrtransf}). Because the minimum of the action is at $r=v$ we also expand the action around this value:
\bq
r_i = v+\eta_i \;.
\eq
The final form of the discrete action is:
\bqa
S &=& \Delta^d \sum_{i_1,\ldots,i_d=0}^{N-1} \Bigg( \h\left(\nabla\eta_i\right)^2 + \h\left(\nabla w_i\right)^2 \nonumber\\
& & \phantom{\Delta^d \sum_{i_1,\ldots,i_d=0}^{N-1} \Bigg(} -\sum_{n=2}^\infty \frac{(-1)^n}{(2n)!} \frac{\Delta^{2n-2}}{v^{2n-2}} \left( \left(\nabla_1w_i\right)^{2n} + \ldots + \left(\nabla_dw_i\right)^{2n} \right) \nonumber\\
& & \phantom{\Delta^d \sum_{i_1,\ldots,i_d=0}^{N-1} \Bigg(} -2\sum_{n=1}^\infty \frac{(-1)^n}{(2n)!} \frac{\Delta^{2n-2}}{v^{2n-2}} \left( \eta_i\left(\nabla_1w_i\right)^{2n} + \ldots + \eta_i\left(\nabla_dw_i\right)^{2n} \right) \nonumber\\
& & \phantom{\Delta^d \sum_{i_1,\ldots,i_d=0}^{N-1} \Bigg(} -\sum_{n=1}^\infty \frac{(-1)^n}{(2n)!} \frac{\Delta^{2n-2}}{v^{2n-2}} \left( \left(\nabla_1\eta_i\right)\left(\nabla_1w_i\right)^{2n} + \ldots + \left(\nabla_d\eta_i\right)\left(\nabla_dw_i\right)^{2n} \right) \nonumber\\
& & \phantom{\Delta^d \sum_{i_1,\ldots,i_d=0}^{N-1} \Bigg(} -\sum_{n=1}^\infty \frac{(-1)^n}{(2n)!} \frac{\Delta^{2n-2}}{v^{2n-2}} \left( \eta_i^2\left(\nabla_1w_i\right)^{2n} + \ldots + \eta_i^2\left(\nabla_dw_i\right)^{2n} \right) \nonumber\\
& & \phantom{\Delta^d \sum_{i_1,\ldots,i_d=0}^{N-1} \Bigg(} -\sum_{n=1}^\infty \frac{(-1)^n}{(2n)!} \frac{\Delta^{2n-2}}{v^{2n-2}} \left( \eta_i\left(\nabla_1\eta_i\right)\left(\nabla_1w_i\right)^{2n} + \ldots + \eta_i\left(\nabla_d\eta_i\right)\left(\nabla_dw_i\right)^{2n} \right) \nonumber\\
& & \phantom{\Delta^d \sum_{i_1,\ldots,i_d=0}^{N-1} \Bigg(} +V\left((v+\eta_i)\cos{w_i\over v}, (v+\eta_i)\sin{w_i\over v}\right) \Bigg) \label{discractioncompl}
\eqa

Note that this expression is still exact, as long as we keep all the terms in the sums coming from expanding the cosines. Our complete discrete path integral $P$ (defined in (\ref{genericpathint})), formulated in terms of the polar fields, now looks like:\index{path integral on lattice}
\bq
P = \int_{-v}^\infty \left(\prod_{i=0}^{N-1} (v+\eta_i)d\eta_i\right) \int_{-v\pi}^{v\pi} \left(\prod_{i=0}^{N-1} dw_i\right) \; O \; \exp\left(-{1\over\hbar}S\right) \;,
\eq
where the action $S$ is given by (\ref{discractioncompl}), $O$ is some product of the $\f$-fields:
\bq
O \equiv \left(v+\eta_{j_1}\right)\cos{w_{j_1}\over v} \cdots \left(v+\eta_{j_m}\right)\cos{w_{j_m}\over v} \cdot \left(v+\eta_{k_1}\right)\sin{w_{k_1}\over v} \cdots \left(v+\eta_{k_n}\right)\sin{w_{k_n}\over v} \;,
\eq
and $i$ still is a shorthand notation for $i_1,\ldots,i_d$. Here $j_1,\ldots,j_m$ and $k_1,\ldots,k_n$ are discrete space-time coordinates. (By $j_1$ we actually mean $j_{1 \; 1},\ldots,j_{1 \; d}$, as in (\ref{shorthand}).) 

The product
\bq
\prod_{i=0}^{N-1} (v+\eta_i) = \prod_{i=0}^{N-1} r_i
\eq
is the \ind{Jacobian} from the transformation to polar fields. This Jacobian factor can be recast in the following form:
\bq
\left(\prod_{i=0}^{N-1} (v+\eta_i)\right) = \exp\left(\sum_{i=0}^{N-1} \ln(v+\eta_i)\right)
\eq

Also the domain of integration for the $w$-fields can be extended from $[-v\pi,v\pi]$ to $\langle-\infty,\infty\rangle$, because the whole integrand is periodic in the $w$-fields. Finally we can also extend the lower integration boundary for the $\eta$-fields from $-v$ to $-\infty$, this shift will only have non-perturbative effects.

So we have brought our path integral to the form
\bq
P = \int_{-\infty}^\infty \left(\prod_{i=0}^{N-1} d\eta_i\right) \int_{-\infty}^{\infty} \left(\prod_{i=0}^{N-1} dw_i\right) \; O \; \exp\left(-{1\over\hbar}S + \sum_{i=0}^{N-1} \ln(v+\eta_i)\right) \;.
\eq
This is a normal path integral on a lattice\index{path integral on lattice}, in the sense that it has the same form of a path integral in terms of Cartesian fields. Such a path integral we can calculate in the ordinary way, with perturbation theory. The action of this path integral does have an infinite number of vertices however, because the discrete action (\ref{discractioncompl}) has an infinite number of interaction terms and also because the expansion of the logarithm coming from the Jacobian has an infinite number of terms. But it is still an \emph{exact} expression, because we keep \emph{all} terms.

Now we have to write down the (discrete) Feynman rules for the action we have found. To write down the momentum-space Feynman rules we must first transform the configuration-space fields $\eta$ and $w$ to momentum-space fields $\tilde{\eta}$ and $\tilde{w}$. In the continuum such a transformation is given by:\index{Fourier transform}
\bq
\tilde{\eta}(k) = \int d^dx \; \eta(x) \; e^{ik\cdot x} \;,
\eq
and similar for the $\tilde{w}$-field. The discrete analogue of this formula is:\index{Fourier transform!discrete}
\bq \label{discrFourier}
\tilde{\eta}_{k_1,\ldots,k_d} = \Delta^d \sum_{i_1,\ldots,i_d=0}^{N-1} \eta_{i_1,\ldots,i_d} \; \exp\left({2\pi i\over L}\left(k_1(\Delta i_1-L/2)+\ldots+k_d(\Delta i_d-L/2)\right)\right) \;,
\eq
where we have used that the continuous momentum is related to the discrete momentum as
\bq \label{discrcontmom}
k_{\mathrm{cont}} = \frac{2\pi k_{\mathrm{discr}}}{L} \;.
\eq
The inverse transformation of (\ref{discrFourier}) is
\bq \label{invdiscrFourier}
\eta_{i_1,\ldots,i_d} = {1\over L^d} \sum_{k_1,\ldots,k_d=-N/2}^{N/2-1} \tilde{\eta}_{k_1,\ldots,k_d} \; \exp\left(-{2\pi i\over L}\left(k_1(\Delta i_1-L/2)+\ldots+k_d(\Delta i_d-L/2)\right)\right) \;,
\eq
and similar for the $w$-field. To see that this is indeed the inverse transformation of (\ref{discrFourier}) one can use the identity
\bq
{1\over N^d} \sum_{i_1,\ldots,i_d=0}^{N-1} \exp\left({2\pi i\over N}(k_1i_1+\ldots+k_di_d)\right) = \delta_{k_1,0\;\mathrm{mod}\;N} \cdots \delta_{k_d,0\;\mathrm{mod}\;N} \;.
\eq

From the relation (\ref{discrcontmom}) we can see that when we take $L\rightarrow\infty$ the momenta become a continuous set. Their domain is still finite however. Because the discrete momenta are between $-N/2$ and $N/2-1$, the continuous momenta are in the domain
\bq
k_{\mathrm{cont}} \in \left\langle -{\pi\over\Delta}, {\pi\over\Delta} \right\rangle \;.
\eq
The finiteness of this domain reflects the discreteness of space-time. From now on we shall understand that we have taken the limit $L\rightarrow\infty$, such that all sums over momenta become integrals. But of course $\Delta$ is still finite.

By using (\ref{invdiscrFourier}), in the limit $L\rightarrow\infty$, we can now express the discrete action (\ref{discractioncompl}) in terms of the momentum-space fields $\tilde{\eta}$ and $\tilde{w}$. From this action one can then read of the discrete, momentum-space Feynman rules. Notice that we did not specify the potential $V$, so we will not include the Feynman rules coming from this part of the action. This potential $V$ will also determine the masses for the $\eta$- and $w$-field. We shall keep these masses general, the upcoming proof for the conjecture will not depend on the explicit form of the potential $V$ and the masses. In the Feynman rules below we will neither include the Feynman rules from the Jacobian, the proof of the conjecture will also not depend on the exact form of these vertices.

The discrete, momentum-space Feynman rules are then:\index{Feynman rules!discrete}
{\allowdisplaybreaks\bqa
\begin{picture}(100, 20)(0, 17)
\Line(20, 20)(80, 20)
\end{picture}
&\leftrightarrow& \frac{\hbar}{{2d\over\Delta^2}-{2\over\Delta^2}\cos\Delta k_1-\ldots-{2\over\Delta^2}\cos\Delta k_d+m_\eta^2} \nonumber\\
\begin{picture}(100, 20)(0, 17)
\DashLine(20, 20)(80, 20){5}
\end{picture}
&\leftrightarrow& \frac{\hbar}{{2d\over\Delta^2}-{2\over\Delta^2}\cos\Delta k_1-\ldots-{2\over\Delta^2}\cos\Delta k_d+m_w^2} \nonumber\\
\begin{picture}(100, 40)(0, 17)
\Line(20, 20)(50, 20)
\DashLine(50, 20)(80, 0){5}
\DashLine(50, 20)(80, 40){5}
\Text(20, 26)[]{$p$}
\Text(84, 32)[]{$k^{(1)}$}
\Text(84, 10)[]{$k^{(2)}$}
\end{picture}
&\leftrightarrow& -\frac{1}{\hbar v}{1\over\Delta^2} \bigg[ \left(e^{-i\Delta p_1}+1\right)\left(e^{-i\Delta k^{(1)}_1}-1\right)\left(e^{-i\Delta k^{(2)}_1}-1\right) \nonumber\\
& & \phantom{-\frac{1}{\hbar v}{1\over\Delta^2} \bigg[} + \ldots + \nonumber\\
& & \phantom{-\frac{1}{\hbar v}{1\over\Delta^2} \bigg[} \left(e^{-i\Delta p_d}+1\right)\left(e^{-i\Delta k^{(1)}_d}-1\right)\left(e^{-i\Delta k^{(2)}_d}-1\right) \bigg] \nonumber\\
&\vdots& \nonumber\\
\begin{picture}(100, 40)(0, 17)
\Line(20, 20)(50, 20)
\DashLine(50, 20)(80, 0){5}
\DashLine(50, 20)(80, 40){5}
\DashCArc(50,20)(30,-20,20){2}
\Text(20, 26)[]{$p$}
\Text(84, 48)[]{$k^{(1)}$}
\Text(84, -8)[]{$k^{(2n)}$}
\end{picture}
&\leftrightarrow& \frac{(-1)^n}{\hbar v^{2n-1}}{1\over\Delta^2} \bigg[ \left(e^{-i\Delta p_1}+1\right)\left(e^{-i\Delta k^{(1)}_1}-1\right) \cdots \left(e^{-i\Delta k^{(2n)}_1}-1\right) \nonumber\\
& & \phantom{\frac{(-1)^n}{\hbar v^{2n-1}}{1\over\Delta^2} \bigg[} + \ldots + \nonumber\\
& & \phantom{\frac{(-1)^n}{\hbar v^{2n-1}}{1\over\Delta^2} \bigg[} \left(e^{-i\Delta p_d}+1\right)\left(e^{-i\Delta k^{(1)}_d}-1\right) \cdots \left(e^{-i\Delta k^{(2n)}_d}-1\right) \bigg] \nonumber\\
&\vdots& \nonumber\\
\begin{picture}(100, 40)(0, 17)
\Line(20,0)(50,20)
\Line(20,40)(50,20)
\DashLine(50, 20)(80, 0){5}
\DashLine(50, 20)(80, 40){5}
\Text(20, 32)[]{$p$}
\Text(20, 10)[]{$q$}
\Text(84, 32)[]{$k^{(1)}$}
\Text(84, 10)[]{$k^{(2)}$}
\end{picture}
&\leftrightarrow& -\frac{1}{\hbar v^2}{1\over\Delta^2} \bigg[ \left(e^{-i\Delta p_1}+e^{-i\Delta q_1}\right)\left(e^{-i\Delta k^{(1)}_1}-1\right)\left(e^{-i\Delta k^{(2)}_1}-1\right) \nonumber\\
& & \phantom{-\frac{1}{\hbar v^2}{1\over\Delta^2} \bigg[} + \ldots + \nonumber\\
& & \phantom{-\frac{1}{\hbar v^2}{1\over\Delta^2} \bigg[} \left(e^{-i\Delta p_d}+e^{-i\Delta q_d}\right)\left(e^{-i\Delta k^{(1)}_d}-1\right)\left(e^{-i\Delta k^{(2)}_d}-1\right) \bigg] \nonumber\\
&\vdots& \nonumber\\
\begin{picture}(100, 40)(0, 17)
\Line(20,0)(50,20)
\Line(20,40)(50,20)
\DashLine(50, 20)(80, 0){5}
\DashLine(50, 20)(80, 40){5}
\DashCArc(50,20)(30,-20,20){2}
\Text(20, 32)[]{$p$}
\Text(20, 10)[]{$q$}
\Text(84, 48)[]{$k^{(1)}$}
\Text(84, -8)[]{$k^{(2n)}$}
\end{picture}
&\leftrightarrow& \frac{(-1)^n}{\hbar v^{2n}}{1\over\Delta^2} \bigg[ \left(e^{-i\Delta p_1}+e^{-i\Delta q_1}\right)\left(e^{-i\Delta k^{(1)}_1}-1\right) \cdots \left(e^{-i\Delta k^{(2n)}_1}-1\right) \nonumber\\
& & \phantom{\frac{(-1)^n}{\hbar v^{2n}}{1\over\Delta^2} \bigg[} + \ldots + \nonumber\\
& & \phantom{\frac{(-1)^n}{\hbar v^{2n}}{1\over\Delta^2} \bigg[} \left(e^{-i\Delta p_d}+e^{-i\Delta q_d}\right)\left(e^{-i\Delta k^{(1)}_d}-1\right) \cdots \left(e^{-i\Delta k^{(2n)}_d}-1\right) \bigg] \nonumber\\
&\vdots& \nonumber\\
\begin{picture}(100, 40)(0, 17)
\DashLine(20,0)(80,40){5}
\DashLine(20,40)(80,0){5}
\Text(20, 32)[]{$k^{(1)}$}
\Text(20, 10)[]{$k^{(2)}$}
\Text(84, 32)[]{$k^{(3)}$}
\Text(84, 10)[]{$k^{(4)}$}
\end{picture}
&\leftrightarrow& \frac{1}{\hbar v^2}{1\over\Delta^2} \bigg[ \left(e^{-i\Delta k^{(1)}_1}-1\right) \cdots \left(e^{-i\Delta k^{(4)}_1}-1\right) \nonumber\\
& & \phantom{\frac{1}{\hbar v^2}{1\over\Delta^2} \bigg[} + \ldots + \nonumber\\
& & \phantom{\frac{1}{\hbar v^2}{1\over\Delta^2} \bigg[} \left(e^{-i\Delta k^{(1)}_d}-1\right) \cdots \left(e^{-i\Delta k^{(4)}_d}-1\right) \bigg] \nonumber\\
&\vdots& \nonumber\\
\begin{picture}(100, 40)(0, 17)
\DashLine(50,20)(80,40){5}
\DashLine(50,20)(80,0){5}
\DashCArc(50,20)(30,50,310){2}
\Text(84, 48)[]{$k^{(1)}$}
\Text(84, -8)[]{$k^{(2n)}$}
\end{picture}
&\leftrightarrow& \frac{(-1)^n}{\hbar v^{2n-2}}{1\over\Delta^2} \bigg[ \left(e^{-i\Delta k^{(1)}_1}-1\right) \cdots \left(e^{-i\Delta k^{(2n)}_1}-1\right) \nonumber\\
& & \phantom{\frac{(-1)^n}{\hbar v^{2n-2}}{1\over\Delta^2} \bigg[} + \ldots + \nonumber\\
& & \phantom{\frac{(-1)^n}{\hbar v^{2n-2}}{1\over\Delta^2} \bigg[} \left(e^{-i\Delta k^{(1)}_d}-1\right) \cdots \left(e^{-i\Delta k^{(2n)}_d}-1\right) \bigg] \nonumber\\
&\vdots& \label{discrFeynman}
\eqa}
Here all the (continuum) momenta are counted incoming.
Together with these Feynman rules for the propagator and the vertices we have the rule that every internal momentum should be integrated over from $-\pi/\Delta$ to $\pi/\Delta$.

Now all these vertices can be written in a more convenient form. By combining all the complex exponentials (i.e.\ writing out all products) and using momentum conservation at the vertex one will notice that for each exponential also its complex conjugate occurs. They can be combined into a cosine, the same cosine that occurs in the discrete propagator. In this way we can write the vertex expressions above in terms of the denominator of the propagator. As a shorthand notation we denote the denominators by $\Pi_k$ and $\bar{\Pi}_k$:
\bqa
\Pi_k &\equiv& {2d\over\Delta^2}-{2\over\Delta^2}\cos{2\pi k_1\over N}-\ldots-{2\over\Delta^2}\cos{2\pi k_d\over N}+m_w^2 \nonumber\\
\bar{\Pi}_k &\equiv& {2d\over\Delta^2}-{2\over\Delta^2}\cos{2\pi k_1\over N}-\ldots-{2\over\Delta^2}\cos{2\pi k_d\over N}+m_\eta^2
\eqa
To write the vertices into this more convenient form we also have to define an operator $P$. We denote the set consisting of the $j^\mathrm{th}$ components of the momenta $k^{(1)},\ldots,k^{(2n)}$ by $\{k_j\}$:
\bq
\{k_j\} \equiv \{k_j^{(1)},k_j^{(2)},\ldots,k_j^{(2n)}\} \;.
\eq
Then the operator $P_i$ working on $\{k_j\}$ returns the sum of $i$ momenta chosen from the set $\{k_j\}$. There are ${2n\choose i}$ ways to choose $i$ momenta from a set of $2n$ momenta, so there are also ${2n\choose i}$ different operators $P_i$. For example:
\bq
P_2 \{k_j\} = k_j^{(1)} + k_j^{(2)} \;.
\eq
With these notations we can write the vertex expressions as follows.
{\allowdisplaybreaks\bqa
\begin{picture}(100, 40)(0, 17)
\Line(20, 20)(50, 20)
\DashLine(50, 20)(80, 0){5}
\DashLine(50, 20)(80, 40){5}
\Text(20, 26)[]{$p$}
\Text(84, 32)[]{$k^{(1)}$}
\Text(84, 10)[]{$k^{(2)}$}
\end{picture}
&\leftrightarrow& -\frac{1}{\hbar v}\left( \Pi_{k^{(1)}} + \Pi_{k^{(2)}} - \bar{\Pi}_p - 2m_w^2 + m_\eta^2 \right) \nonumber\\
&\vdots& \nonumber\\
\begin{picture}(100, 40)(0, 17)
\Line(20, 20)(50, 20)
\DashLine(50, 20)(80, 0){5}
\DashLine(50, 20)(80, 40){5}
\DashCArc(50,20)(30,-20,20){2}
\Text(20, 26)[]{$p$}
\Text(84, 48)[]{$k^{(1)}$}
\Text(84, -8)[]{$k^{(2n)}$}
\end{picture}
&\leftrightarrow& \frac{(-1)^n}{\hbar v^{2n-1}} \Bigg( -\bar{\Pi}_p + \sum_{P_1} \Pi_{P_1\{k\}} + \sum_{P_1} \Pi_{p+P_1\{k\}} + \nonumber\\
& & \phantom{\frac{(-1)^n}{\hbar v^{2n-1}} \Bigg(} -\sum_{P_2} \bar{\Pi}_{P_2\{k\}} - \sum_{P_2} \bar{\Pi}_{p+P_2\{k\}} + \ldots + \nonumber\\
& & \phantom{\frac{(-1)^n}{\hbar v^{2n-1}} \Bigg(} (-1)^n\sum_{P_{n-1}} \Pi_{P_{n-1}\{k\}} + (-1)^n\sum_{P_{n-1}} \Pi_{p+P_{n-1}\{k\}} + \nonumber\\
& & \phantom{\frac{(-1)^n}{\hbar v^{2n-1}} \Bigg(} \h(-1)^{n+1}\sum_{P_n} \bar{\Pi}_{P_n\{k\}} + \h(-1)^{n+1}\sum_{P_n} \bar{\Pi}_{p+P_n\{k\}} + \nonumber\\
& & \phantom{\frac{(-1)^n}{\hbar v^{2n-1}} \Bigg(} -2^{2n-1}m_w^2 - \left(1-2^{2n-1}\right)m_\eta^2 \Bigg) \nonumber\\
&\vdots& \nonumber\\
\begin{picture}(100, 40)(0, 17)
\Line(20,0)(50,20)
\Line(20,40)(50,20)
\DashLine(50, 20)(80, 0){5}
\DashLine(50, 20)(80, 40){5}
\Text(20, 32)[]{$p$}
\Text(20, 10)[]{$q$}
\Text(84, 32)[]{$k^{(1)}$}
\Text(84, 10)[]{$k^{(2)}$}
\end{picture}
&\leftrightarrow& -\frac{1}{\hbar v^2}\bigg( -\bar{\Pi}_p - \bar{\Pi}_q + \nonumber\\
& & \phantom{-\frac{1}{\hbar v^2}\bigg(} \h\Pi_{p+k^{(1)}} + \h\Pi_{p+k^{(2)}} + \h\Pi_{q+k^{(1)}} + \h\Pi_{q+k^{(2)}} + \nonumber\\
& & \phantom{-\frac{1}{\hbar v^2}\bigg(} -2m_w^2 + 2m_\eta^2 \bigg) \nonumber\\
&\vdots& \nonumber\\
\begin{picture}(100, 40)(0, 17)
\Line(20,0)(50,20)
\Line(20,40)(50,20)
\DashLine(50, 20)(80, 0){5}
\DashLine(50, 20)(80, 40){5}
\DashCArc(50,20)(30,-20,20){2}
\Text(20, 32)[]{$p$}
\Text(20, 10)[]{$q$}
\Text(84, 48)[]{$k^{(1)}$}
\Text(84, -8)[]{$k^{(2n)}$}
\end{picture}
&\leftrightarrow& \frac{(-1)^n}{\hbar v^{2n}}\Bigg( -\bar{\Pi}_p - \bar{\Pi}_q + \sum_{P_1} \Pi_{p+P_1\{k\}} + \sum_{P_1} \Pi_{q+P_1\{k\}} + \nonumber\\
& & \phantom{\frac{(-1)^n}{\hbar v^{2n}}\Bigg(} -\sum_{P_2} \bar{\Pi}_{p+P_2\{k\}} - \sum_{P_2} \bar{\Pi}_{q+P_2\{k\}} + \ldots + \nonumber\\
& & \phantom{\frac{(-1)^n}{\hbar v^{2n}}\Bigg(} (-1)^n\sum_{P_{n-1}} \Pi_{p+P_{n-1}\{k\}} + (-1)^n\sum_{P_{n-1}} \Pi_{q+P_{n-1}\{k\}} + \nonumber\\
& & \phantom{\frac{(-1)^n}{\hbar v^{2n}}\Bigg(} \h(-1)^{n+1}\sum_{P_n} \bar{\Pi}_{p+P_n\{k\}} + \h(-1)^{n+1}\sum_{P_n} \bar{\Pi}_{q+P_n\{k\}} + \nonumber\\
& & \phantom{\frac{(-1)^n}{\hbar v^{2n}}\Bigg(} - 2^{2n-1}m_w^2 + 2^{2n-1}m_\eta^2 \Bigg) \nonumber\\
&\vdots& \nonumber\\
\begin{picture}(100, 40)(0, 17)
\DashLine(20,0)(80,40){5}
\DashLine(20,40)(80,0){5}
\Text(20, 32)[]{$k^{(1)}$}
\Text(20, 10)[]{$k^{(2)}$}
\Text(84, 32)[]{$k^{(3)}$}
\Text(84, 10)[]{$k^{(4)}$}
\end{picture}
&\leftrightarrow& \frac{1}{\hbar v^2}\bigg( \Pi_{k^{(1)}} + \Pi_{k^{(2)}} + \Pi_{k^{(3)}} + \Pi_{k^{(4)}} + \nonumber\\
& & \phantom{\frac{1}{\hbar v^2}\bigg(} -\h\bar{\Pi}_{k^{(1)}+k^{(2)}} - \h\bar{\Pi}_{k^{(1)}+k^{(3)}} - \h\bar{\Pi}_{k^{(1)}+k^{(4)}} + \nonumber\\
& & \phantom{\frac{1}{\hbar v^2}\bigg(} -\h\bar{\Pi}_{k^{(2)}+k^{(3)}} - \h\bar{\Pi}_{k^{(2)}+k^{(4)}} - \h\bar{\Pi}_{k^{(3)}+k^{(4)}} - 4m_w^2 + 3m_\eta^2 \bigg) \nonumber\\
&\vdots& \nonumber\\
\begin{picture}(100, 40)(0, 17)
\DashLine(50,20)(80,40){5}
\DashLine(50,20)(80,0){5}
\DashCArc(50,20)(30,50,310){2}
\Text(84, 48)[]{$k^{(1)}$}
\Text(84, -8)[]{$k^{(2n)}$}
\end{picture}
&\leftrightarrow& \frac{(-1)^n}{\hbar v^{2n-2}}\Bigg( \sum_{P_1} \Pi_{P_1\{k\}} - \sum_{P_2} \bar{\Pi}_{P_2\{k\}} + \ldots + \nonumber\\
& & \phantom{\frac{(-1)^n}{\hbar v^{2n-2}}\Bigg(} (-1)^n\sum_{P_{n-1}} \Pi_{P_{n-1}\{k\}} + \h(-1)^{n+1}\sum_{P_n} \bar{\Pi}_{P_n\{k\}} + \nonumber\\
& & \phantom{\frac{(-1)^n}{\hbar v^{2n-2}}\Bigg(} -2^{2n-2}m_w^2 - \left(1-2^{2n-2}\right)m_\eta^2 \Bigg) \nonumber\\ \label{discrFeynmanrules}
&\vdots&
\eqa}
Notice that the bars are always placed on the terms with an operator $P_i$ with $i$ even. For the sake of the argument it is convenient to place these bars in this way. Whether or not a bar is placed on the last term, with $P_n$, thus depends on $n$, whether $n$ is even or odd. Above the bars are placed as if $n$ were even, but it should be clear how they should be placed when $n$ is odd.

Now notice that these vertex rules look identical to the rules one would use when doing a $d$-dimensional continuum calculation. For example, in the continuum the 3-vertex would be:
\vspace{-17pt}
\bq
\begin{picture}(100, 40)(0, 17)
\Line(20, 20)(50, 20)
\DashLine(50, 20)(80, 0){5}
\DashLine(50, 20)(80, 40){5}
\Text(20, 26)[]{$p$}
\Text(84, 32)[]{$k^{(1)}$}
\Text(84, 10)[]{$k^{(2)}$}
\end{picture}
\leftrightarrow \frac{2}{\hbar v} k^{(1)}\cdot k^{(2)}
\eq\\
To simplify the dot-product in a $d$-dimensional calculation one would write it as
\bqa
\frac{2}{\hbar v} k^{(1)}\cdot k^{(2)} &=& -\frac{1}{\hbar v}\bigg( \left(\left(k^{(1)}\right)^2+m_w^2\right) + \left(\left(k^{(2)}\right)^2+m_w^2\right) + \nonumber\\
& & \phantom{-\frac{1}{\hbar v}\bigg(} -\left(\left(k^{(1)}+k^{(2)}\right)^2+m_\eta^2\right) - 2m_w^2 + m_\eta^2 \bigg) \;,
\eqa
which corresponds exactly to the discrete vertex expression given in (\ref{discrFeynmanrules}) for the 3-vertex. So when one uses the continuum rules to rewrite dot-products of momenta, as in the 3-vertex example above, one is actually doing a \emph{correct} calculation, although one is doing a continuum calculation.

Another rule that one uses in a continuum calculation is that it is allowed to shift the loop momenta. Also in a discrete calculation this is allowed, because of the periodicity of the discrete propagators and vertex expressions.

What then goes wrong in a continuum calculation? There is one more rule that one uses in a continuum calculation that we have not mentioned up to now. This rule is:
\bq \label{badrule}
{1\over2\pi}\int d^dk \; \frac{k_i}{k^2+m^2} = 0 \;.
\eq
This rule, however, is \emph{not} correct in a discrete calculation. For example, in dimension 1, we have to realize that by $k$ and $k^2+m^2$ we actually mean:
\bqa
k &\leftrightarrow& {i\over\Delta}\left(e^{-i\Delta k}-1\right) \nonumber\\
k^2+m^2 &\leftrightarrow& {2\over\Delta^2}-{2\over\Delta^2}\cos(\Delta k)+m^2
\eqa
So by the integral above we actually mean:
\bq
{1\over2\pi}\int dk \; \frac{k}{k^2+m^2} \leftrightarrow {1\over2\pi}\int_{-\pi/\Delta}^{\pi/\Delta} dk \; \frac{{i\over\Delta}\left(e^{-i\Delta k}-1\right)}{{2\over\Delta^2}-{2\over\Delta^2}\cos(\Delta k)+m^2} = -\h i + \mathcal{O}(\Delta)
\eq
This shows that the rule (\ref{badrule}) is \emph{not} correct to use. The only instances that one would use the rule (\ref{badrule}) is when a $\Pi$ (or $\bar{\Pi}$) is left in the numerator, and \emph{cannot} cancel anything in the denominator anymore.

Below we shall show that all such terms, where a $\Pi$ (or $\bar{\Pi}$) remains in the numerator, cancel when one adds \emph{all} diagrams for a certain Green's function. To this end it is convenient to split up the vertices in (\ref{discrFeynmanrules}) as follows:
{\allowdisplaybreaks\bqa
\begin{picture}(100, 40)(0, 17)
\Line(20, 20)(50, 20)
\DashLine(50, 20)(80, 0){5}
\DashLine(50, 20)(80, 40){5}
\DashCArc(50,20)(30,-20,20){2}
\Vertex(40,20){3}
\Text(20, 26)[]{$p$}
\Text(84, 48)[]{$k^{(1)}$}
\Text(84, -8)[]{$k^{(2n)}$}
\end{picture}
&\leftrightarrow& \frac{(-1)^n}{\hbar v^{2n-1}} \left( -\bar{\Pi}_p \right)  \nonumber\\[20pt]
\begin{picture}(100, 40)(0, 17)
\Line(20, 20)(50, 20)
\DashLine(50, 20)(80, 0){5}
\DashLine(50, 20)(80, 40){5}
\DashCArc(50,20)(30,-20,20){2}
\Vertex(58.3,25.5){3}
\Text(20, 26)[]{$p$}
\Text(84, 48)[]{$k^{(1)}$}
\Text(84, -8)[]{$k^{(2n)}$}
\end{picture}
&\leftrightarrow& \frac{(-1)^n}{\hbar v^{2n-1}} \Pi_{k^{(1)}} \nonumber\\[20pt]
&\vdots& \nonumber\\
\begin{picture}(100, 40)(0, 17)
\Line(20, 20)(50, 20)
\DashLine(50, 20)(80, 0){5}
\DashLine(50, 20)(80, 40){5}
\DashCArc(50,20)(30,-20,20){2}
\Vertex(58.3,14.5){3}
\Text(20, 26)[]{$p$}
\Text(84, 48)[]{$k^{(1)}$}
\Text(84, -8)[]{$k^{(2n)}$}
\end{picture}
&\leftrightarrow& \frac{(-1)^n}{\hbar v^{2n-1}} \Pi_{k^{(2n)}} \nonumber\\[20pt]
\begin{picture}(100, 40)(0, 17)
\Line(20, 20)(50, 20)
\DashLine(50, 20)(80, 0){5}
\DashLine(50, 20)(80, 40){5}
\DashCArc(50,20)(30,-20,20){2}
\Vertex(50,20){3}
\Text(20, 26)[]{$p$}
\Text(84, 48)[]{$k^{(1)}$}
\Text(84, -8)[]{$k^{(2n)}$}
\end{picture}
&\leftrightarrow& \frac{(-1)^n}{\hbar v^{2n-1}} \Bigg( \sum_{P_1} \Pi_{p+P_1\{k\}} - \sum_{P_2} \bar{\Pi}_{P_2\{k\}} - \sum_{P_2} \bar{\Pi}_{p+P_2\{k\}} + \ldots + \nonumber\\
& & \phantom{\frac{(-1)^n}{\hbar v^{2n-1}} \Bigg(} (-1)^n\sum_{P_{n-1}} \Pi_{P_{n-1}\{k\}} + (-1)^n\sum_{P_{n-1}} \Pi_{p+P_{n-1}\{k\}} + \nonumber\\
& & \phantom{\frac{(-1)^n}{\hbar v^{2n-1}} \Bigg(} \h(-1)^{n+1}\sum_{P_n} \bar{\Pi}_{P_n\{k\}} + \h(-1)^{n+1}\sum_{P_n} \bar{\Pi}_{p+P_n\{k\}} \Bigg) \nonumber\\
\begin{picture}(100, 40)(0, 17)
\Line(20, 20)(50, 20)
\DashLine(50, 20)(80, 0){5}
\DashLine(50, 20)(80, 40){5}
\DashCArc(50,20)(30,-20,20){2}
\Text(20, 26)[]{$p$}
\Text(84, 48)[]{$k^{(1)}$}
\Text(84, -8)[]{$k^{(2n)}$}
\end{picture}
&\leftrightarrow& \frac{(-1)^n}{\hbar v^{2n-1}} \left( -2^{2n-1}m_w^2 - \left(1-2^{2n-1}\right)m_\eta^2 \right) \nonumber\\[20pt]
&\vdots& \nonumber\\
\begin{picture}(100, 40)(0, 17)
\Line(20,0)(50,20)
\Line(20,40)(50,20)
\DashLine(50, 20)(80, 0){5}
\DashLine(50, 20)(80, 40){5}
\DashCArc(50,20)(30,-20,20){2}
\Vertex(41.7,25.5){3}
\Text(20, 32)[]{$p$}
\Text(20, 10)[]{$q$}
\Text(84, 48)[]{$k^{(1)}$}
\Text(84, -8)[]{$k^{(2n)}$}
\end{picture}
&\leftrightarrow& \frac{(-1)^n}{\hbar v^{2n}} \left( -\bar{\Pi}_p \right) \nonumber\\[20pt]
\begin{picture}(100, 40)(0, 17)
\Line(20,0)(50,20)
\Line(20,40)(50,20)
\DashLine(50, 20)(80, 0){5}
\DashLine(50, 20)(80, 40){5}
\DashCArc(50,20)(30,-20,20){2}
\Vertex(41.7,14.5){3}
\Text(20, 32)[]{$p$}
\Text(20, 10)[]{$q$}
\Text(84, 48)[]{$k^{(1)}$}
\Text(84, -8)[]{$k^{(2n)}$}
\end{picture}
&\leftrightarrow& \frac{(-1)^n}{\hbar v^{2n}} \left( -\bar{\Pi}_q \right) \nonumber\\[20pt]
\begin{picture}(100, 40)(0, 17)
\Line(20,0)(50,20)
\Line(20,40)(50,20)
\DashLine(50, 20)(80, 0){5}
\DashLine(50, 20)(80, 40){5}
\DashCArc(50,20)(30,-20,20){2}
\Vertex(50,20){3}
\Text(20, 32)[]{$p$}
\Text(20, 10)[]{$q$}
\Text(84, 48)[]{$k^{(1)}$}
\Text(84, -8)[]{$k^{(2n)}$}
\end{picture}
&\leftrightarrow& \frac{(-1)^n}{\hbar v^{2n}}\Bigg( \sum_{P_1} \Pi_{p+P_1\{k\}} + \sum_{P_1} \Pi_{q+P_1\{k\}} + \nonumber\\
& & \phantom{\frac{(-1)^n}{\hbar v^{2n}}\Bigg(} -\sum_{P_2} \bar{\Pi}_{p+P_2\{k\}} - \sum_{P_2} \bar{\Pi}_{q+P_2\{k\}} + \ldots + \nonumber\\
& & \phantom{\frac{(-1)^n}{\hbar v^{2n}}\Bigg(} (-1)^n\sum_{P_{n-1}} \Pi_{p+P_{n-1}\{k\}} + (-1)^n\sum_{P_{n-1}} \Pi_{q+P_{n-1}\{k\}} + \nonumber\\
& & \phantom{\frac{(-1)^n}{\hbar v^{2n}}\Bigg(} \h(-1)^{n+1}\sum_{P_n} \bar{\Pi}_{p+P_n\{k\}} + \h(-1)^{n+1}\sum_{P_n} \bar{\Pi}_{q+P_n\{k\}} \Bigg) \nonumber\\
\begin{picture}(100, 40)(0, 17)
\Line(20,0)(50,20)
\Line(20,40)(50,20)
\DashLine(50, 20)(80, 0){5}
\DashLine(50, 20)(80, 40){5}
\DashCArc(50,20)(30,-20,20){2}
\Text(20, 32)[]{$p$}
\Text(20, 10)[]{$q$}
\Text(84, 48)[]{$k^{(1)}$}
\Text(84, -8)[]{$k^{(2n)}$}
\end{picture}
&\leftrightarrow& \frac{(-1)^n}{\hbar v^{2n}} \left( -2^{2n-1}m_w^2 + 2^{2n-1}m_\eta^2 \right) \nonumber\\[20pt]
&\vdots& \nonumber\\
\begin{picture}(100, 40)(0, 17)
\DashLine(50,20)(80,40){5}
\DashLine(50,20)(80,0){5}
\DashCArc(50,20)(30,50,310){2}
\Vertex(58.3,25.5){3}
\Text(84, 48)[]{$k^{(1)}$}
\Text(84, -8)[]{$k^{(2n)}$}
\end{picture}
&\leftrightarrow& \frac{(-1)^n}{\hbar v^{2n-2}} \Pi_{k^{(1)}} \nonumber\\[20pt]
\begin{picture}(100, 40)(0, 17)
\DashLine(50,20)(80,40){5}
\DashLine(50,20)(80,0){5}
\DashCArc(50,20)(30,50,310){2}
\Vertex(58.3,14.5){3}
\Text(84, 48)[]{$k^{(1)}$}
\Text(84, -8)[]{$k^{(2n)}$}
\end{picture}
&\leftrightarrow& \frac{(-1)^n}{\hbar v^{2n-2}} \Pi_{k^{(2n)}} \nonumber\\[20pt]
\begin{picture}(100, 40)(0, 17)
\DashLine(50,20)(80,40){5}
\DashLine(50,20)(80,0){5}
\DashCArc(50,20)(30,50,310){2}
\Vertex(50,20){3}
\Text(84, 48)[]{$k^{(1)}$}
\Text(84, -8)[]{$k^{(2n)}$}
\end{picture}
&\leftrightarrow& \frac{(-1)^n}{\hbar v^{2n-2}}\Bigg( -\sum_{P_2} \bar{\Pi}_{P_2\{k\}} + \ldots + \nonumber\\
& & \phantom{\frac{(-1)^n}{\hbar v^{2n-2}}\Bigg(} (-1)^n\sum_{P_{n-1}} \Pi_{P_{n-1}\{k\}} + \h(-1)^{n+1}\sum_{P_n} \bar{\Pi}_{P_n\{k\}} \Bigg) \nonumber\\
\begin{picture}(100, 40)(0, 17)
\DashLine(50,20)(80,40){5}
\DashLine(50,20)(80,0){5}
\DashCArc(50,20)(30,50,310){2}
\Text(84, 48)[]{$k^{(1)}$}
\Text(84, -8)[]{$k^{(2n)}$}
\end{picture}
&\leftrightarrow& \frac{(-1)^n}{\hbar v^{2n-2}} \left( -2^{2n-2}m_w^2 - \left(1-2^{2n-2}\right)m_\eta^2 \right) \nonumber\\[20pt]
&\vdots&
\eqa}

Having written the vertices in this form it is clear where the problem terms in a certain diagram come from. They come from a vertex with a dot in the center or two vertices connected by a line with two dots. That a dotted vertex is a source can immediately be seen from the vertex expressions above. A line with two dots and momentum $k$ flowing through it gets two $\Pi_k$'s in the numerator, from the vertices, and only one $\Pi_k$ in the denominator, from the propagator.

It is now easy to derive the following \ind{recursion relation}, valid for $n\geq2$:
\vspace{-17pt}
{\allowdisplaybreaks\bqa
& & \begin{picture}(100, 40)(0, 17)
\DashLine(10,40)(40,20){5}
\DashLine(10,0)(40,20){5}
\DashLine(60,20)(90,40){5}
\DashLine(60,20)(90,0){5}
\Line(40,20)(60,20)
\Vertex(44,20){2}
\Vertex(56,20){2}
\DashCArc(60,20)(30,-28,28){2}
\Text(92,20)[l]{$2n-2$}
\end{picture} \hspace{40pt} +
\begin{picture}(100, 40)(0, 17)
\DashLine(10,40)(40,20){5}
\DashLine(10,0)(40,20){5}
\DashLine(10,26.7)(40,20){5}
\DashLine(10,13.3)(40,20){5}
\DashLine(60,20)(90,40){5}
\DashLine(60,20)(90,0){5}
\Line(40,20)(60,20)
\Vertex(44,20){2}
\Vertex(56,20){2}
\DashCArc(60,20)(30,-28,28){2}
\Text(92,20)[l]{$2n-4$}
\end{picture} \hspace{40pt} + \ldots + \nonumber\\[10pt]
& & \begin{picture}(100, 40)(0, 17)
\DashLine(10,40)(40,20){5}
\DashLine(10,0)(40,20){5}
\DashLine(60,20)(90,40){5}
\DashLine(60,20)(90,0){5}
\Line(40,20)(60,20)
\Vertex(44,20){2}
\Vertex(56,20){2}
\DashCArc(60,20)(30,-28,28){2}
\DashCArc(40,20)(30,152,208){2}
\Text(8,20)[r]{$e(n)$}
\Text(92,20)[l]{$2n-e(n)$}
\end{picture} \hspace{50pt} + \nonumber\\[10pt]
& & \begin{picture}(100, 40)(0, 17)
\DashLine(10,40)(40,20){5}
\DashLine(10,20)(40,20){5}
\DashLine(10,0)(40,20){5}
\DashLine(60,20)(90,40){5}
\DashLine(60,20)(90,0){5}
\DashLine(40,20)(60,20){5}
\Vertex(44,20){2}
\Vertex(56,20){2}
\DashCArc(60,20)(30,-28,28){2}
\Text(92,20)[l]{$2n-3$}
\end{picture} \hspace{40pt} +
\begin{picture}(100, 40)(0, 17)
\DashLine(10,40)(40,20){5}
\DashLine(10,30)(40,20){5}
\DashLine(10,20)(40,20){5}
\DashLine(10,10)(40,20){5}
\DashLine(10,0)(40,20){5}
\DashLine(60,20)(90,40){5}
\DashLine(60,20)(90,0){5}
\DashLine(40,20)(60,20){5}
\Vertex(44,20){2}
\Vertex(56,20){2}
\DashCArc(60,20)(30,-28,28){2}
\Text(92,20)[l]{$2n-5$}
\end{picture} \hspace{40pt} + \ldots + \nonumber\\[10pt]
& & \begin{picture}(100, 40)(0, 17)
\DashLine(10,40)(40,20){5}
\DashLine(10,0)(40,20){5}
\DashLine(60,20)(90,40){5}
\DashLine(60,20)(90,0){5}
\DashLine(40,20)(60,20){5}
\Vertex(44,20){2}
\Vertex(56,20){2}
\DashCArc(60,20)(30,-28,28){2}
\DashCArc(40,20)(30,152,208){2}
\Text(92,20)[l]{$e(n)+1$}
\Text(8,20)[r]{$2n-e(n)-1$}
\end{picture} \hspace{50pt} + \nonumber\\[20pt]
& & \begin{picture}(100, 40)(0, 17)
\DashLine(50,20)(80,40){5}
\DashLine(50,20)(80,0){5}
\Vertex(50,20){3}
\DashCArc(50,20)(30,50,310){2}
\Text(18,20)[r]{$2n$}
\end{picture} = 0
\eqa}\\[10pt]
In these diagrams it is understood that the outgoing legs should be connected in \emph{all} possible ways. $e(n)$ Is defined as:
\bq
e(n) \equiv \left\{ \begin{array}{ll}
n & \textrm{if $n$ is even} \\
n-1 & \textrm{if $n$ is odd}
\end{array} \right.
\eq
Notice that the third and fourth line in the recursion relation above are not there when $n=2$, these diagrams simply do not exist.

We also have the following two recursion relations:\index{recursion relation}
{\allowdisplaybreaks\bqa
& & \begin{picture}(100, 40)(0, 17)
\DashLine(10,40)(40,20){5}
\DashLine(10,0)(40,20){5}
\DashLine(60,20)(90,40){5}
\DashLine(60,20)(90,0){5}
\Line(40,20)(60,20)
\Line(40,20)(90,50)
\Vertex(44,20){2}
\Vertex(56,20){2}
\DashCArc(60,20)(30,-28,28){2}
\Text(92,20)[l]{$2n-2$}
\end{picture} \hspace{40pt} + \ldots + \hspace{30pt}
\begin{picture}(100, 40)(0, 17)
\DashLine(10,40)(40,20){5}
\DashLine(10,0)(40,20){5}
\DashLine(60,20)(90,40){5}
\DashLine(60,20)(90,0){5}
\Line(40,20)(60,20)
\Line(40,20)(90,50)
\Vertex(44,20){2}
\Vertex(56,20){2}
\DashCArc(60,20)(30,-28,28){2}
\DashCArc(40,20)(30,152,208){2}
\Text(8,20)[r]{$e(n)$}
\Text(92,20)[l]{$2n-e(n)$}
\end{picture} \hspace{50pt} + \nonumber\\[20pt]
& & \begin{picture}(100, 40)(0, 17)
\DashLine(10,40)(40,20){5}
\DashLine(10,0)(40,20){5}
\DashLine(60,20)(90,40){5}
\DashLine(60,20)(90,0){5}
\Line(40,20)(60,20)
\Line(60,20)(90,50)
\Vertex(44,20){2}
\Vertex(56,20){2}
\DashCArc(60,20)(30,-28,28){2}
\Text(92,20)[l]{$2n-2$}
\end{picture} \hspace{40pt} + \ldots + \hspace{30pt}
\begin{picture}(100, 40)(0, 17)
\DashLine(10,40)(40,20){5}
\DashLine(10,0)(40,20){5}
\DashLine(60,20)(90,40){5}
\DashLine(60,20)(90,0){5}
\Line(40,20)(60,20)
\Line(60,20)(90,50)
\Vertex(44,20){2}
\Vertex(56,20){2}
\DashCArc(60,20)(30,-28,28){2}
\DashCArc(40,20)(30,152,208){2}
\Text(8,20)[r]{$e(n)$}
\Text(92,20)[l]{$2n-e(n)$}
\end{picture} \hspace{50pt} + \nonumber\\[20pt]
& & \begin{picture}(100, 40)(0, 17)
\DashLine(10,20)(40,20){5}
\DashLine(60,20)(90,40){5}
\DashLine(60,20)(90,0){5}
\DashLine(40,20)(60,20){5}
\Line(40,20)(90,50)
\Vertex(44,20){2}
\Vertex(56,20){2}
\DashCArc(60,20)(30,-28,28){2}
\Text(92,20)[l]{$2n-1$}
\end{picture} \hspace{40pt} + \ldots + \hspace{80pt}
\begin{picture}(100, 40)(0, 17)
\DashLine(10,40)(40,20){5}
\DashLine(10,0)(40,20){5}
\DashLine(60,20)(90,40){5}
\DashLine(60,20)(90,0){5}
\DashLine(40,20)(60,20){5}
\Line(40,20)(90,50)
\Vertex(44,20){2}
\Vertex(56,20){2}
\DashCArc(60,20)(30,-28,28){2}
\DashCArc(40,20)(30,152,208){2}
\Text(92,20)[l]{$e(n)+1$}
\Text(8,20)[r]{$2n-e(n)-1$}
\end{picture} \hspace{50pt} + \nonumber\\[20pt]
& & \begin{picture}(100, 40)(0, 17)
\DashLine(10,40)(40,20){5}
\DashLine(10,20)(40,20){5}
\DashLine(10,0)(40,20){5}
\DashLine(60,20)(90,40){5}
\DashLine(60,20)(90,0){5}
\DashLine(40,20)(60,20){5}
\Line(60,20)(90,50)
\Vertex(44,20){2}
\Vertex(56,20){2}
\DashCArc(60,20)(30,-28,28){2}
\Text(92,20)[l]{$2n-3$}
\end{picture} \hspace{40pt} + \ldots + \hspace{80pt}
\begin{picture}(100, 40)(0, 17)
\DashLine(10,40)(40,20){5}
\DashLine(10,0)(40,20){5}
\DashLine(60,20)(90,40){5}
\DashLine(60,20)(90,0){5}
\DashLine(40,20)(60,20){5}
\Line(60,20)(90,50)
\Vertex(44,20){2}
\Vertex(56,20){2}
\DashCArc(60,20)(30,-28,28){2}
\DashCArc(40,20)(30,152,208){2}
\Text(92,20)[l]{$e(n)+1$}
\Text(8,20)[r]{$2n-e(n)-1$}
\end{picture} \hspace{50pt} + \nonumber\\[30pt]
& & \begin{picture}(100, 40)(0, 17)
\DashLine(50,20)(80,40){5}
\DashLine(50,20)(80,0){5}
\Line(50,20)(80,50)
\DashLine(50,20)(20,40){5}
\Vertex(50,20){3}
\DashCArc(50,20)(30,156,316){2}
\Text(18,20)[r]{$2n$}
\end{picture} = 0
\eqa}\\[20pt]
In this recursion relation $n\geq2$.

{\allowdisplaybreaks\bqa
& & \begin{picture}(100, 40)(0, 17)
\DashLine(10,40)(40,20){5}
\DashLine(10,0)(40,20){5}
\DashLine(60,20)(90,40){5}
\DashLine(60,20)(90,0){5}
\Line(40,20)(60,20)
\Line(40,20)(90,60)
\Line(60,20)(90,50)
\Vertex(44,20){2}
\Vertex(56,20){2}
\DashCArc(60,20)(30,-28,28){2}
\Text(92,20)[l]{$2n-2$}
\end{picture} \hspace{40pt} + \ldots + \hspace{30pt}
\begin{picture}(100, 40)(0, 17)
\DashLine(10,40)(40,20){5}
\DashLine(10,0)(40,20){5}
\DashLine(60,20)(90,40){5}
\DashLine(60,20)(90,0){5}
\Line(40,20)(60,20)
\Line(40,20)(90,60)
\Line(60,20)(90,50)
\Vertex(44,20){2}
\Vertex(56,20){2}
\DashCArc(60,20)(30,-28,28){2}
\DashCArc(40,20)(30,152,208){2}
\Text(8,20)[r]{$e(n)$}
\Text(92,20)[l]{$2n-e(n)$}
\end{picture} \hspace{50pt} + \nonumber\\[30pt]
& & \begin{picture}(100, 40)(0, 17)
\DashLine(10,40)(40,20){5}
\DashLine(10,0)(40,20){5}
\DashLine(60,20)(90,40){5}
\DashLine(60,20)(90,0){5}
\Line(40,20)(60,20)
\Line(40,20)(90,50)
\Line(60,20)(90,60)
\Vertex(44,20){2}
\Vertex(56,20){2}
\DashCArc(60,20)(30,-28,28){2}
\Text(92,20)[l]{$2n-2$}
\end{picture} \hspace{40pt} + \ldots + \hspace{30pt}
\begin{picture}(100, 40)(0, 17)
\DashLine(10,40)(40,20){5}
\DashLine(10,0)(40,20){5}
\DashLine(60,20)(90,40){5}
\DashLine(60,20)(90,0){5}
\Line(40,20)(60,20)
\Line(40,20)(90,50)
\Line(60,20)(90,60)
\Vertex(44,20){2}
\Vertex(56,20){2}
\DashCArc(60,20)(30,-28,28){2}
\DashCArc(40,20)(30,152,208){2}
\Text(8,20)[r]{$e(n)$}
\Text(92,20)[l]{$2n-e(n)$}
\end{picture} \hspace{50pt} + \nonumber\\[30pt]
& & \begin{picture}(100, 40)(0, 17)
\DashLine(10,20)(40,20){5}
\DashLine(60,20)(90,40){5}
\DashLine(60,20)(90,0){5}
\DashLine(40,20)(60,20){5}
\Line(40,20)(90,60)
\Line(60,20)(90,50)
\Vertex(44,20){2}
\Vertex(56,20){2}
\DashCArc(60,20)(30,-28,28){2}
\Text(92,20)[l]{$2n-1$}
\end{picture} \hspace{40pt} + \ldots + \hspace{80pt}
\begin{picture}(100, 40)(0, 17)
\DashLine(10,40)(40,20){5}
\DashLine(10,0)(40,20){5}
\DashLine(60,20)(90,40){5}
\DashLine(60,20)(90,0){5}
\DashLine(40,20)(60,20){5}
\Line(40,20)(90,60)
\Line(60,20)(90,50)
\Vertex(44,20){2}
\Vertex(56,20){2}
\DashCArc(60,20)(30,-28,28){2}
\DashCArc(40,20)(30,152,208){2}
\Text(92,20)[l]{$e(n)+1$}
\Text(8,20)[r]{$2n-e(n)-1$}
\end{picture} \hspace{50pt} + \nonumber\\[30pt]
& & \begin{picture}(100, 40)(0, 17)
\DashLine(10,20)(40,20){5}
\DashLine(60,20)(90,40){5}
\DashLine(60,20)(90,0){5}
\DashLine(40,20)(60,20){5}
\Line(40,20)(90,50)
\Line(60,20)(90,60)
\Vertex(44,20){2}
\Vertex(56,20){2}
\DashCArc(60,20)(30,-28,28){2}
\Text(92,20)[l]{$2n-1$}
\end{picture} \hspace{40pt} + \ldots + \hspace{80pt}
\begin{picture}(100, 40)(0, 17)
\DashLine(10,40)(40,20){5}
\DashLine(10,0)(40,20){5}
\DashLine(60,20)(90,40){5}
\DashLine(60,20)(90,0){5}
\DashLine(40,20)(60,20){5}
\Line(40,20)(90,50)
\Line(60,20)(90,60)
\Vertex(44,20){2}
\Vertex(56,20){2}
\DashCArc(60,20)(30,-28,28){2}
\DashCArc(40,20)(30,152,208){2}
\Text(92,20)[l]{$e(n)+1$}
\Text(8,20)[r]{$2n-e(n)-1$}
\end{picture} \hspace{50pt} + \nonumber\\[40pt]
& & \begin{picture}(100, 40)(0, 17)
\DashLine(50,20)(80,40){5}
\DashLine(50,20)(80,0){5}
\Line(50,20)(80,50)
\Line(50,20)(80,60)
\DashLine(50,20)(20,40){5}
\Vertex(50,20){3}
\DashCArc(50,20)(30,156,316){2}
\Text(18,20)[r]{$2n$}
\end{picture} = 0
\eqa}\\[10pt]
In this recursion relation $n\geq1$. The first two lines are not there when $n=1$.

With these recursion relations it is easy to see that the problem terms always cancel in the complete set of diagrams for a certain Green's function. If somewhere in a diagram an internal line with two dots occurs, then at this same point in the diagram also the dotted vertex can occur. These diagrams then sum up to zero.

So finally we have proven that the problem terms cancel in the complete set of diagrams for a certain Green's function. If they cancel out anyway it is also correct to treat these problem terms like one would in the continuum. Of course one makes a mistake for each problem term, but these mistakes cancel out again in the complete set of diagrams. So there is nothing wrong with taking the continuum limit right from the start and doing a $d$-dimensional continuum calculation.

Notice that it is no problem to add the vertices coming from the potential $V$ and the Jacobian to this argument. These vertices give no problem terms themselves. They can be combined with the dotted vertices, but the problem terms always come from a \emph{clear}, \emph{separated} part of the diagram, either two vertices connected by a line with two dots, or a dotted vertex.

Now we know it is correct to take the \ind{continuum limit} $\Delta\rightarrow0$ straightaway in the discrete Feynman rules (\ref{discrFeynmanrules}). If we do this it is easy to see that only the $\eta ww$- and $\eta\eta ww$-vertex do not vanish. All the other vertices go to zero, as can be seen from their expressions in (\ref{discrFeynmanrules}), but also, and much quicker, from (\ref{discrFeynman}), because they all have one or more factors of $\Delta$ in front when the exponentials are expanded.

So, finally we are left with the Feynman rules:\index{Feynman rules!continuum}
{\allowdisplaybreaks\bqa
\begin{picture}(100, 20)(0, 17)
\Line(20, 20)(80, 20)
\end{picture}
&\leftrightarrow& \frac{\hbar}{k^2+m_\eta^2} \nonumber\\
\begin{picture}(100, 20)(0, 17)
\DashLine(20, 20)(80, 20){5}
\end{picture}
&\leftrightarrow& \frac{\hbar}{k^2+m_w^2} \nonumber\\
\begin{picture}(100, 40)(0, 17)
\Line(20, 20)(50, 20)
\DashLine(50, 20)(80, 0){5}
\DashLine(50, 20)(80, 40){5}
\Text(80,30)[c]{$k_1$}
\Text(80,10)[c]{$k_2$}
\end{picture}
&\leftrightarrow& \frac{2}{\hbar v} k_1\cdot k_2 \nonumber\\
\begin{picture}(100, 40)(0, 17)
\Line(20,0)(50,20)
\Line(20,40)(50,20)
\DashLine(50,20)(80,40){5}
\DashLine(50,20)(80,0){5}
\Text(80,30)[c]{$k_1$}
\Text(80,10)[c]{$k_2$}
\end{picture}
&\leftrightarrow& \frac{2}{\hbar v^2} k_1\cdot k_2 \label{continuousFeynmanrules}
\eqa}\\
plus the Feynman rules coming from the potential $V$ and the Jacobian. These are exactly the Feynman rules that one would have read off from the continuum action:
\bqa
S &=& \int d^dx \left( \h\left(\nabla r\right)^2 + \h\left(\nabla w\right)^2 + {1\over v}\eta\left(\nabla w\right)^2 + {1\over2v^2}\eta^2\left(\nabla w\right)^2 + V\left(r\cos{w\over v},r\sin{w\over v}\right) \right) \nonumber\\
&=& \int d^dx \left( \h\left(\nabla r\right)^2 + \h{r^2\over v^2}\left(\nabla w\right)^2 + V\left(r\cos{w\over v},r\sin{w\over v}\right) \right) \label{continuumaction}
\eqa

Recapitulating the proof, we have done the following. The basis of our proof is the \emph{discrete} path integral in terms of the Cartesian fields $\f_1$ and $\f_2$. In this path integral on the lattice it is completely legitimate to transform to polar fields. After this transformation we get a very big, complicated action. Looking at the discrete vertex expressions we find how we can simplify these expressions, and how we can let them cancel against propagators in a certain diagram. We notice that these rules are exactly the same as in a $d$-dimensional continuum calculation. All the rules that we would use in a continuum calculation appear to be valid in a calculation on the lattice as well, except for one: rule (\ref{badrule}). The terms where we would need to use this rule can then be shown to cancel in the complete set of diagrams, by using the three recursion relations. So by using the incorrect rule (\ref{badrule}) we actually make a mistake, but all these mistakes cancel in the complete set of diagrams. Thus we know that \emph{all} the rules that we use in $d$-dimensional continuum calculation are \emph{also} valid in a correct, discrete calculation. This means we might as well take the continuum limit directly in the discrete Feynman rules (\ref{discrFeynman}). Then these Feynman rules simplify to (\ref{continuousFeynmanrules}), and we have proven that a $d$-dimensional continuum calculation with the action (\ref{continuumaction}) is \emph{correct}.

\subsection{An Example}

To see explicitly how the mechanism described in the previous section works we consider an example. Consider the 1-loop $\eta$-propagator. There are two types of diagrams (We do not include vertices from the potential $V$ and the Jacobian, because such vertices will never give problem terms.):
\begin{center}
\begin{picture}(70,40)(0,18)
\Line(0,20)(20,20)
\DashCArc(35,20)(15,-180,180){3}
\Line(50,20)(70,20)
\end{picture} \quad and \quad
\begin{picture}(70,40)(0,18)
\Line(0,5)(70,5)
\DashCArc(35,20)(15,-90,270){3}
\end{picture}
\end{center}\vspace{10pt}
Here, dots should still be put on the lines or in the vertices. There are a lot of diagrams, but it is easy to see that there are only two diagrams that contain problem terms. These are:
\vspace{-17pt}
\bqa
\begin{picture}(70,40)(0,18)
\Line(0,20)(20,20)
\DashCArc(35,20)(15,-180,180){3}
\Line(50,20)(70,20)
\Vertex(48,27.5){2}
\Vertex(22,27.5){2}
\end{picture} &=& {1\over(2\pi)^d} \int_{-\pi/\Delta}^{\pi/\Delta} d^dk \; \left(-{1\over\hbar v}\right) \Pi_k \left(-{1\over\hbar v}\right) \Pi_k \frac{\hbar}{\Pi_k} \frac{\hbar}{\Pi_{p-k}} \nonumber\\
&=& {1\over v^2} {1\over(2\pi)^d} \int_{-\pi/\Delta}^{\pi/\Delta} d^dk \; \frac{\Pi_{p-k}}{\Pi_k} \nonumber\\
\begin{picture}(70,40)(0,18)
\Line(0,5)(70,5)
\DashCArc(35,20)(15,-90,270){3}
\Vertex(35,5){3}
\end{picture} &=& \h {1\over(2\pi)^d} \int_{-\pi/\Delta}^{\pi/\Delta} d^dk \; \left(-{1\over\hbar v^2}\right) \nonumber\\
& & \phantom{\h {1\over(2\pi)^d} \int_{-\pi/\Delta}^{\pi/\Delta} d^dk \;} \left( \h\left(\Pi_{p-k}+\Pi_{p+k}\right) + \h\left(\Pi_{p+k}+\Pi_{p-k}\right) \right) \frac{\hbar}{\Pi_k} \nonumber\\
&=& -{1\over v^2} {1\over(2\pi)^d} \int_{-\pi/\Delta}^{\pi/\Delta} d^dk \; \frac{\Pi_{p-k}}{\Pi_k}
\eqa
Indeed these diagrams cancel, as is guaranteed by the recursion relations derived in the previous section. The other diagrams, contributing to this $\eta$-propagator at 1-loop order, have their dots in other places, or have vertices without dots, that can also come from the potential $V$. These diagrams can never have a problem term. And thus the whole 1-loop propagator is free of problem terms, and the continuum limit could have been taken right from the start.

\subsection{The Jacobian and $w$-Loops}

In the previous sections it has become clear that it is allowed to work with the continuum Feynman rules (\ref{continuousFeynmanrules}), as the conjecture states. Together with these Feynman rules we have of course the rules from the arbitrary potential $V$, and the rules from the Jacobian. From the discrete calculation it is easy to see that the Jacobian can indeed be rewritten as:
\bqa
\prod_x r(x) &=& \prod_x \exp\left(-{1\over\hbar}\left(-\hbar\ln r(x)\right)\right) \nonumber\\
&=& \exp\left(-{1\over\hbar} \sum_x \left(-\hbar\ln r(x)\right)\right) \nonumber\\
&\rightarrow& \exp\left(-{1\over\hbar}{1\over\Delta^d} \int d^dx \left(-\hbar\ln r(x)\right)\right) \nonumber\\
&\rightarrow& \exp\left(-{1\over\hbar} \left[{1\over(2\pi)^d}\int d^dk\right] \int d^dx \left(-\hbar\ln r(x)\right)\right) \nonumber\\
&=& \exp\left(-{1\over\hbar} \left[{1\over(2\pi)^d}\int d^dk\right] \int d^dx \left(-\hbar \sum_{n=1}^\infty \frac{(-1)^{n+1}}{n} \left({\eta(x)\over v}\right)^n \right)\right) \;, \nonumber\\
\eqa
as the conjecture states. The Feynman rules coming from this Jacobian are:
\vspace{-17pt}
\bqa
\begin{picture}(100, 40)(0, 17)
\Line(20, 20)(50, 20)
\Vertex(50, 20){3}
\end{picture}
&\leftrightarrow& {1\over v} \; I \nonumber\\
\begin{picture}(100, 40)(0, 17)
\Line(20, 20)(80, 20)
\Vertex(50, 20){3}
\end{picture}
&\leftrightarrow& -{1\over v^2} \; I \nonumber\\
&\vdots& \nonumber\\
\begin{picture}(100, 40)(0, 17)
\Line(20, 20)(50, 20)
\Line(50, 20)(80, 40)
\Line(50, 20)(80, 0)
\DashCArc(50,20)(30,-30,30){2}
\Vertex(50, 20){3}
\Text(82,20)[l]{n}
\end{picture}
&\leftrightarrow& {(-1)^{n+1}(n-1)!\over v^n} \; I
\eqa\\[5pt]
Here the standard integral $I$ has been defined earlier, in (\ref{standintI}).

We see that all these vertices give strange integrals $I$. In the case of the shifted toy model and the arctangent toy model we already saw that these integrals $I$ always cancelled against identical terms coming from $w$-loops. We shall now prove this in general.

Consider a $w$-loop with only $\eta ww$- and $\eta\eta ww$-vertices (as given in (\ref{continuousFeynmanrules})) on it. So the diagrams we are calculating can only have external $\eta$-legs. Such a diagram would look like:
\begin{center}
\begin{picture}(80,80)(0,0)
\DashCArc(40,40)(20,0,360){5}
\Line(60,40)(80,40)
\Line(40,60)(36,78)
\Line(40,60)(44,78)
\Line(20,40)(2,44)
\Line(20,40)(2,36)
\Line(40,0)(40,20)
\DashCArc(40,40)(35,10,75){2}
\DashCArc(40,40)(35,105,165){2}
\DashCArc(40,40)(35,195,260){2}
\DashCArc(40,40)(35,280,350){2}
\end{picture}
\end{center}
Now these diagrams have a part which is going to cancel the $I$-integrals from the Jacobian. This part is exactly the worst divergent part of the diagrams above. To calculate this worst divergent part the masses and incoming momenta can be neglected.

In this case it is easy to write down the generating functional for such diagrams. This generating functional is defined as:
\vspace{-25pt}
\bq
Z(x) = \sum_{n=1}^\infty \quad
\SetScale{0.7}
\begin{picture}(56,56)(0,25)
\DashCArc(40,40)(20,0,360){5}
\Line(60,40)(80,40)
\Line(40,60)(36,78)
\Line(40,60)(44,78)
\Line(20,40)(2,44)
\Line(20,40)(2,36)
\Line(40,0)(40,20)
\DashCArc(40,40)(35,10,75){2}
\DashCArc(40,40)(35,105,165){2}
\DashCArc(40,40)(35,195,260){2}
\DashCArc(40,40)(35,280,350){2}
\end{picture} \quad \frac{x^n}{n!} \;,
\SetScale{1}
\eq\\
where the diagram symbolizes \emph{all} 1-loop diagrams of this type with $n$ outgoing $\eta$-lines.

We denote the number of $\eta ww$-vertices in the $w$-loop by $n_3$ and the number of $\eta\eta ww$-vertices by $n_4$. Then the generating functional $Z(x)$ for diagrams of this type is given by:
\bqa
Z(x) &=& {1\over(2\pi)^d} \int d^dk \underset{n_3+n_4>0}{\sum_{n_3,n_4=0}^\infty} \left({1\over2!}\right)^{n_3} \left({1\over2!2!}\right)^{n_4} {1\over n_3!} {1\over n_4!} \left({2\over\hbar v}(-k^2)\right)^{n_3} \left({2\over\hbar v^2}(-k^2)\right)^{n_4} \nonumber\\
& & \phantom{{1\over(2\pi)^d} \int d^dk \underset{n_3+n_4>0}{\sum_{n_3,n_4=0}^\infty}} \left(\frac{\hbar}{k^2}\right)^{n_3+n_4} (2n_3+2n_4)!! (n_3+2n_4)! \frac{x^{n_3+2n_4}}{(n_3+2n_4)!} \nonumber\\
&=& -I \; \ln\left(1+{x\over v}\right) \nonumber\\
&=& I \; \sum_{n=1}^\infty \frac{(-1)^n(n-1)!}{v^n} \frac{x^n}{n!}
\eqa
So we can read off that a $w$-loop with $n$ outgoing $\eta$-lines has a worst divergent part given by:
\bq
\frac{(-1)^n(n-1)!}{v^n} \; I
\eq
This exactly cancels the vertices from the Jacobian. In any diagram, wherever a dotted vertex from the Jacobian with $n$ legs occurs, also a $w$-loop with $n$ outgoing legs can occur, and their part that contains the standard integral $I$ cancels!

\subsection{The Dimensional Regularization Scheme}

In the case that one uses the \ind{dimensional regularization scheme} one has that:
\bq
{1\over(2\pi)^d} \int d^dk \; \frac{1}{(k^2)^m} = 0 \qquad \forall \; m \;,
\eq
which means that also our standard integral\index{standard integrals} $I$ becomes zero:
\bq
I = 0 \;.
\eq
This means that in the dimensional regularization scheme it becomes even easier to work with a path integral in polar field variables. In this case one can also completely forget about the Jacobian one gets from the transformation. Also one can ignore the integrals $I$ that are generated by $w$-loops.

In this thesis we will keep everything general however, and not specify a regularization scheme.

\section{The 1-Dimensional Case}

In section \ref{proofconj} we have proven the conjecture for a general model with two fields in $d$ space-time dimensions. This conjecture, which is promoted to a theorem by now, enables us to actually calculate things via the path integral in terms polar fields for any $d$-dimensional model. In a $d$-dimensional model the only analytical computations we can do in practice are continuum calculations. Analytical discrete calculations, i.e.\ calculations on the lattice, are in practice much too hard to do. That is why we have \emph{not} bothered to simplify the \emph{discrete} $d$-dimensional path integral, hoping to do a discrete calculation with this simplified form.

In \emph{one} dimension analytical discrete calculations \emph{are} sometimes possible (as we will see in the next section). Therefore it is convenient to have a reasonably simple, \emph{discrete} path integral in terms of polar fields for the case $d=1$. It is this path integral that we shall derive in this section.

By deriving this path integral we shall also make contact with the literature on quantum mechanical (i.e.\ 1-dimensional) path integrals in terms of polar fields \cite{Lee,Edwards,Peak}.

Our starting point will again be the discrete Feynman rules (\ref{discrFeynman}), now specified to $d=1$ however. Also, for the sake of the argument, we will split up the vertices as given below. The 1-dimensional Feynman rules are then:
{\allowdisplaybreaks\bqa
\begin{picture}(100, 20)(0, 17)
\Line(20, 20)(80, 20)
\end{picture}
&\leftrightarrow& \frac{\hbar}{{2\over\Delta^2}-{2\over\Delta^2}\cos{\Delta k}+m_\eta^2} \nonumber\\
\begin{picture}(100, 20)(0, 17)
\DashLine(20, 20)(80, 20){5}
\end{picture}
&\leftrightarrow& \frac{\hbar}{{2\over\Delta^2}-{2\over\Delta^2}\cos{\Delta k}+m_w^2} \nonumber\\
\begin{picture}(100, 40)(0, 17)
\Line(20, 20)(50, 20)
\DashLine(50, 20)(80, 0){5}
\DashLine(50, 20)(80, 40){5}
\DashCArc(50,20)(30,-20,20){2}
\Text(20, 26)[]{$p$}
\Text(84, 48)[]{$k^{(1)}$}
\Text(84, -8)[]{$k^{(2n)}$}
\end{picture}
&\leftrightarrow& \frac{2(-1)^n}{\hbar v^{2n-1}}{1\over\Delta^2} \left[ \left(e^{-i\Delta k^{(1)}}-1\right) \cdots \left(e^{-i\Delta k^{(2n)}}-1\right) \right] \nonumber\\[20pt]
\begin{picture}(100, 40)(0, 17)
\ArrowLine(20, 20)(50, 20)
\DashLine(50, 20)(80, 0){5}
\DashLine(50, 20)(80, 40){5}
\DashCArc(50,20)(30,-20,20){2}
\Text(20, 26)[]{$p$}
\Text(84, 48)[]{$k^{(1)}$}
\Text(84, -8)[]{$k^{(2n)}$}
\end{picture}
&\leftrightarrow& \frac{(-1)^n}{\hbar v^{2n-1}}{1\over\Delta^2} \left[ \left(e^{-i\Delta p}-1\right)\left(e^{-i\Delta k^{(1)}}-1\right) \cdots \left(e^{-i\Delta k^{(2n)}}\right) \right] \nonumber\\[20pt]
\begin{picture}(100, 40)(0, 17)
\Line(20,0)(50,20)
\Line(20,40)(50,20)
\DashLine(50, 20)(80, 0){5}
\DashLine(50, 20)(80, 40){5}
\DashCArc(50,20)(30,-20,20){2}
\Text(20, 32)[]{$p$}
\Text(20, 10)[]{$q$}
\Text(84, 48)[]{$k^{(1)}$}
\Text(84, -8)[]{$k^{(2n)}$}
\end{picture}
&\leftrightarrow& \frac{2(-1)^n}{\hbar v^{2n}}{1\over\Delta^2} \left[ \left(e^{-i\Delta k^{(1)}}-1\right) \cdots \left(e^{-i\Delta k^{(2n)}}-1\right) \right] \nonumber\\[20pt]
\begin{picture}(100, 40)(0, 17)
\Line(20,0)(50,20)
\ArrowLine(20,40)(50,20)
\DashLine(50, 20)(80, 0){5}
\DashLine(50, 20)(80, 40){5}
\DashCArc(50,20)(30,-20,20){2}
\Text(20, 32)[]{$p$}
\Text(20, 10)[]{$q$}
\Text(84, 48)[]{$k^{(1)}$}
\Text(84, -8)[]{$k^{(2n)}$}
\end{picture}
&\leftrightarrow& \frac{(-1)^n}{\hbar v^{2n}}{1\over\Delta^2} \left[ \left(e^{-i\Delta p}-1\right)\left(e^{-i\Delta k^{(1)}}-1\right) \cdots \left(e^{-i\Delta k^{(2n)}_1}-1\right) \right] \nonumber\\[20pt]
\begin{picture}(100, 40)(0, 17)
\DashLine(50,20)(80,40){5}
\DashLine(50,20)(80,0){5}
\DashCArc(50,20)(30,50,310){2}
\Text(84, 48)[]{$k^{(1)}$}
\Text(84, -8)[]{$k^{(2n)}$}
\end{picture}
&\leftrightarrow& \frac{(-1)^n}{\hbar v^{2n-2}}{1\over\Delta^2} \left[ \left(e^{-i\Delta k^{(1)}}-1\right) \cdots \left(e^{-i\Delta k^{(2n)}}-1\right) \right]
\eqa}\\

Looking at these vertex expressions we notice that only the vertices
\vspace{-17pt}
\bq \label{survivingvertices}
\begin{picture}(100, 40)(0,17)
\Line(20, 20)(50, 20)
\DashLine(50, 20)(80, 0){5}
\DashLine(50, 20)(80, 40){5}
\end{picture} \textrm{and}
\begin{picture}(100, 40)(0,17)
\Line(20,0)(50,20)
\Line(20,40)(50,20)
\DashLine(50, 20)(80, 0){5}
\DashLine(50, 20)(80, 40){5}
\end{picture}
\eq\\
have a finite continuum limit, \emph{all} the other vertices go to zero when $\Delta$ is sent to zero in the Feynman rules. First we are going to consider \emph{all} diagrams which have at least one of these vertices that vanish in the continuum limit. The only way these vertices can survive a continuum limit in a complete diagram is when there occur loops that give a $1/\Delta$.

First consider 1-loop diagrams. All 1-loop diagrams can be built from the \ind{vacuum diagram}
\vspace{-17pt}
\bq
\begin{picture}(50, 40)(0,17)
\CArc(25,20)(15,0,360)
\end{picture}
\eq\\
By attaching legs we can build any 1-loop diagram from these. Having an $\eta$-line in the loop will never give a $1/\Delta$, no matter which vertices we use. If the whole loop is a $w$-line this loop can give $1/\Delta$'s. If we construct a diagram from this vacuum graph with at least one of the vertices that go to zero in the continuum limit, one can verify easily that either the whole diagram goes to zero in the continuum limit \emph{or} diagrams cancel among each other in the complete set of graphs for a certain process, such that the whole process is zero.

The same thing can now be done on 2-loop level. Here we can construct all diagrams from the vacuum graphs\index{vacuum diagram}
\vspace{-17pt}
\bq
\begin{picture}(100, 40)(0,17)
\CArc(35,20)(15,0,360)
\CArc(65,20)(15,-180,180)
\end{picture},
\begin{picture}(100, 40)(0,17)
\Line(30,20)(50,20)
\Line(50,20)(70,20)
\CArc(50,20)(20,0,360)
\end{picture} \textrm{and} \quad
\begin{picture}(100, 40)(0,17)
\Line(40,20)(60,20)
\CArc(25,20)(15,0,360)
\CArc(75,20)(15,-180,180)
\end{picture}
\eq\\
One can see that the only diagrams surviving the continuum limit and containing at least one of the vertices that vanish in the continuum limit can be constructed from the following vacuum graphs by \emph{only} attaching lines with the vertices (\ref{survivingvertices}), because these vertices do not give additional powers of $\Delta$.
\vspace{-17pt}
\bq \label{twoloopvacgraphs}
\begin{picture}(100, 40)(0,17)
\DashCArc(35,20)(15,0,360){5}
\DashCArc(65,20)(15,-180,180){5}
\end{picture},
\begin{picture}(100, 40)(0,17)
\DashCArc(35,20)(15,0,360){5}
\CArc(65,20)(15,-180,90)
\ArrowArc(65,20)(15,90,180)
\end{picture},
\begin{picture}(100, 40)(0,17)
\Line(30,20)(50,20)
\ArrowLine(50,20)(70,20)
\DashCArc(50,20)(20,0,360){5}
\end{picture} \textrm{and}
\begin{picture}(100, 40)(0,17)
\ArrowLine(50,20)(30,20)
\ArrowLine(50,20)(70,20)
\DashCArc(50,20)(20,0,360){5}
\end{picture}
\eq\\
For the vacuum graphs we have the following expressions, excluding vertex constants and symmetry factors. Only the discrete loop integration is done and the worst behavior in $\Delta$ is kept.
\vspace{-17pt}
\bqa
\begin{picture}(100, 40)(0,17)
\DashCArc(35,20)(15,0,360){5}
\DashCArc(65,20)(15,-180,180){5}
\end{picture} &=& \frac{\hbar^2}{\Delta^2} \nonumber\\
\begin{picture}(100, 40)(0,17)
\DashCArc(35,20)(15,0,360){5}
\CArc(65,20)(15,-180,90)
\ArrowArc(65,20)(15,90,180)
\end{picture} &=& -\h\frac{\hbar^2}{\Delta} \nonumber\\
\begin{picture}(100, 40)(0,17)
\Line(30,20)(50,20)
\ArrowLine(50,20)(70,20)
\DashCArc(50,20)(20,0,360){5}
\end{picture} &=& -\h\frac{\hbar^3}{\Delta} \nonumber\\
\begin{picture}(100, 40)(0,17)
\ArrowLine(50,20)(30,20)
\ArrowLine(50,20)(70,20)
\DashCArc(50,20)(20,0,360){5}
\end{picture} &=& \frac{\hbar^3}{\Delta^2}
\eqa\\

We now construct all 1PI diagrams from these vacuum graphs by attaching lines with the vertices (\ref{survivingvertices}). Because we can only attach lines with these vertices we can only get external $\eta$-lines. We shall now calculate the generating functional of all the diagrams that can be constructed in this way:
\vspace{-17pt}
\bq
Z(x) = \sum_{n=0}^\infty 
\begin{picture}(60, 40)(0,17)
\Line(30,20)(55,40)
\Line(30,20)(57.7,36)
\Line(30,20)(55,0)
\DashCArc(30,20)(25,-34,26){2}
\Text(57,20)[l]{n}
\GCirc(30,20){15}{0.5}
\end{picture} \quad {1\over n!} x^n
\eq

Now this $Z$, for the first vacuum graph, is given by:
\bqa
Z &=& {1\over8}{\hbar^2\over\Delta^2} \sum_{m_3,m_4,n_3,n_4=0}^\infty \left( {\Delta^2\over\hbar v^2} + {2\Delta^2\over\hbar v^3}x + {2\Delta^2\over\hbar v^4}\h x^2 \right) \hbar^{m_3+m_4+n_3+n_4} \nonumber\\
& & \phantom{\h{\hbar^2\over\Delta^2}{1\over4} \sum_{m_3,m_4,n_3,n_4=0}^\infty} \left(-\frac{2}{\hbar v}\right)^{m_3+n_3} \left(-\frac{2}{\hbar v^2}\right)^{m_4+n_4} \left({1\over2!}\right)^{m_3+n_3} \left({1\over2!2!}\right)^{m_4+n_4} \nonumber\\
& & \phantom{\h{\hbar^2\over\Delta^2}{1\over4} \sum_{m_3,m_4,n_3,n_4=0}^\infty} {1\over m_3!} {1\over n_3!} {1\over m_4!} {1\over n_4!} (2m_3+2m_4)!! (2n_3+2n_4)!! x^{m_3+n_3+2n_4+2m_4} \nonumber\\
&=& {1\over8}{\hbar\over v^2} \frac{1}{\left(1+{x\over v}\right)^2} \nonumber\\
&=& {\hbar\over8} \sum_{n=0}^\infty \frac{(-1)^n (n+1)!}{v^{n+2}} {1\over n!} x^n \label{genfunctvacblob}
\eqa
Here $1/8$ is the symmetry factor of the vacuum graph, $m_3$ and $m_4$ denote respectively the number of $\eta ww$- and $\eta\eta ww$-vertices in the left loop and $n_3$ and $n_4$ denote the number of $\eta ww$- and $\eta\eta ww$-vertices in the right loop. Now we can read off
\bq \label{Leeterm}
\begin{picture}(60, 40)(0,17)
\Line(30,20)(55,40)
\Line(30,20)(57.7,36)
\Line(30,20)(55,0)
\DashCArc(30,20)(25,-34,26){2}
\Text(57,20)[l]{n}
\GCirc(30,20){15}{0.5}
\end{picture}  \quad = {\hbar\over8} \frac{(-1)^n (n+1)!}{v^{n+2}}
\eq\\
for the first vacuum graph.

Also the contributions from the three other vacuum blobs can be constructed in the same way. Their generating functions appear to cancel each other. The reason for this shall become clear below. For now the \emph{only} 2-loop contribution we get is (\ref{Leeterm}).

The $n$-leg diagrams that we find in (\ref{Leeterm}) are exactly the vertices one would get from a term
\bq \label{Leeterm2}
-{\hbar^2\over8}{1\over r_i^2}
\eq
in the action. This can easily be seen by substituting $r=v+\eta$ in this term and expanding it in $\eta$. This means, up to 2-loop level, one can discard the vertices that go to zero in the continuum limit and replace them by the vertices from (\ref{Leeterm2}). In the action this means one is left with
\bq
S = \Delta \sum_{i=0}^{N-1} \left( \h\frac{(r_{i+1}-r_i)^2}{\Delta^2} + \h {r_i^2\over v^2}\frac{(w_{i+1}-w_i)^2}{\Delta^2} - \frac{\hbar^2}{8r_i^2} \right) \;.
\eq

Disregarding 3- and higher-loop level we have now proven that our 1-dimensional discrete path integral $P$ (defined in (\ref{genericpathint})) is equal to:\index{path integral on lattice!1-dimensional}
\bqa
P &=& \int_{-\infty}^{\infty} dr_0 \ldots dr_{N-1} \; r_0 \ldots r_{N-1} \; \int_{-\infty}^{\infty} dw_0 \ldots dw_{N-1} \; O \nonumber\\
& & \qquad \exp\Bigg( -\frac{1}{\hbar} \Delta \sum_{i=0}^{N-1} \bigg( \h\frac{(r_{i+1}-r_i)^2}{\Delta^2} + \h {r_i^2\over v^2}\frac{(w_{i+1}-w_i)^2}{\Delta^2} - \frac{\hbar^2}{8r_i^2} + \nonumber\\
& & \phantom{\qquad \exp\Bigg( -\frac{1}{\hbar} \Delta \sum_{i=0}^{N-1} \bigg(} V\left(r_i\cos{w_i\over v},r_i\sin{w_i\over v}\right) \bigg) \Bigg) \;. \label{1dimpathintLee}
\eqa

This is a form that one can also find in the literature. This same path integral is derived by Lee \cite{Lee} in chapter 19, formula (19.49). Also Edwards et al.\ \cite{Edwards} and Peak et al.\ \cite{Peak} find a term (\ref{Leeterm2}). However they start with the discrete path integral in terms of Cartesian fields, transform to polar fields and actually perform the angular integration. Only then they find the term (\ref{Leeterm2}). We have presented a more general proof of this term here, like Lee \cite{Lee}.

Up to now we have \emph{not} proven that at 3- and higher-loop level there are diagrams, containing at least on of the vertices that vanish in the continuum limit, that can \emph{not} be built also from vertices from the term (\ref{Leeterm2}). We shall not prove this in this thesis. In this thesis we are mostly interested in $d$-dimensional models, and the conjecture needed to compute via polar fields in these models has already been fully proven. It should be clear however that, to have agreement with the literature, the path integral (\ref{1dimpathintLee}) is correct, up to all orders. So, although we cannot prove it at this point, there are \emph{no} diagrams at 3- and higher-loop order that cannot also be constructed with only the term (\ref{Leeterm2}).

\subsubsection{Lee's Proof}

The strictly 1-dimensional (i.e.\ quantum mechanical) derivation of (\ref{1dimpathintLee}) by Lee \cite{Lee} is based on how quantum mechanical path integrals are mostly derived in elementary textbooks. Note that Lee derives the Minkowskian version of (\ref{1dimpathintLee}). Here we will shortly sketch Lee's proof.

One starts with a certain amplitude
\bq
\langle x',y' | x,y \rangle
\eq
that one wants to calculate, where $|x,y\rangle$ is a state in the Heisenberg picture. Here we have a two-dimensional space-time, the analog of this in our quantum field theory is the two-dimensional field-space with fields $\f_1$ and $\f_2$. To derive the path integral one makes time discrete and at each discrete time point $t$ one inserts the unit operator
\bq
\int dx dy \; |x,y,t\rangle\langle x,y,t| \;.
\eq
(Here $|x,y,t\rangle$ is a state in the Schr\"odinger picture.) Working this out one finds the path integral in Cartesian coordinates.

One might also have inserted the unit operator
\bq \label{unitopppol}
\int dr d\theta \; |r,\theta,t\rangle_P \; _P\langle r,\theta,t|
\eq
at each discrete time point $t$. Here the subscript $P$ denotes that we are dealing with polar states, which are related to the Cartesian states as:
\bq
|r,\theta,t\rangle_P = \sqrt{r}|r\cos\theta,r\sin\theta\rangle \;.
\eq
We need the $\sqrt{r}$ to have the proper normalization for the polar states. Only \emph{with} this $\sqrt{r}$ we have that the operator (\ref{unitopppol}) is a unit operator.

At some point in the derivation towards the path integral we have to let the Lagrangian density
\bq
\mathcal{L} = \h\hat{\pi}_x^2 + \h\hat{\pi}_y^2 - V(x,y)
\eq
operate on the polar states. Here $\hat{\pi}_x$ and $\hat{\pi}_x$ are the canonical momenta conjugate to $x$ and $y$. We know that these canonical momenta operate on the Cartesian states as:
\bqa
\hat{\pi}_x &=& -i\hbar \frac{\partial}{\partial x} \;, \nonumber\\
\hat{\pi}_y &=& -i\hbar \frac{\partial}{\partial y} \;.
\eqa
If we then define the canonical momenta conjugate to $r$ and $\theta$ as
\bqa
\hat{\pi}_r &=& -i\hbar \frac{\partial}{\partial r} \;, \nonumber\\
\hat{\pi}_\theta &=& -i\hbar \frac{\partial}{\partial\theta} \;,
\eqa
we have
\bq
\hat{\pi}^2_x + \hat{\pi}^2_y = \hat{\pi}^2_r + \frac{1}{r^2} \hat{\pi}^2_w - \frac{\hbar^2}{4r^2} \;.
\eq
Here we see the emergence of the term (\ref{Leeterm2}) in this derivation of the path integral in polar fields. This roughly sketches how (\ref{1dimpathintLee}) can also be derived in this way. 

Notice that the term (\ref{Leeterm2}) is only found in a 1-dimensional argument. In $d$ dimensions there is no hope that all the vertices that vanish in the continuum can be replaced by a simple term like (\ref{Leeterm2}).

\subsubsection{Another Way to Derive (\ref{1dimpathintLee})}

Another way to derive the discrete 1-dimensional path integral (\ref{1dimpathintLee}) is by first using the conjecture. So one computes all diagrams in a $d$-dimensional way in the continuum, with the simple Feynman rules from the continuum action, then one lets $d\rightarrow1$. To obtain the discrete version of these diagrams one has to know what the difference is between calculating in the continuum and calculating in the discrete. This difference, we know, comes from the problem terms. If we formulate the (continuum) $\eta ww$ and $\eta\eta ww$-vertex with the dots, as we did in section \ref{proofconj}, then we have a clear source of problem terms. Because in this case we \emph{only} have the $\eta ww$- and $\eta\eta ww$-vertex we also \emph{only} have the recursion relation:
\bqa
& & \begin{picture}(100, 40)(0, 17)
\DashLine(10,20)(90,20){5}
\Line(40,20)(10,50)
\Line(60,20)(90,50)
\Vertex(44,20){2}
\Vertex(56,20){2}
\end{picture} +
\begin{picture}(100, 40)(0, 17)
\DashLine(10,20)(90,20){5}
\Line(40,20)(90,50)
\Line(60,20)(10,50)
\Vertex(44,20){2}
\Vertex(56,20){2}
\end{picture} +
\begin{picture}(100, 40)(0, 17)
\DashLine(10,20)(90,20){5}
\Line(50,20)(90,50)
\Line(50,20)(10,50)
\Vertex(50,20){3}
\end{picture} = 0 \label{specialrecrel}
\eqa\\[10pt]
This recursion relation ensures that all problem terms from a $w$-line with two dots cancel against the dotted part of the $\eta\eta ww$-vertex. So the problem terms from $w$-lines are never going to give a difference between a discrete and continuum calculation. All we have to do is find the problem terms coming from $\eta$-lines with two dots.

Now, as in the previous (partial) derivation of (\ref{1dimpathintLee}) we can build all diagrams from the vacuum graphs. At 1-loop order there is \emph{no} difference between a continuum and discrete calculation. At 2-loop order the only problem terms come from the vacuum graph
\vspace{-17pt}
\bq \label{dotgraph}
\begin{picture}(100, 40)(0,17)
\Line(30,20)(70,20)
\DashCArc(50,20)(20,0,360){5}
\Vertex(40,20){3}
\Vertex(60,20){3}
\end{picture}
\eq\\
Now we can understand why, in the other derivation of (\ref{1dimpathintLee}), the generating functionals from the last three vacuum graphs in (\ref{twoloopvacgraphs}) cancelled. Only the first graph in (\ref{twoloopvacgraphs}) corresponds to the vacuum graph above. This correspondence can be seen by pinching the dotted $\eta$-line in the vacuum graph above. The last three vacuum graphs in (\ref{twoloopvacgraphs}) correspond to problem terms from dotted $w$-lines or a dotted $\eta\eta ww$-vertex. These cancel among each other because of the recursion relation (\ref{specialrecrel}).

Now the difference between a continuum and discrete calculation of the graph (\ref{dotgraph}) can be calculated. Also the generating functional of diagrams where we connect any number of $\eta$-legs via the $\eta ww$- and $\eta\eta ww$-vertices can be calculated. The result of this generating functional is given by (\ref{genfunctvacblob}). In this way we find that, to compensate for the differences that we get by doing a discrete instead of a continuum calculation, we have to introduce the term (\ref{Leeterm2}) in the action again.

Also in this way of deriving (\ref{1dimpathintLee}) we do not know how to show that 3- and higher-loop diagrams give no new differences.

\section{A 1-Dimensional Illustration} \label{1dimill}

In the previous section we have shown that, in the case of \emph{one} dimension, the path integral in polar field variables is given by (\ref{1dimpathintLee}). In this section we give a specific example of how (\ref{1dimpathintLee}) can be used in a discrete calculation. We will calculate $\langle r(x)\rangle$ and $\langle\f_1(0)\f_1(x)\rangle$ in the arctangent toy model through this formula and compare the results with the results we can find in a Cartesian calculation. The Cartesian results for some $\f$-Green's-functions have already been found in section \ref{arctanCartsect}.

\subsection{$\langle r(x)\rangle$}

According to (\ref{1dimpathintLee}) $\langle r(x)\rangle$ is given by:
\bqa
\langle r(x) \rangle &=& {1\over Z(0)} \int_{-\infty}^{\infty} dr_0 \ldots dr_{N-1} \; r_0 \ldots r_{N-1} \; \int_{-\infty}^{\infty} dw_0 \ldots dw_{N-1} \; r_j \nonumber\\
& & \qquad \exp\Bigg( -\frac{1}{\hbar} \Delta \sum_{i=0}^{N-1} \bigg( \h\frac{(r_{i+1}-r_i)^2}{\Delta^2} + \h {r_i^2\over v^2}\frac{(w_{i+1}-w_i)^2}{\Delta^2} - \frac{\hbar^2}{8r_i^2} + \nonumber\\
& & \phantom{\qquad \exp\Bigg( -\frac{1}{\hbar} \Delta \sum_{i=0}^{N-1} \bigg(} \frac{m^2}{4v^2} (r_i^2-v^2)^2 + \frac{m^2v^2}{2r_i^2} w_i^2 \bigg) \Bigg) \;. \label{r1dim}
\eqa
Here $j$ is the discrete coordinate corresponding to $x$:
\bq
x = -{L\over2} + j\Delta \;,
\eq
and $Z(0)$ is given by
\bqa
Z(0) &=& \int_{-\infty}^{\infty} dr_0 \ldots dr_{N-1} \; r_0 \ldots r_{N-1} \; \int_{-\infty}^{\infty} dw_0 \ldots dw_{N-1} \nonumber\\
& & \qquad \exp\Bigg( -\frac{1}{\hbar} \Delta \sum_{i=0}^{N-1} \bigg( \h\frac{(r_{i+1}-r_i)^2}{\Delta^2} + \h {r_i^2\over v^2}\frac{(w_{i+1}-w_i)^2}{\Delta^2} - \frac{\hbar^2}{8r_i^2} + \nonumber\\
& & \phantom{\qquad \exp\Bigg( -\frac{1}{\hbar} \Delta \sum_{i=0}^{N-1} \bigg(} \frac{m^2}{4v^2} (r_i^2-v^2)^2 + \frac{m^2v^2}{2r_i^2} w_i^2 \bigg) \Bigg) \;.
\eqa

First we shall focus our attention on the $w$-part in (\ref{r1dim}). This part we call $Z_w(0)$, it is defined as:
\bq
Z_w(0) \equiv \int_{-\infty}^{\infty} dw_0 \ldots dw_{N-1} \; \exp \left( -\frac{1}{\hbar} \Delta \sum_{i=0}^{N-1} \left( \frac{1}{2} {r_i^2\over v^2} \frac{(w_{i+1}-w_i)^2}{\Delta^2} + \frac{m^2v^2}{2r_i^2} w_i^2 \right) \right).
\eq
This $Z_w(0)$ is a quantity depending on all $r$'s. It is this dependence we have to know before we can do the $r$-integrals.

We will now try to find this dependence by finding a set of $N$ differential equations which are satisfied by $Z_w(0)$. Then we will try to solve these differential equations. If we are successful we will have fixed $Z_w(0)$ up to a constant, which is unimportant. Such a set of differential equations is readily found. If we let 
\bq
-\hbar \frac{1}{\Delta} \frac{\partial}{\partial r_j^2}, \quad j = 0,\ldots,N-1
\eq
operate on $Z_w(0)$ we get:
\bqa
-\hbar \frac{1}{\Delta} \frac{\partial}{\partial r_j^2} Z_w(0) &=& \int_{-\infty}^{\infty} dw_0 \ldots dw_{N-1} \left( \frac{1}{2v^2} \frac{(w_{j+1}-w_j)^2}{\Delta^2} - \frac{m^2v^2}{2r_j^4} w_j^2 \right) \nonumber\\
& & \qquad \exp \left( -\frac{1}{\hbar} \Delta \sum_{i=0}^{N-1} \left( \frac{1}{2} {r_i^2\over v^2} \frac{(w_{i+1}-w_i)^2}{\Delta^2} + \frac{m^2v^2}{2r_i^2} w_i^2 \right) \right) \nonumber\\
&=& \frac{1}{2v^2} Z_w(0) \left\langle \frac{(w_{j+1}-w_j)^2}{\Delta^2} \right\rangle_w - Z_w(0) \frac{m^2v^2}{2r_j^4} \langle w_j^2 \rangle_w \;.
\eqa
This differential equation is satisfied by $Z_w(0)$ for $j =0,\ldots,N-1$. Here the $w$-average ($\langle\ldots\rangle_w$) of some product of fields $\alpha$ is given by:
\bq
\langle \alpha \rangle_w \equiv \frac{1}{Z_w(0)} \int_{-\infty}^{\infty} dw_0 \ldots dw_{N-1} \; \alpha \; \exp \left( -\frac{1}{\hbar} \Delta \sum_{i=0}^{N-1} \left( \frac{1}{2} {r_i^2\over v^2} \frac{(w_{i+1}-w_i)^2}{\Delta^2} + \frac{m^2v^2}{2r_i^2} w_i^2 \right) \right) \;.
\eq

Before we can proceed we have to know the quantities
\bq
\left\langle \frac{(w_{j+1}-w_j)^2}{\Delta^2} \right\rangle_w \quad \textrm{and} \quad \langle w_j^2 \rangle_w \;.
\eq
So first we will have to calculate the discrete $w$-propagator $\langle w_jw_k\rangle_w$.

\subsubsection{The Discrete $w$-Propagator}

The discrete $w$-propagator is given by:
\bqa
\langle w_jw_k \rangle_w &\equiv& \frac{1}{Z_w(0)} \int_{-\infty}^{\infty} dw_0 \ldots dw_{N-1} \; w_jw_k \nonumber\\
& & \qquad \exp \left( -\frac{1}{\hbar} \Delta \sum_{i=0}^{N-1} \left( \frac{1}{2} {r_i^2\over v^2} \frac{(w_{i+1}-w_i)^2}{\Delta^2} + \frac{m^2v^2}{2r_i^2} w_i^2 \right) \right) \;.
\eqa
We will find the $w$-propagator through the discrete Schwinger-Dyson equations. To find these Schwinger-Dyson equations we first have to know the Schwinger-Symanzik equation for $Z_w(J)$ in discrete form. $Z_w(J)$ is defined by
\bq
Z_w(J) \equiv \int_{-\infty}^{\infty} dw_0 \ldots dw_{N-1} \exp \left( -\frac{1}{\hbar} \Delta \sum_{i=0}^{N-1} \left( \frac{1}{2} {r_i^2\over v^2} \frac{(w_{i+1}-w_i)^2}{\Delta^2} + \frac{m^2v^2}{2r_i^2} w_i^2 - J_iw_i \right) \right) \;.
\eq
and the \ind{Schwinger-Symanzik equation} it satisfies is
\bq
\left( \frac{1}{\Delta} \left.\frac{\partial S}{\partial w_j}\right|_{w_j = \hbar \frac{1}{\Delta} \frac{\partial}{\partial J_j}} - J_j \right) Z_w(J) = 0 \;.
\eq
The part with the partial derivative of $S$ can be written out. When doing this it is convenient to introduce the following discrete derivative definitions:
\bq \label{discreteprimes}
\alpha'_j \equiv \frac{\alpha_{j+1}-\alpha_j}{\Delta}, \quad \alpha''_j \equiv \frac{-2\alpha_{j+1}+\alpha_j+\alpha_{j+2}}{\Delta^2}
\eq
where $\alpha$ can be any quantity with a label $j$.

The Schwinger-Symanzik equation becomes:
\bqa
& & \Bigg( -r_j^2 \left( \hbar \frac{1}{\Delta} \frac{\partial}{\partial J_{j-1}} \right)'' + \Delta (r'_{j-1})^2 \left( \hbar \frac{1}{\Delta} \frac{\partial}{\partial J_{j-1}} \right)' - 2r_jr'_{j-1} \left( \hbar \frac{1}{\Delta} \frac{\partial}{\partial J_{j-1}} \right)' + \nonumber\\
& & \qquad \frac{m^2v^4}{r_j^2} \left( \hbar \frac{1}{\Delta} \frac{\partial}{\partial J_j} \right) - J_j \Bigg) Z_w(J) = 0 \label{SchwSymill}
\eqa
Now, to obtain the \ind{Schwinger-Dyson equations} for the $w$-propagator, let $\hbar \frac{1}{\Delta} \frac{\partial}{\partial J_k}$ work on both sides of the Schwinger-Symanzik equation (\ref{SchwSymill}) and put all $J$'s to zero. If we also divide by $Z_w(0)$ we get:
\bq \label{SchwDysill}
r_j^2 \langle w''_{j-1}w_k \rangle_w - \Delta (r'_{j-1})^2 \langle w'_{j-1}w_k \rangle_w + 2r_jr'_{j-1} \langle w'_{j-1}w_k \rangle_w - \frac{m^2v^4}{r_j^2} \langle w_jw_k \rangle_w = -\hbar \frac{v^2}{\Delta} \delta_{jk} 
\eq

Now we have to find a solution that satisfies this discrete Schwinger-Dyson equation. Because later we are only interested in the continuum limit of our final result, this solution only has to satisfy the Schwinger-Dyson equation up to order $\Delta^0$. The following solution does the job:
\bqa
\langle w_jw_k \rangle_w &=& \frac{\hbar}{2m} \exp \left( -mv^2\Delta \sum_{i=j+1}^{k} \frac{1}{r_{i-1}^2} \right) \quad \textrm{if $k>j$} \nonumber\\
\langle w_jw_k \rangle_w &=& \frac{\hbar}{2m} \quad \textrm{if $k=j$} \nonumber\\
\langle w_jw_k \rangle_w &=& \frac{\hbar}{2m} \exp \left( -mv^2\Delta \sum_{i=k+1}^{j} \frac{1}{r_{i-1}^2} \right) \quad \textrm{if $k<j$}.
\eqa
To verify this one has to treat the cases $k<j-1$, $k=j-1$, $k=j$, $k=j+1$ and $k>j+1$ separately. As an example we will verify the case $k>j+1$ below. In this case, the average $\langle w'_{j-1}w_k \rangle_w$ becomes:
\bqa
\langle w'_{j-1}w_k \rangle_w &=& \frac{1}{\Delta} \frac{\hbar}{2m} \left( \exp \left( -mv^2\Delta \sum_{i=j+1}^{k} \frac{1}{r_{i-1}^2} \right) - \exp \left( -mv^2\Delta \sum_{i=j}^{k} \frac{1}{r_{i-1}^2} \right) \right) \nonumber\\
&=& \frac{1}{\Delta} \frac{\hbar}{2m} \exp \left( -mv^2\Delta \sum_{i=j+1}^{k} \frac{1}{r_{i-1}^2} \right) \left( 1- e^{-mv^2\Delta \frac{1}{r_{j-1}^2}} \right) \nonumber\\
&=& \langle w_jw_k \rangle_w \frac{1}{\Delta} \left( mv^2\Delta \frac{1}{r_{j-1}^2} + \mathcal{O}(\Delta^2) \right) \nonumber\\
&=& \langle w_jw_k \rangle_w \left( mv^2\frac{1}{r_j^2} + 2mv^2\frac{1}{r_j^3}\Delta r'_{j-1} + \mathcal{O}(\Delta) \right) \;.
\eqa
Remember that $r'_j$ is of order $\frac{1}{\sqrt{\Delta}}$ in one dimension. The average $\langle w''_{j-1}w_k \rangle_w$ becomes:
\bqa
\langle w''_{j-1}w_k \rangle_w &=& \frac{1}{\Delta^2} \frac{\hbar}{2m} \Bigg( -2 \exp \left( -mv^2\Delta \sum_{i=j+1}^{k} \frac{1}{r_{i-1}^2} \right) + \exp \left( -mv^2\Delta \sum_{i=j+2}^{k} \frac{1}{r_{i-1}^2} \right) + \nonumber\\
& & \phantom{\frac{1}{\Delta^2} \frac{\hbar}{2m} \Bigg(} \exp \left( -mv^2\Delta \sum_{i=j}^{k} \frac{1}{r_{i-1}^2} \right) \Bigg) \nonumber\\
&=& \langle w_jw_k \rangle_w \frac{1}{\Delta^2} \left( -2 + e^{mv^2\Delta \frac{1}{r_j^2}} + e^{-mv^2\Delta \frac{1}{r_{j-1}^2}} \right) \nonumber\\
&=& \langle w_jw_k \rangle_w \frac{1}{\Delta^2} \bigg( mv^2\Delta \frac{1}{r_j^2} + \frac{1}{2} m^2v^4\Delta^2 \frac{1}{r_j^4} - mv^2\Delta \frac{1}{r_{j-1}^2} + \frac{1}{2} m^2v^4\Delta^2 \frac{1}{r_{j-1}^4} + \nonumber\\
& & \phantom{\langle w_jw_k \rangle_w \frac{1}{\Delta^2} \bigg(} \mathcal{O}(\Delta^3) \bigg) \nonumber\\
&=& \langle w_jw_k \rangle_w \frac{1}{\Delta^2} \bigg( mv^2\Delta \frac{1}{r_j^2} + m^2v^4\Delta^2 \frac{1}{r_j^4} \nonumber\\
& & \phantom{\langle w_jw_k \rangle_w \frac{1}{\Delta^2} \bigg(} mv^2\Delta \left( -\frac{1}{r_j^2} - 2 \frac{1}{r_j^3} (r_j-r_{j-1}) - 3 \frac{1}{r_j^4} (r_j-r_{j-1})^2 \right) + \nonumber\\
& & \phantom{\langle w_jw_k \rangle_w \frac{1}{\Delta^2} \bigg(} \mathcal{O}(\Delta^{5/2}) \bigg) \nonumber\\
&=& \langle w_jw_k \rangle_w \left( -\frac{2mv^2}{r_j^3} r'_{j-1} - \frac{3mv^2\Delta}{r_j^4} (r'_{j-1})^2 + \frac{m^2v^4}{r_j^4} + \mathcal{O}(\Delta^{\frac{1}{2}}) \right) \;.
\eqa
These results can now be substituted in the Schwinger-Dyson equation (\ref{SchwDysill}) and we see that it is indeed satisfied. 

Now we have our desired discrete solution and we can continue calculating $Z_w(0)$.

\subsubsection{The Calculation Of $Z_w(0)$}

Now that we know the discrete $w$-propagator we can use this result in the differential equations for $Z_w(0)$. In these differential equations the two following quantities occur.
\bqa
\left\langle \frac{(w_{j+1}-w_j)^2}{\Delta^2} \right\rangle_w &=& \frac{1}{\Delta^2} ( \langle w_j^2 \rangle_w + \langle w_{j+1}^2 \rangle_w - 2 \langle w_jw_{j+1} \rangle_w ) \nonumber\\
&=& \frac{1}{\Delta^2} \left( \frac{\hbar}{m} - \frac{\hbar}{m} e^{-mv^2\Delta \frac{1}{r_j^2}} \right) \nonumber\\
&=& \frac{\hbar v^2}{\Delta} \frac{1}{r_j^2} - \frac{\hbar mv^4}{2} \frac{1}{r_j^4} + \mathcal{O}(\Delta) \nonumber\\
\langle w_j^2 \rangle_w &=& \frac{\hbar}{2m}
\eqa
With these results the differential equations for $Z_w(0)$ become:
\bq
-\hbar \frac{1}{\Delta} \frac{\partial}{\partial r_j^2} Z_w(0) = \frac{1}{2} Z_w(0) \Big( \frac{\hbar}{\Delta} \frac{1}{r_j^2} - \hbar mv^2 \frac{1}{r_j^4} + \mathcal{O}(\Delta) \Big)
\eq
or
\bq
\frac{\partial}{\partial r_j^2} Z_w(0) = -\frac{1}{2r_j^2} Z_w(0) + \frac{\Delta mv^2}{2r_j^4} Z_w(0) + \mathcal{O}(\Delta^2) Z_w(0).
\eq

We now have a set of $N$ uncoupled differential equations, which are easily solved:
\bqa
\frac{1}{Z_w(0)} \frac{\partial}{\partial r_j^2} Z_w(0) &=& -\frac{1}{2r_j^2} + \frac{\Delta mv^2}{2r_j^4} + \mathcal{O}(\Delta^2) 
\nonumber\\
\frac{\partial}{\partial r_j^2} \ln Z_w(0) &=& \frac{\partial}{\partial r_j^2} \Big( -\frac{1}{2} \ln r_j^2 - \frac{\Delta mv^2}{2r_j^2} + \mathcal{O}(\Delta^2) \Big) \nonumber\\
\ln Z_w(0) &=& \ln r_j^{-1} - \frac{\Delta mv^2}{2r_j^2} + \mathcal{O}(\Delta^2) + C \nonumber\\
Z_w(0) &=& \frac{1}{r_j} e^{-\frac{\Delta mv^2}{2r_j^2} + \mathcal{O}(\Delta^2) + C}
\eqa
Here $C$ can depend on $r_i$ with $i\neq j$. The complete $Z_w(0)$ is then easily constructed:
\bq
Z_w(0) = C' \left( \prod_{i=0}^{N-1} \frac{1}{r_i} \right) \exp \left( -\Delta \sum_{i=0}^{N-1} \left( \frac{mv^2}{2r_i^2} + \mathcal{O}(\Delta) \right) \right)
\eq
We see that in the continuum limit the term of order $\Delta$ between the brackets in the exponential is unimportant, this verifies our earlier statement that we only have to know the discrete $w$-propagator up to order $\Delta^0$. So the important part of $Z_w(0)$ is:
\bq \label{solZw0}
Z_w(0) = \left( \prod_{i=0}^{N-1} \frac{1}{r_i} \right) \exp \left( -\Delta \sum_{i=0}^{N-1} \frac{mv^2}{2r_i^2} \right).
\eq

\subsubsection{The Calculation Of $\langle r(x)\rangle$}

Now that we know $Z_w(0)$ we can actually calculate $\langle r(x) \rangle$. Remember that $\langle r(x) \rangle$ is given by:
\bqa
\langle r(x) \rangle &=& \frac{1}{Z(0)} \int_{-\infty}^{\infty} dr_0 \ldots dr_{N-1} r_0 \cdots r_{N-1} \int_{-\infty}^{\infty} dw_0 \ldots dw_{N-1} \nonumber\\
& & \quad \exp \Bigg( -\frac{1}{\hbar} \Delta \sum_{i=0}^{N-1} \bigg( \frac{1}{2} \frac{(r_{i+1}-r_i)^2}{\Delta^2} + \frac{1}{2} {r_i^2\over v^2} \frac{(w_{i+1}-w_i)^2}{\Delta^2} - \frac{\hbar^2}{8r_i^2} + \nonumber\\
& & \phantom{\quad \exp \Bigg(} \frac{m^2}{4v^2} (r_i^2-v^2)^2 + \frac{m^2v^2}{2r_i^2} w_i^2 \bigg) \Bigg) \; r_j \nonumber\\
&=& \frac{1}{Z(0)} \int_{-\infty}^{\infty} dr_0 \ldots dr_{N-1} r_0 \ldots r_{N-1} \nonumber\\
& & \quad \exp \left( -\frac{1}{\hbar} \Delta \sum_{i=0}^{N-1} \left( \frac{1}{2} \frac{(r_{i+1}-r_i)^2}{\Delta^2} - \frac{\hbar^2}{8r_i^2} + \frac{m^2}{4v^2} (r_i^2-v^2)^2 \right) \right) \; Z_w(0) \; r_j \;. \nonumber\\
\eqa
Substituting our solution (\ref{solZw0}) for $Z_w(0)$ in here gives:
\bqa
\langle r(x) \rangle &=& \frac{1}{Z(0)} \int_{-\infty}^{\infty} dr_0 \ldots dr_{N-1} \nonumber\\
& & \quad \exp \left( -\frac{1}{\hbar} \Delta \sum_{i=0}^{N-1} \left( \frac{1}{2} \frac{(r_{i+1}-r_i)^2}{\Delta^2} - \frac{\hbar^2}{8r_i^2} + \frac{m^2}{4v^2} (r_i^2-v^2)^2 + \frac{\hbar mv^2}{2r_i^2} \right) \right) \; r_j \;. \nonumber\\
\eqa
What we have here is a normal path integral in $r$. This means we can do this path integral in the ordinary way by using Feynman diagrams. At this point also the continuum limit can be taken in the action, since we have no derivative couplings anymore. The continuum form of the action is:
\bq
S = \int dx \left( \frac{1}{2} (r'(x))^2 + \frac{m^2}{4v^2} (r^2(x)-v^2)^2 - \frac{\hbar^2}{8r^2(x)} + \frac{\hbar mv^2}{2r^2(x)} \right) \;.
\eq

First we should find the minimum of the action. This minimum is not exactly at $r=v$, but is shifted a little because of the terms $Z_w(0)$ has introduced in the action. The minimum $r_m$ satisfies:
\bq \label{l8}
\frac{m^2}{v^2} (r_m^2-v^2) r_m + \frac{\hbar^2}{4r_m^3} - \frac{\hbar mv^2}{r_m^3} = 0 \;.
\eq
Again we want to know $r_m$ up to order $\hbar^2$. The $r_m$ that satisfies (\ref{l8}) up to order $\hbar^2$ is:
\bq \label{l11}
r_m = v + \frac{\hbar}{2mv} - \frac{5\hbar^2}{4m^2v^3} \;.
\eq
The expectation value of $r(0)$ becomes:
\bq
\langle r(0) \rangle = v + \frac{\hbar}{2mv} - \frac{5\hbar^2}{4m^2v^3} + \langle \eta(0) \rangle + \mathcal{O}(\hbar^3) \;.
\eq

Now we should expand the action around the minimum $r_m$ and read off the Feynman rules. To get $\langle \eta(0) \rangle$ correct up to order $\hbar^2$ we have to keep all terms of order $\hbar^{5/2}$ and lower in the action. After the expansion around the minimum the action becomes:
\bq
S = \int dx \Big( \frac{1}{2} (\eta'(x))^2 + \frac{\hbar m}{2} - \frac{3\hbar^2}{8v^2} + m^2 \eta^2 + \frac{3\hbar m}{v^2} \eta^2 + \frac{m^2}{v} \eta^3 - \frac{3\hbar m}{2v^3} \eta^3 + \frac{m^2}{4v^2} \eta^4 \Big).
\eq
The Feynman rules are:
\bqa
\begin{picture}(100, 20)(0, 17)
\Line(20, 20)(80, 20)
\end{picture}
&\leftrightarrow& \frac{\hbar}{k^2+\mu^2} \nonumber\\
\begin{picture}(100, 20)(0, 17)
\Line(20, 20)(80, 20)
\Vertex(50, 20){3}
\end{picture}
&\leftrightarrow& -\frac{6m}{v^2} \nonumber\\
\begin{picture}(100, 40)(0, 17)
\Line(20, 20)(50, 20)
\Line(50, 20)(80, 0)
\Line(50, 20)(80, 40)
\end{picture}
&\leftrightarrow& -\frac{6m^2}{v} \frac{1}{\hbar} \nonumber\\
\begin{picture}(100, 40)(0, 17)
\Line(20, 20)(50, 20)
\Line(50, 20)(80, 0)
\Line(50, 20)(80, 40)
\Vertex(50, 20){3}
\end{picture}
&\leftrightarrow& \frac{9\hbar m}{v^3} \frac{1}{\hbar} \nonumber\\
\begin{picture}(100, 40)(0, 17)
\Line(20, 0)(80, 40)
\Line(20, 40)(80, 0)
\end{picture}
&\leftrightarrow& -\frac{6m^2}{v^2} \frac{1}{\hbar} \\\nonumber
\eqa
With these Feynman rules we have to compute $\langle \eta(0) \rangle$ up to order $\hbar^2$. We get the following contributions.
{\allowdisplaybreaks\bqa
\begin{picture}(80, 40)(0, 18)
\Line(20, 20)(40, 20)
\Oval(50, 20)(10, 10)(0)
\end{picture}
&=& -\frac{3}{4 \sqrt{2}} \frac{\hbar}{vm} \nonumber\\
\begin{picture}(80, 40)(0, 18)
\Line(20, 20)(40, 20)
\Oval(50, 20)(10, 10)(0)
\Vertex(60, 20){3}
\end{picture}
&=& \frac{9}{8 \sqrt{2}} \frac{\hbar^2}{v^3m^2} \nonumber\\
\begin{picture}(80, 40)(0, 18)
\Line(20, 20)(40, 20)
\Oval(50, 20)(10, 10)(0)
\Vertex(30, 20){3}
\end{picture}
&=& \frac{9}{4 \sqrt{2}} \frac{\hbar^2}{v^3m^2} \nonumber\\
\begin{picture}(80, 40)(0, 18)
\Line(20, 20)(40, 20)
\Oval(50, 20)(10, 10)(0)
\Vertex(40, 20){3}
\end{picture}
&=& \frac{9}{8 \sqrt{2}} \frac{\hbar^2}{v^3m^2} \nonumber\\
\begin{picture}(80, 40)(0, 18)
\Line(10, 20)(30, 20)
\Line(30, 20)(53, 28)
\Line(30, 20)(53, 12)
\Oval(60, 30)(8, 8)(0)
\Oval(60, 10)(8, 8)(0)
\end{picture}
&=& -\frac{27}{64} \frac{\hbar^2}{v^3m^2} \nonumber\\
\begin{picture}(80, 40)(0, 18)
\Line(10, 20)(20, 20)
\Oval(30, 20)(10, 10)(0)
\Line(40, 20)(50, 20)
\Oval(60, 20)(10, 10)(0)
\end{picture}
&=& -\frac{27}{64} \frac{\hbar^2}{v^3m^2} \nonumber\\
\begin{picture}(80, 40)(0, 18)
\Line(10, 20)(40, 20)
\Oval(55, 20)(15, 15)(0)
\Line(55, 5)(55, 35)
\end{picture}
&=& -\frac{3}{16} \frac{\hbar^2}{v^3m^2} \nonumber\\
\begin{picture}(80, 40)(0, 18)
\Line(10, 20)(30, 20)
\Oval(40, 20)(10, 10)(0)
\Oval(60, 20)(10, 10)(0)
\end{picture}
&=& \frac{9}{64} \frac{\hbar^2}{v^3m^2} \nonumber\\
\begin{picture}(80, 40)(0, 18)
\Line(10, 20)(50, 20)
\Oval(30, 30)(10, 10)(0)
\Oval(60, 20)(10, 10)(0)
\end{picture}
&=& \frac{9}{32} \frac{\hbar^2}{v^3m^2} \nonumber\\
\begin{picture}(80, 40)(0, 18)
\Line(10, 20)(70, 20)
\Oval(55, 20)(15, 15)(0)
\end{picture}
&=& \frac{1}{16} \frac{\hbar^2}{v^3m^2} \\\nonumber
\eqa}
Summing up these contributions we get:
\bq \label{l12}
\langle \eta(0) \rangle = \frac{\hbar}{vm} \Big( -\frac{3}{4 \sqrt{2}} \Big) + \frac{\hbar^2}{v^3m^2} \Big( -\frac{35}{64} + \frac{9}{2 \sqrt{2}} \Big) + \mathcal{O}(\hbar^3).
\eq
For $\langle r(0) \rangle$ we finally get:
\bq \label{1dimrpolar}
\langle r(0) \rangle = \frac{\hbar}{vm} \Big( \frac{1}{2} -\frac{3}{4 \sqrt{2}} \Big) + \frac{\hbar^2}{v^3m^2} \Big( -\frac{115}{64} + \frac{9}{2 \sqrt{2}} \Big) + \mathcal{O}(\hbar^3).
\eq

From our Cartesian results in section \ref{arctanCartsect} $\langle r(x)\rangle$ can easily be found via the formula:
\bqa
\langle r(x)\rangle &=& \left\langle \sqrt{(v+\eta_1(x))^2+\eta_2^2(x)} \right\rangle \nonumber\\
&=& v + \langle \eta_1(x) \rangle + \frac{1}{2v} \langle \eta_2^2(x) \rangle - \frac{1}{2v^2} \langle \eta_1(x)\eta_2^2(x) \rangle + \frac{1}{2v^3} \langle \eta_1^2(x)\eta_2^2(x) \rangle + \nonumber\\
& & -\frac{1}{8v^3} \langle \eta_2^4(x) \rangle + \mathcal{O}(\hbar^3)
\eqa
Substituting the results from section \ref{arctanCartsect}, specified to the case $d=1$, indeed gives the same result as (\ref{1dimrpolar}). Here we used the 1-dimensional results for the standard integrals given in appendix \ref{appstandint}.

\subsection{The $\f_1$-Propagator}

As a last illustration of the 1-dimensional path integral in terms of polar fields we will calculate the $\f_1$-propagator. We have:
\bqa
\langle \f_1(x) \f_1(y) \rangle &=& \langle r(x) r(y) \cos(w(x)) \cos(w(y)) \rangle \nonumber\\
&=& \langle r(x) r(y) \rangle - \frac{1}{2} \langle r(x) r(y) w^2(x) \rangle - \frac{1}{2} \langle r(x) r(y) w^2(y) \rangle + \nonumber\\
& & \frac{1}{24} \langle r(x) r(y) w^4(x) \rangle + \frac{1}{24} \langle r(x) r(y) w^4(y) \rangle + \frac{1}{4} \langle r(x) r(y) w^2(x) w^2(y) \rangle + \nonumber\\[5pt]
& & \mathcal{O}(\hbar^3).
\eqa

We calculate the six averages that occur one by one.
\bq
\langle r(x) r(y) \rangle = r_m^2 + 2r_m \langle \eta(0) \rangle + \langle \eta(x) \eta(y) \rangle
\eq
Here $r_m$ is, as in the calculation of $\langle r(0) \rangle$, given by (\ref{l11}). The quantity $\langle \eta(0) \rangle$ is already known from earlier calculations, it is given by (\ref{l12}). So we only have to calculate $\langle \eta(x) \eta(y) \rangle$. For this average we get the following contributions.
\bqa
\begin{picture}(80, 40)(0, 18)
\Line(10, 20)(70, 20)
\end{picture}
&=& \frac{\hbar}{2 \sqrt{2} m} e^{-\mu |x-y|} \nonumber\\
\begin{picture}(80, 40)(0, 18)
\Line(10, 20)(70, 20)
\Vertex(40, 20){3}
\end{picture}
&=& \frac{\hbar^2}{v^2m^2} \Big( -\frac{3}{4 \sqrt{2}} e^{-\mu |x-y|} - \frac{3}{4} m|x-y| e^{-\mu |x-y|} \Big) \nonumber\\
\begin{picture}(80, 40)(0, 18)
\Line(10, 20)(70, 20)
\Oval(40, 30)(10, 10)(0)
\end{picture}
&=& \frac{\hbar^2}{v^2m^2} \Big( -\frac{3}{32} e^{-\mu |x-y|} - \frac{3}{16 \sqrt{2}} m|x-y| e^{-\mu |x-y|} \Big) \nonumber\\
\begin{picture}(80, 40)(0, 18)
\Line(10, 20)(25, 20)
\Line(55, 20)(70, 20)
\Oval(40, 20)(15, 15)(0)
\end{picture}
&=& \frac{\hbar^2}{v^2m^2} \Big( \frac{1}{16} e^{-\mu |x-y|} + \frac{1}{16} e^{-2\mu |x-y|} + \frac{3}{8 \sqrt{2}} m|x-y| e^{-\mu |x-y|} \Big)  \nonumber\\
\begin{picture}(80, 40)(0, 18)
\Line(10, 10)(70, 10)
\Line(40, 10)(40, 20)
\Oval(40, 30)(10, 10)(0)
\end{picture}
&=& \frac{\hbar^2}{v^2m^2} \Big( \frac{9}{32} e^{-\mu |x-y|} + \frac{9}{16 \sqrt{2}} m|x-y| e^{-\mu |x-y|} \Big) \nonumber\\
\begin{picture}(80, 40)(0, 18)
\Line(10, 20)(25, 20)
\Line(55, 20)(70, 20)
\Oval(32, 20)(7, 7)(0)
\Oval(48, 20)(7, 7)(0)
\end{picture}
&=& \frac{\hbar^2}{v^2m^2} \frac{9}{32} \\\nonumber
\eqa
Summing these contributions gives:
\bqa
\langle \eta(x) \eta(y) \rangle &=& \frac{\hbar}{m} \frac{1}{2 \sqrt{2}} e^{-\mu |x-y|} + \frac{\hbar^2}{v^2m^2} \Big( \frac{9}{32} + \frac{1}{4} e^{-\mu |x-y|} - \frac{3}{4 \sqrt{2}} e^{-\mu |x-y|} + \nonumber\\
& & \frac{1}{16} e^{-2\mu |x-y|} - \frac{3}{4} m|x-y| e^{-\mu |x-y|} + \frac{3}{4 \sqrt{2}} m|x-y| e^{-\mu |x-y|} \Big).
\eqa
Now $\langle r(x) r(y) \rangle$ becomes:
\bqa
\langle r(x) r(y) \rangle &=& v^2 + \frac{\hbar}{m} \Big( 1 - \frac{3}{2 \sqrt{2}} + \frac{1}{2 \sqrt{2}} e^{-\mu |x-y|} \Big) + \nonumber\\
& & \frac{\hbar^2}{v^2m^2} \Big( -\frac{49}{16} + \frac{33}{4 \sqrt{2}} + \frac{1}{4} e^{-\mu |x-y|} - \frac{3}{4 \sqrt{2}} e^{-\mu |x-y|} + \nonumber\\
& & \frac{1}{16} e^{-2\mu |x-y|} - \frac{3}{4} m|x-y| e^{-\mu |x-y|} + \frac{3}{4 \sqrt{2}} m|x-y| e^{-\mu |x-y|} \Big). \label{l13}
\eqa

In the next five averages that we have to calculate, before we find our final result for the $\f_1$-propagator, also $w$'s occur. These $w$'s have to be taken into account in the $w$-path integral that we perform first in all our calculations. Fortunately we know the $w$-propagator already, so that this result can be used in these calculations. We get:
\bq
-\frac{1}{2} \langle r(x) r(y) w^2(x) \rangle = -\frac{1}{2} \frac{\hbar}{2mv^2} \langle r(x) r(y) \rangle.
\eq
Now we can substitute our result (\ref{l13}) up to order $\hbar$ in here and find:
\bq
-\frac{1}{2} \langle r(x) r(y) w^2(x) \rangle = \frac{\hbar}{m} \Big( -\frac{1}{4} \Big) + \frac{\hbar^2}{v^2m^2} \Big( -\frac{1}{4} + \frac{3}{8 \sqrt{2}} - \frac{1}{8 \sqrt{2}} e^{-\mu |x-y|} \Big).
\eq
In the same way we find:
\bq
-\frac{1}{2} \langle r(x) r(y) w^2(y) \rangle = \frac{\hbar}{m} \Big( -\frac{1}{4} \Big) + \frac{\hbar^2}{v^2m^2} \Big( -\frac{1}{4} + \frac{3}{8 \sqrt{2}} - \frac{1}{8 \sqrt{2}} e^{-\mu |x-y|} \Big).
\eq

In the next three averages products of four $w$'s occur. Because the $w$-path integral is Gaussian we can write this into a sum of products of averages of two $w$'s:
\bqa
\langle w^4(x) \rangle_w &=& 3 \langle w^2(x) \rangle_w \langle w^2(x) \rangle_w = \frac{3\hbar^2}{4m^2v^4} \nonumber\\
\langle w^4(y) \rangle_w &=& 3 \langle w^2(y) \rangle_w \langle w^2(y) \rangle_w = \frac{3\hbar^2}{4m^2v^4} \nonumber\\
\langle w^2(x) w^2(y) \rangle_w &=& 2 \langle w(x) w(y) \rangle_w \langle w(x) w(y) \rangle_w + \langle w^2(x) \rangle_w \langle w^2(y) \rangle_w \nonumber\\
&=& \frac{\hbar^2}{2m^2v^4} e^{-2mv^2 | \int_x^y dx' \frac{1}{r^2(x')} |} + \frac{\hbar^2}{4m^2v^4}.
\eqa

For our last three averages we get:
\bqa
\frac{1}{24} \langle r(x) r(y) w^4(x) \rangle &=& \frac{\hbar^2}{32m^2v^4} \langle r(x) r(y) \rangle = \frac{\hbar^2}{v^2m^2} \frac{1}{32} \nonumber\\
\frac{1}{24} \langle r(x) r(y) w^4(y) \rangle &=& \frac{\hbar^2}{32m^2v^4} \langle r(x) r(y) \rangle = \frac{\hbar^2}{v^2m^2} \frac{1}{32}
\eqa
\bqa
\frac{1}{4} \langle r(x) r(y) w^2(x) w^2(y) \rangle &=& \frac{\hbar^2}{16m^2v^4} \langle r(x) r(y) \rangle + \frac{\hbar^2}{8m^2v^4} \langle r(x) r(y) e^{-2mv^2 | \int_x^y dx' \frac{1}{r^2(x')} |} \rangle \nonumber\\
&=& \frac{\hbar^2}{m^2v^2} \frac{1}{16} + \frac{\hbar^2}{8m^2v^4} \langle r_m^2 e^{-2mv^2 | \int_x^y dx' \frac{1}{r_m^2} |} \rangle \nonumber\\
&=& \frac{\hbar^2}{m^2v^2} \frac{1}{16} + \frac{\hbar^2}{m^2v^2} \frac{1}{8} e^{-2m|x-y|}.
\eqa

Finally we get for the $\f_1$-propagator:
\bqa
\langle \f_1(x) \f_1(y) \rangle &=& v^2 + \frac{\hbar}{m} \Big( \frac{1}{2} - \frac{3}{2 \sqrt{2}} + \frac{1}{2 \sqrt{2}} e^{-\mu |x-y|} \Big) + \nonumber\\
& & \frac{\hbar^2}{v^2m^2} \Big( -\frac{55}{16} + \frac{9}{\sqrt{2}} + \frac{1}{4} e^{-\mu |x-y|} - \frac{1}{\sqrt{2}} e^{-\mu |x-y|} + \frac{1}{16} e^{-2\mu |x-y|} + \nonumber\\
& & \frac{1}{8} e^{-2m |x-y|} - \frac{3}{4} m|x-y| e^{-\mu |x-y|} + \frac{3}{4 \sqrt{2}} m|x-y| e^{-\mu |x-y|} \Big).
\eqa
And again this matches the result we find in section \ref{arctanCartsect}, specified to $d=1$. (To calculate the standard integrals use appendix \ref{appstandint} again.)


\chapter{The $N=2$ LSM: The Path-Integral Approach II}\label{chapN2sigmamodelpathII}

The action\index{action!$N=2$ LSM} of the Euclidean $N=2$ linear sigma model is given by:\index{$N=2$ linear sigma model!path-integral approach}
\bq \label{polaractionN2}
S = \int d^dx \left( {1\over2}\left(\nabla\varphi_1(x)\right)^2 + {1\over2}\left(\nabla\varphi_2(x)\right)^2 + {\mu\over4v^2}\left( \varphi_1^2(x)+\varphi_2^2(x)-v^2 \right)^2 \right) \;.
\eq

In chapter \ref{chapN2sigmamodelpathI} we already calculated the Green's functions of this model by naively calculating Green's functions around each of the minima and then integrating over all minima. It was not at all clear that this was the correct thing to do, especially because in this approach one has to do perturbation theory around each of the minima. Each time we expand around one of these minima we pretend the ring of minima is actually an infinite line. So in this way we ignore the damping in the $\eta_2$-direction (i.e.\ tangential direction), which is there because of the $\eta_2^4$-term. This damping effect is lost in perturbation theory because the exponential of $\eta_2^4$ is expanded and not all terms are kept.

Also, by integrating over all minima we implicitly assume that the minima do not communicate, which is not true at all.

In this chapter we will calculate the same Green's functions via the \ind{path integral in polar field variables}. These polar variables are the natural variables for a model with $O(2)$-symmetry. The action in terms of polar field variables will \emph{not} depend on the angular field $w$, but only on $\nabla w$. Therefore we have that:
\bqa
w &=& \mathcal{O}(1) \nonumber\\
\nabla w &=& \mathcal{O}(\sqrt{\hbar})
\eqa
The first relation merely states that all points on the ring of minima have an equal weight in the path integral. This means it is also incorrect to expand around $w=0$, for which we would have to assume that $w$ is small. This expansion is what we did in chapter \ref{chapN2sigmamodelpathI}. From the second relation we see that it \emph{is} correct to expand in $\nabla w$, because $\nabla w$ is small.

Because the action in terms of polar fields does not depend on $w$ there is also \emph{no} need to expand around $w=0$ in the formalism in terms of polar fields. In this way we avoid doing perturbation theory in $w$, which was the big problem of chapter \ref{chapN2sigmamodelpathI}.

In this chapter also the effective potential of the $N=2$ linear sigma model will be calculated via the path integral in terms of polar fields.

\section{Green's Functions}

According to the \ind{conjecture} from the previous chapter the path integral in terms of polar field variables\index{path integral in polar field variables} for this model is given by
\bqa
& & \langle\f_1(x_1)\cdots\f_1(x_m)\f_2(y_1)\cdots\f_2(y_n)\rangle = \nonumber\\[5pt]
& & \hspace{50pt} {1\over Z(0)} \int_{-\infty}^{\infty} \mathcal{D}r \int_{-\infty}^{\infty} \mathcal{D}\theta \; \exp\left(-{1\over\hbar} I \int d^dx \left(-\hbar\ln r(x)\right)\right) \cdot \nonumber\\
& & \hspace{50pt} \phantom{{1\over Z(0)} \int_{-\infty}^{\infty} \mathcal{D}r \int_{-\infty}^{\infty} \mathcal{D}w \;} r(x_1)\cos(\theta(x_1)) \cdots r(x_m)\cos(\theta(x_m)) \cdot \nonumber\\
& & \hspace{50pt} \phantom{{1\over Z(0)} \int_{-\infty}^{\infty} \mathcal{D}r \int_{-\infty}^{\infty} \mathcal{D}w \;} r(y_1)\sin(\theta(y_1)) \cdots r(y_n)\sin(\theta(y_n)) \cdot \nonumber\\
& & \hspace{50pt} \phantom{{1\over Z(0)} \int_{-\infty}^{\infty} \mathcal{D}r \int_{-\infty}^{\infty} \mathcal{D}w \;} \exp\left(-{1\over\hbar}S(r,\theta)\right) \;,
\eqa
provided we perform the calculation in a $d$-dimensional way. Here $Z(0)$ given by
\bq
Z(0) = \int_{-\infty}^{\infty} \mathcal{D}r \int_{-\infty}^{\infty} \mathcal{D}\theta \; \exp\left(-{1\over\hbar} I \int d^dx \left(-\hbar\ln r(x)\right)\right) \exp\left(-{1\over\hbar}S(r,\theta)\right) \;,
\eq
and $S(r,\theta)$ given by
\bq
S(r,\theta) = \int d^dx \left( {1\over2}\left(\nabla r(x)\right)^2 + {1\over2}r^2(x)\left(\nabla\theta(x)\right)^2 + {\mu\over4v^2}\left( r^2(x)-v^2 \right)^2 \right) \;.
\eq

Because we are dealing with a $d$-dimensional model divergences will arise and we must renormalize the fields, masses and coupling constants. First we rewrite the action in the form
\bq
S(r,\theta) = \int d^dx \left( {1\over2}\left(\nabla r(x)\right)^2 + {1\over2}r^2(x)\left(\nabla\theta(x)\right)^2 - \h\mu r^2(x) + {\lambda\over24} r^4(x) \right) \;,
\eq
where
\bq
\lambda = {6\mu\over v^2} \;,
\eq
as in the rest of this thesis. The fields, masses and coupling constants are renormalized in the same way as in chapter \ref{chapN2sigmamodelcan} and \ref{chapN2sigmamodelpathI}:
\bqa
\varphi_i^R &=& {1\over\sqrt{Z}} \varphi_i \quad (i=1,2) \nonumber\\
\mu^R &=&\mu Z - \delta_\mu \nonumber\\
\lambda^R &=& \lambda Z^2 - \delta_{\lambda}
\eqa
In terms of polar variables the field renormalization means:
\bq
r^R = {1\over\sqrt{Z}} r \;,
\eq
the $\theta$-field is \emph{not} renormalized. We also define a new angular field as
\bq
w(x) \equiv v^R \theta(x) \;,
\eq
where
\bq
v^R = \sqrt{6\mu^R\over\lambda^R} \;.
\eq

Making these substitutions in the action (\ref{polaractionN2}) we get (also defining $\delta_Z \equiv Z-1$):
\bqa
S &=& \int d^dx \Bigg( {1\over2}\left(\nabla r^R\right)^2 + {1\over2}{\left(r^R\right)^2\over \left(v^R\right)^2}\left(\nabla w\right)^2 + {\left(\mu^R\right)^2\over4\left(v^R\right)^2}\left( \left(r^R\right)^2-\left(v^R\right)^2 \right)^2 + \nonumber\\
& & \phantom{\int d^dx \Bigg(} {1\over2} \delta_Z \left(\nabla r^R\right)^2 + {1\over2} \delta_Z {\left(r^R\right)^2\over \left(v^R\right)^2}\left(\nabla w\right)^2 - {1\over2} \delta_\mu \left(r^R\right)^2 + {1\over24} \delta_{\lambda} \left(r^R\right)^4 \Bigg) \;. \label{renormS}
\eqa

From here on we shall suppress the $R$-superscripts, understanding that we \emph{always} work with renormalized fields, masses and coupling constants.

Notice that the counter terms have nothing to do with the transformation to polar fields, both in a Cartesian and polar formulation we have the \emph{same} counter terms.

To do perturbation theory we expand around the minimum of the first line (i.e.\ the classical part) of the renormalized action:
\bq
r(x) = v + \eta(x) \;.
\eq

Remember that we also have to include the Feynman rules from the Jacobian. The procedure of renormalization does not change these rules.

The Feynman rules (in momentum space) up to order $\hbar^{5/2}$ are:\index{Feynman rules!$N=2$ LSM}
{\allowdisplaybreaks\bqa
\begin{picture}(100, 20)(0, 17)
\Line(20, 20)(80, 20)
\end{picture}
&\leftrightarrow& \frac{\hbar}{k^2 + m^2} \nonumber\\
\begin{picture}(100, 20)(0, 17)
\DashLine(20, 20)(80, 20){5}
\end{picture}
&\leftrightarrow& \frac{\hbar}{k^2} \nonumber\\
\begin{picture}(100, 40)(0, 17)
\Line(20, 20)(50, 20)
\DashLine(50, 20)(80, 0){5}
\DashLine(50, 20)(80, 40){5}
\Text(80, 10)[]{$q$}
\Text(80, 32)[]{$p$}
\end{picture}
&\leftrightarrow& \frac{2}{\hbar v} \; p\cdot q \nonumber\\
\begin{picture}(100, 40)(0, 17)
\Line(20, 0)(50, 20)
\Line(20, 40)(50, 20)
\DashLine(50, 20)(80, 0){5}
\DashLine(50, 20)(80, 40){5}
\Text(80, 10)[]{$q$}
\Text(80, 32)[]{$p$}
\end{picture}
&\leftrightarrow& \frac{2}{\hbar v^2} \; p\cdot q \nonumber\\
\begin{picture}(100, 40)(0, 17)
\Line(20, 20)(50, 20)
\Line(50, 20)(80, 0)
\Line(50, 20)(80, 40)
\end{picture}
&\leftrightarrow& -\frac{3m^2}{\hbar v} \nonumber\\
\begin{picture}(100, 40)(0, 17)
\Line(20, 40)(80, 0)
\Line(20, 0)(80, 40)
\end{picture}
&\leftrightarrow& -\frac{3m^2}{\hbar v^2} \nonumber\\
\begin{picture}(100, 40)(0, 17)
\Line(20, 20)(50, 20)
\Vertex(50,20){5}
\end{picture}
&\leftrightarrow& {v\over\hbar} \delta_\mu - {1\over6} {v^3\over\hbar} \delta_{\lambda} \nonumber\\[5pt]
\begin{picture}(100, 40)(0, 17)
\Line(20, 20)(80, 20)
\Vertex(50,20){5}
\end{picture}
&\leftrightarrow& -{1\over\hbar} \; p^2 \; \delta_Z + {1\over\hbar} \delta_\mu - {1\over2} {v^2\over\hbar} \delta_{\lambda} \nonumber\\[5pt]
\begin{picture}(100, 40)(0, 17)
\DashLine(20, 20)(80, 20){5}
\Vertex(50,20){5}
\end{picture}
&\leftrightarrow& -{1\over\hbar} \; p^2 \; \delta_Z \nonumber\\[5pt]
\begin{picture}(100, 40)(0, 17)
\Line(20, 20)(50, 20)
\DashLine(50, 20)(80, 0){5}
\DashLine(50, 20)(80, 40){5}
\Vertex(50,20){5}
\Text(80, 10)[]{$q$}
\Text(80, 32)[]{$p$}
\end{picture}
&\leftrightarrow& \frac{2}{\hbar v} \; p\cdot q \; \delta_Z \nonumber\\
\begin{picture}(100, 40)(0, 17)
\Line(20, 0)(50, 20)
\Line(20, 40)(50, 20)
\DashLine(50, 20)(80, 0){5}
\DashLine(50, 20)(80, 40){5}
\Vertex(50,20){5}
\Text(80, 10)[]{$q$}
\Text(80, 32)[]{$p$}
\end{picture}
&\leftrightarrow& \frac{2}{\hbar v^2} \; p\cdot q \; \delta_Z \nonumber\\
\begin{picture}(100, 40)(0, 17)
\Line(20, 20)(50, 20)
\Line(50, 20)(80, 0)
\Line(50, 20)(80, 40)
\Vertex(50,20){5}
\end{picture}
&\leftrightarrow& -\frac{v}{\hbar} \delta_{\lambda} \nonumber\\
\begin{picture}(100, 40)(0, 17)
\Line(20, 40)(80, 0)
\Line(20, 0)(80, 40)
\Vertex(50,20){5}
\end{picture}
&\leftrightarrow& -\frac{1}{\hbar} \delta_{\lambda} \nonumber\\
\begin{picture}(100, 40)(0, 17)
\Line(20, 20)(50, 20)
\Vertex(50, 20){3}
\end{picture}
&\leftrightarrow& {1\over v} \; I \nonumber\\
\begin{picture}(100, 40)(0, 17)
\Line(20, 20)(80, 20)
\Vertex(50, 20){3}
\end{picture}
&\leftrightarrow& -{1\over v^2} \; I \nonumber\\
\begin{picture}(100, 40)(0, 17)
\Line(20, 20)(50, 20)
\Line(50, 20)(80, 40)
\Line(50, 20)(80, 0)
\Vertex(50, 20){3}
\end{picture}
&\leftrightarrow& {2\over v^3} \; I
\eqa}
\\[10pt]
Here we have defined $\mu=\h m^2$ as in chapters \ref{chapN2sigmamodelcan} and \ref{chapN2sigmamodelpathI}. Also all indicated momenta flow into the vertex. The counter-term vertices have been indicated by a big dot in the vertex, the vertices from the \ind{Jacobian} have been indicated by a small dot.

\subsection{$\eta$- And $w$-Green's-Functions} \label{O2etaw}

Now we can compute all the $\eta$- and $w$-Green's-functions.\index{Green's functions!$N=2$ LSM}
\bqa
\langle \tilde{\eta} \rangle |_{\hbar} &=& \quad
\begin{picture}(50,40)(0,18)
\Line(0,20)(20,20)
\DashCArc(35,20)(15,-180,180){3}
\end{picture} \quad + \quad
\begin{picture}(50,40)(0,18)
\Line(0,20)(20,20)
\CArc(35,20)(15,-180,180)
\end{picture} \quad + \quad
\begin{picture}(50,40)(0,18)
\Line(0,20)(35,20)
\Vertex(35,20){5}
\end{picture} \quad + \quad
\begin{picture}(50,40)(0,18)
\Line(0,20)(35,20)
\Vertex(35,20){3}
\end{picture} \nonumber\\[25pt]
&=& -{3\over2} {\hbar\over v} \; I(0,m) + {v\over m^2} \; \delta_\mu|_{\hbar} - {1\over6} {v^3\over m^2} \; \delta_{\lambda}|_{\hbar}
\eqa
{\allowdisplaybreaks\bqa
\langle \tilde{\eta}(p) \tilde{\eta}(q) \rangle_\mathrm{c} |_{\hbar^2} &=& \quad
\begin{picture}(70,40)(0,18)
\Line(0,20)(20,20)
\DashCArc(35,20)(15,-180,180){3}
\Line(50,20)(70,20)
\end{picture} \quad + \quad
\begin{picture}(70,40)(0,18)
\Line(0,20)(20,20)
\CArc(35,20)(15,-180,180)
\Line(50,20)(70,20)
\end{picture} \quad + \quad
\begin{picture}(70,40)(0,18)
\Line(0,5)(70,5)
\Line(35,5)(35,15)
\GCirc(35,25){10}{0.7}
\end{picture} \quad + \nonumber\\
& & \quad \begin{picture}(70,40)(0,18)
\Line(0,5)(70,5)
\DashCArc(35,20)(15,-90,270){3}
\end{picture} \quad + \quad
\begin{picture}(70,40)(0,18)
\Line(0,5)(70,5)
\CArc(35,20)(15,-90,270)
\end{picture} \quad + \quad
\begin{picture}(70,40)(0,18)
\Line(0,20)(70,20)
\Vertex(35,20){5}
\end{picture} \quad + \nonumber\\
& & \quad \begin{picture}(70,40)(0,18)
\Line(0,20)(70,20)
\Vertex(35,20){3}
\end{picture} \nonumber\\[30pt]
&=& -{\hbar^2\over v^2} \frac{p^2}{(p^2+m^2)^2} \; I(0,0) + \h {\hbar^2\over v^2} \frac{(p^2)^2}{(p^2+m^2)^2} \; I(0,0,p,0) \nonumber\\
& & +3{\hbar^2m^2\over v^2} \frac{1}{(p^2+m^2)^2} \; I(0,m) + {9\over2} {\hbar^2m^4\over v^2} \frac{1}{(p^2+m^2)^2} \; I(0,m,p,m) \nonumber\\
& & -\hbar \frac{p^2}{(p^2+m^2)^2} \; \delta_Z|_{\hbar} - 2\hbar \frac{1}{(p^2+m^2)^2} \; \delta_\mu|_{\hbar}
\eqa}
\bqa
\langle \tilde{w}(p) \tilde{w}(q) \rangle_\mathrm{c} |_{\hbar^2} &=& \quad
\begin{picture}(70,40)(0,18)
\DashLine(0,20)(20,20){3}
\DashCArc(35,20)(15,-180,0){3}
\CArc(35,20)(15,0,180)
\DashLine(50,20)(70,20){3}
\end{picture} \quad + \quad
\begin{picture}(70,40)(0,18)
\DashLine(0,5)(70,5){3}
\Line(35,5)(35,15)
\GCirc(35,25){10}{0.7}
\end{picture} \quad + \quad
\begin{picture}(70,40)(0,18)
\DashLine(0,5)(70,5){3}
\CArc(35,20)(15,-90,270)
\end{picture} \quad + \nonumber\\
& & \quad \begin{picture}(70,40)(0,18)
\DashLine(0,20)(70,20){3}
\Vertex(35,20){5}
\end{picture} \nonumber\\[30pt]
&=& -{\hbar^2\over v^2} \frac{p^2+m^2}{(p^2)^2} \; I(0,0) + {\hbar^2\over v^2} \frac{5p^2+m^2}{(p^2)^2} \; I(0,m) \nonumber\\
& & +{\hbar^2\over v^2} \frac{(p^2+m^2)^2}{(p^2)^2} \; I(0,0,p,m) \nonumber\\
& & -\hbar \frac{1}{p^2} \; \delta_Z|_{\hbar} - 2{\hbar\over m^2} \frac{1}{p^2} \; \delta_\mu|_{\hbar} + {1\over3}{\hbar v^2\over m^2} \frac{1}{p^2} \; \delta_{\lambda}|_{\hbar}
\eqa
{\allowdisplaybreaks\bqa
\langle \tilde{\eta} \rangle|_{\hbar^2} &=& \quad
\begin{picture}(70,40)(0,18)
\Line(0,20)(20,20)
\GCirc(55,20){15}{0.7}
\DashCArc(37,20)(17,50,180){3}
\DashCArc(37,20)(17,180,310){3}
\end{picture} \quad + \quad
\begin{picture}(70,40)(0,18)
\Line(0,20)(20,20)
\GCirc(55,20){15}{0.7}
\CArc(37,20)(17,50,180)
\CArc(37,20)(17,180,310)
\end{picture} \quad + \quad
\quad \begin{picture}(70,40)(0,18)
\Line(0,20)(40,20)
\GCirc(55,20){15}{0.7}
\CArc(20,30)(10,-90,270)
\end{picture} \quad + \nonumber\\[10pt]
& & \quad \begin{picture}(70,40)(0,18)
\Line(0,20)(40,20)
\GCirc(55,20){15}{0.7}
\DashCArc(20,30)(10,-90,270){3}
\end{picture} \quad + \quad
\begin{picture}(70,40)(0,18)
\Line(0,20)(25,20)
\Line(25,20)(60,33)
\Line(25,20)(60,7)
\GCirc(60,33){10}{0.7}
\GCirc(60,7){10}{0.7}
\end{picture} \quad + \quad
\begin{picture}(70,40)(0,18)
\Line(0,20)(70,20)
\DashCArc(50,20)(20,-180,180){3}
\end{picture} \quad + \nonumber\\[10pt]
& & \quad \begin{picture}(70,40)(0,18)
\Line(0,20)(70,20)
\CArc(50,20)(20,-180,180)
\end{picture} \quad + \quad
\begin{picture}(70,40)(0,18)
\Line(0,20)(50,20)
\Vertex(50,20){5}
\end{picture} \quad + \quad
\begin{picture}(70,40)(0,18)
\Line(0,20)(40,20)
\GCirc(55,20){15}{0.7}
\Vertex(20,20){5}
\end{picture} \quad + \nonumber\\[10pt]
& & \quad \begin{picture}(70,40)(0,18)
\Line(0,20)(40,20)
\CArc(55,20)(15,0,360)
\Vertex(40,20){5}
\end{picture} \quad + \quad
\begin{picture}(70,40)(0,18)
\Line(0,20)(40,20)
\DashCArc(55,20)(15,-180,180){3}
\Vertex(40,20){5}
\end{picture} \quad + \quad
\begin{picture}(70,40)(0,18)
\Line(0,20)(40,20)
\CArc(55,20)(15,-180,180)
\Vertex(40,20){3}
\end{picture} \quad \nonumber\\[30pt]
&=& -{1\over4} {\hbar^2\over v^3} \; I(0,0)^2 + \h {\hbar^2\over v^3} \; I(0,0) I(0,m) - {3\over2} {\hbar^2m^2\over v^3} \; I(0,0) I(0,m,0,m) \nonumber\\
& & +{\hbar^2m^2\over v^3} \; D_{00m} + {3\over2} {\hbar^2m^2\over v^3} \; D_{mmm} - {3\over4} {\hbar^2m^4\over v^3} \; B_{m00} - {27\over4} {\hbar^2m^4\over v^3} \; B_{mmm} \nonumber\\
& & -{9\over2} {\hbar^2m^2\over v^3} \; I(0,m) I(0,m,0,m) - {9\over8} {\hbar^2\over v^3} \; I(0,m)^2 \nonumber\\
& & +{3\over2} {\hbar\over v} \; I(0,m) \; \delta_Z|_{\hbar} + {3\over2} {\hbar\over m^2v} \; I(0,m) \; \delta_\mu|_{\hbar} - {1\over4} {\hbar v\over m^2} \; I(0,m) \; \delta_{\lambda}|_{\hbar} \nonumber\\
& & -{3\over2} {\hbar m^2\over v} \; I(0,m,0,m) \; \delta_Z|_{\hbar} + 3 {\hbar\over v} \; I(0,m,0,m) \; \delta_\mu|_{\hbar} \nonumber\\
& & -\h {v\over m^4} \; \left(\delta_\mu|_{\hbar}\right)^2 + {1\over24} {v^5\over m^4} \; \left(\delta_{\lambda}|_{\hbar}\right)^2 - {1\over6} {v^3\over m^4} \; \delta_\mu|_{\hbar} \delta_{\lambda}|_{\hbar} + {v\over m^2} \; \delta_\mu|_{\hbar^2} - {1\over6} {v^3\over m^2} \; \delta_{\lambda}|_{\hbar^2} \nonumber\\
\eqa}
\bqa
\langle \tilde{\eta}(p) \tilde{w}(q_1) \tilde{w}(q_2) \rangle_{\mathrm{c}} |_{\hbar^2} &=& \quad
\begin{picture}(70,40)(0,18)
\Line(0,20)(35,20)
\DashLine(35,20)(70,35){3}
\DashLine(35,20)(70,5){3}
\end{picture} \nonumber\\[20pt]
&=& 2{\hbar^2\over v} {1\over p^2+m^2} {1\over q_1^2} {1\over q_2^2} \; q_1\cdot q_2 \; (2\pi)^d \delta^d(p+q_1+q_2)
\eqa

\subsection{The $\f$-Green's-Functions}

The path integral for the $\f_1$-vacuum-expectation value is given by:\index{Green's functions!$N=2$ LSM}
\bqa
\langle\f_1(x)\rangle &=& {1\over Z(0)} \int_{-\infty}^{\infty} \mathcal{D}r \int_{-\infty}^{\infty} \mathcal{D}w \; \exp\left(-{1\over\hbar} I \int d^dx \left(-\hbar\ln r(x)\right)\right) \cdot \nonumber\\
& & \phantom{{1\over Z(0)} \int_{-\infty}^{\infty} \mathcal{D}r \int_{-\infty}^{\infty} \mathcal{D}w \;} r(x)\cos(w(x)/v) \exp\left(-{1\over\hbar}S(r,w)\right) \;,
\eqa
where $S$ is given by (\ref{renormS}). This action \emph{only} depends on $\nabla w$, which reflects the $O(2)$-invariance of the $N=2$ linear sigma model. This means that if we shift \emph{all} $w$-fields (i.e.\ the $w$-fields at \emph{all} space-time points) by the same amount the action does \emph{not} change. Also the path-integral measure does not change. So we can show:
\bq
\langle r(x)\cos(w(x)/v)\rangle = \langle r(x)\cos(w(x)/v+\pi)\rangle = -\langle r(x)\cos(w(x)/v)\rangle \;,
\eq
such that
\bq
\langle\f_1(x)\rangle = 0 \;.
\eq
For the same reason we have that
\bq
\langle\f_2(x)\rangle = 0 \;.
\eq
Notice that we have been able to show this through a non-perturbative argument. This is the great merit of a calculation via the path integral in terms of polar fields.

The $\f_1$- and $\f_2$-propagator can be calculated in a similar way:
\bqa
\langle\f_1(0)\f_1(x)\rangle &=& {1\over Z(0)} \int_{-\infty}^{\infty} \mathcal{D}r \int_{-\infty}^{\infty} \mathcal{D}w \; \exp\left(-{1\over\hbar} I \int d^dx \left(-\hbar\ln r(x)\right)\right) \cdot \nonumber\\
& & \phantom{{1\over Z(0)} \int_{-\infty}^{\infty} \mathcal{D}r \int_{-\infty}^{\infty} \mathcal{D}w \;} r(0)r(x) \cos(w(0)/v)\cos(w(x)/v) \cdot \nonumber\\
& & \phantom{{1\over Z(0)} \int_{-\infty}^{\infty} \mathcal{D}r \int_{-\infty}^{\infty} \mathcal{D}w \;} \exp\left(-{1\over\hbar}S(r,w)\right) \nonumber\\
&=& {1\over Z(0)} \int_{-\infty}^{\infty} \mathcal{D}r \int_{-\infty}^{\infty} \mathcal{D}w \; \exp\left(-{1\over\hbar} I \int d^dx \left(-\hbar\ln r(x)\right)\right) \; r(0)r(x) \cdot \nonumber\\
& & \phantom{{1\over Z(0)} \int_{-\infty}^{\infty} \mathcal{D}r \int_{-\infty}^{\infty} \mathcal{D}w \;} \Bigg( \h\cos\left(\frac{w(0)-w(x)}{v}\right) + \nonumber\\
& & \phantom{{1\over Z(0)} \int_{-\infty}^{\infty} \mathcal{D}r \int_{-\infty}^{\infty} \mathcal{D}w \; \Bigg(} \h\cos\left(\frac{w(0)+w(x)}{v}\right) \Bigg) \cdot \nonumber\\
& & \phantom{{1\over Z(0)} \int_{-\infty}^{\infty} \mathcal{D}r \int_{-\infty}^{\infty} \mathcal{D}w \;} \exp\left(-{1\over\hbar}S(r,w)\right) \nonumber\\
&=& \h{1\over Z(0)} \int_{-\infty}^{\infty} \mathcal{D}r \int_{-\infty}^{\infty} \mathcal{D}w \; \exp\left(-{1\over\hbar} I \int d^dx \left(-\hbar\ln r(x)\right)\right) \cdot \nonumber\\
& & \phantom{\h{1\over Z(0)} \int_{-\infty}^{\infty} \mathcal{D}r \int_{-\infty}^{\infty} \mathcal{D}w \;} r(0)r(x) \; \cos\left(\frac{w(0)-w(x)}{v}\right) \cdot \nonumber\\
& & \phantom{\h{1\over Z(0)} \int_{-\infty}^{\infty} \mathcal{D}r \int_{-\infty}^{\infty} \mathcal{D}w \;} \exp\left(-{1\over\hbar}S(r,w)\right) \nonumber\\
&=& \h \left\langle r(0)r(x) \; \cos\left(\frac{w(0)-w(x)}{v}\right) \right\rangle \nonumber\\
\langle\f_2(0)\f_2(x)\rangle &=& \langle\f_1(0)\f_1(x)\rangle \nonumber\\
\langle\f_1(0)\f_2(x)\rangle &=& 0
\eqa
Here we could discard the cosine of the sum $w(0)+w(x)$ because this cosine is \emph{not} invariant under a global shift of the $w$-field, i.e.\ this cosine is not $O(2)$-invariant.

The cosine of the difference of two $w$-fields can now be expanded, because a difference of two $w$'s can always be written as an integral over $\nabla w$, which \emph{is} small. (Remember $\nabla w=\mathcal{O}(\sqrt{\hbar})$.) 

Then the $\eta$- and $w$-Green's-functions we have calculated in the previous section can be used to find:
{\allowdisplaybreaks\bqa
\langle \varphi_1(0) \varphi_1(x) \rangle &=& {1\over2} v^2 - {1\over2} \langle w^2 \rangle + {1\over2} \langle w(0) w(x) \rangle + {1\over24v^2} \langle w^4 \rangle + {3\over24v^2} \langle w^2(0) w^2(x) \rangle + \nonumber\\
& & -{1\over12v^2} \langle w^3(0) w(x) \rangle - {1\over12v^2} \langle w(0) w^3(x) \rangle + v \langle \eta \rangle - {1\over2v} \langle \eta w^2 \rangle + \nonumber\\
& & -{1\over4v} \langle \eta(0) w^2(x) \rangle - {1\over4v} \langle \eta(x) w^2(0) \rangle + {1\over2v} \langle \eta(0) w(0) w(x) \rangle + \nonumber\\
& & {1\over2v} \langle \eta(x) w(0) w(x) \rangle + {1\over2} \langle \eta(0) \eta(x) \rangle - {1\over4v^2} \langle \eta(0) \eta(x) w^2(0) \rangle + \nonumber\\
& & -{1\over4v^2} \langle \eta(0) \eta(x) w^2(x) \rangle + {1\over2v^2} \langle \eta(0) \eta(x) w(0) w(x) \rangle + \mathcal{O}\left(\hbar^3\right) \nonumber\\[5pt]
&=& +\h v^2 \nonumber\\
& & -\h \hbar \; I(0,0) + \h \hbar \; A_0(x) - {3\over2} \hbar \; I(0,m) + \h \hbar \; A_m(x) \nonumber\\
& & +{v^2\over m^2} \; \delta_\mu|_{\hbar} - {1\over6} {v^4\over m^2} \; \delta_{\lambda}|_{\hbar} \nonumber\\
& & +\h {\hbar^2m^2\over v^2} \; I(0,0) I(0,0,0,0) - \h {\hbar^2m^2\over v^2} \; I(0,m) I(0,0,0,0) \nonumber\\
& & -{3\over2} {\hbar^2m^2\over v^2} \; I(0,0) I(0,m,0,m) - {9\over2} {\hbar^2m^2\over v^2} \; I(0,m) I(0,m,0,m) \nonumber\\
& & +\h {\hbar^2m^2\over v^2} \; D_{00m} + {3\over2} {\hbar^2m^2\over v^2} \; D_{mmm} - \h {\hbar^2m^4\over v^2} \; B_{00m} - {3\over4} {\hbar^2m^4\over v^2} \; B_{m00} \nonumber\\
& & -{27\over4} {\hbar^2m^4\over v^2} \; B_{mmm} \nonumber\\
& & -\h {\hbar^2m^2\over v^2} \; I(0,0) C_{00}(x) + \h {\hbar^2m^2\over v^2} \; I(0,m) C_{00}(x) \nonumber\\
& & +\h {\hbar^2m^2\over v^2} \; I(0,0) C_{mm}(x) + {3\over2} {\hbar^2m^2\over v^2} \; I(0,m) C_{mm}(x) \nonumber\\
& & +\h {\hbar^2m^4\over v^2} \; B_{00m}(x) + {1\over4} {\hbar^2m^4\over v^2} \; B_{m00}(x) + {9\over4} {\hbar^2m^4\over v^2} \; B_{mmm}(x) \nonumber\\
& & +\h \hbar \; I(0,0) \; \delta_Z|_{\hbar} - \h \hbar \; A_0(x) \; \delta_Z|_{\hbar} + {3\over2} \hbar \; I(0,m) \; \delta_Z|_{\hbar} - \h \hbar \; A_m(x) \; \delta_Z|_{\hbar} \nonumber\\
& & -{3\over2} \hbar m^2 \; I(0,m,0,m) \; \delta_Z|_{\hbar} + 3 \hbar \; I(0,m,0,m) \; \delta_\mu|_{\hbar} \nonumber\\
& & +\h \hbar m^2 \; C_{mm}(x) \; \delta_Z|_{\hbar} - \hbar \; C_{mm}(x) \; \delta_\mu|_{\hbar} \nonumber\\
& & +{1\over18}{v^6\over m^4} \; \left(\delta_{\lambda}|_{\hbar}\right)^2 - {1\over3}{v^4\over m^4} \; \delta_\mu|_{\hbar} \; \delta_{\lambda}|_{\hbar} + {v^2\over m^2} \; \delta_\mu|_{\hbar^2} - {1\over6}{v^4\over m^2} \; \delta_{\lambda}|_{\hbar^2} \nonumber\\[5pt]
& & +\mathcal{O}\left(\hbar^3\right) \label{symm}
\eqa}

If we compare this result to the result for the $\f_1$- and $\f_2$-propagator (\ref{phi1propchap7}), obtained in chapter \ref{chapN2sigmamodelpathI}, we find that both results agree. So, although it was far from obvious that the simple calculation done in chapter \ref{chapN2sigmamodelpathI} was correct, the result agrees with the proper calculation done in this chapter.

\subsection{Schwinger-Dyson Check}

We can check the result (\ref{symm}) by substituting it in the Schwinger-Dyson equations of the $N=2$ linear sigma model. This check is most conveniently done on the level of the unrenormalized action.

The Schwinger-Dyson equations for the propagator can be derived through the Schwinger-Symanzik equations\index{Schwinger-Symanzik equation}:
\bqa
\left[ \frac{\partial S}{\partial \varphi_1(x)}|_{\varphi_i=\hbar\frac{\partial}{\partial J_i}} - J_1(x) \right] Z\left(J_1, J_2\right) &=& 0 \nonumber\\
\left[ \frac{\partial S}{\partial \varphi_2(x)}|_{\varphi_i=\hbar\frac{\partial}{\partial J_i}} - J_2(x) \right] Z\left(J_1, J_2\right) &=& 0
\eqa
Substituting the unrenormalized action of our $N=2$ linear sigma model, operating on both sides of the first Schwinger-Symanzik equation with $\frac{\partial}{\partial J_1(0)}$ and finally putting all sources to zero we find the Schwinger-Dyson equation for the propagator:\index{Schwinger-Dyson equations}
\bq \label{SD}
\left( \nabla^2+{1\over2}m^2 \right) \langle \varphi_1(0) \varphi_1(x) \rangle - {m^2\over2v^2} \langle \varphi_1(0) \varphi_1^3(x) \rangle - {m^2\over2v^2} \langle \varphi_1(0) \varphi_1(x) \varphi_2^2(x) \rangle = -\hbar \delta^d(x)
\eq

Now we will check the result (\ref{symm}). First we have to know the 4-points Green's functions however (not including counter terms).
\bqa
& & \langle \varphi_1(0) \varphi_1^3(x) \rangle + \langle \varphi_1(0) \varphi_1(x) \varphi_2^2(x) \rangle = \nonumber\\[10pt]
& & \hspace{30pt} \langle (v+\eta(0)) (v+\eta(x))^3 \; \cos w(0)/v \; \cos w(x)/v \rangle = \nonumber\\
& & \hspace{30pt} {1\over2} \langle (v+\eta(0)) (v+\eta(x))^3 \; \cos\left((w(0)-w(x))/v\right) \rangle = \nonumber\\
& & \hspace{30pt} {1\over2}v^4 + 2v^3 \langle \eta \rangle - {1\over2}v^2 \langle w^2 \rangle + {1\over2}v^2 \langle w(0) w(x) \rangle + {3\over2}v^2 \langle \eta^2 \rangle + {3\over2}v^2 \langle \eta(0) \eta(x) \rangle + \nonumber\\
& & \hspace{30pt} {1\over2}v \langle \eta^3 \rangle + {3\over2}v \langle \eta(0) \eta^2(x) \rangle - {3\over4}v \langle \eta(x) w^2(0) \rangle + {3\over2}v \langle \eta(x) w(0) w(x) \rangle - v \langle \eta w^2 \rangle + \nonumber\\
& & \hspace{30pt} {1\over2}v \langle \eta(0) w(0) w(x) \rangle - {1\over4}v \langle \eta(0) w^2(x) \rangle + {1\over24} \langle w^4 \rangle -{1\over12} \langle w^3(0) w(x) \rangle + \nonumber\\
& & \hspace{30pt} {1\over8} \langle w^2(0) w^2(x) \rangle - {1\over12} \langle w(0) w^3(x) \rangle - {3\over4} \langle \eta^2(x) w^2(0) \rangle + {3\over2} \langle \eta^2(x) w(0) w(x) \rangle \nonumber\\
& & \hspace{30pt} -{3\over4} \langle \eta^2 w^2 \rangle - {3\over4} \langle \eta(0) \eta(x) w^2(0) \rangle + {3\over2} \langle \eta(0) \eta(x) w(0) w(x) \rangle - {3\over4} \langle \eta(0) \eta(x) w^2(x) \rangle + \nonumber\\
& & \hspace{30pt} {1\over2} \langle \eta(0) \eta^3(x) \rangle + \mathcal{O}(\hbar^3)
\eqa
Substituting the results from section \ref{O2etaw} gives:
\bqa
& & \langle \varphi_1(0) \varphi_1^3(x) \rangle + \langle \varphi_1(0) \varphi_1(x) \varphi_2^2(x) \rangle = \nonumber\\
& & \hspace{30pt} +{1\over2}v^4 \nonumber\\
& & \hspace{30pt} -{3\over2}\hbar v^2 \; I(0,m) + {3\over2}\hbar v^2 \; A_m(x) - \h\hbar v^2 \; I(0,0) + {1\over2}\hbar v^2 \; A_0(x) \nonumber\\
& & \hspace{30pt} -\hbar^2 \; I(0,m) A_0(x) - 3\hbar^2 \; I(0,m) A_m(x) + \hbar^2 \; I(0,0) A_0(x) - \hbar^2 \; I(0,0) A_m(x) \nonumber\\
& & \hspace{30pt} +{9\over2}\hbar^2m^2 \; I(0,m) C_{mm}(x) - {9\over2}\hbar^2m^2 \; I(0,m) I(0,m,0,m) + {1\over2}\hbar^2m^2 \; I(0,m) C_{00}(x) \nonumber\\
& & \hspace{30pt} -{3\over2}\hbar^2m^2 \; I(0,0) I(0,m,0,m) + \h\hbar^2m^2 \; I(0,0) I(0,0,0,0) \nonumber\\
& & \hspace{30pt} -\h\hbar^2m^2 \; I(0,m) I(0,0,0,0) - \h\hbar^2m^2 \; I(0,0) C_{00}(x) + {3\over2}\hbar^2m^2 \; I(0,0) C_{mm}(x) \nonumber\\
& & \hspace{30pt} +{1\over2}\hbar^2m^2 \; D_{00m} - {3\over4}\hbar^2m^4 \; B_{m00} - {27\over4}\hbar^2m^4 \; B_{mmm} + {3\over2}\hbar^2m^2 \; D_{mmm} \nonumber\\
& & \hspace{30pt} -{1\over2}\hbar^2m^4 \; B_{00m} - \hbar^2m^2 \; D_{00m}(x) + {1\over2}\hbar^2m^4 \; B_{00m}(x) - {1\over2}\hbar^2m^2 \; D_{m00}(x) \nonumber\\
& & \hspace{30pt} +{3\over4}\hbar^2m^4 \; B_{m00}(x) + {27\over4}\hbar^2m^4 \; B_{mmm}(x) - {9\over2}\hbar^2m^2 \; D_{mmm}(x)
\eqa

Now, substituting this and the propagator (\ref{symm}) with all the counter terms set to zero in the Schwinger-Dyson equation (\ref{SD}) we find that the equation is satisfied.

\subsection{The Canonical $\f_1$-Propagator}

From our path integral in terms of polar field variables we can also recover the $\f_1$-propagator one would find in the canonical approach. To this end we have to ignore the fact that
\bq
\left\langle\cos\left(\frac{w(x)+w(y)}{v}\right)\right\rangle = 0 \;.
\eq
Instead we have to expand both cosines around $w=0$, although this is actually incorrect in the path-integral approach. Expanding the cosines around $w=0$ here corresponds to doing perturbation theory around one minimum, where we also ignore the damping in the $\eta_2$-direction and replace the ring by an infinite line. In this case we obtain:
\bqa
\langle \varphi_1(0) \varphi_1(x) \rangle_\mathrm{c} &=& \langle \varphi_1(0) \varphi_1(x) \rangle - \langle \varphi_1 \rangle^2 \nonumber\\
&=& \langle \eta(0) \eta(x) \rangle_{\mathrm{c}} - {1\over v} \langle \eta(0) w^2(x) \rangle_{\mathrm{c}} + {1\over4v^2} \langle w(0) w(x) \rangle^2 \nonumber\\
& & -{1\over v^2} \langle \eta(0) \eta(x) \rangle \langle w^2 \rangle + \mathcal{O}\left(\hbar^3\right) \nonumber\\[5pt]
&=& +\hbar \; A_m(x) \nonumber\\
& & +{\hbar^2m^2\over v^2} \; I(0,0) C_{mm}(x) + 3{\hbar^2m^2\over v^2} \; I(0,m) C_{mm}(x) + \h {\hbar^2m^4\over v^2} \; B_{m00}(x) \nonumber\\
& & +{9\over2} {\hbar^2m^4\over v^2} \; B_{mmm}(x) \nonumber\\
& & -\hbar \; A_m(x) \; \delta_Z|_{\hbar} + \hbar m^2 \; C_{mm}(x) \; \delta_Z|_{\hbar} - 2\hbar \; C_{mm}(x) \; \delta_\mu|_{\hbar} \nonumber\\[5pt]
& & +\mathcal{O}\left(\hbar^3\right) \label{phi1propcanonical}
\eqa
This propagator agrees with the $\eta_1$-propagator we found in the canonical approach (\ref{eta1propcan}).

In this $\f_1$-propagator we can substitute the counter terms (\ref{counttermsN2}) that we found in chapter \ref{chapN2sigmamodelcan}. Then this result will satisfy the renormalization conditions from chapter \ref{chapN2sigmamodelcan}.

Also this propagator can be substituted in the Schwinger-Dyson equation, together with the result for $\langle \varphi_1(0) \varphi_1^3(x) \rangle + \langle \varphi_1(0) \varphi_1(x) \varphi_2^2(x) \rangle$ in this approach, where we expand around $w=0$. These results also satisfy the Schwinger-Dyson equation. This demonstrates that \emph{both} the canonical and path-integral approach give proper solutions to the Schwinger-Dyson equations of the $N=2$ linear sigma model.

Now we can also clearly see the \emph{difference} between results from the canonical and path-integral approach.\index{difference canonical and path-integral approach} (Compare (\ref{phi1propcanonical}) to (\ref{symm}).)

\section{1-Dimensional Calculation}

In this section we shall calculate $\langle r(x)\rangle$ and the $\f_1$-propagator for the 1-dimensional $N=2$ linear sigma model using the 1-dimensional path integral in terms of polar fields (\ref{1dimpathintLee}). In section \ref{1dimill} we already used this formula to compute some Green's functions for the arctangent toy model. We saw there that to compute Green's functions we first have to know $Z_w(0)$ and we calculated $Z_w(0)$ for the case of the arctangent toy model. From this result we can easily get $Z_w(0)$ for the $N=2$ linear sigma model. By looking at the definition of $Z_w(0)$ we see that we can get $Z_w(0)$ for the sigma model by putting $m=0$ in the $Z_w(0)$ for the arctangent toy model. So we find the simple result:
\bq
Z_w(0) \sim \prod_{i=0}^{N-1} \frac{1}{r_i^2} \;.
\eq

In this strictly 1-dimensional calculation we shall \emph{not} renormalize.

\subsection{$\langle r(x)\rangle$}

For $\langle r(x)\rangle$ we thus find:
\bqa
\langle r(x) \rangle &=& \frac{1}{Z(0)} \int_{-\infty}^{\infty} dr_0 \ldots dr_{N-1} \nonumber\\
& & \quad \exp \left( -\frac{1}{\hbar} \Delta \sum_{i=0}^{N-1} \left( \frac{1}{2} \frac{(r_{i+1}-r_i)^2}{\Delta^2} - \frac{\hbar^2}{8r_i^2} + \frac{\mu}{4v^2} (r_i^2-v^2)^2 \right) \right) \; r_j \;. \label{O2rxav}
\eqa
This is a normal one-dimensional path integral that we can calculate in the continuum. The minimum of the action is at $r_m$, where $r_m$ is defined by:
\bq
\frac{\hbar^2}{4r_m^3} + \frac{\mu}{v^2} (r_m^2-v^2) r_m = 0.
\eq
We can calculate $r_m$ up to a certain order in $\hbar$. We want our final answer of the expectation value of $r(x)$ up to order $\hbar^2$. Then we should also know $r_m$ up to this order. This is readily calculated:
\bq
r_m = v - \frac{1}{8\mu v^3} \hbar^2 + \mathcal{O}(\hbar^3).
\eq
Now we use the saddle-point method and expand the action in (\ref{O2rxav}) around $r=r_m$:
\bq
r(x) = r_m + \eta(x).
\eq
In the action we can discard terms of order $\hbar^3$ or higher.  In the continuum limit the action becomes:
\bq
S(\eta) = \int dx \left( \frac{1}{2} (\eta')^2 + \mu \eta^2 + \frac{\mu}{v} \eta^3 + \frac{\mu}{4v^2} \eta^4 - \frac{\hbar^2}{8v^2} + \mathcal{O}(\hbar^3) \right) \;.
\eq
Notice that the term of order $\hbar^{2\frac{1}{2}}$ ($\hbar^2\eta$) exactly drops out of the action.

Finally we get for $\langle r(x) \rangle$:
\bq \label{l3}
\langle r(x) \rangle = v - \frac{\hbar^2}{8\mu v^3} + \langle\eta(x)\rangle + \mathcal{O}(\hbar^3)
\eq
The last term gives the following contributions (We define again $\mu\equiv\h m^2$.):
{\allowdisplaybreaks\bqa
\begin{picture}(80, 40)(0, 18)
\Line(20, 20)(40, 20)
\Oval(50, 20)(10, 10)(0)
\end{picture}
&=& -\frac{3}{4} \frac{\hbar}{vm} \nonumber\\
\begin{picture}(80, 40)(0, 18)
\Line(10, 20)(30, 20)
\Line(30, 20)(52, 28)
\Line(30, 20)(52, 12)
\Oval(60, 30)(8, 8)(0)
\Oval(60, 10)(8, 8)(0)
\end{picture}
&=& -\frac{27}{32} \frac{\hbar^2}{v^3m^2} \nonumber\\
\begin{picture}(80, 40)(0, 18)
\Line(10, 20)(20, 20)
\Oval(30, 20)(10, 10)(0)
\Line(40, 20)(50, 20)
\Oval(60, 20)(10, 10)(0)
\end{picture}
&=& -\frac{27}{32} \frac{\hbar^2}{v^3m^2} \nonumber\\
\begin{picture}(80, 40)(0, 18)
\Line(10, 20)(40, 20)
\Oval(55, 20)(15, 15)(0)
\Line(55, 5)(55, 35)
\end{picture}
&=& -\frac{3}{8} \frac{\hbar^2}{v^3m^2} \nonumber\\
\begin{picture}(80, 40)(0, 18)
\Line(10, 20)(30, 20)
\Oval(40, 20)(10, 10)(0)
\Oval(60, 20)(10, 10)(0)
\end{picture}
&=& \frac{9}{32} \frac{\hbar^2}{v^3m^2} \nonumber\\
\begin{picture}(80, 40)(0, 18)
\Line(10, 20)(50, 20)
\Oval(30, 30)(10, 10)(0)
\Oval(60, 20)(10, 10)(0)
\end{picture}
&=& \frac{9}{16} \frac{\hbar^2}{v^3m^2} \nonumber\\
\begin{picture}(80, 40)(0, 18)
\Line(10, 20)(70, 20)
\Oval(55, 20)(15, 15)(0)
\end{picture}
&=& \frac{1}{8} \frac{\hbar^2}{v^3m^2}
\eqa}\\
Summing everything we get for $\langle r(x) \rangle$:
\bq \label{onedimresultr}
\langle r(x) \rangle = v - \frac{3}{4} \frac{\hbar}{vm} - \frac{43}{32} \frac{\hbar^2}{v^3m^2} + \mathcal{O}(\hbar^3) \;.
\eq

Does this agree with the $d$-dimensional continuum calculation? In this $d$-dimensional calculation $\langle r(x) \rangle$ is given by:
\bq
\langle r(x) \rangle = v + \langle\eta(x)\rangle \;.
\eq
We have computed the $d$-dimensional result for $\langle\eta(x)\rangle$ in section \ref{O2etaw}. To compare we first have to set all counter terms in the $d$-dimensional result from section \ref{O2etaw} to zero, because we did not renormalize in the 1-dimensional case. Also we have to realize that the $d$-dimensional result contains infrared divergent loop integrals. These \ind{infrared divergences} have to be regularized. The most straightforward way to regularize them is to introduce a mass $\varepsilon$ for the $w$-field. However then we would have to do the computation of $\langle\eta\rangle$ in section \ref{O2etaw} all over again. In $\langle\eta\rangle$ from section \ref{O2etaw} all terms of order $\varepsilon$ are absent, we need those terms now. For example a term
\bq
\varepsilon \; I(0,\varepsilon)
\eq
\emph{will} give a finite contribution in $d=1$, it gives $1/2$ plus terms of order $\varepsilon$.

A more convenient way to regularize the infrared divergences is to use dimensional regularization.\index{dimensional regularization scheme} In this scheme we have, for $d=1$:
\bqa
I(0,0) &=& 0 \nonumber\\
D_{00m} &=& {1\over2m^4} \nonumber\\
B_{m00} &=& {1\over m^6}
\eqa
Substituting these results and the results for the other standard integrals for $d=1$ (see appendix \ref{appstandint}) in $\langle\eta\rangle$ from section \ref{O2etaw} we find indeed the same result as (\ref{onedimresultr}).

The results of the standard integrals above can be found conveniently by the \ind{Mellin-Barnes technique}. This technique is studied extensively in the literature, see the book by Smirnov \cite{Smirnov} and articles by Bollini et al.\ \cite{Bollini}, Smirnov \cite{SmirnovArt}, Tausk \cite{Tausk}, Czakon \cite{Czakon} and Anastasiou et al.\ \cite{Anastasiou}.

\subsection{The $\f_1$-Propagator}

Now we wish to calculate the $\varphi_1$-propagator. For this we need to know $\langle \cos w_{j_1} \cos w_{j_2} \rangle_w$. We shall use the following formula:
\bq
\cos w_{j_1} \cos w_{j_2} = \frac{1}{2} \big( \cos(w_{j_1}-w_{j_2}) + \cos(w_{j_1}+w_{j_2}) \big).
\eq
Now when we take the average over $w$ the last term will give zero since it is not invariant under a global shift of the $w$'s. We get:
\bqa
\langle \cos w_{j_1} \cos w_{j_2} \rangle_w &=& \frac{1}{2} \langle \cos(w_{j_1}-w_{j_2}) \rangle_w \nonumber\\
&=& \frac{1}{2} \sum_{n=0}^{\infty} \frac{1}{(2n)!} (-1)^n \langle (w_{j_1}-w_{j_2})^{2n} \rangle_w.
\eqa
We should calculate the quantities $\langle (w_{j_1}-w_{j_2})^{2n} \rangle_w$. Now, as can easily be seen from the path integral the variables $w_{i+1}-w_i$ have a Gaussian distribution. So the variables $w_{j_1}-w_{j_2}$, which are sums of these variables, also have a Gaussian distribution. So we can use the Wick expansion for the averages $\langle (w_{j_1}-w_{j_2})^{2n} \rangle_w$:
\bqa
\langle (w_{j_1}-w_{j_2})^{2n} \rangle_w &=& (2n-1)!! \langle (w_{j_1}-w_{j_2})^2 \rangle_w^n \nonumber\\
&=& \frac{(2n-1)!}{2^{n-1}(n-1)!} \langle (w_{j_1}-w_{j_2})^2 \rangle_w^n.
\eqa
Now we know $\langle (w_{j_1}-w_{j_2})^2 \rangle_w$:
\bqa
\langle (w_{j_1}-w_{j_2})^2 \rangle_w &=& \Delta^2 \sum_{k=j_1}^{j_2-1} \sum_{l=j_1}^{j_2-1} \langle w'_lw'_k \rangle_w \nonumber\\
&=& \sum_{k=j_1}^{j_2-1} \sum_{l=j_1}^{j_2-1} \hbar \Delta \frac{1}{r_{l+1}^2} \delta_{lk} \nonumber\\
&=& \sum_{k=j_1+1}^{j_2} \hbar \Delta \frac{1}{r_k^2}.
\eqa
Here the prime denotes the discrete derivative again, as defined in (\ref{discreteprimes}). Now we finally get:
\bqa
\langle \cos w_{j_1} \cos w_{j_2} \rangle_w &=& \frac{1}{2} \sum_{n=0}^{\infty} \frac{1}{(2n)!} (-1)^n \frac{(2n-1)!}{2^{n-1}(n-1)!} \left( \sum_{k=j_1+1}^{j_2} \frac{\hbar \Delta}{r_k^2} \right)^n \nonumber\\
&=& \frac{1}{2} \sum_{n=0}^{\infty} \frac{1}{n!} \left( -\frac{1}{2}\sum_{k=j_1+1}^{j_2} \frac{\hbar \Delta}{r_k^2} \right)^n \nonumber\\
&=& \frac{1}{2} \exp \left( -\sum_{k=j_1+1}^{j_2} \frac{\hbar \Delta}{2r_k^2} \right) \;.
\eqa

Now the complete propagator becomes:
\bqa
\langle \varphi_1(x_1) \varphi_1(x_2) \rangle &=& \h\frac{1}{Z(0)} \int_{-\infty}^{\infty} dr_0 \ldots dr_{N-1} \; r_{j_1}r_{j_2} \nonumber\\
& & \qquad \exp \left( -\frac{1}{\hbar} \Delta \sum_{i=0}^{N-1} \left( \frac{1}{2} \frac{(r_{i+1}-r_i)^2}{\Delta^2} - \frac{\hbar^2}{8r_i^2} + \frac{\mu}{4v^2} (r_i^2-v^2)^2 \right) \right) \nonumber\\
& & \qquad \exp \left( -\sum_{i=j_1+1}^{j_2} \frac{\hbar\Delta}{2r_i^2} \right) \;. \label{l6}
\eqa
We expand again around $r=r_m$. Also the last exponent in (\ref{l6}) should be expanded:
\bqa
& & \exp \left( -\sum_{i=j_1+1}^{j_2} \frac{\hbar\Delta}{2r_i^2} \right) = \exp \left( -\sum_{i=j_1+1}^{j_2} \frac{\hbar\Delta}{2} \frac{1}{r_m^2+2r_m\eta_i+\eta_i^2} \right) \nonumber\\
&=& \exp \left( -\frac{\hbar\Delta}{2} \sum_{i=j_1+1}^{j_2} \left( \frac{1}{r_m^2} - \frac{1}{r_m^3} 2\eta_i + \frac{1}{r_m^4} 3\eta_i^2 + \ldots \right) \right) \nonumber\\
&=& e^{-\frac{\hbar}{2r_m^2} |x_1-x_2|} \left( 1 + \frac{\hbar\Delta}{2} \sum_{i=j_1+1}^{j_2} \Big( \frac{1}{r_m^3} 2\eta_i - \frac{1}{r_m^4} 3\eta_i^2 + \ldots \Big) + \ldots \right) \;. \label{l4}
\eqa
Remember that we only wish to calculate everything up to order $\hbar^2$, so the terms which are indicated by the dots in (\ref{l4}) are unimportant to us. For the $\f_1$-propagator we get:
\bqa
& & \langle \varphi_1(x_1) \varphi_1(x_2) \rangle = \nonumber\\[5pt]
& & \qquad \frac{1}{2} e^{-\frac{\hbar}{2r_m^2} |x_1-x_2|} \cdot \nonumber\\
& & \qquad \frac{ \int \mathcal{D}\eta \; (r_m+\eta(x_1))(r_m+\eta(x_2)) \Big( 1 + \frac{\hbar}{2} \int_{x_1}^{x_2} dx \big( \frac{2}{r_m^3} \eta(x) - \frac{3}{r_m^4} \eta^2(x) \big) \Big) \; e^{-\frac{1}{\hbar} S}}{\int \mathcal{D}\eta \; e^{-\frac{1}{\hbar} S}} \nonumber\\
& & \qquad +\mathcal{O}(\hbar^3) \label{l5}
\eqa
with
\bq
S = \int dx \left( \frac{1}{2} (\eta'(x))^2 + \mu \eta^2(x) + \frac{\mu}{v} \eta^3(x) + \frac{\mu}{4v^2} \eta^4(x) - \frac{\hbar^2}{8v^2} \right) \;.
\eq

The fraction in this expression can be computed in the ordinary way by using Feynman diagrams. In all $r_m$'s occurring in (\ref{l5}) there are $\hbar$'s. We should also expand all of these around $\hbar=0$.
\bq
e^{-\frac{\hbar}{2r_m^2} |x_1-x_2|} = e^{-\frac{\hbar}{2v^2} |x_1-x_2|} \big( 1 + \mathcal{O}(\hbar^3) \big)
\eq
We will not expand the exponential on the right-hand side around $\hbar=0$ but leave it as an overall factor. If we would expand this exponential as well it would no longer be directly apparent that the propagator goes to zero if the distance between $x_1$ and $x_2$ becomes large. All the rest we will expand up to order $\hbar^2$.
\bqa
& & \frac{ \int \mathcal{D}\eta \; (r_m+\eta(x_1))(r_m+\eta(x_2)) \Big( 1 + \frac{\hbar}{2} \int_{x_1}^{x_2} dx \big( \frac{2}{r_m^3} \eta(x) - \frac{3}{r_m^4} \eta^2(x) \big) \Big) \; e^{-\frac{1}{\hbar} S}}{\int \mathcal{D}\eta \; e^{-\frac{1}{\hbar} S}} \nonumber\\
&=& r_m^2 + r_m^2 \frac{\hbar}{2} \int_{x_1}^{x_2} dx \frac{ \int \mathcal{D}\eta \big( \frac{2}{r_m^3} \eta(x) - \frac{3}{r_m^4} \eta^2(x) \big) \; e^{-\frac{1}{\hbar} S}}{\int \mathcal{D}\eta \; e^{-\frac{1}{\hbar} S}} + \nonumber\\
& & r_m \frac{ \int \mathcal{D}\eta \; (\eta(x_1)+\eta(x_2)) \Big( 1 + \frac{\hbar}{2} \int_{x_1}^{x_2} dx \big( \frac{2}{r_m^3} \eta(x) - \frac{3}{r_m^4} \eta^2(x) \big) \Big) \; e^{-\frac{1}{\hbar} S}}{\int \mathcal{D}\eta \; e^{-\frac{1}{\hbar} S}} + \nonumber\\ 
& & \frac{ \int \mathcal{D}\eta \; \eta(x_1)\eta(x_2) \Big( 1 + \frac{\hbar}{2} \int_{x_1}^{x_2} dx \big( \frac{2}{r_m^3} \eta(x) - \frac{3}{r_m^4} \eta^2(x) \big) \Big) \; e^{-\frac{1}{\hbar} S}}{\int \mathcal{D}\eta \; e^{-\frac{1}{\hbar} S}} + \mathcal{O}(\hbar^3) \label{l7}\\ 
&=& v^2 - \frac{\hbar^2}{4\mu v^2} + \frac{\hbar}{v} \int_{x_1}^{x_2} dx 
\begin{picture}(80, 40)(0, 18)
\Line(20, 20)(40, 20)
\GOval(50, 20)(10, 10)(0){0.8}
\Text(20, 28)[]{$x$}
\end{picture}
- \frac{3\hbar}{2v^2} \int_{x_1}^{x_2} dx
\begin{picture}(80, 40)(0, 18)
\Line(20, 20)(60, 20)
\GOval(40, 20)(8, 8)(0){0.8}
\Text(20, 28)[]{$x$}
\Text(60, 28)[]{$x$}
\end{picture} + \nonumber\\
& & v \bigg(
\begin{picture}(80, 40)(0, 18)
\Line(20, 20)(40, 20)
\GOval(50, 20)(10, 10)(0){0.8}
\Text(20, 28)[]{$x_1$}
\end{picture} +
\begin{picture}(80, 40)(0, 18)
\Line(20, 20)(40, 20)
\GOval(50, 20)(10, 10)(0){0.8}
\Text(20, 28)[]{$x_2$}
\end{picture}
+ \nonumber\\
& & \frac{\hbar}{2} \int_{x_1}^{x_2} dx \frac{2}{v^3}
\begin{picture}(80, 40)(0, 18)
\Line(20, 20)(60, 20)
\Text(20, 28)[]{$x_1$}
\Text(60, 28)[]{$x$}
\end{picture}
+ \frac{\hbar}{2} \int_{x_1}^{x_2} dx \frac{2}{v^3}
\begin{picture}(80, 40)(0, 18)
\Line(20, 20)(60, 20)
\SetPFont{Helvetica}{6}
\Text(20, 28)[]{$x_2$}
\Text(60, 28)[]{$x$}
\end{picture}
\bigg) + \nonumber\\
& & \begin{picture}(80, 40)(0, 18)
\Line(20, 20)(60, 20)
\GOval(40, 20)(8, 8)(0){0.8}
\Text(20, 28)[]{$x_1$}
\Text(60, 28)[]{$x_2$}
\end{picture}
+ \mathcal{O}(\hbar^3) \nonumber\\\nonumber\\
&=& v^2 + \hbar \left( -\frac{3}{2m} + \frac{1}{2m} e^{-m |x_1-x_2|} \right) + \nonumber\\
& & \hbar^2 \bigg( -\frac{9}{8m^2v^2} - \frac{1}{2m^2v^2} e^{-m |x_1-x_2|} + \frac{1}{8m^2v^2} e^{-2m |x_1-x_2|} + \nonumber\\
& & \frac{3}{4m v^2} |x_1-x_2| e^{-m |x_1-x_2|} - \frac{3}{2m v^2} |x_1-x_2| \bigg) + \mathcal{O}(\hbar^3)
\eqa

Finally we get for the complete $\varphi_1$-propagator:
\bqa
\langle \varphi_1(x_1) \varphi_1(x_2) \rangle &=& \frac{1}{2} e^{-\frac{\hbar}{2v^2} |x_1-x_2|} \bigg( v^2 + \hbar \left( -\frac{3}{2m} + \frac{1}{2m} e^{-m |x_1-x_2|} \right) + \nonumber\\
& & \hbar^2 \bigg( -\frac{9}{8m^2v^2} - \frac{1}{2m^2v^2} e^{-m |x_1-x_2|} + \frac{1}{8m^2v^2} e^{-2m |x_1-x_2|} + \nonumber\\
& & \phantom{\hbar^2 \bigg(} \frac{3}{4m v^2} |x_1-x_2| e^{-m |x_1-x_2|} - \frac{3}{2m v^2} |x_1-x_2| \bigg) + \mathcal{O}(\hbar^3) \bigg) \;. \nonumber\\ \label{onedimfprop}
\eqa

Does this result agree with the $d$-dimensional continuum calculation? It does, however first we have to expand the exponential
\bq
e^{-\frac{\hbar}{2v^2} |x_1-x_2|} \;,
\eq
because the $d$-dimensional result (\ref{symm}) is a perturbative series in $\hbar$, whereas the exponential above is not. Secondly we have to set all counter terms in (\ref{symm}) to zero. Thirdly we have to realize that (\ref{symm}) contains \ind{infrared divergences}. These are most conveniently regularized again in the \ind{dimensional regularization scheme}. In this scheme we have:
\bqa
I(0,0) &=& 0 \nonumber\\
A_0(x) &=& -\h|x| \nonumber\\
I(0,0,0,0) &=& 0 \nonumber\\
D_{00m} &=& {1\over2m^4} \nonumber\\
B_{00m} &=& -{1\over m^6} \nonumber\\
B_{m00} &=& {1\over m^6} \nonumber\\
C_{00}(x) &=& {1\over12}|x|^3 \nonumber\\
B_{00m}(x) &=& {1\over24m^6}\left( -m^3|x|^3 - 6m|x|e^{-m|x|} - 24e^{-m|x|} - 18m|x| \right) \nonumber\\
B_{m00}(x) &=& {1\over8 m^6}\left( 2m^2|x|^2 + 8 \right)
\eqa
These integrals can again easily be calculated with the \ind{Mellin-Barnes technique}. Substituting these results in (\ref{symm}) gives indeed the result (\ref{onedimfprop}).

There is a difference between (\ref{onedimfprop}) and (\ref{symm}) however. In (\ref{onedimfprop}) we see that the propagator goes to zero for large $|x_1-x_2|$. Only when we expand the exponential that causes this behavior we recover the result (\ref{symm}) we get from the $d$-dimensional continuum calculation specified to $d=1$. How do we recover the exponential from the $d$-dimensional continuum calculation?

\subsection{Recovering (\ref{onedimfprop})}

To recover the 1-dimensional result (\ref{onedimfprop}) completely from the $d$-dimensional continuum result (\ref{symm}) we actually also need (parts of) higher order terms in (\ref{symm}). It will appear that the momentum-space version of (\ref{symm}) has terms like:
\bq \label{oneoverps}
{\hbar\over p^2} \;, {\hbar^3\over p^4} \;, {\hbar^5\over p^6} \;, \textrm{etc.} \;,
\eq
when these higher order terms are included. (Here we have calculated in the dimensional regularization scheme.) For $d=1$ these terms are exactly what we get when transforming the exponential
\bq \label{expon1d}
e^{-\frac{\hbar}{2v^2} |x_1-x_2|} \;,
\eq
to momentum space (in the dimensional regularization scheme). In this way we can in principle recover the result (\ref{onedimfprop}) from the $d$-dimensional result (\ref{symm}), although in practice we can never compute the terms of arbitrarily high  order.

Considering the one-dimensional result, and knowing that we have to re-sum part of it to obtain the true propagator, which drops to zero nicely for large distances, a good question would be: Should we also re-sum in $d$ dimensions? If this were the case, then our result (\ref{symm}) would still be incomplete, in the sense that we also need higher order terms to see the true physics. Fortunately this is not the case. The reason is as follows.

In one dimension the terms (\ref{oneoverps}) \emph{all} come from the $\psi_2$-propagator. (Here we mean the $\psi_2$-propagator from chapter \ref{chapN2sigmamodelpathI}.) These terms generate a mass for the $\psi_2$-particle. In configuration space this is mirrored by the exponential (\ref{expon1d}), from which we can read off a mass
\bq
\frac{\hbar}{2v^2} \;.
\eq
Now the \ind{Goldstone theorem} tells us that if we have spontaneous symmetry breaking, the \ind{Goldstone boson} is massless, and stays massless at \emph{all} orders in perturbation theory. In our case $\psi_2$ is the Goldstone boson in the canonical approach. For $d=1$ we do \emph{not} have SSB (see \cite{Coleman,Jackiw}), so the Goldstone theorem does \emph{not} forbid that a mass is generated for the $\psi_2$-particle. In fact, we have seen that a mass \emph{is} generated. For $d>2$ we \emph{do} have SSB in the canonical approach, and the Goldstone theorem forbids that a mass is generated. As a consequence the terms
\bq
{\hbar^3\over p^4} \;, {\hbar^5\over p^6} \;, \textrm{etc.}
\eq
do \emph{not} occur in the $\psi_2$-propagator. And thus there is nothing to re-sum, and the result (\ref{symm}) can be considered complete in the sense that higher order terms will \emph{not} cause the propagator to drop to zero for large distances. This means we are not missing any important physics by only having the $\f_1$- and $\f_2$-propagator up to order $\hbar^2$.

The case $d=2$ is a special case, see also \cite{Coleman,Jackiw}.

\section{The Effective Potential}

We can also calculate the effective potential\index{effective potential!$N=2$ LSM} via the path integral in terms of polar fields. To this end we introduce source terms in the renormalized action:
\bqa
S &=& \int d^dx \; \bigg( \h\left(\nabla\f_1\right)^2 + \h\left(\nabla\f_2\right)^2 - \h\mu\left(\f_1^2+\f_2^2\right) + {\lambda\over24}\left(\f_1^2+\f_2^2\right)^2 - J_1\f_1 - J_2\f_2 + \nonumber\\
& & \phantom{\int d^dx \; \bigg(} \h\delta_Z\left(\nabla\f_1\right)^2 + \h\delta_Z\left(\nabla\f_2\right)^2 - \h\delta_{\mu}\left(\f_1^2+\f_2^2\right) + {\delta_{\lambda}\over24}\left(\f_1^2+\f_2^2\right)^2 \bigg) \;.
\eqa

In chapter \ref{chapN2sigmamodelpathI} we already computed the effective potential for the $N=2$ linear sigma model. However there we had to discard the term (\ref{discpsi2term}) to avoid ending up with an expression that contained infrared divergences at order $\hbar$. Also we could not find the interpolation of the effective potential between small $J$ (order $\hbar$) and $J$ of order 1 ($\hbar^0$). In this section we shall see if we can do a better job by calculating in terms of polar fields.

According to the \ind{conjecture} the action in terms of polar fields is
\bqa
S &=& \int d^dx \; \bigg( \h\left(\nabla r\right)^2 + \h{r^2\over v^2}\left(\nabla w\right)^2 - \h\mu r^2 + {\lambda\over24}r^4 - J_1r\cos{w\over v} - J_2\sin{w\over v} + \nonumber\\
& & \phantom{\int d^dx \; \bigg(} \h\delta_Z\left(\nabla\f_1\right)^2 + \h\delta_Z\left(\nabla\f_2\right)^2 - \h\delta_{\mu}r^2 + {\delta_{\lambda}\over24}r^4 \bigg) \;,
\eqa
provided we calculate in a $d$-dimensional way in the continuum.

Introducing
\bqa
J_1 &\equiv& J\cos\beta \nonumber\\
J_2 &\equiv& J\sin\beta
\eqa
the minimum of the first line of the action, i.e.\ the classical action, is given by:
\bq
r = {2v\over\sqrt{3}} \sin\left(\alpha+{\pi\over3}\right) \;, \quad w = v\beta \;,
\eq
with
\bq
{2\mu v\over3\sqrt{3}} \sin 3\alpha = J \;.
\eq

Expanding the action around the minimum,
\bqa
r(x) &=& \bar{v} + \eta(x) \;, \qquad \bar{v} \equiv {2v\over\sqrt{3}} \sin\left(\alpha+{\pi\over3}\right) \nonumber\\
w(x) &=& v\beta + \bar{w} \;,
\eqa
we find:
\bqa
S &=& \int d^dx \; \bigg( \h\left(\nabla\eta\right)^2 + \left(-\h\mu+{3\over2}{\mu\bar{v}^2\over v^2}\right)\eta^2 + \h{\bar{v}^2\over v^2}\left(\nabla\bar{w}\right)^2 + \nonumber\\
& & \phantom{\int d^dx \; \bigg(} {\bar{v}\over v^2}\eta\left(\nabla\bar{w}\right)^2 + \h{1\over v^2}\eta^2\left(\nabla\bar{w}\right)^2 + {\mu\bar{v}\over v^2}\eta^3 + {\mu\over4v^2}\eta^4 + \nonumber\\
& & \phantom{\int d^dx \; \bigg(} -J\bar{v}\cos{\bar{w}\over v} + J\eta\left(1-\cos{\bar{w}\over v}\right) - \h\mu\bar{v}^2 + {1\over4}{\mu\bar{v}^4\over v^2} + \nonumber\\
& & \phantom{\int d^dx \; \bigg(} \h\delta_Z\left(\nabla\eta\right)^2 + \h\delta_Z{\bar{v}^2\over v^2}\left(\nabla\bar{w}\right)^2 + \delta_Z{\bar{v}\over v^2}\eta\left(\nabla\bar{w}\right)^2 + \h\delta_Z{1\over v^2}\eta^2\left(\nabla\bar{w}\right)^2 + \nonumber\\
& & \phantom{\int d^dx \; \bigg(} -\h\delta_{\mu}\bar{v}^2 + {\bar{v}^4\over24}\delta_\lambda + \nonumber\\
& & \phantom{\int d^dx \; \bigg(} \left(-\delta_\mu\bar{v}+{1\over6}\bar{v}^3\delta_\lambda\right)\eta + \left(-\h\delta_\mu+{1\over4}\bar{v}^2\delta_\lambda\right)\eta^2 + {1\over6}\bar{v}\delta_\lambda\eta^3 + {1\over24}\delta_\lambda\eta^4 \bigg) \;. \nonumber\\
\eqa
According to the conjecture the generating functional is now given by:
\bq
Z(J_1,J_2) = \int \mathcal{D}\eta \int \mathcal{D}\bar{w} \; \exp\left(-{1\over\hbar} I \int d^dx \left(-\hbar\ln\left(\bar{v}+\eta\right)\right)\right) \; e^{-{1\over\hbar}S}
\eq

As can easily be seen from the action we have \emph{one} minimum for $J\neq0$, whereas we have a ring of minima for $J=0$. This means that for $J$ of order 1 it is correct to expand the cosine of $\bar{w}/v$. In that case we recover the effective potential from the canonical approach. This is expected because for large $J$ it is correct to take into account only \emph{one} minimum. For small $J$, i.e.\ $J$ of order $\hbar$ things are a bit more difficult. For such small $J$ there is strictly speaking still one minimum, but the ring is so flat that it is incorrect to ignore the other points. We know this because when $J$ becomes really zero the other points in the ring start to play an important role. What we can do is the following. We write the generating functional as:
\bqa
Z(J_1,J_2) &=& \int \mathcal{D}\eta \int \mathcal{D}\bar{w} \; \exp\left(-{1\over\hbar} I \int d^dx \left(-\hbar\ln\left(\bar{v}+\eta\right)\right)\right) \nonumber\\
& & \phantom{\int \mathcal{D}\eta \int \mathcal{D}\bar{w} \;} \exp\left({1\over\hbar}\int d^dx \; J\left(\bar{v}+\eta\right)\cos{\bar{w}\over v}\right) \; e^{-{1\over\hbar}S'}
\eqa
with $S'$ given by:
\bqa
S' &=& \int d^dx \; \bigg( \h\left(\nabla\eta\right)^2 + \left(-\h\mu+{3\over2}{\mu\bar{v}^2\over v^2}\right)\eta^2 + \h{\bar{v}^2\over v^2}\left(\nabla\bar{w}\right)^2 + \nonumber\\
& & \phantom{\int d^dx \; \bigg(} {\bar{v}\over v^2}\eta\left(\nabla\bar{w}\right)^2 + \h{1\over v^2}\eta^2\left(\nabla\bar{w}\right)^2 + {\mu\bar{v}\over v^2}\eta^3 + {\mu\over4v^2}\eta^4 + \nonumber\\
& & \phantom{\int d^dx \; \bigg(} J\eta - \h\mu\bar{v}^2 + {1\over4}{\mu\bar{v}^4\over v^2} + \nonumber\\
& & \phantom{\int d^dx \; \bigg(} \h\delta_Z\left(\nabla\eta\right)^2 + \h\delta_Z{\bar{v}^2\over v^2}\left(\nabla\bar{w}\right)^2 + \delta_Z{\bar{v}\over v^2}\eta\left(\nabla\bar{w}\right)^2 + \h\delta_Z{1\over v^2}\eta^2\left(\nabla\bar{w}\right)^2 + \nonumber\\
& & \phantom{\int d^dx \; \bigg(} -\h\delta_{\mu}\bar{v}^2 + {\bar{v}^4\over24}\delta_\lambda + \nonumber\\
& & \phantom{\int d^dx \; \bigg(} \left(-\delta_\mu\bar{v}+{1\over6}\bar{v}^3\delta_\lambda\right)\eta + \left(-\h\delta_\mu+{1\over4}\bar{v}^2\delta_\lambda\right)\eta^2 + {1\over6}\bar{v}\delta_\lambda\eta^3 + {1\over24}\delta_\lambda\eta^4 \bigg) \;. \nonumber\\
\eqa
Note that in the new action $S'$ only $\nabla\bar{w}$ occurs. Now focus on the part that we pulled out of the action:
\bqa
& & \exp\left({1\over\hbar}\int d^dx \; J\left(\bar{v}+\eta(x)\right)\cos{\bar{w}(x)\over v}\right) \nonumber\\
& & \qquad = \exp\left({1\over\hbar}\int d^dx \; Jr(x)\cos{\bar{w}(x)\over v}\right) \nonumber\\
& & \qquad = \sum_{n=0}^\infty {1\over n!} \left( {1\over\hbar}\int d^dx \; Jr(x)\cos{\bar{w}(x)\over v} \right)^n \nonumber\\
& & \qquad = \sum_{n=0}^\infty {1\over n!} {J^n\over\hbar^n} \int d^dx_1\cdots d^dx_n \; r(x_1)\cdots r(x_n) \cos{\bar{w}(x_1)\over v}\cdots\cos{\bar{w}(x_n)\over v}
\eqa
We are going to combine the cosines into a sum of single cosines. Because the action $S'$ only depends on $\nabla\bar{w}$ only the $O(2)$-invariant cosines are going to survive in the path integral. This means, when combining the cosines, all cosines with an unequal number of $+\bar{w}$'s and $-\bar{w}$'s are going to vanish under the path integral.
\bqa
& & \exp\left({1\over\hbar}\int d^dx \; J\left(\bar{v}+\eta(x)\right)\cos{\bar{w}(x)\over v}\right) \nonumber\\
& & \qquad = \sum_{n=0}^\infty {1\over(2n)!} {J^{2n}\over\hbar^{2n}} \int d^dx_1\cdots d^dx_{2n} \; r(x_1)\cdots r(x_{2n}) \cos{\bar{w}(x_1)\over v}\cdots\cos{\bar{w}(x_{2n})\over v} \nonumber\\
& & \qquad = \sum_{n=0}^\infty {1\over(2n)!} {J^{2n}\over\hbar^{2n}} \int d^dx_1\cdots d^dx_{2n} \; r(x_1)\cdots r(x_{2n}) {2n-1\choose n} {1\over2^{2n-1}} \cdot \nonumber\\
& & \hspace{50pt} \cos{1\over v}\left(\bar{w}(x_1)+\bar{w}(x_2)+\ldots+\bar{w}(x_n)-\bar{w}(x_{n+1})-\bar{w}(x_{n+2})-\ldots-\bar{w}(x_{2n})\right) \nonumber\\
& & \qquad = \sum_{n=0}^\infty {1\over(n!)^2} \left({J\over2\hbar}\right)^{2n} \int d^dx_1\cdots d^dx_{2n} \; r(x_1)\cdots r(x_{2n}) \cdot \nonumber\\
& & \hspace{50pt} \cos{1\over v}\left(\bar{w}(x_1)+\bar{w}(x_2)+\ldots+\bar{w}(x_n)-\bar{w}(x_{n+1})-\bar{w}(x_{n+2})-\ldots-\bar{w}(x_{2n})\right) \nonumber\\
\eqa

Now we can expand the cosine, because it contains only differences of two $\bar{w}$'s. Such differences can be written as an integral over $\nabla\bar{w}$. From the path integral it can be seen that $\nabla\bar{w}$ is small (of order $\sqrt{\hbar}$), such that it is indeed correct to expand the cosine. Keeping the first and second term from the expansion of the cosine we find:
\bqa
& & \exp\left({1\over\hbar}\int d^dx \; J\left(\bar{v}+\eta(x)\right)\cos{\bar{w}(x)\over v}\right) \nonumber\\
& & \qquad = \sum_{n=0}^\infty {1\over(n!)^2} \left({J\over2\hbar}\right)^{2n} \left(\int d^dx \; r(x)\right)^{2n} + \nonumber\\
& & \qquad \phantom{=} -\h\sum_{n=0}^\infty {1\over(n!)^2} \left({J\over2\hbar}\right)^{2n} \int d^dx_1\cdots d^dx_{2n} \; r(x_1)\cdots r(x_{2n}) \cdot \nonumber\\
& & \hspace{70pt} {1\over v^2}\left(\bar{w}(x_1)+\ldots+\bar{w}(x_n)-\bar{w}(x_{n+1})-\ldots-\bar{w}(x_{2n})\right)^2 \nonumber\\
& & \qquad = \sum_{n=0}^\infty {1\over(n!)^2} \left({J\over2\hbar}\right)^{2n} \left(\int d^dx \; r(x)\right)^{2n} + \nonumber\\
& & \qquad \phantom{=} -\h\sum_{n=0}^\infty {1\over(n!)^2} \left({J\over2\hbar}\right)^{2n} \int d^dx_1\cdots d^dx_{2n} \; r(x_1)\cdots r(x_{2n}) \cdot \nonumber\\
& & \hspace{70pt} {1\over v^2} n\left( \bar{w}(x_1) - \bar{w}(x_2) \right)^2 \nonumber\\
& & \qquad = I_0\left({J\over\hbar}\int d^dx \; r(x)\right) + \nonumber\\
& & \qquad \phantom{=} -{J\over4\hbar v^2} \frac{\int d^dx \int d^dy \; r(x)r(y)\left(\bar{w}(x)-\bar{w}(y)\right)^2}{\int d^dx \; r(x)} \; I_1\left({J\over\hbar}\int d^dx \; r(x)\right)
\eqa
Here $I_n(x)$ is a modified Bessel function of the first kind.

We can see clearly here that, if $J$ is of order 1, the first and second term are of the same magnitude (both order 1). This means that when $J$ is of order 1 we need \emph{all} terms of the expansion of the cosine. This mirrors the fact that for $J$ of order 1 there is \emph{one} clear minimum and the cosine plays a crucial role in determining where this minimum is located. Keeping all the terms of the expansion of the cosine is very hard in an actual computation, so the formula above is not very convenient to find the generating functional for $J$ of order 1.

It is convenient for $J$ of order $\hbar$ however, in this case we see that the first term above is of order 1, while the second term is of order $\hbar$. This means discarding the higher order term seems to be a good approximation. Discarding these terms means the generating functional is correct up to order $\hbar$. Also discarding other terms of higher order than $\hbar$ we find for the generating functional:
\bqa
Z(J_1,J_2) &=& \int \mathcal{D}\eta \int \mathcal{D}\bar{w} \; \exp\left(-{1\over\hbar} I \int d^dx \left(-\hbar\ln\left(\bar{v}+\eta\right)\right)\right) \; e^{-{1\over\hbar}S'} \nonumber\\
& & \qquad \Bigg[ I_0\left({J\over\hbar}\int d^dx \; (\bar{v}+\eta(x))\right) + \nonumber\\
& & \phantom{\qquad \Bigg[} -{J\over4\hbar v^2} \frac{\int d^dx \int d^dy \; (\bar{v}+\eta(x))(\bar{v}+\eta(y))\left(\bar{w}(x)-\bar{w}(y)\right)^2}{\int d^dx \; (\bar{v}+\eta(x))} \cdot \nonumber\\
& & \phantom{\qquad \Bigg[} I_1\left({J\over\hbar}\int d^dx \; (\bar{v}+\eta(x))\right) \Bigg] \nonumber\\
&=& \int \mathcal{D}\eta \int \mathcal{D}\bar{w} \; \exp\left(-{1\over\hbar} I \int d^dx \left(-\hbar\ln\left(\bar{v}+\eta\right)\right)\right) \; e^{-{1\over\hbar}S'} \nonumber\\
& & \qquad \Bigg[ I_0\left({Jv\Omega\over\hbar}\right) + I_1\left({Jv\Omega\over\hbar}\right) \; {J\over\hbar}\int d^dx \left({J\over2\mu}+\eta\right) + \nonumber\\
& & \phantom{\qquad \Bigg[} \left( I_2\left({Jv\Omega\over\hbar}\right) + {\hbar\over Jv\Omega} I_1\left({Jv\Omega\over\hbar}\right) \right) \h{J^2\over\hbar^2} \int d^dx \int d^dy \; \eta(x)\eta(y) \nonumber\\
& & \phantom{\qquad \Bigg[} -I_1\left({Jv\Omega\over\hbar}\right) {J\over4\hbar v\Omega} \int d^dx \int d^dy \; \left(\bar{w}(x)-\bar{w}(y)\right)^2 \Bigg]
\eqa
Here we have also expanded $\bar{v}$,
\bq
\bar{v} = v + {J\over2\mu} + \mathcal{O}(\hbar^2) \;,
\eq
and $\Omega$ denotes the space-time volume. Also in the action $S'$ terms of higher order than $\hbar^2$ should be discarded. (There is a $1/\hbar$ in front of the action.) In the Jacobian we should discard all terms of order higher than $\hbar$.

From the formula above one could in principle calculate the generating functional, and from it the $\f_1$- and $\f_2$-expectation-value, all up to order $\hbar$. The expectation values are given by:
\bqa
\langle\f_1\rangle(J_1,J_2) &=& {\hbar\over\Omega}\frac{\partial}{\partial J_1} \ln Z(J_1,J_2) \nonumber\\
&=& {\hbar\over\Omega} \cos\beta \frac{\partial}{\partial J} \ln Z(J_1,J_2) \nonumber\\
\langle\f_2\rangle(J_1,J_2) &=& {\hbar\over\Omega}\frac{\partial}{\partial J_2} \ln Z(J_1,J_2) \nonumber\\
&=& {\hbar\over\Omega} \sin\beta \frac{\partial}{\partial J} \ln Z(J_1,J_2)
\eqa
Notice that the generating functional does not depend on the direction of the source $\beta$. From these expectation values one can then find the effective potential.

However, when calculating the generating functional one encounters infrared divergences again. The reason is the same as in chapter \ref{chapN2sigmamodelpathI}. The formula above is only valid for small $J$, whereas we saw already in the canonical approach that to avoid the infrared singularities we need \emph{all} $n$-points Green's functions. So we also need to know $Z(J_1,J_2)$ for \emph{all} $J$, which is very hard, as we saw above. In chapter \ref{chapN2sigmamodelpathI} we could find a result (up to order $\hbar$) without infrared divergences by discarding the term (\ref{discpsi2term}), which caused the infrared divergences at order $\hbar$. In the formula above it is not clear what we can do to avoid the infrared divergences.

It is however easy to find the $\f_1$- and $\f_2$-expectation-values at lowest order from the formula for $Z$ above. We find:
\bqa
\langle\f_1\rangle(J_1,J_2) &=& v\cos\beta \; \frac{I_1\left({Jv\Omega\over\hbar}\right)}{I_0\left({Jv\Omega\over\hbar}\right)} \nonumber\\
\langle\f_2\rangle(J_1,J_2) &=& v\sin\beta \; \frac{I_1\left({Jv\Omega\over\hbar}\right)}{I_0\left({Jv\Omega\over\hbar}\right)} \;,
\eqa
and
\bq \label{sqrtphi}
\sqrt{\langle \f_1 \rangle^2+\langle \f_2 \rangle^2} = v\frac{I_1\left({\Omega vJ\over\hbar}\right)}{I_0\left({\Omega vJ\over\hbar}\right)} \;,
\eq
which agrees with the results from chapter \ref{chapN2sigmamodelpathI}.

The important thing however, even though we have not been able to explicitly calculate the effective potential up to order $\hbar$ in this approach, or find the interpolating form of the effective potential between the cases $J=\mathcal{O}(1)$ and $J=\mathcal{O}(\hbar)$, is that the effective potential we find is flat at the origin. This means we find the \ind{Maxwell construction} of the effective potential from the canonical approach. And we find a \emph{convex} effective potential, as it should be in the path-integral approach.

From the formula (\ref{sqrtphi}) one can also clearly see that it matters for the 1-point Green's function, or tadpole, in which order the limits $J\rightarrow0$ and $\Omega\rightarrow\infty$ are taken. If we first take $J\rightarrow0$, and then $\Omega\rightarrow\infty$, then (\ref{sqrtphi}) becomes zero. This corresponds to the path-integral approach.

If we first take $\Omega\rightarrow\infty$, and only then $J\rightarrow0$, then (\ref{sqrtphi}) becomes $v$, which corresponds to the canonical approach.


\chapter{Conclusions}

The most fundamental theory of nature known at present day is the `Standard Model'. This theory agrees very well with experimental results. All particles that are predicted in the Standard Model have also been detected in experiments, except for one: the Higgs boson. The existence of this Higgs boson in the Standard Model is derived within this model via what we call `the canonical approach'.

In the canonical approach one takes a classical field theory and quantizes it by imposing certain commutation or anti-commutation relations on the fields. The particle content of the theory is found by solving the time independent Schr\"odinger equation. One can find the vacuum state, i.e.\ the lowest energy state, via this equation, and one can build a whole Fock space on this vacuum state. The time evolution of the states is governed by the time evolution operator. Via this time evolution operator one can derive the Schwinger-Dyson equations. These equations tell one about the probability amplitudes for certain physical processes.

In the Higgs sector of the Standard Model the time independent Schr\"odinger equation is too hard to actually solve. Therefore one postulates some properties of the vacuum state, inspired by the classical lowest energy state. For example, one \emph{assumes} that the vacuum expectation value of the Higgs field is non-zero, after which one can construct the Fock space. This assumption is also very important when solving the Schwinger-Dyson equations. These Schwinger-Dyson equations can be solved iteratively. In this way one obtains a perturbative series for the Green's functions of the theory. Assuming that the vacuum expectation value of the Higgs field is non-zero one finds the Green's functions of the canonical approach. This canonical approach is completely self-consistent.

Another formulation of quantum field theory is the so-called path-integral formulation. The path integral is merely a solution to the Schwinger-Dyson equations, like the perturbative series mentioned above. For ordinary theories the path-integral formulation is just another formulation of the theory, it gives the \emph{same} physical results. The Green's functions in both formulations come out to be the same.

However, in theories for which the canonical approach predicts spontaneous symmetry breaking, it appears that both formulations of the same quantum field theory do \emph{not} yield identical results. This was the central topic of this thesis. We have calculated Green's functions for two such theories, for which the canonical approach predicts SSB. Surprisingly it appeared that, indeed, the path-integral approach gives very different Green's functions than the canonical approach.

For example, the effective potential in the canonical approach is \emph{not} convex, although one can derive, via the path-integral formulation that an effective potential should \emph{always} be convex. This is known as the convexity problem. However, it is not really a problem, because the convexity is derived in the path-integral formulation of the theory. If we accept that the canonical approach and the path-integral approach are \emph{different}, then the problem is resolved.

The first model we have studied is the Euclidean version of the $N=1$ linear sigma model. There we clearly saw that the Green's functions from the canonical approach and the path-integral approach are \emph{different}. The canonical Green's functions can be obtained from the path integral by, for some reason, only taking into account \emph{one} minimum. We also saw that the divergences in both approaches are identical, meaning that one can use the same counter terms in both approaches to make everything finite. We also calculated the effective potential in the path-integral approach and found it to be well-defined and convex, as it should be on grounds of general arguments.

We also studied the path-integral approach to this model where we now fix the paths in the path integral at some point in time over all of space. In this case we saw that we reproduce the canonical Green's functions. Thus it is possible to get the physics from the canonical approach from a path-integral approach, however in order to obtain this we have to fix the paths. Also in this case the divergences are the same as in the canonical approach. We also found the (alternative) effective potential, and found it to be convex.

The second model that we studied was the $N=2$ Euclidean linear sigma model. Here we saw again that the Green's functions obtained in the canonical and path-integral approach are very different. Divergences are identical in both approaches again. In chapter \ref{chapN2sigmamodelpathI} we first tried a naive approach, making some questionable steps, to take into account all minima of the path integral. In chapter \ref{chapN2sigmamodelpathII} we performed a more rigorous calculation of the path integral, based on the path integral in terms of polar fields. Results appeared to be the same. Also we obtained the effective potential of the $N=2$ LSM within the path-integral approach and found it to be convex.

With all these calculations we have established that, in the case of a theory which exhibits SSB in the canonical approach, the path-integral approach gives \emph{different} Green's functions, which may indicate \emph{different} physics. This brings up some interesting questions related to the Higgs sector of the Standard Model. The prediction of the Higgs particle and its interaction are all based on the \emph{canonical} approach. What if we treat the Higgs sector of the Standard Model not in the canonical way, but instead via the path integral? What would the phenomenology of such an approach be? Could we build a theory without a Higgs particle in this way, or could we explain why the Higgs particle has not been found up to now?

The first step to answer these questions would be to look at the phenomenology of the $N=1$ LSM and the $N=2$ LSM. This is still unknown territory, which marks the end of this thesis, but hopefully also the beginning of a new quest to resolve the mysteries surrounding `the holy grail of particle physics'.


\appendix
\chapter{Standard Integrals}\label{appstandint}

Throughout this thesis we have introduced the following \ind{standard integrals}:
\bqa
& & I\left(q_1,m_1,q_2,m_2,\ldots,q_n,m_n\right) \equiv \nonumber\\
& & \qquad {1\over(2\pi)^d} \int d^dk \; \frac{1}{\left(k+q_1\right)^2+m_1^2} \frac{1}{\left(k+q_2\right)^2+m_2^2} \cdots \frac{1}{\left(k+q_n\right)^2+m_n^2}
\eqa
\bqa
D_{m_1m_2m_3} &\equiv& {1\over(2\pi)^{2d}} \int d^dk \; d^dl \; \frac{1}{k^2+m_1^2} \; \frac{1}{l^2+m_2^2} \; \frac{1}{(k-l)^2+m_3^2} \\
B_{m_1m_2m_3} &\equiv& {1\over(2\pi)^{2d}} \int d^dk \; d^dl \; \frac{1}{(k^2+m_1^2)^2} \; \frac{1}{l^2+m_2^2} \; \frac{1}{(k-l)^2+m_3^2} \\
A_m(x) &\equiv& {1\over(2\pi)^d} \int d^dk \; e^{ik\cdot x} \; \frac{1}{k^2+m^2} \\
C_{m_1m_2}(x) &\equiv& {1\over(2\pi)^d} \int d^dk \; e^{ik\cdot x} \; \frac{1}{k^2+m_1^2} \; \frac{1}{k^2+m_2^2} \\
D_{m_1m_2m_3}(x) &\equiv& {1\over(2\pi)^{2d}} \int d^dk \; d^dl \; e^{ik\cdot x} \; \frac{1}{k^2+m_1^2} \; \frac{1}{l^2+m_2^2} \; \frac{1}{(k-l)^2+m_3^2} \\
B_{m_1m_2m_3}(x) &\equiv& {1\over(2\pi)^{2d}} \int d^dk \; d^dl \; e^{ik\cdot x} \; \frac{1}{(k^2+m_1^2)^2} \; \frac{1}{l^2+m_2^2} \; \frac{1}{(k-l)^2+m_3^2} \\
I &\equiv& {1\over(2\pi)^d} \int d^dk
\eqa

In this appendix we list several results for standard integrals, which are used in our computations throughout this thesis.

\section{$d=1$}

For $d=1$ the standard integrals that we come across in calculations are not divergent. Therefore it is not necessary to introduce a regularization scheme. The following results are used in this thesis:\index{standard integrals!$d=1$}

\bqa
I(0,m) &=& {1\over2m} \\
I(0,m_1,0,m_2) &=& \frac{1}{2m_1m_2(m_1+m_2)} \\
I(0,m,0,m) &=& {1\over4m^3} \\
I(0,m,0,m,0,m) &=& {3\over16m^5} \\
I(0,m,0,m,0,m,0,m) &=& {5\over32m^7} \\
I(0,m,p,m) &=& {1\over m}{1\over p^2+4m^2} \\
A_m(x) &=& {1\over2m} e^{-m|x|} \\
D_{m_1m_2m_3} &=& \frac{1}{4m_1m_2m_3(m_1+m_2+m_3)} \\
D_{mmm} &=& {1\over12m^4} \\
B_{m_1m_2m_3} &=& \frac{2m_1+m_2+m_3}{8m_1^3m_2m_3(m_1+m_2+m_3)^2} \\
B_{mmm} &=& {1\over18m^6} \\
C_{m_1m_2}(x) &=& \frac{1}{2m_1m_2(m_1^2-m_2^2)}\left( m_1e^{-m_2|x|} - m_2e^{-m_1|x|} \right) \\
D_{m_1m_2m_3}(x) &=& \frac{-m_1e^{-(m_2+m_3)|x|}+(m_2+m_3)e^{-m_1|x|}}{4m_1m_2m_3(2m_2m_3-m_1^2+m_2^2+m_3^2)} \\
B_{m_1m_2m_3}(x) &=& \frac{1}{8m_1^3m_2m_3(m_1+m_2+m_3)^2(m_1-m_2-m_3)^2} \nonumber\\
& & \Big( 2m_1^3 e^{-(m_2+m_3)|x|} + \nonumber\\
& & \phantom{\Big(} (m_2^3+m_3^3+3m_2^2m_3+3m_2m_3^2+\nonumber\\
& & \phantom{\Big( (} \quad -3m_1^2m_2-3m_1^2m_3) e^{-m_1|x|} + \nonumber\\
& & \phantom{\Big(} (m_2^3+m_3^3+3m_2^2m_3+3m_2m_3^2+\nonumber\\
& & \phantom{\Big( (} \quad -m_1^2m_2-m_1^2m_3) m_1|x|e^{-m_1|x|} \Big)
\eqa

\section{$d=4$}

For $d=4$ some of the standard integrals that we encounter are divergent, and we have to introduce a regularization scheme. We will calculate these integrals in the \ind{dimensional regularization scheme}. The following results are used in this thesis:\index{standard integrals!$d=4$}

\bqa
I(0,m) &=& {m^2\over8\pi^2}{1\over d-4} + {m^2\over16\pi^2}\left(-1-\ln4\pi+\ln m^2+\gamma\right) \\
I(0,m,0,m) &=& -{1\over8\pi^2}{1\over d-4} + {1\over16\pi^2}\left(\ln4\pi-\ln m^2-\gamma\right) \\
I(0,m,0,m,0,m) &=& {1\over(4\pi)^2}{1\over2m^2} \\
I(0,m,0,m,0,m,0,m) &=& {1\over(4\pi)^2}{1\over6m^4}
\eqa
\bq 
I(0,m,p,m) = -{1\over8\pi^2}{1\over d-4} + {1\over16\pi^2}\left(\ln4\pi-\ln m^2-\gamma\right) + {1\over16\pi^2} f\left({p^2\over m^2}\right)
\eq
with
\bq
f(x) = \sum_{n=1}^{\infty} (-1)^n{n!(n-1)!\over(1+2n)!}x^n
\eq

\printindex

\pagestyle{plain}

\chapter*{Summary}

\addcontentsline{toc}{chapter}{\numberline{}Summary}

A quantum field theory, like the Standard Model, can be set up in \emph{two} ways. The first, and mostly used, way is by \emph{canonical} quantization. This means one takes a classical field theory and imposes commutation or anti-commutation relations on the canonical fields and their conjugated momenta. In this way one can eventually find the Feynman rules of the theory and calculate probabilities for scattering processes.

The second way to set up a quantum field theory is by postulating a \emph{path integral}. In this case the path integral determines the Green's functions, which give the probabilities for actual (physical) scattering processes again.

In principle both ways of setting up the QFT give identical results. So both ways are merely different formalisms: similarly, quantum mechanics can be formulated via the Schr\"odinger equation \emph{and} via the Feynman path integral.

For some theories however, the canonical approach and the path-integral approach do \emph{not} yield identical results. This is the case for theories that exhibit spontaneous symmetry breaking in the canonical approach. In these models, although both approaches satisfy the same fundamental Schwinger-Dyson equations, results are different. This difference has been the main topic of this thesis, and has been investigated for two models: the Euclidean $N=1$ linear sigma model and the Euclidean $N=2$ linear sigma model.

A manifestation of the difference between the canonical and path-integral approach can also be found in the literature. For some theories exhibiting spontaneous symmetry breaking it was known that the effective potential calculated in the canonical approach is non-convex, whereas from the path integral one can prove that such an effective potential is \emph{always} convex\footnote{In this thesis we have only proven this for Euclidean theories, however in the literature one finds proofs also for Minkowskian theories, see e.g.\ \cite{Haymaker}.}. This contradiction is known as the convexity problem. A soon as one realizes that the canonical and path-integral results do not have to be the same this convexity problem is resolved.

The convexity problem is discussed thoroughly, and also resolved, in the literature. However a clear discussion of the difference between the canonical and path-integral formalism does not exist. It is not clear whether the different results from both approaches also indicate different physics, i.e.\ different probabilities for actual scattering processes.

In this thesis we have investigated both approaches for the case of two simple Euclidean quantum field theories. In the case of the $N=1$ linear sigma model we have presented the canonical approach (chapter \ref{chapN1sigmamodelcan}) and the path-integral approach (chapter \ref{chapN1sigmamodelpath}). Results for the Green's functions and the effective potential are clearly different. The effective potential calculated from the path-integral approach is nicely convex, as it should be, whereas the effective potential from the canonical approach is not. In the case of the $N=1$ linear sigma model we have also considered a path-integral formalism where we fix the paths all over space to have a given value at some time (chapter \ref{chapN1sigmamodelfix}). In this case the Green's functions came out to be the same as in the canonical approach. But to obtain this we had to introduce the somewhat artificial path-fixing constraint.

In the case of the $N=2$ linear sigma model we have also presented the canonical (chapter \ref{chapN2sigmamodelcan}) and the path-integral approach (chapters \ref{chapN2sigmamodelpathI} and \ref{chapN2sigmamodelpathII}). Here the calculations in the path-integral approach are more complicated than in the $N=1$ linear sigma model, because the minima of the bare potential form a continuous set. To this end we had to investigate also how one can formulate a path integral in terms of polar field variables (chapter \ref{chappathintpol}). In the end we demonstrated again that the Green's functions from the canonical and path-integral approach are different.

Although we have established that, in general, canonical and path-integral results (i.e.\ Green's functions) differ, in the case of a theory that exhibits spontaneous symmetry breaking, it is not yet clear what this means for the physics of both approaches. Of course the physics of the canonical approach is known, but the physics of the path-integral approach requires more research. Several exciting questions remain within this approach: are the Green's functions of a Minkowskian quantum field theory in the path-integral approach also different than in the canonical approach? Are the Green's functions for a theory with \emph{local} gauge invariance also different in both approaches? Is the physics of the Standard Model different if we formulate this model via the path integral? Is there something like the Higgs mechanism in this approach? Is there even a Higgs particle in this approach? The present thesis can be considered a first step towards finding answers to these fundamental questions.

\chapter*{Samenvatting}

\addcontentsline{toc}{chapter}{\numberline{}Samenvatting}

Een quantumveldentheorie, zoals het Standaard Model, kan op \emph{twee} manieren opgezet worden. De eerste, en meest gebruikte, manier gaat via \emph{canonieke} quantizatie. Dit betekent dat men een klassieke veldentheorie neemt en commutatie- of anticommutatierelaties oplegt aan de canonieke velden en hun geconjugeerde momenta. Op deze manier construeert men de Feynmanregels van de theorie en kan men waarschijnlijkheden voor verstrooiingsprocessen berekenen.

De tweede manier om een quantumveldentheorie op te zetten is door het postuleren van een \emph{padintegraal}. In dit geval bepaalt de padintegraal de Greense functies, welke op hun beurt de waarschijnlijkheden voor fysische verstrooiingsprocessen geven.

In principe geven beide manieren van het opzetten van een quantumveldentheorie identieke resultaten. Beide manieren zijn slechts andere formalismen, net zoals quantummechanica geformuleerd kan worden via de Schr\"odingervergelijking \emph{en} de Feynman-padintegraal.

Voor sommige theorie\"en geven het canonieke en padintegraalformalisme echter \emph{niet} dezelfde resultaten. Dit is het geval voor theorie\"en waarin spontane symmetriebreking optreedt. In deze modellen, ondanks dat beide formalismen voldoen aan dezelfde fundamentele Schwinger-Dysonvergelijkingen, verschillen de resultaten. Dit verschil is het hoofdonderwerp van dit proefschrift, en is onderzocht voor twee modellen: het Euclidische $N=1$ lineaire sigmamodel en het Euclidische $N=2$ lineaire sigmamodel.

Een manifestatie van het verschil tussen het canonieke en het padintegraalformalisme is al bekend in de literatuur. Voor sommige theorie\"en die spontane symmetriebreking vertonen is het bekend dat de effectieve potentiaal berekend via het canonieke formalisme niet convex is, terwijl men via het padintegraalformalisme kan bewijzen dat deze potentiaal \emph{altijd} convex moet zijn\footnote{In dit proefschrift wordt dit slechts bewezen voor Euclidische theorie\"en, in de literatuur vindt men ook bewijzen voor Minkowskitheorie\"en, zie bijv. \cite{Haymaker}.}. In de literatuur staat deze tegenstelling bekend als het convexiteitsprobleem. Echter wanneer men accepteert dat het canonieke en het padintegraalformalisme simpelweg niet hetzelfde zijn, dan verdwijnt ook deze tegenstelling.

Het convexiteitsprobleem wordt uitgebreid besproken, en opgelost, in de literatuur. Een heldere discussie over het verschil tussen het canonieke en padintegraalformalisme ontbreekt echter. Het is niet duidelijk of de verschillende resultaten uit beide formalismen ook verschillende fysische overgangswaarschijnlijkheden als gevolg zullen hebben. 

In dit proefschrift zijn de twee formalismen voor twee simpele Euclidische quantumveldentheorie\"en onderzocht. In het geval van het $N=1$ lineaire sigmamodel hebben we het canonieke formalisme (hoofdstuk \ref{chapN1sigmamodelcan}) en het padintegraalformalisme (hoofdstuk \ref{chapN1sigmamodelpath}) gepresenteerd. De resultaten voor de Greense functies en de effectieve potentiaal zijn duidelijk verschillend. De effectieve potentiaal berekend vanuit het padintegraalformalisme is netjes convex, zoals het hoort, terwijl de effectieve potentiaal van het canonieke formalisme deze eigenschap niet heeft. In het geval van het $N=1$ lineaire sigmamodel hebben we ook een padintegraalformalisme beschouwd waarbij we de paden in de padintegraal over de gehele ruimte vasthouden op een bepaalde waarde, op een bepaald tijdstip (hoofdstuk \ref{chapN1sigmamodelfix}). In dit geval zijn de Greense functies wel hetzelfde als in het canonieke formalisme. Maar om dit te bereiken moeten we wel de zojuist genoemde kunstmatige beperking van paden inbouwen.

In het geval van het $N=2$ lineaire sigmamodel hebben we ook het canonieke (hoofdstuk \ref{chapN2sigmamodelcan}) en het padintegraalformalisme (hoofdstukken \ref{chapN2sigmamodelpathI} en \ref{chapN2sigmamodelpathII}) gepresenteerd. In dit geval waren de berekeningen in het padintegraalformalisme gecompliceerder dan in het $N=1$-geval, omdat de minima van de naakte potentiaal een continue set vormen. Hiertoe hebben we onderzocht hoe men een padintegraal in termen van polaire velden kan formuleren (hoofdstuk \ref{chappathintpol}). Uiteindelijk kon ook voor dit $N=2$ lineaire sigma model aangetoond worden dat de Greense functies in beide formalismen anders zijn.

Ondanks dat we aangetoond hebben dat, in het algemeen, canonieke en padintegraalresultaten (i.e.\ Greense functies) niet gelijk zijn voor het geval van een theorie met spontane symmetriebreking, is het onduidelijk wat dit betekent voor fysische resultaten van beide formalismen. De fysische resulaten van het canonieke formalisme zijn natuurlijk bekend, maar verder onderzoek zal moeten uitwijzen wat de fysische resultaten van het padintegraalformalisme zijn. Een aantal spannende vragen blijven bestaan binnen dit formalisme: zijn ook de Greense functies van een Minkowskische quantumveldentheorie anders in beide formalismen? Zijn de Greense functies van een theorie met \emph{lokale} ijkinvariantie anders in beide formalismen? Zijn de fysische resultaten van het Standaard Model anders als we dit model via het padintegraalformalisme opzetten? Bestaat er zoiets als het Higgsmechanisme in het padintegraalformalisme? En bestaat er een Higgsdeeltje in dit formalisme? Dit proefschrift is bedoeld als een eerste stap op weg naar het vinden van antwoorden op deze fundamentele vragen.

\chapter*{Curriculum Vitae}

\addcontentsline{toc}{chapter}{\numberline{}Curriculum Vitae}

Marcel van Kessel werd op 19 maart 1980 geboren te Tegelen. In 1998 behaalde hij het VWO eindexamen aan het Stedelijk College Den Hulster te Venlo. In 1998 begon hij aan de studie natuurkunde aan de Katholieke Universiteit Nijmegen. In 2003 begon hij zijn afstudeeronderzoek onder leiding van Prof.dr. R. Kleiss. In 2004 rondde hij dit afstudeeronderzoek succesvol af, en behaalde hij het doctoraal examen in de theoretische natuurkunde.

Van 1 april 2004 tot 30 september 2008 was hij werkzaam als onderzoeker-in-opleiding, ook onder leiding van Prof.dr. R. Kleiss aan de Radboud Universiteit Nijmegen. Het onderzoek gedaan in deze periode heeft geleid tot dit proefschrift.

\chapter*{Acknowledgements}

\addcontentsline{toc}{chapter}{\numberline{}Acknowledgements}

I would like to thank everybody who made a contribution to this thesis and the research within.

First of all I would like to thank my supervisor Ronald Kleiss, who thought with me those four and a half years, and who always had an answer to my questions, though never the straightforward answer I expected or hoped for. He taught me that this is the way science is done, that an answer is found by hard labor instead of sitting down with a cup of strong coffee and thinking. His fresh insights and his unique way of approaching a problem were crucial for the research within this thesis. I liked very much the way he worked and thought with me, instead of above me, and I enjoyed the time we spent in front of his very big (almost too big for his office) old blackboard. Especially in the first two years we spent hours and hours in front of it, sadly in the last two years other duties of Ronald called for more and more time.

I also want to thank Ronald for introducing me to his Greek friend Ernestos Argyres. I have had the pleasure of working with him since the beginning of my PhD research. We must have spent many hours on the phone, and I imagine that the university is not at all too happy with their phone bill since I met him. I wish to thank Ernestos for the numerous discussions we have had and for showing me that I really like working together with other people while doing research. I learned from Ernestos that science is also about knowing what has been done in the past, and not just doing everything yourself. (This may sound obvious, but is quite an insight for a pigheaded person like myself.) Also I would like to thank Ernestos for his hospitality during the (much too short) times I spent in Athens. What better place is there to gain inspiration than the origin of western civilization? Also I enjoyed very much the times we went sightseeing in Greece. 

Also I wish to thank Ernestos for invoking Costas Papadopoulos in our problems and discussions. I would like to thank Costas for all his contributions to our research, one of them being the bold conjecture from chapter \ref{chappathintpol}. (In our daily discussions this conjecture was known as `the Greek conjecture'.) From which epiphany he got it I do not know, and I did not believe it at first, but after a lot of hard labor it has been proven. From Costas I learned that sometimes one has to be bold and just make a pragmatic statement. I also want to thank Costas for his hospitality while I was in Athens.

Finally I would like to express my gratitude to Wim Beenakker and Tom Rijken, who were confronted with my urgent questions and problems whenever Ronald was not there or had no time. Also thanks to my colleagues at the department, who made working there very nice from a social point of view. Especially in the last one and a half year the department has been growing and has been blessed with new people who organize a lot of new things, making the department more and more attractive.

\end{document}

%% file: titlesimple.tex




\begin{center}
\vspace*{1cm}

{\LARGE\bf The Path-Integral Approach to Spontaneous Symmetry Breaking}\\

\end{center}


\clearpage \thispagestyle{empty}

%

\vspace*{21.6cm}


\noindent
The front cover shows quantum mechanics (symbolized by Schr\"odinger's cat) playing with the Higgs-field in the Mexican hat potential. The back cover shows the Maxwell construction of the Mexican hat potential, which is obtained in the path-integral approach to the $N=2$ linear sigma model.

I wish to thank Pepper the cat for her cooperation in the making of the cover.


\normalsize


%% file: title.tex




\begin{center}
\vspace*{1cm}

{\LARGE\bf The Path-Integral Approach to Spontaneous Symmetry Breaking}\\

\vspace{3.5cm} {
{\large \textsc{Een wetenschappelijke proeve op het gebied der Natuurwetenschappen, Wiskunde en Informatica}}\\
\vspace{2.0cm}
{\large \textsc{Proefschrift}} \\
\vspace{2.0cm}
\textsc{ter verkrijging van de graad van doctor}\\
\textsc{aan de Radboud Universiteit Nijmegen}\\
\textsc{op gezag van de rector magnificus prof. mr. S.C.J.J. Kortmann,}\\
\textsc{volgens besluit van het College van Decanen}\\
\textsc{in het openbaar te verdedigen op dinsdag 3 februari 2009}\\
\textsc{om 15.30 uur precies}\\
\vspace{2cm}
\textsc{door}  \\
\vspace{2cm}
{\bf Marcel Theodorus Maria van Kessel}   \\
\vspace{1cm}
\textsc{geboren op 19 maart 1980} \\
\textsc{te Tegelen}
}
\end{center}


\clearpage \thispagestyle{empty}

\vspace{5cm} 
\begin{tabbing}
Manuscriptcommissie:kkkkk\=Prof. Mr. M.W.J.M. Demarteaukkkkk\=    \kill
Promotores: \>Prof. dr. R.H.P. Kleiss \\
            \>Prof. dr. E.N. Argyres (NCSR Demokritos Athens) \\
\\
\\
Manuscriptcommissie: \> Prof. dr. S. de Jong \\
                     \> Prof. dr. E. Laenen (NIKHEF Amsterdam) \\
                     \> Prof. dr. C. Papadopoulos (NCSR Demokritos Athens) \\

\end{tabbing}

\vspace{14.6cm}

\epsfig{file=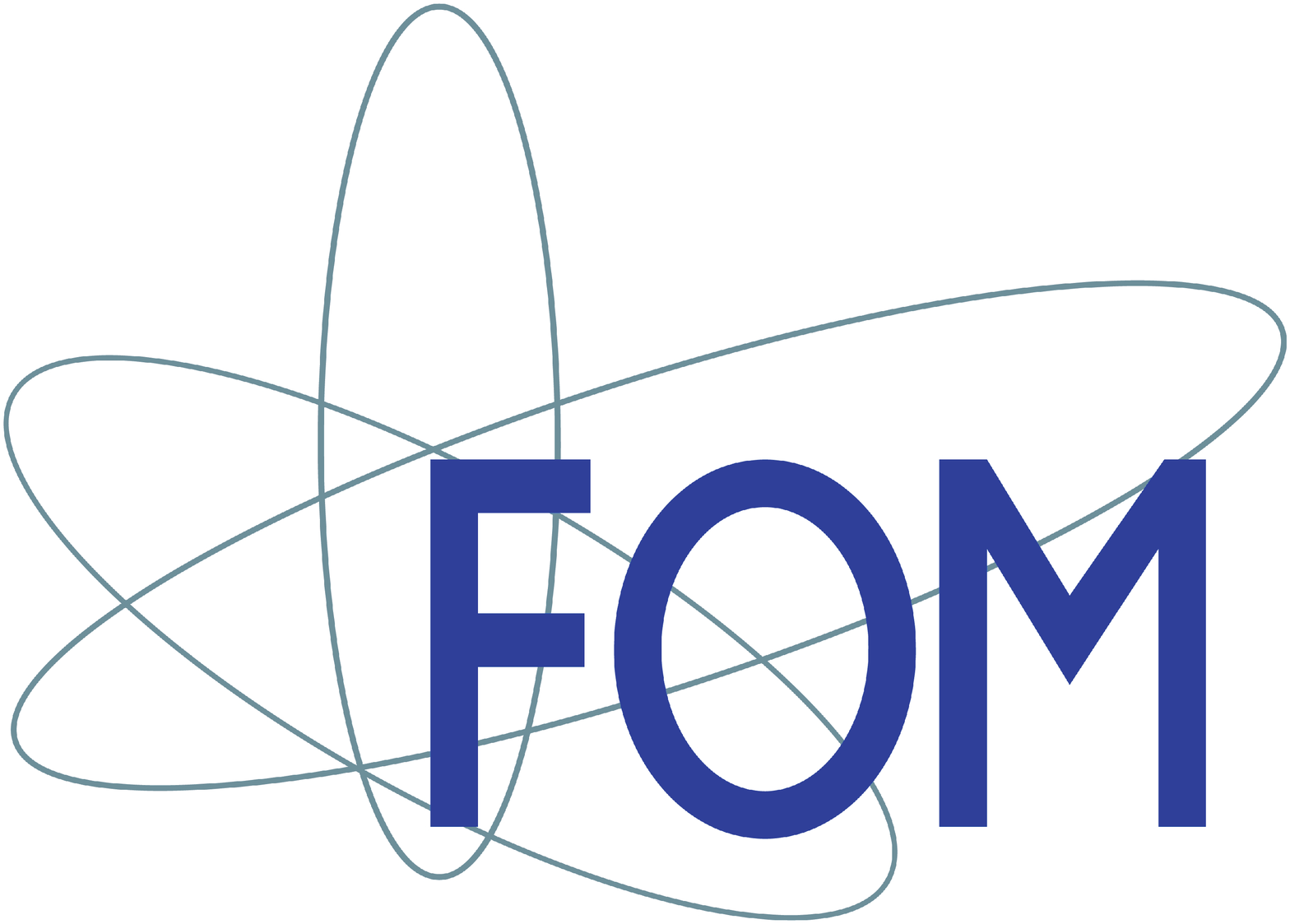,width=4cm}\\
\noindent
Het werk beschreven in dit proefschrift maakt deel uit van het onderzoeksprogramma van de Stichting voor Fundamenteel Onderzoek der Materie (FOM), die financieel wordt gesteund door de Nederlandse Organisatie voor Wetenschappelijk Onderzoek (NWO).\\
\\
\noindent
ISBN 978-90-9023799-2


\normalsize
